\documentclass[11pt,a4paper]{article}

\usepackage[dvipsnames]{xcolor} 
\usepackage{tikz}
\usetikzlibrary{calc,arrows,backgrounds}
\usepackage{amsmath,mathtools}   
\usepackage{amsfonts}
\usepackage{amssymb}
\usepackage{lscape}
\usepackage{afterpage}
\usepackage{setspace}
\usepackage{multicol}
\usepackage{changepage}
\usepackage{simplewick}
\usepackage{youngtab}
\usepackage{pgfmath}
\usepackage[title]{appendix}
\usepackage[margin=1in]{geometry}
\newcommand{\kk}{\mathfrak}

\newcommand{\bs}{\boldsymbol}
\newcommand{\THICC}{ultra thick}
\newcommand{\wt}{\widetilde}
\newcommand{\sz}{\scriptsize}
\newcommand{\Q}{\textsf{Q}}
\newcommand{\es}{\mathcal{S}}
\newcommand{\Hg}{\mathcal{H}}
\newcommand{\dimh}{\dim_{\mathbb{H}}}

\begin{document}
\def\O{\bar{\mathcal{O}}}

\begin{flushright}
LTH 1199
\end{flushright}

\begin{center}
{\Large\textbf{\boldmath$D_n$ Dynkin quiver moduli spaces} }

Jamie Rogers and Radu Tatar

\textit{Department of Mathematics, University of Liverpool, \\ Liverpool, L69 7ZL, United Kingdom}

jamie.rogers@liv.ac.uk, rtatar@liv.ac.uk
\end{center}

\begin{abstract}
We study $3d$ $\mathcal{N}=4$ quiver gauge theories with gauge nodes forming a $D_n$ Dynkin diagram and their relation to nilpotent varieties in $\kk{so}_{2n}$. The class of good $D_n$ Dynkin quivers is completely characterised and the moduli space singularity structure fully determined for all such theories. The class of good $D_n$ Dynkin quivers is denoted $D_\nu^\mu(n)_p$ where $n \geq 2$ is an integer, $\nu$ and $\mu$ are integer partitions and $p \in \{ \textrm{even},  \textrm{odd}\}$ denotes membership of one of two broad subclasses. Small subclasses of these quivers are known to realise some $\mathfrak{so}_{2n}$ nilpotent varieties with their moduli space branches. We fully determine which $\mathfrak{so}_{2n}$ nilpotent varieties are realisable as $D_n$ Dynkin quiver moduli spaces and which are not. Quiver addition is introduced and is used to give large subclasses of $D_n$ Dynkin quivers poset structure. The partial ordering is determined by inclusion relations for the moduli space branches. The resulting Hasse diagrams are used to both classify $D_n$ Dynkin quivers and determine the moduli space singularity structure for an arbitrary good theory. The poset constructions and local moduli space analyses are complemented throughout by explicit checks utilising moduli space dimension matching. 
\end{abstract}

\tableofcontents

\section{Introduction and summary}
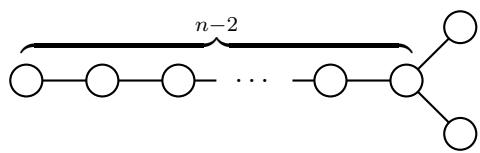
\begin{figure}[h]
	\begin{center}
		\begin{tikzpicture}
		\draw[thick] (0,0) -- (2.5,0)  (3.5,0) -- (5,0) -- (5.7,0.7) (5,0) -- (5.7,-0.7);
		
		\draw[ thick, fill=white] (0,0) circle (6pt);
		\draw[ thick, fill=white] (1,0) circle (6pt);
		\draw[ thick, fill=white] (2,0) circle (6pt);
		\draw[ thick, fill=white] (4,0) circle (6pt);
		\draw[ thick, fill=white] (5,0) circle (6pt);
		\begin{scope}[xshift=5cm]
		\draw[thick, fill=white] (45:1) circle (6pt);
		\draw[ thick, fill=white] (-45:1) circle (6pt);
		\end{scope}
		
		\draw (2.5,0.6) node {$\overbrace{~\quad\qquad\qquad\qquad\qquad\qquad\qquad}^{n-2}$}
		
		(3,0) node {$\dots$}
		;
		\end{tikzpicture}
		
		
	\end{center}
	\caption{The Dynkin diagram for $D_n$.}
	\label{Dynkin}
\end{figure}

The moduli space of vacua of a supersymmetric quantum field theory is both a cornerstone to our physical insight and a richly structured geometric object in its own right. Over the years, the study of moduli spaces for theories with eight supercharges, especially $\mathcal{N}=4$ theories in three dimensions, \cite{Intril} - \cite{Boer}, has been the subject of much attention. An important class of $3d$ $\mathcal{N}=4$ gauge theories are those whose field content can be represented by a quiver \cite{jima}. This work considers good (in the sense of \cite{WitGia}) $3d$ $\mathcal{N}=4$ quiver gauge theories whose shape is that of a $D_n$ Dynkin diagram as in figure \ref{Dynkin}. These $3d$ $\mathcal{N}=4$ $D_n$ Dynkin quiver gauge theories are herein referred to simply as $D_n$ Dynkin quivers. Note that in the literature, `$D_n$ quiver' is often used to refer to quivers that would be more properly called $\wt{D}_n$ Dynkin quivers\footnote{With gauge nodes given by the affine $D_n$ Dynkin diagram shown in figure \ref{DynkinwtD}.}, for example in \cite{Kasputin}.
The specifics of the connections between quiver gauge theories and the nilpotent cones of classical and exceptional Lie algebras, \cite{CollMc} - \cite{Ryo} have been the subject of numerous papers in recent years \cite{Crem} - \cite{excep}. Work has often involved using $3d$ $\mathcal{N}=4$ quiver gauge theories as tools with which to construct the varieties which also arise as nilpotent varieties of Lie algebras. An important question then arises as to which nilpotent varieties are constructable at all using the quiver gauge theory approach. In this work we answer this question for the construction of $\kk{so}_{2n}$ nilpotent varieties by $D_n$ Dynkin quivers. The varieties realised by $D_n$ quivers which do not realise $\kk{so}_{2n}$ nilpotent varieties are also fully locally analysed using techniques developed from those in \cite{Rogers}.

Characterisation of the local moduli space structure allows the partial ordering of classes of quiver gauge theories, opening a wealth of new insights. For example, the moduli space branches of linear ($A_n$ Dynkin) quivers are exactly subvarieties of the nilpotent cone of $\kk{sl}_n$. An important class of subvarieties of the nilpotent cone of a Lie algebra are the closures of nilpotent orbits. The nilpotent cone is the union of finitely many orbits and the closures of these orbits arise naturally as moduli space branches of a subclass of good $A_n$ Dynkin quivers. The nilpotent orbits of $\kk{sl}_n$ can be classified using integer partitions (the standard text is \cite{CollMc}). The set of integer partitions is partially ordered in a natural way, interpreted in the set of nilpotent orbit closures as arising from a simple inclusion relation. The partial ordering can be realised in the subclass of $A_n$ Dynkin quivers which realise the orbit closures as moduli space branches.

Another set of important subvarieties of the nilpotent cone consists of the intersections of the parts of the algebra that are \textit{transverse} to a given orbit, called Slodowy slices, with the cone itself. These are also classified using integer partitions in the case of $\kk{sl}_n$. A generic nilpotent subvariety of the nilpotent cone of $\kk{sl}_n$ arises by intersecting an orbit closure and a Slodowy slice and can be characterised by two partitions of the integer $n$. There are restrictions as to which pairs define a variety, namely one partition must \textit{dominate} the other. Every nilpotent variety of $\kk{sl}_n$ is realised as the moduli space of a good $A_n$ Dynkin quiver gauge theory and every good $A_n$ Dynkin quiver realises a nilpotent variety of some $\kk{sl}_n$ algebra with its moduli space branches. A good $A_n$ Dynkin quiver gauge theory can therefore be classified by two integer partitions of the same magnitude. This class in called $T_{\rho^t}^\sigma(SU(n))$ in the literature where $\rho$ and $\sigma$ are two partitions of $n$ defining a nilpotent variety\footnote{Formally there are an infinite number of pairs of integer partitions for any given linear quiver gauge theory. This is because the nilpotent cones and subvarieties of smaller algebras can appear fully as nilpotent subvarieties of larger algebras ad infinitum. In context it can vary whether one considers the theory to be defined using the smallest magnitude such pair or whether a theory is defined using partitions of a specific magnitude.}. These partitions can also be interpreted in terms of the linking number of branes in a type IIB Hanany-Witten configuration whose low energy dynamics are described by that $3d$ field theory \cite{hanwit}. This interpretation was used extensively in \cite{Rogers}, however in this work we move away from these descriptions so as to inform the study of quiver gauge theories without such interpretations.

The extension of these ideas to other Dynkin quiver gauge theories has only been partially successful. Of most immediate interest is the other family of classic algebras whose Dynkin diagram is simply laced, namely $\kk{so}_{2n}$, with the $D_n$ Dynkin diagram, figure \ref{Dynkin}. The nilpotent varieties of $\kk{so}_{2n}$ can be characterised once again using a pair of integer partitions, however this time there is a restriction to those partitions of $2n$ in which even parts occur an even number of times. This restriction subsequently encapsulates a number of difficulties which arise for nilpotent varieties of $\kk{so}_{2n}$, discussed in \cite{CollMc} and reviewed here in section 2. Only a relatively small number of low-rank $D_n$ Dynkin quivers have been found to realise $\kk{so}_{2n}$ nilpotent varieties as moduli space branches, for example \cite{Kalveks}, \cite{kalveks2}. 

\vspace{1mm}

This work presents the full analysis of the singularity structure of the Higgs and Coulomb branches of the moduli space of vacua for good $D_n$ Dynkin quivers. A classification of all good $D_n$ Dynkin quivers is constructed. The subsequent comparison of the singularity structure for the nilpotent cone of $\kk{so}_{2n}$ and of the moduli space varieties shows that most $D_n$ Dynkin quivers do \textit{not} realise $\kk{so}_{2n}$ nilpotent varieties with their moduli space branches, and most nilpotent varieties of $\kk{so}_{2n}$ are \textit{not} realised as $D_n$ Dynkin quiver moduli space branches. While the local analysis is proving fruitful, a technique to match and discern the global structure from a complete local analysis would prove an influential tool in the investigation of global moduli space descriptions. 

The primary technique employed here is \textit{quiver addition}. Much like the very closely related quiver subtraction introduced in \cite{QuivSub}, quiver addition concerns the singularities that appear as transverse slices to the moduli space of a smaller quiver inside the moduli space of a dominant, larger quiver. In quiver subtraction, the two quivers are known and the quiver for the slice between them is determined via subtraction of the smaller quiver from the larger. This is a more sophisticated version of the Kraft-Procesi transition developed in \cite{KPT} and \cite{KPTC} and which was used extensively in \cite{Rogers} for circular ($\wt{A}_n$ Dynkin) quivers. Quiver addition is the reverse of this process. Given a class of quiver gauge theories, for example balanced $D_n$ Dynkin quivers, and a set of known quivers corresponding to minimal singularities, one identifies which singularities can appear as the difference between a known smaller quiver and some larger quiver of the same class. The singularities are then `added' to the smaller quiver and the class is built from the ground up. Starting with the smallest (lowest rank) quiver for the class, one obtains the set of quivers which can be reached via quiver addition from that starting quiver with poset structure corresponding to the inclusion relations of moduli space branches which are the natural result of implementing addition in this manner. In the case of balanced $D_n$ Dynkin quivers, for example, the Hasse diagram, built via quiver addition, has nodes corresponding to all balanced $D_n$ Dynkin quivers and so the poset structure of the entire class is established. Whereas the poset structure of $A_n$ Dynkin quivers was initially established `from the top down' with reference to knowledge of the global structure ($\kk{sl}_n$ nilpotent cones), quiver addition constitutes an entirely `bottom-up' construction of moduli space singularity structure, once usable singularities are established.

The main result is that the moduli space singularity structure for a generic good $D_n$ Dynkin quiver is composed of a sequence of nilpotent varieties of \textit{$\kk{sl}_n$} between which there is a \textit{traversing structure} of transverse slices of either $D_k$ or $A_k \cup A_k$ singularities, along with other specific singularities which may arise under necessary systematic editing procedures which are fully determined. At the level of the quiver this can be interpreted as a moduli space singularity structure reflecting the fact that the $D_n$ Dynkin diagram is mostly a line of unitary gauge nodes, making the appearance of similar structures to the linear case not surprising. From this point of view, the $\kk{so}_{2n}$ varieties that get realised are only those where this Hasse diagram construction coincides with some substructure of the Hasse diagram for nilpotent varieties of $\kk{so}_{2n}$. A full characterisation of good $D_n$ Dynkin quivers requires two completely separate quiver addition constructions which never connect with one another. The theories are therefore divided into two broad subclasses depending on which of the constructions they belong to. These subclasses are called \textit{even} and \textit{odd} (denoted with a $p\in\{e,o\}$) in reference to the difference in the flavour content of the two end nodes. Within each subclass, a precise theory can be characterised using the integer $n$ and two integer partitions, $\nu$ and $\mu$ whose magnitudes are not necessarily either equal to $n$ or equal to one another, nevertheless they obey restrictions determined by $p$. The full class of good $D_n$ Dynkin quivers is denoted as $D_\nu^\mu(n)_p$.

In section 2 the nilpotent varieties of $\kk{so}_{2n}$ are discussed starting with some background on their classification using a restricted set of partitions of integers. All subvarieties of the \textit{maximal special slice} for $\kk{so}_{2n}$ are identified. These are all of the nilpotent varieties of $\kk{so}_{2n}$ which get realised as $D_n$ Dynkin quiver moduli space branches. The precise $D_n$ Dynkin quivers which realise the subvarieties of this slice are then identified. The proof that these are all of the $D_n$ Dynkin quivers realising $\kk{so}_{2n}$ nilpotent varieties is a natural result of the classification and moduli space analysis of all good $D_n$ Dynkin quivers. Section 3 contains the full classification and moduli space singularity structure analysis for balanced and good $D_n$ Dynkin quivers. Using the aforementioned quiver addition techniques, the Hasse diagram for \textit{balanced} $D_n$ Dynkin quivers is constructed. These theories are shown to necessarily be of \textit{even} type and all good, even $D_n$ Dynkin quivers are then constructed as differences of two balanced quivers. While one Hasse diagram for even quivers would be sufficient, it is more informative to construct two Hasse diagrams dealing with two subtly different types of even theory. We then cover good quivers of odd type that do not arise as the difference of balanced quivers. An alternative Hasse diagram illustrating their poset structure is constructed. From these Hasse diagrams the classification using $n$, $\nu$, $\mu$, and $p$ is straightforward and is proved to contain all good $D_n$ Dynkin quivers. Explicit non-trivial checks of the constructions are performed throughout by matching the expected dimension of the moduli space branches of a theory with the dimension of the variety described by the Hasse diagrams. The theories from section 2 which realise nilpotent varieties of $\kk{so}_{2n}$ are also identified within this larger discussion. Section 4 contains conclusions and discussion of future directions of interest.

\section{Dynkin quivers and nilpotent varieties in Lie algebras}
The $D_n$ Dynkin quivers whose moduli space branches realise nilpotent varieties of $\kk{so}_{2n}$ have been studied recently, for example in \cite{Kalveks}, \cite{Sperling} and \cite{kalveks2}. We present an a full determination of which $D_n$ Dynkin quivers realise $\kk{so}_{2n}$ nilpotent varieties. The proof that this analysis is complete follows from the characterisation of the singularity structure of the moduli space branches of a generic, good $D_n$ Dynkin quiver given in section 3. In this section we first review the essential tools and building blocks useful for the rest of the analysis. The classic text for the study of nilpotent orbits in classical and exceptional Lie algebras is \cite{CollMc}. The study of their singularity structure was performed in  \cite{KP2}--\cite{fu}.

\subsection{Nilpotent varieties in $\kk{so}_{2n}$}
We will work with three types of nilpotent varieties in Lie algebras, closures of nilpotent orbits, transverse slices to these orbits\footnote{Also called Slodowy slices to distinguish them from transverse slices defined in other ways.}, and intersections between the two. Nilpotent orbits are the conjugacy classes of nilpotent elements under the action of an associated Lie group. There is a subtlety for $\kk{so}_{2n}$ depending on whether the action is with the group $SO(2n)$ or $O(2n)$. The nilpotent orbits of classical algebras are characterised by a (potentially restricted) set of integer partitions. The orbits in $\kk{so}_{2n}$ are associated to a restricted set of integer partitions of magnitude $2n$. Slodowy slices are defined as transverse to these orbits and so are also classified using those same integer partitions. Integer partitions will play a central role throughout this work so the rudiments of their study are repeated here. 

\subsubsection{Partitions}
A partition $\mu$, of magnitude $N$, is a weakly decreasing tuple of non-negative integers (parts), $\mu = (m_1,...,m_{N}) $ with $ m_1 \geq \dots \geq m_{N} \geq 0$, such that $\sum_{i=1}^N m_i = N$. Partitions are usually written in exponential notation where each part is labelled with its multiplicity and parts with multiplicity zero are dropped. The set of partitions of $N$, denoted $\mathcal{P}(N)$, has a natural dominance ordering whereby for two partitions $\mu$ and $\nu$ we write $\mu > \nu$ if
\begin{equation}\label{partord}
\sum_{i=1}^k \mu_i \geq \sum_{i=1}^k \nu_i,
\end{equation}
for all $1\leq k\leq N$. For $N\geq6$ this is a \textit{partial} ordering of $\mathcal{P}(N)$ which is illustrated with a \textit{Hasse diagram}. A Hasse diagram is a graph of nodes and edges where nodes represent partitions, dominant partitions' nodes are placed higher, and nodes are joined with an edge if they are \textit{adjacent}. $\mu$ and $\nu$ are adjacent if there is no partition $\rho$ such that $\mu > \rho > \nu$. In the language of posets (partially ordered sets) adjacent domination is a covering relation. Examples of Hasse diagrams for sets of integer partitions are given in appendix \ref{appendix}.
Given $\mu$, one constructs the \textit{transpose}, $\mu^t$, by taking the difference between the $i^\textrm{th}$ and $i+1^\textrm{th}$ parts of $\mu$ to be the multiplicity of $i$ in $\mu^t$. The transpose will again be a partition of $N$. The one-to-one nature of transposition on $\mathcal{P}(N)$ manifests in the Hasse diagrams as top-bottom reflection symmetry. Alternatively,  a transposition is a reflection of the Young tableau representation of a partition in the NW-SE diagonal. 

\begin{figure}
	\begin{center}
		\begin{tikzpicture}[scale = 0.65]
		\begin{scope}[yshift=3cm]
		\draw[ultra thick] (-1, 1.5) -- (3.5,1.5);
		
		\draw (0,2.25) node {\scriptsize{Hasse}}
		(0,1.85) node {\scriptsize{diagram}}
		(2.5,2.05) node {\scriptsize{Partition}}
		(1.25,3.3) node {{$\mathcal{P}_+(6)$}};
		\end{scope}
		
		\draw[fill=black] (0,0) circle (4pt);
		\draw[fill=black] (0,1) circle (4pt);
		\draw[fill=black] (0,2) circle (4pt);
		\draw[fill=black] (0,3) circle (4pt);
		\draw[fill=black] (0,4) circle (4pt);
		
		\draw[\THICC] (0,0) -- (0,4);
		
		\draw (2.5,0) node {$(1^6)$};
		\draw (2.5,1) node {$(2^2,1^2)$};
		\draw (2.5,2) node {$(3,1^3)$};
		\draw (2.5,3) node {$(3^2)$};
		\draw (2.5,4) node {$(5,1)$};
		
		\draw (0,-2.27) node {};
		
		\end{tikzpicture}$\qquad~~~~~~$
		\begin{tikzpicture}[scale=0.65]
		\begin{scope}[yshift=6cm]
		\draw[ultra thick] (-1, 1.5) -- (4.5,1.5);
		
		\draw (0,2.25) node {\scriptsize{Hasse}}
		(0,1.85) node {\scriptsize{diagram}}
		(3.5,2.05) node {\scriptsize{Partition}}
		(1.75,3.3) node {{$\mathcal{P}_+(8)$}};
		\end{scope}
		
		\draw[\THICC] (0,0) -- (0,1) -- (1,2) -- (0,3) -- (0,4) -- (1,5) -- (0,6) -- (0,7) (0,6) -- (-1,5) -- (0,4) (0,3) -- (-1,2) -- (0,1);
		
		\draw[fill=black] (0,0) circle (4pt);
		\draw[fill=black] (0,1) circle (4pt);
		\draw[fill=black] (0,3) circle (4pt);
		\draw[white, fill=white] (0,3) circle (1.8pt);
		\draw[fill=black] (0,4) circle (4pt);
		\draw[fill=black] (0,6) circle (4pt);
		\draw[fill=black] (0,7) circle (4pt);
		
		\draw[fill=black] (-1,2) circle (4pt);
		\draw[fill=black] (1,2) circle (4pt);
		\draw[fill=black] (-1,5) circle (4pt);
		\draw[fill=black] (1,5) circle (4pt);
		
		\begin{scope}[xshift = 3.5cm]
		\draw[fill=black] (0,0) node {$(1^8)$};
		\draw[fill=black] (0,1) node {$(2^2,1^4)$};
		\draw[fill=black] (0,3) node {$(3,2^2,1)$};
		\draw[fill=black] (0,4) node {$(3^2,1^2)$};
		\draw[fill=black] (0,6) node {$(5,3)$};
		\draw[fill=black] (0,7) node {$(7,1)$};
		
		\draw[fill=black] (-1,2) node {$(2^4)$};
		\draw[fill=black] (1,2) node {$(3,1^5)$};
		\draw[fill=black] (-1,5) node {$(4^2)$};
		\draw[fill=black] (1,5) node {$(5,1^3)$};
		
		\end{scope}

		\end{tikzpicture}
	\end{center}
	\caption{Hasse diagrams for restricted sets of partitions $\mathcal{P}_+(6)$ and $\mathcal{P}_+(8)$. Hollow nodes ($(3,2^2,1) \in \mathcal{P}_+(8)$) are \textit{non-special}, see discussion.}
	\label{Hasseex}
\end{figure}
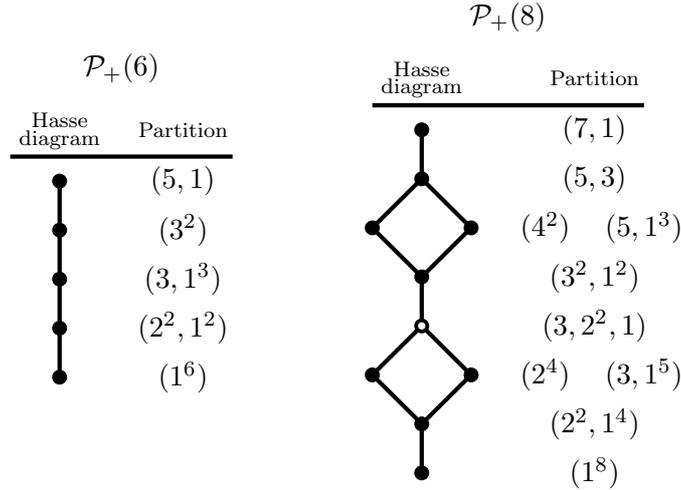

The nilpotent orbits in $\kk{so}_{2n}$ are in one-to-one correspondence with the restricted set of partitions of $2n$ where \emph{even} parts occur with \emph{even} multiplicity (including zero). This set is written $\mathcal{P}_+(2n)$. The dominance ordering can once again be used to give $\mathcal{P}_+(2n)$ poset structure and hence construct a Hasse diagram. Some example Hasse diagrams are given in figure \ref{Hasseex} and appendix \ref{appendix}.
However there are complications in defining transpose in $\mathcal{P}_+(2n)$ because $\mu \in \mathcal{P}_+(2n)$ doesn't imply that $\mu^t \in \mathcal{P}_+(2n)$. Since transpose maps $\mathcal{P}_+(2n) \rightarrow \mathcal{P}(2n)$, one defines another map called the \textit{D-collapse} which maps $\mathcal{P}(2n) \rightarrow \mathcal{P}_+(2n)$. The D-collapse takes a partition $\sigma$ to the largest partition in $\mathcal{P}_+(2n)$ that is equal to, or dominated by, $\sigma$. Clearly $\sigma_D = \sigma$ for $\sigma \in \mathcal{P}_+(2n)$. If $\sigma \notin \mathcal{P}_+(2n)$ then (at least one) even part, say $\sigma_j = 2r$, must have odd multiplicity. In this case, take the final $2r$ to $2r-1$ and take the largest part $\sigma_i < 2r-1$ to $\sigma_i+1$. Repeat this process until the resulting partition is in $\mathcal{P}_+(2n)$. D-collapse is many-to-one.

The \textit{Lusztig-Spaltenstein} map, $d_{LS}$, is transposition followed by D-collapse. $d_{LS}$ is a many-to-one (due to the D-collapse) map $\mathcal{P}_+(2n) \rightarrow \mathcal{P}_+(2n)$. Note that $d_{LS}^3 = d_{LS}$. A partition is called \textit{special} if $d_{LS}^2$ is the identity on the partition. Non-special partitions' nodes are drawn hollow in the Hasse diagram.

\noindent\textbf{Notable partitions in}
{\boldmath$\mathcal{P}_+(2n)$} $\qquad$ There are several important partitions to highlight. For every $n \geq 4$ there are unique, special partitions which are the highest, next-to-highest, lowest and next-to lowest partitions in the set. These are $(2n-1,1)$, $(2n-3,3)$, $(1^{2n})$ and $(2^2,1^{2n-4})$ respectively. For $n\geq 4$ there are the highest and lowest non-special partitions. These always take the form $(2n-5,2^2,1)$ and $(3,2^2,1^{2n-7})$ respectively (these coincide for $n=4$).

\subsubsection{Nilpotent orbits}
Nilpotency of an element of a Lie algebra is preserved under the action of its corresponding Lie group. The conjugacy class of a nilpotent element of a Lie algebra under this action is called a nilpotent orbit. There are finitely many nilpotent orbits in a Lie algebra. Nilpotent orbits in $\kk{so}_{2n}$ are in one-to-one correspondence with the restricted set $\mathcal{P}_+(2n)$ of partitions of $2n$ in which \emph{even} parts occur with \emph{even} multiplicity. A complication arises for \textit{very even} partitions, consisting of only even parts. Under the action of $SO(2n)$ these partitions are associated to \textit{two} orbits, but under the action of $O(2n)$ these orbits combine into a single orbit. For example there are twelve nilpotent orbits in $\kk{so}_8$ under the action of $SO(2n)$, these are  $\mathcal{O}_{(7,1)}$, $\mathcal{O}_{(5,3)}$,$\mathcal{O}_{(4^2)}^{I}$, $\mathcal{O}_{(4^2)}^{II}$, $\mathcal{O}_{(5,1^3)}$,  $\mathcal{O}_{(3^2,1^2)}$, $\mathcal{O}_{(3,2^2,1)}$, $\mathcal{O}_{(2^4)}^I$,$\mathcal{O}_{(2^4)}^{II}$, $\mathcal{O}_{(3,1^5)}$, $\mathcal{O}_{(2^2,1^4)}$ and  $\mathcal{O}_{(1^8)}$. Under the action of $O(2n)$ there are ten nilpotent orbits, all of the non-very-even orbits and the two orbits $\mathcal{O}^I_{(4^2)}\cup\mathcal{O}^{II}_{(4^2)}$ and $\mathcal{O}^I_{(2^4)}\cup\mathcal{O}^{II}_{(2^4)}$. For the rest of this work these unions of orbits will be what is meant by $\mathcal{O}_\lambda$ for very even $\lambda$.

The \textit{closure} of a nilpotent orbit $\mathcal{O}_\mu$ is the union of the orbit with all of the orbits labelled with partitions dominated by $\mu$,
\begin{equation}
\O_\mu = \bigcup_{\nu \leq \mu} \mathcal{O}_\nu.
\end{equation}
A partial ordering of these closures can be defined by their inclusion relations. This partial order has the same structure as that for the partitions to which the orbits are related. Writing a general partition $\mu = ((2n)^{r_{2n}}, \dots , 3^{r_3},2^{r_2} , 1^{r_1},0^{r_0})$, the closures of nilpotent orbits in $\kk{so}_{2n}$ are algebraic varieties of quaternionic dimension 
\begin{equation}\label{dimorb}
\dim_{\mathbb{H}}(\O_\mu) = \displaystyle\frac{1}{2} \Big{(}2n^2 - n - \frac{1}{2} \sum_i (\mu_i^t)^2 + \frac{1}{2}\sum_{i \textrm{ odd}}r_i \Big{)}.
\end{equation} 

Nilpotent orbit closures for adjacent partitions form \textit{minimal degenerations}. In \cite{KP2}--\cite{KP3}, Kraft and Procesi determined $\textrm{Sing}(\O_{\mu}, \O_{\nu})$ for all minimal degenerations in the classical Lie algebras\footnote{The singularity of a point $x$ in a variety $\mathcal{X}$ and the singularity of a point $y$ in a variety $\mathcal{Y}$ are \textit{smoothly equivalent} if there exists a variety $\mathcal{Z}$ with a point $z\in \mathcal{Z}$ and morphisms $\phi: \mathcal{Z} \rightarrow \mathcal{X}$ and $\psi:\mathcal{Z} \rightarrow \mathcal{Y}$ which are smooth and for which $\phi(z) =x$ and $\psi(z)=y$. The equivalence class of singularities smoothly equivalent to the singularity of a pointed variety $(\mathcal{X},x)$ is denoted $\textrm{Sing}(\mathcal{X}, x)$. In $\O_\mu$, the smooth equivalence class for the singularity of an element in $\mathcal{O}_\nu$, where $\O_\nu \subset \O_\mu$ is a minimal degeneration, is independent of the element and so is denoted $\textrm{Sing}(\O_{\mu}, \O_{\nu})$.}. The edges of the Hasse diagram for nilpotent orbit closures are labelled with these singularities. For example, the diagrams in figure \ref{Hasseex} become figure \ref{Hasseex2}. All the Hasse diagrams in appendix \ref{appendix} are labelled with these singularities.
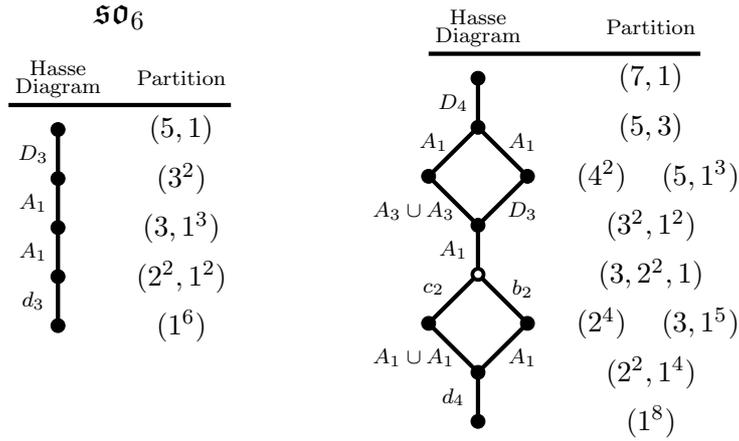
\begin{figure}
	\begin{center}
		\begin{tikzpicture}[scale = 0.65]
		\begin{scope}[yshift=3cm]
		\draw[ultra thick] (-1, 1.5) -- (3.5,1.5);
		
		\draw (0,2.25) node {\scriptsize{Hasse}}
		(0,1.85) node {\scriptsize{Diagram}}
		(2.5,2.05) node {\scriptsize{Partition}}
		(1.25,3.3) node {\Large{$\kk{so}_6$}};
		\end{scope}
		
		\draw[fill=black] (0,0) circle (4pt);
		\draw[fill=black] (0,1) circle (4pt);
		\draw[fill=black] (0,2) circle (4pt);
		\draw[fill=black] (0,3) circle (4pt);
		\draw[fill=black] (0,4) circle (4pt);
		
		\draw[\THICC] (0,0) -- (0,4);
		
		\draw (2.5,0) node {$(1^6)$};
		\draw (2.5,1) node {$(2^2,1^2)$};
		\draw (2.5,2) node {$(3,1^3)$};
		\draw (2.5,3) node {$(3^2)$};
		\draw (2.5,4) node {$(5,1)$};
		
		\draw (-0.5,3.5) node {\scriptsize$D_3$};
		\draw (-0.5,2.5) node {\scriptsize$A_1$};
		\draw (-0.5,1.5) node {\scriptsize$A_1$};
		\draw (-0.5,0.5) node {\scriptsize$d_3$};
		
		\draw (0,-2.27) node {};
		
		\end{tikzpicture}$\qquad~~~~~~$
		\begin{tikzpicture}[scale=0.65]
		\begin{scope}[yshift=6cm]
		\draw[ultra thick] (-1, 1.5) -- (4.5,1.5);
		
		\draw (0,2.25) node {\scriptsize{Hasse}}
		(0,1.85) node {\scriptsize{Diagram}}
		(3.5,2.05) node {\scriptsize{Partition}}
		(1.75,3.3) node {\Large{$\kk{so}_8$}};
		\end{scope}
		
		\draw[\THICC] (0,0) -- (0,1) -- (1,2) -- (0,3) -- (0,4) -- (1,5) -- (0,6) -- (0,7) (0,6) -- (-1,5) -- (0,4) (0,3) -- (-1,2) -- (0,1);
		
		\draw[fill=black] (0,0) circle (4pt);
		\draw[fill=black] (0,1) circle (4pt);
		\draw[fill=black] (0,3) circle (4pt);
		\draw[white, fill=white] (0,3) circle (1.8pt);
		\draw[fill=black] (0,4) circle (4pt);
		\draw[fill=black] (0,6) circle (4pt);
		\draw[fill=black] (0,7) circle (4pt);
		
		\draw[fill=black] (-1,2) circle (4pt);
		\draw[fill=black] (1,2) circle (4pt);
		\draw[fill=black] (-1,5) circle (4pt);
		\draw[fill=black] (1,5) circle (4pt);
		
		\begin{scope}[xshift = 3.5cm]
		\draw[fill=black] (0,0) node {$(1^8)$};
		\draw[fill=black] (0,1) node {$(2^2,1^4)$};
		\draw[fill=black] (0,3) node {$(3,2^2,1)$};
		\draw[fill=black] (0,4) node {$(3^2,1^2)$};
		\draw[fill=black] (0,6) node {$(5,3)$};
		\draw[fill=black] (0,7) node {$(7,1)$};
		
		\draw[fill=black] (-1,2) node {$(2^4)$};
		\draw[fill=black] (1,2) node {$(3,1^5)$};
		\draw[fill=black] (-1,5) node {$(4^2)$};
		\draw[fill=black] (1,5) node {$(5,1^3)$};
		
		\end{scope}
		
		\draw (-0.5,0.5) node {\scriptsize$d_4$};
		\draw (-0.5,3.5) node {\scriptsize$A_1$};
		\draw (-0.5,6.5) node {\scriptsize$D_4$};
		
		\draw (-1.3,1.3) node {\scriptsize$A_1 \cup A_1$};
		\draw (0.9,1.3) node {\scriptsize$A_1$};
		
		\draw (-0.9,2.7) node {\scriptsize$c_2$};
		\draw (0.9,2.7) node {\scriptsize$b_2$};
		
		\draw (-1.3,4.3) node {\scriptsize$A_3 \cup A_3$};
		\draw (0.9,4.3) node {\scriptsize$D_3$};
		
		\draw (-0.9,5.7) node {\scriptsize$A_1$};
		\draw (0.9,5.7) node {\scriptsize$A_1$};

		\end{tikzpicture}
	\end{center}
	\caption{The example Hasse diagrams from figure \ref{Hasseex} with the edges labelled with  $\textrm{Sing}(\O_{\mu}, \O_{\nu})$ for the orbits they join as identified by Kraft and Procesi. The labels $A_n$ and $D_n$ refer to simple, or Du Val, singularities, and the labels $a_n$, $b_n$, $c_n$, $d_n$, refer to the closures of the minimal nilpotent orbits of the corresponding algebras. }
	\label{Hasseex2}
\end{figure}

Orbits associated to non-special partitions are themselves called non-special. Nilpotent orbit closures can be thought of as consisting of a \textit{run} of edges and nodes on a Hasse diagram from a node $\mathcal{O}_\mu$ down to the bottom. Orbit closures which only have special nodes in their Hasse sub-diagram play an important role in our discussion. These are exactly the closures of those orbits with \textit{height} two or less. The height of a nilpotent orbit in $\kk{so}_{2n}$ is \cite{height}
\begin{equation}
\textrm{ht}(\mathcal{O}_\mu) = \begin{cases} m_1+m_2-2, &~~ m_2 \geq m_1-1 \\ 2m_1 - 4, &~~ m_2 \leq m_1-2 \end{cases}.
\end{equation}
The $m_i$ here are from the non-exponential notation for $\mu$. Low-height orbit closures are important because ht$(\mathcal{O}_\mu) \leq 2$ nilpotent orbits always have a realisation as the Coulomb branch of a $D_n$ Dynkin quiver, \cite{QuivSub}.

\subsubsection{Slodowy slices}
Given a nilpotent element $X \in \mathcal{O}_{\nu} \subset \kk{so}_{2n}$, the \textit{Slodowy slice} to $X$ is 
\begin{equation}
\mathcal{S}_X :=X + \ker(\textrm{ad}(Y))
\end{equation}
where $Y \in \mathcal{N} \subset \kk{so}_{2n}$ is a nilpotent element associated to $X$ inside an $\kk{sl}_2$ triple (\cite{CollMc}, 3.2). This triple is unique up to conjugacy and so this defines a transverse slice to the orbit $\O_\nu$ denoted $\mathcal{S}_\nu$. Recall that from the point of view of the Hasse diagram the closure of a nilpotent orbit, $\O_\nu$, can be considered as a \textit{run} from the bottom node up to the node $\nu$. Restricting the Slodowy slice to nilpotent elements by intersecting the variety with the nilpotent cone, $\mathcal{N} = \O_{(2n-1,1)}$, gives a variety which corresponds to a run from the node $\nu$ up to the top of the Hasse diagram. From here on a slice written $\mathcal{S}_\nu$ means the intersection of the full slice with the nilpotent cone. 

Writing $\nu$ in exponential notation with exponents $t_i$, the Slodowy slice $\mathcal{S}_\nu$ is a hyperK{\"a}hler singular variety of quaternionic dimension
\begin{equation}\label{dimslice}
\dim_{\mathbb{H}}(\mathcal{S}_\nu) = \frac{1}{2}\Big{(} \frac{1}{2} \sum_i(\nu_i^t)^2  -\frac{1}{2}\sum_{i \textrm{ odd}} t_i - n \Big{)}.
\end{equation} 

Once again those varieties whose Hasse diagrams contain only special nodes will play an important role in our discussion. Slodowy slices that correspond to runs at the top of the Hasse diagram which contain no non-special nodes shall be referred to as \textit{special slices}. These special slices (and their subvarieties) will have realisations as the Higgs branches of $D_n$ Dynkin quivers.

There always exists a largest special slice in a nilpotent cone. This is due to the presence of a highest non-special partition whose node must be avoided. This non-special node is $(2n-5,2^2,1)$. For the Hasse diagram for a slice $\mathcal{S}_\nu$ to avoid containing this node, $\nu$ must not be dominated by $(2n-5,2^2,1)$. The lowest partition not dominated by $(2n-5,2^2,1)$ always takes the form $(n-1^2,1^2)$. Therefore the \textit{maximal special slice} in an algebra $\kk{so}_{2n}$ is always $\mathcal{S}_{(n-1^2,1^2)} \in \kk{so}_{2n}$. It is this variety and its subvarieties that have realisations as the Higgs branches of Dynkin quivers. Note that 
\begin{equation}\label{height}
\textrm{ht}(d_{LS}((n-1^2,1^2))) = \begin{cases} \textrm{ht}((3^2,2^{n-3})) = 4 > 2, & n \textrm{ odd} \\ \textrm{ht}((3^2,2^{n-4},1^2)) = 4 > 2, & n \textrm{ even}, \end{cases} \end{equation}
so while all the Dynkin quivers which realise nilpotent orbit closures as their Coulomb branches also realise Slodowy slices with their Higgs branches, the reverse doesn't necessarily follow.

Happily, Hasse diagrams pertaining to maximal special slices in $\kk{so}_{2n}$ take a regular form which allows them to be written generally. It transpires that there are two forms for the Hasse diagrams, $\mathcal{S}_{(2m-1^2,1^2)} \in \kk{so}_{4m}$ or $\mathcal{S}_{(2m^2,1^2)} \in \kk{so}_{4m+2}$. The Hasse diagrams are given in figure \ref{MSSHasse}.
\begin{figure}
	\begin{center}
		\begin{tikzpicture}[scale=0.65]
		\begin{scope}[yshift=8cm]
		\draw[ultra thick] (-1, 1.5) -- (9.5,1.5);
		
		\draw (0,2.25) node {\scriptsize{Hasse}}
		(0,1.85) node {\scriptsize{Diagram}}
		(8,2.05) node {\scriptsize{Partition}}
		(4,2.3) node {$\mathcal{S}_{(2m-1^2,1^2)} \in \kk{so}_{4m}$};
		\end{scope}
		
		\draw[\THICC] (-1,3.3) -- (-1,1) -- (0,0) -- (1,1) -- (1,3.3) (0,9) -- (0,8) -- (-1,7) -- (-1,4.7) (0,8) -- (1,7) -- (1,4.7)
		
		(-1,2) -- (1,1) (-1,3) -- (1,2) (1,3) -- (0.7,3.15)
		
		(-1,7) -- (1,6) (-1,6) -- (1,5) (-1,5) -- (-0.7,4.85)
		;
		
		\draw[fill=black] (0,0) circle (4pt)
		(0,8) circle (4pt)
		(0,9) circle (4pt);
		\foreach \i in {1,2,3,5,6,7}{\draw[fill=black] (-1,\i) circle (4pt)
			(1,\i) circle (4pt);}
		
		\draw (0,4.1) node {$\vdots$};
		
		\draw (-0.5,8.5) node {\sz$D_{2m}$}
		(-1.4,7.7) node {\sz$D_{2m-2}$}
		(-1.8,6.5) node {\sz$D_{2m-4}$}
		(-1.8,5.5) node {\sz$D_{2m-6}$}
		
		(-1.6,2.5) node {\sz$D_{4}$}
		(-1.6,1.5) node {\sz$A_1$}
		
		(0.1,1.75) node[rotate=-25] {\sz$A_{2m-3}$}
		(0.1,2.75) node[rotate=-25] {\sz$A_{2m-5}$}
		
		(0.1,5.75) node[rotate=-25] {\sz$A_{5}$}
		(0.1,6.75) node[rotate=-25] {\sz$A_{3}$}
		
		(1,7.7) node {\sz$A_{1}$}
		(1.8,6.5) node {\sz$D_{2m-1}$}
		(1.8,5.5) node {\sz$D_{2m-3}$}
		
		(1.6,2.5) node {\sz$D_{7}$}
		(1.6,1.5) node {\sz$D_5$}
		
		(1.1,0.3) node {\sz$D_{3}$}
		(-2.05,0.3) node {\sz$A_{2m-1} \cup A_{2m-1}$}
		
		(14,0.3) node {\sz$~$}
		;

		\begin{scope}[xshift = 8cm]
		\draw[fill=black] (0,0) node {$(2m-1^2,1^2)$};
		
		\draw[fill=black] (-2.5,1) node {$(2m^2)$};
		\draw[fill=black] (2.5,1) node {$(2m+1,2m-3,1^2)$};
		
		\draw[fill=black] (-2.5,2) node {$(2m+1,2m-1)$};
		\draw[fill=black] (2.5,2) node {$(2m+3,2m-5,1^2)$};
		
		\draw[fill=black] (-2.5,3) node {$(2m+3,2m-3)$};
		\draw[fill=black] (2.5,3) node {$(2m+5,2m-7,1^2)$};
		
		\draw[fill=black] (0,9) node {$(4m-1,1)$};
		
		\draw[fill=black] (0,8) node {$(4m-3,3)$};
		
		\draw[fill=black] (-2.5,7) node {$(4m-5,5)$};
		\draw[fill=black] (2.5,7) node {$(4m-3,1^3)$};
		
		\draw[fill=black] (-2.5,6) node {$(4m-7,7)$};
		\draw[fill=black] (2.5,6) node {$(4m-5,3,1^2)$};
		
		\draw[fill=black] (-2.5,5) node {$(4m-9,9)$};
		\draw[fill=black] (2.5,5) node {$(4m-7,5,1^2)$};
		
		\draw (0,4.1) node {$\vdots$};
		\end{scope}
		\end{tikzpicture}
		
		$~$
		
		$~$
		
		\begin{tikzpicture}[scale=0.65]
		\begin{scope}[yshift=8cm]
		\draw[ultra thick] (-1, 1.5) -- (9.5,1.5);
		
		\draw (0,2.25) node {\scriptsize{Hasse}}
		(0,1.85) node {\scriptsize{Diagram}}
		(8,2.05) node {\scriptsize{Partition}}
		(4,2.3) node {$\mathcal{S}_{(2m^2,1^2)} \in \kk{so}_{4m+2}$};
		\end{scope}
		
		
		\draw[\THICC] (0,0) -- (0,-1) (-1,3.3) -- (-1,1) -- (0,0) -- (1,1) -- (1,3.3) (0,9) -- (0,8) -- (-1,7) -- (-1,4.7) (0,8) -- (1,7) -- (1,4.7)
		
		(-1,2) -- (1,1) (-1,3) -- (1,2) (1,3) -- (0.7,3.15)
		
		(-1,7) -- (1,6) (-1,6) -- (1,5) (-1,5) -- (-0.7,4.85)
		;
		
		\draw[fill=black] (0,-1) circle (4pt)
		(0,0) circle (4pt)
		(0,8) circle (4pt)
		(0,9) circle (4pt);
		\foreach \i in {1,2,3,5,6,7}{\draw[fill=black] (-1,\i) circle (4pt)
			(1,\i) circle (4pt);}
		
		\draw (0,4.1) node {$\vdots$};
		
		\draw (-0.7,8.5) node {\sz$D_{2m+1}$}
		(-1.4,7.7) node {\sz$D_{2m-1}$}
		(-1.8,6.5) node {\sz$D_{2m-3}$}
		(-1.8,5.5) node {\sz$D_{2m-5}$}
		
		(-1.6,2.5) node {\sz$D_{5}$}
		(-1.6,1.5) node {\sz$D_3$}
		
		(0.1,1.75) node[rotate=-25] {\sz$A_{2m-3}$}
		(0.1,2.75) node[rotate=-25] {\sz$A_{2m-5}$}
		
		(0.1,5.75) node[rotate=-25] {\sz$A_{5}$}
		(0.1,6.75) node[rotate=-25] {\sz$A_{3}$}
		
		(1,7.7) node {\sz$A_{1}$}
		(1.8,6.5) node {\sz$D_{2m}$}
		(1.8,5.5) node {\sz$D_{2m-2}$}
		
		(1.6,2.5) node {\sz$D_{8}$}
		(1.6,1.5) node {\sz$D_6$}
		
		(1.1,0.3) node {\sz$D_{4}$}
		(-1.3,0.3) node {\sz$A_{2m-1}$}
		
		(-0.5,-0.5) node {\sz$A_1$}
		;
		
		\begin{scope}[xshift = 8cm]
		\draw[fill=black] (0,-1) node {$(2m^2,1^2)$};
		
		\draw[fill=black] (0,0) node {$(2m+1,2m-1,1^2)$};
		
		\draw[fill=black] (-2.5,1) node {$(2m+1^2)$};
		\draw[fill=black] (2.5,1) node {$(2m+3,2m-3,1^2)$};
		
		\draw[fill=black] (-2.5,2) node {$(2m+3,2m-1)$};
		\draw[fill=black] (2.5,2) node {$(2m+5,2m-5,1^2)$};
		
		\draw[fill=black] (-2.5,3) node {$(2m+5,2m-3)$};
		\draw[fill=black] (2.5,3) node {$(2m+7,2m-7,1^2)$};
		
		\draw[fill=black] (0,9) node {$(4m+1,1)$};
		
		\draw[fill=black] (0,8) node {$(4m-1,3)$};
		
		\draw[fill=black] (-2.5,7) node {$(4m-3,5)$};
		\draw[fill=black] (2.5,7) node {$(4m-1,1^3)$};
		
		\draw[fill=black] (-2.5,6) node {$(4m-5,7)$};
		\draw[fill=black] (2.5,6) node {$(4m-3,3,1^2)$};
		
		\draw[fill=black] (-2.5,5) node {$(4m-7,9)$};
		\draw[fill=black] (2.5,5) node {$(4m-5,5,1^2)$};
		
		\draw (0,4.1) node {$\vdots$};
		\end{scope}
		\end{tikzpicture}
	\end{center}
	\caption{The Hasse diagrams for the maximal special slices, $\mathcal{S}_{(2m-1^2,1^2)} \in \kk{so}_{4m}$ and $\mathcal{S}_{(2m^2,1^2)} \in \kk{so}_{4m+2}$. The Higgs branches of the $D_n$ Dynkin quivers in this section will be subvarieties appearing as runs in these Hasse diagrams.}
	\label{MSSHasse}
\end{figure}
The nilpotent varieties which appear as the Higgs branches of $D_n$ Dynkin quivers are subvarieties of the maximal special slices and so their Hasse diagrams appear as subdiagrams of figure \ref{MSSHasse}.

Some obvious subvarieties of the maximal special slice are all the Slodowy slices $\es_\nu$ where $\nu>(2m-1^2,1^2)$ (resp. $(2m^2,1^2)$). These Slodowy slices are the varieties which appear as the Higgs branches of the balanced Dynkin quivers considered in \cite{kalveks2}. Further subvarieties can be found by considering slice and orbit-closure intersections.

\subsubsection{Intersections}
The intersection of a Slodowy slice with a closure of a nilpotent orbit is a subvariety which corresponds to some run of nodes in the Hasse diagram. An intersection $\O_\mu \cap \es_\nu$ corresponds to a run from a node $\mu$ down to a node $\nu$ for $\mu>\nu$. The subvarieties of the maximal special slice are those varieties $\O_\mu \cap \es_\nu$ where $\mu>\nu\geq (2m-1^2,1^2)$ (resp. $(2m^2,1^2)$). Intersections are hyperK{\"a}hler varieties of quaternionic dimension
\begin{equation}\label{nilpvardim}
\dim_{\mathbb{H}}(\O_\mu \cap \mathcal{S}_\nu) = 
\displaystyle \frac{1}{2}\Big{(} \frac{1}{2} \sum_i(\nu_i^t)^2 - \frac{1}{2}\sum_i(\mu_i^t)^2 +\frac{1}{2}\sum_{i \textrm{ odd}} r_i - \frac{1}{2}\sum_{i \textrm{ odd}} t_i \Big{)},
\end{equation}
when $\mu$ and $\nu$ are written using the aforementioned exponential notation. Note that this equation generalises (\ref{dimorb}) and (\ref{dimslice}), in which the nilpotent orbits' dimensions are obtained by setting $\nu$ to be the minimal partition, and the Slodowy slices' dimensions are obtained by setting $\mu$ to be the maximal partition.

To characterise the general subvarieties of the maximal special slices note that the partitions in each maximal special slice fall into a small number of general forms. In $\kk{so}_{4m}$ the partitions are all of the form $\psi_j = (4m-(2j+1),2j+1)$ or $\varphi_j = (4m-(2j+3),2j+1,1^2)$, for $0\leq j\leq m-1$, or $(2m^2)$. There are therefore five general forms for subvarieties $\mathcal{V} \subseteq \es_{(2m-1^2,1^2)} \subset \kk{so}_{4m}$,
\begin{equation}\label{so4vars}
\mathcal{V} = \begin{cases}
\O_{\psi_j} \cap \es_{\psi_k} ~&~ \textrm{for } k>j \\
\O_{\psi_j} \cap \es_{\varphi_k} ~&~ \textrm{for } k>j-1 \geq 0 \\
\O_{\varphi_j} \cap \es_{\varphi_k} ~&~ \textrm{for } k>j \\
\O_{\psi_j} \cap \es_{(2m^2)} & \\
\O_{(2m^2)} \cap \es_{(2m-1^2,1^2)} = A_{2m-1} \cup A_{2m-1}. &
\end{cases}
\end{equation}
In $\kk{so}_{4m+2}$ the maximal special slice partitions all take the form $\psi_j' = (4m-(2j-1),2j+1)$ or $\varphi_j' = (4m-(2j+1), 2j+1,1^2)$, for $0\leq j \leq m$, or $(2m^2,1^2)$. There are therefore five general forms for subvarieties $\mathcal{V} \subseteq \es_{(2m^2,1^2)} \subset \kk{so}_{4m+2}$,
\begin{equation}\label{so42vars}
\mathcal{V} = \begin{cases}
\O_{\psi_j'} \cap \es_{\psi_k'} ~&~ \textrm{for } k>j \\
\O_{\psi_j'} \cap \es_{\varphi_k'} ~&~ \textrm{for } k>j-1 \geq 0 \\
\O_{\varphi_j'} \cap \es_{\varphi_k'} ~&~ \textrm{for } k>j \\
\O_{\psi_j'} \cap \es_{(2m^2,1^2)} & \\
\O_{\varphi_j'} \cap \es_{(2m^2,1^2)}. &
\end{cases}
\end{equation}
There are some varieties that appear in both maximal special slices. These are identified by looking for identical Hasse subdiagrams in the two maximal special slice Hasse diagrams. Hasse subdiagrams shared by both diagrams in figure \ref{MSSHasse} are: the individual singularities of types $A_l$ and $D_l$, and varieties with Hasse diagrams that take the form of a chain of $D_l$ singularities for odd or even $l$. Casting these shared subvarieties in the notation above gives the equalities
\begin{equation}\label{matchingvar}
\begin{split}
\O_{\psi_j} \cap \es_{\psi_k} & = \O_{\varphi_j'} \cap \es_{\varphi_k'} \\
\O_{\varphi_j} \cap \es_{\varphi_k} & = \O_{\psi_{j+1}'} \cap \es_{\psi_{k+1}'} \\
\O_{\psi_j} \cap \es_{(2m^2)} & = \O_{\varphi_j'} \cap \es_{(2m^2,1^2)}.
\end{split}
\end{equation}  
The varieties in (\ref{so4vars}) and (\ref{so42vars}) are all of the nilpotent subvarieties of $\kk{so}_{2n}$ which get realised as the Higgs branches of $D_n$ Dynkin quivers. This will be confirmed in section 3 when the singularity structure of all good $D_n$ Dynkin quivers is determined and these are the only $\kk{so}_{2n}$ nilpotent varieties to arise.

\subsection{Nilpotent varieties as Dynkin quiver Higgs branches}
\begin{figure}
	\begin{center}
		\begin{tikzpicture}[scale = 0.8]
		\def\q{7}
		\def\k{0.7071}
		\def\size{\sz}
		
		\draw[thick] (0,0) -- (4,0) -- (4.7,0.7) (4,0) -- (4.7,-0.7) (0,0) -- (0,1)
		(1,0) -- (1,1)
		(2,0) -- (2,1)
		(3,0) -- (3,1)
		(4,0) -- (5,0)
		
		(4+\k,\k) -- (5+\k,\k)
		(4+\k,-\k) -- (5+\k,-\k)
		;
		
		\draw[thick, fill=white] (0,0) circle (\q pt)
		(1,0) circle (\q pt)
		(2,0) circle (\q pt)
		(3,0) circle (\q pt)
		(4,0) circle (\q pt)
		(4+\k,\k) circle (\q pt)
		(4+\k,-\k) circle (\q pt)
		;
		
		\draw[thick, fill=white] 
		(0 -\q/30,1 -\q/30) rectangle (0 +\q/30,1 +\q/30)
		(1 -\q/30,1 -\q/30) rectangle (1 +\q/30,1 +\q/30)
		(2 -\q/30,1 -\q/30) rectangle (2 +\q/30,1 +\q/30)
		(3 -\q/30,1 -\q/30) rectangle (3 +\q/30,1 +\q/30)
		(5 -\q/30,0 -\q/30) rectangle (5 +\q/30,0 +\q/30)
		(5+\k -\q/30,\k -\q/30) rectangle (5+\k +\q/30,\k +\q/30)
		(5+\k -\q/30,-\k -\q/30) rectangle (5+\k +\q/30,-\k +\q/30)
		;

		\draw (0,1.5) node {\size $f_1$}
		(1,1.5) node {\size $f_2$}
		(2,1.5) node {\size $f_3$}
		(3,1.5) node {\size $f_{n-3}$}
		
		(5.8,0) node  {\size $f_{n-2}$}
		(5.8 +\k,0+\k) node  {\size $f'$}
		(5.8+\k,0-\k) node  {\size $f''$}

		(0,-0.5) node {\size $g_1$}
		(1,-0.5) node {\size $g_2$}
		(2,-0.5) node {\size $g_3$}
		(3,-0.5) node {\size $g_{n-3}$}
		(3.75,0.45) node {\size $g_{n-2}$}
		
		(4+\k,1.25) node {\size $g'$}
		(4+\k,-1.25) node {\size $g''$}
		;
		
		\draw[fill=white, white] (1.5,-1.5) rectangle (2.5,2);
		
		\draw (2,0) node {$\dots$};
		
		\end{tikzpicture}
	\end{center}
\caption{A general $3d$ $\mathcal{N}=4$ unitary $D_n$ Dynkin quiver. A circular gauge node labelled $g$ represents a $U(g)$ gauge group and carries an adjoint $U(g)$ vectormultiplet. Square nodes labelled $f$ represent global flavour symmetry. Edges between circular nodes are bifundamental hypermultiplets and edges between a circular and a square node are fundamental hypermultiplets. The gauge nodes and bifundamental hypermultiplets form the Dynkin diagram so a $D_n$ Dynkin quiver has $n$ gauge nodes in total and a gauge group $U(g') \times U(g'') \times \prod_{i=1}^{n-2}U(g_i)$.}
\label{Dynkinquiv}
\end{figure}
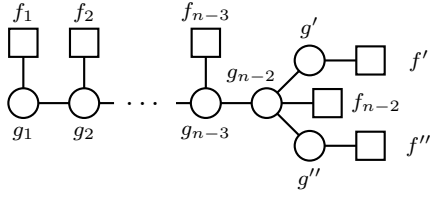

The $\kk{so}_{2n}$ nilpotent varieties in (\ref{so4vars}) and (\ref{so42vars}) are realised as the Higgs branches of $D_n$ Dynkin quivers. A general $D_n$ Dynkin quiver is given in figure \ref{Dynkinquiv}.

The moduli space branches of $D_n$ Dynkin quivers have been discussed before in numerous contexts. Their capacity to realise closures of  $\kk{so}_{2n}$ nilpotent orbits of (characteristic) height ht$(\O_\rho) \leq 2$ as their \textit{Coulomb} branches was considered in \cite{Kalveks}, \cite{Sperling}, \cite{QuivSub}, to realise $\kk{so}_{2n}$ Slodowy slices as their Higgs branches in \cite{kalveks2}, and in the context of brane constructions for which they are IR descriptions in, for example, \cite{Kasputin}. The discussion in this section will concern the Dynkin quivers which realise nilpotent varieties appearing as a subvariety of the maximal special slice as their Higgs branches. The $D_n$ Dynkin quivers with ht$(\O_{\rho}) \leq 2$ nilpotent orbit Coulomb branches are a subset of those which realise the Slodowy slices as Higgs branches thanks to (\ref{height}) and so the discussion automatically generalises the Coulomb branch results in the same manner. 

A central tool to write down these quivers is the Kraft-Procesi transition \cite{KPT}, \cite{KPTC}, and its generalization, quiver subtraction, \cite{QuivSub}. These processes identify and remove the transverse slice structure of the moduli space branches. This work introduces the reverse procedure, \textit{quiver addition} whereby transverse slices are `added' onto known moduli spaces by the introduction of appropriate fields at the level of the quiver. To demonstrate the technique's effectiveness, quiver addition will be used to find the appropriate form for quivers realising Slodowy slices, then quiver subtraction will be used to identify quivers for subvarieties of the maximal special slice. In order to perform the addition the quivers for the singularities labelling the Hasse diagrams must be determined.

The maximal special slice $\mathcal{S}_{(2m-1^2,1^2)} \in \kk{so}_{4m}$ has a Hasse diagram given in figure \ref{MSSHasse}. It is clear that there are only three types of minimal degenerations needed in order to construct these varieties, $D_l$, $A_{2l-1}$ and $A_{2l-1} \cup A_{2l-1}$.

Note that the Hasse diagram for partitions and the Hasse diagram of inclusion relations for Slodowy slices are flipped. The Hasse subdiagram describing the singularity structure of $\mathcal{S}_{(2m-1^2,1^2)}$ places the partition $(4m-1,1)$ at the top, as it is the most dominant partition. However the Higgs branches are Slodowy slices\footnote{Which can be considered to be all the edges and nodes from a given node \textit{up} in the Hasse diagram.} and as such this topmost node corresponds to a theory with \textit{trivial} Higgs branch. If instead one drew a Hasse diagram corresponding to the inclusion relations of the Higgs branches of the quivers, the largest branch would be the slice $\mathcal{S}_{(2m-1^2,1^2)}$ and this theory would be placed at the top. Labelling the edges of this quiver Hasse diagram with the transverse slice between the quiver's Higgs branches then yields a Hasse diagram which is exactly the partition diagram flipped upside down. When adding transverse slices in order to build up Higgs branches one builds the Hasse diagram for the slice from the \textit{top} and the theory with the largest (with respect to the moduli space inclusion relation) Higgs branch will be associated to the node(s) at the \textit{bottom}. This is the convention which preserves the notation for the hierarchy of singularities within the varieties.

The three types of minimal degeneration that are realised as the Higgs branches of the Dynkin quivers are given in figure \ref{sliceQ}. 
\begin{figure}
	\begin{center}
		\begin{tikzpicture}[scale = 0.65]
		\def\q{7}
		\def\k{0.7071}
		\def\size{\sz}
		
		\draw[thick] (-1,0) -- (1.5,0) (2.5,0) -- (4,0) -- (4.7,0.7) (4,0) -- (4.7,-0.7)
		(0,0) -- (0,1)
		
		;
		
		\draw[thick, fill=white] 
		(-1,0) circle (\q pt)
		(0,0) circle (\q pt)
		(1,0) circle (\q pt)
		(3,0) circle (\q pt)
		(4,0) circle (\q pt)
		(4+\k,\k) circle (\q pt)
		(4+\k,-\k) circle (\q pt)
		;
		
		\draw[thick, fill=white] 
		(0 -\q/30,1 -\q/30) rectangle (0 +\q/30,1 +\q/30)
		;

		\draw 
		(0,1.5) node {\size $1$}
		
		
		(-1,-0.5) node {\size $1$}
		(0,-0.5) node {\size $2$}
		(1,-0.5) node {\size $2$}
		(3,-0.5) node {\size $2$}
		(3.75,0.45) node {\size $2$}
		
		(4+\k,1.25) node {\size $1$}
		(4+\k,-1.25) node {\size $1$}
		;
		
		\draw 
		(2,0) node {$\dots$}
		;
		
		\draw 
		(-4,0) node {$\Q_{\Hg}(D_k) ~~~~ =$}
		;

		\begin{scope}[yshift=-4.5cm]
		\draw[thick] (-1,0) -- (1.5,0) 
		(2.5,0) -- (4,0)
		(4,0) -- (4.7,0.7) 
		(-1,0) -- (-1,1) 
		
		(4+\k,\k) -- (5+\k,\k)
		;
		
		\draw[thick, fill=white] 
		(-1,0) circle (\q pt)
		(0,0) circle (\q pt)
		(1,0) circle (\q pt)
		(3,0) circle (\q pt)
		(4,0) circle (\q pt)
		(4+\k,\k) circle (\q pt)
		;
		
		\draw[thick, fill=white] 
		(-1 -\q/30,1 -\q/30) rectangle (-1 +\q/30,1 +\q/30)
		(5+\k -\q/30,\k -\q/30) rectangle (5+\k +\q/30,\k +\q/30)
		;

		\draw 
		(-1,1.5) node {\size $1$}
		
		(5.8 +\k,0+\k) node  {\size $1$}
		
		(-1,-0.5) node {\size $1$}
		(0,-0.5) node {\size $1$}
		(1,-0.5) node {\size $1$}
		(3,-0.5) node {\size $1$}
		(3.75,0.45) node {\size $1$}
		
		(4+\k,1.25) node {\size $1$}
		;
		
		\draw 
		(2,0) node {$\dots$}
		;

		\draw 
		(-6.2,-1) node {$\Q_{\Hg}(A_{k} \cup A_{k}) ~~~~ =$}
		(-2.5,0.5) node {$I$}
		(-2.5,-2.5) node {$II$}
		;
		
		\begin{scope}[yshift=-2cm, yscale=-1]
		\draw[thick] (-1,0) -- (1.5,0) 
		(2.5,0) -- (4,0)
		(4,0) -- (4.7,0.7) 
		(-1,0) -- (-1,1) 
		
		(4+\k,\k) -- (5+\k,\k)
		;
		
		\draw[thick, fill=white] 
		(-1,0) circle (\q pt)
		(0,0) circle (\q pt)
		(1,0) circle (\q pt)
		(3,0) circle (\q pt)
		(4,0) circle (\q pt)
		(4+\k,\k) circle (\q pt)
		;
		
		\draw[thick, fill=white] 
		(-1 -\q/30,1 -\q/30) rectangle (-1 +\q/30,1 +\q/30)
		(5+\k -\q/30,\k -\q/30) rectangle (5+\k +\q/30,\k +\q/30)
		;

		\draw 
		(-1,1.5) node {\size $1$}
		
		(5.8 +\k,0+\k) node  {\size $1$}
		
		(-1,-0.5) node {\size $1$}
		(0,-0.5) node {\size $1$}
		(1,-0.5) node {\size $1$}
		(3,-0.5) node {\size $1$}
		(3.75,0.45) node {\size $1$}
		
		(4+\k,1.25) node {\size $1$}
		;
		
		\draw 
		(2,0) node {$\dots$}
		;
		\end{scope}
		\end{scope}

		\begin{scope}[yshift=-11.5cm]
		\draw[thick] (-1,0) -- (1.5,0) 
		(2.5,0) -- (4,0)
		(-1,0) -- (-1,1) 
		(4,0) -- (4,1)
		
		;
		
		\draw[thick, fill=white] 
		(-1,0) circle (\q pt)
		(0,0) circle (\q pt)
		(1,0) circle (\q pt)
		(3,0) circle (\q pt)
		(4,0) circle (\q pt)
		;
		
		\draw[thick, fill=white] 
		(-1 -\q/30,1 -\q/30) rectangle (-1 +\q/30,1 +\q/30)
		(4 -\q/30,1 -\q/30) rectangle (4 +\q/30,1 +\q/30)
		;

		\draw 
		(-1,1.5) node {\size $1$}
		
		(4,1.5) node  {\size $1$}
		
		(-1,-0.5) node {\size $1$}
		(0,-0.5) node {\size $1$}
		(1,-0.5) node {\size $1$}
		(3,-0.5) node {\size $1$}
		(4,-0.5) node {\size $1$}
		
		;
		
		\draw 
		(2,0) node {$\dots$}
		;

		\draw 
		(-4,0) node {$\Q_{\Hg}(A_k) ~~~~ =$}
		;

		\end{scope}
		
		\end{tikzpicture}
	\end{center}
\caption{The $3d$ $\mathcal{N}=4$ quivers which realise $D_k$, $A_k \cup A_k$ and $A_k$ singularities as their Higgs branches. Each quiver has $k$ nodes in each case. The $A_k$ is familiar from linear quivers. The $D_k$ case has been discussed before, for example \cite{kalveks2}. The $A_k \cup A_k$ singularity is a little more complicated, see discussion.}
\label{sliceQ}
\end{figure}
The top and bottom quivers are familiar, however the middle one is not. In the context of $\kk{so}_{2n}$ nilpotent varieties, the $A_k \cup A_k$ singularity is associated to very even partitions and hence to the subtlety regarding the nilpotent orbits for such partitions discussed previously. However in the broader context of $D_n$ quivers the $A_k\cup A_k$ singularity is associated with the choice that exists due to the equivalence of the two end nodes. Simplistically, swapping the two end nodes doesn't change the field theory for a $D_n$ Dynkin quiver, however in the more concrete diagrammatic context of quiver arithmetic, only one at a time will be realised. There is therefore an important implicit assumption in the remaining discussion: When the flavour content and/or gauge group of the two end nodes is different, there are implicitly \textit{two} $D_n$ Dynkin quivers available, which represent the same field theory. These quivers, when they are both drawn, are labelled $I$ and $II$, following the very even nilpotent orbit naming convention. The $A_k \cup A_k$ singularity exists in two scenarios, firstly when the end nodes differ and one of the flavours on an end node is 1\footnote{And this makes an $A_k$ singularity with a flavour node on the tail.}. The implicit choice that exists due to the difference in end nodes allows the observation that the true singularity is $A_k \cup A_k$, not just $A_k$. Alternatively, if \textit{both} end node flavours are 1 this also corresponds to an $A_k \cup A_k$ singularity as either could be considered as forming an $A_k$ singularity with flavour in the tail of the quiver.

Note that there are $B$ and $C$ type Dynkin quivers which realise as \textit{Coulomb} branches two other minimal degenerations $b_n$ and $c_n$ that appear in $BCD$-type orbit Hasse diagrams. However both involve non-simply laced edges and so do not map onto the topology of the $D_n$ Dynkin quiver. The quiver arithmetic for building nilpotent orbits using Coulomb branches fails when one tries to include these singularities. Since the non-special nodes for the $D_n$ always appear in the Hasse diagram as the end of an edge labelled with one of these singularities, this failure of quiver arithmetic is exactly the restriction that non-special nodes can't be included in the moduli space branch Hasse diagrams. Hence the diagrams are limited to ht$(\O_{\rho})\leq 2$ nilpotent orbits and their subvarieties for Coulomb branches, and to the maximal special slice and its subvarieties for Higgs branches. 

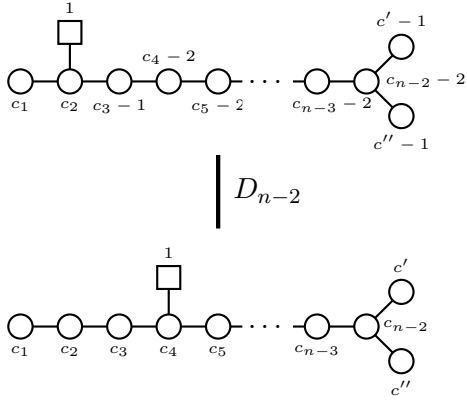
\begin{figure}
	\begin{center}
		\begin{tikzpicture}[scale = 0.65]
		\def\q{7}
		\def\k{0.7071}
		\def\size{\tiny}
		
		\draw[thick] (-3,0) -- (4,0) -- (4.7,0.7) (4,0) -- (4.7,-0.7) 
		(-2,0) -- (-2,1)
		
		;
		
		\draw[thick, fill=white] 
		(-3,0) circle (\q pt)
		(-2,0) circle (\q pt)
		(-1,0) circle (\q pt)
		(0,0) circle (\q pt)
		(1,0) circle (\q pt)
		(2,0) circle (\q pt)
		(3,0) circle (\q pt)
		(4,0) circle (\q pt)
		(4+\k,\k) circle (\q pt)
		(4+\k,-\k) circle (\q pt)
		;
		
		\draw[thick, fill=white] 
		(-2 -\q/30,1 -\q/30) rectangle (-2 +\q/30,1 +\q/30)
		;

		\draw 
		(-2,1.5) node {\size $1$}
		
		

		(-3,-0.5) node {\size $c_1$}
		(-2,-0.5) node {\size $c_2$}
		(-1,-0.5) node {\size $c_3-1$}
		(0,0.5) node {\size $c_4-2$}
		(1,-0.5) node {\size $c_5-2$}
		(3.3,-0.5) node {\size $c_{n-3}-2$}
		(5.2,0) node {\size $c_{n-2}-2$}
		
		(4+\k,1.25) node {\size $c'-1$}
		(4+\k,-1.25) node {\size $c''-1$}
		;
		
		\draw[white,fill=white] (1.5,-1.5) rectangle (2.5,2);
		
		\draw (2,0) node {$\dots$};
		
		\draw[\THICC] (1,-1.5) -- (1,-3);
		
		\draw (2,-2.25) node {$D_{n-2}$};
		
		\begin{scope}[yshift = -5cm]
		\draw[thick] (-3,0) -- (4,0) -- (4.7,0.7) (4,0) -- (4.7,-0.7) 
		(0,0) -- (0,1)
		
		;
		
		\draw[thick, fill=white] 
		(-3,0) circle (\q pt)
		(-2,0) circle (\q pt)
		(-1,0) circle (\q pt)
		(0,0) circle (\q pt)
		(1,0) circle (\q pt)
		(2,0) circle (\q pt)
		(3,0) circle (\q pt)
		(4,0) circle (\q pt)
		(4+\k,\k) circle (\q pt)
		(4+\k,-\k) circle (\q pt)
		;
		
		\draw[thick, fill=white] 
		(0 -\q/30,1 -\q/30) rectangle (0 +\q/30,1 +\q/30)
		;

		\draw 
		
		(0,1.5) node {\size $1$}
		

		(-3,-0.5) node {\size $c_1$}
		(-2,-0.5) node {\size $c_2$}
		(-1,-0.5) node {\size $c_3$}
		(0,-0.5) node {\size $c_4$}
		(1,-0.5) node {\size $c_5$}
		(3,-0.5) node {\size $c_{n-3}$}
		(4.8,0) node {\size $c_{n-2}$}
		
		(4+\k,1.25) node {\size $c'$}
		(4+\k,-1.25) node {\size $c''$}
		;
		
		\draw[white,fill=white] (1.5,-1.5) rectangle (2.5,2);
		
		\draw (2,0) node {$\dots$};

		\end{scope}
		
		\end{tikzpicture}
	\end{center}
\caption{Quiver addition of a $D_{n-2}$ singularity to the Higgs branch of a $D_n$ Dynkin quiver. As per the discussion, the theory with the larger Higgs branch is placed lower and quiver subtraction works from the bottom, up. Reinterpreting the diagram as the \textit{removal} of the singularity via the Higgs mechanism, the behaviour of the flavour becomes clear.}  
\label{removalD}
\end{figure}

\begin{figure}
	\begin{center}
		\begin{tikzpicture}[scale = 0.65]
		\def\q{7}
		\def\k{0.7071}
		\def\size{\tiny}
		\begin{scope}[yshift = 0cm]
		\draw[thick] (-3,0) -- (4,0) -- (4.7,0.7) (4,0) -- (4.7,-0.7) 
		(-2,0) -- (-2,1)
		
		(4+\k,\k) -- (5+\k,\k)
		;
		
		\draw[thick, fill=white] 
		(-3,0) circle (\q pt)
		(-2,0) circle (\q pt)
		(-1,0) circle (\q pt)
		(0,0) circle (\q pt)
		(1,0) circle (\q pt)
		(2,0) circle (\q pt)
		(3,0) circle (\q pt)
		(4,0) circle (\q pt)
		(4+\k,\k) circle (\q pt)
		(4+\k,-\k) circle (\q pt)
		;
		
		\draw[thick, fill=white] 
		(-2 -\q/30,1 -\q/30) rectangle (-2 +\q/30,1 +\q/30)
		(5+\k -\q/30,\k -\q/30) rectangle (5+\k +\q/30,\k +\q/30)
		;

		\draw 
		(-2,1.5) node {\size $1$}
		
		
		(5.6 +\k,0+\k) node  {\size $2$}

		(-3,-0.5) node {\size $c_1$}
		(-2,-0.5) node {\size $c_2$}
		(-1,-0.5) node {\size $c_3-1$}
		(0,0.5) node {\size $c_4-1$}
		(1,-0.5) node {\size $c_5-1$}
		(3.3,-0.5) node {\size $c_{n-3}-1$}
		(5.2,0) node {\size $c_{n-2}-1$}
		
		(4+\k,1.25) node {\size $c'$}
		(4+\k,-1.25) node {\size $c''-1$}
		;
		
		\draw[white,fill=white] (1.5,-1.5) rectangle (2.5,2);
		
		\draw (2,0) node {$\dots$}
			(-5,0) node {$I$};

		\end{scope}
		
		\begin{scope}[yshift = -3.3cm]
		\draw[thick] (-3,0) -- (4,0) -- (4.7,0.7) (4,0) -- (4.7,-0.7) 
		(-2,0) -- (-2,1)
		
		(4+\k,-\k) -- (5+\k,-\k)
		;
		
		\draw[thick, fill=white] 
		(-3,0) circle (\q pt)
		(-2,0) circle (\q pt)
		(-1,0) circle (\q pt)
		(0,0) circle (\q pt)
		(1,0) circle (\q pt)
		(2,0) circle (\q pt)
		(3,0) circle (\q pt)
		(4,0) circle (\q pt)
		(4+\k,\k) circle (\q pt)
		(4+\k,-\k) circle (\q pt)
		;
		
		\draw[thick, fill=white] 
		(-2 -\q/30,1 -\q/30) rectangle (-2 +\q/30,1 +\q/30)
		(5+\k -\q/30,-\k -\q/30) rectangle (5+\k +\q/30,-\k +\q/30)
		;

		\draw 
		(-2,1.5) node {\size $1$}
		
		
		(5.6+\k,0-\k) node  {\size $2$}

		(-3,-0.5) node {\size $c_1$}
		(-2,-0.5) node {\size $c_2$}
		(-1,-0.5) node {\size $c_3-1$}
		(0,0.5) node {\size $c_4-1$}
		(1,-0.5) node {\size $c_5-1$}
		(3.3,-0.5) node {\size $c_{n-3}-1$}
		(5.2,0) node {\size $c_{n-2}-1$}
		
		(4+\k,1.25) node {\size $c'-1$}
		(4+\k,-1.25) node {\size $c''$}
		;
		
		\draw[white,fill=white] (1.5,-1.5) rectangle (2.5,2);
		
		\draw (2,0) node {$\dots$}
		(-5,0) node {$II$};

		\end{scope}
		
		\draw[\THICC] (1,-5) -- (1,-6.5);
		
		\draw (2.75,-5.75) node {$A_{n-3} \cup A_{n-3}$};

		\begin{scope}[yshift = -8.5cm]
		\draw[thick] (-3,0) -- (4,0) -- (4.7,0.7) (4,0) -- (4.7,-0.7) 
		(-1,0) -- (-1,1)
		
		(4+\k,\k) -- (5+\k,\k)
		(4+\k,-\k) -- (5+\k,-\k)
		;
		
		\draw[thick, fill=white] 
		(-3,0) circle (\q pt)
		(-2,0) circle (\q pt)
		(-1,0) circle (\q pt)
		(0,0) circle (\q pt)
		(1,0) circle (\q pt)
		(2,0) circle (\q pt)
		(3,0) circle (\q pt)
		(4,0) circle (\q pt)
		(4+\k,\k) circle (\q pt)
		(4+\k,-\k) circle (\q pt)
		;
		
		\draw[thick, fill=white] 
		(-1 -\q/30,1 -\q/30) rectangle (-1 +\q/30,1 +\q/30)
		(5+\k -\q/30,\k -\q/30) rectangle (5+\k +\q/30,\k +\q/30)
		(5+\k -\q/30,-\k -\q/30) rectangle (5+\k +\q/30,-\k +\q/30)
		;

		\draw 
		(-1,1.5) node {\size $1$}
		
		
		(5.6 +\k,0+\k) node  {\size $1$}
		(5.6+\k,0-\k) node  {\size $1$}

		(-3,-0.5) node {\size $c_1$}
		(-2,-0.5) node {\size $c_2$}
		(-1,-0.5) node {\size $c_3$}
		(0,-0.5) node {\size $c_4$}
		(1,-0.5) node {\size $c_5$}
		(3.3,-0.5) node {\size $c_{n-3}$}
		(5,0) node {\size $c_{n-2}$}
		
		(4+\k,1.25) node {\size $c'$}
		(4+\k,-1.25) node {\size $c''$}
		;
		
		\draw[white,fill=white] (1.5,-1.5) rectangle (2.5,2);
		
		\draw (2,0) node {$\dots$};

		\end{scope}
		
		\end{tikzpicture}
	\end{center}
\caption{Quiver addition of one case of the $A_{n-3} \cup A_{n-3}$ singularity to the Higgs branch of a $D_n$ Dynkin quiver. Reinterpreting the diagram as the \textit{removal} of the singularity via the Higgs mechanism (working up), the behaviour of the flavour is clear. The $A_k \cup A_k$ transition for uneven initial end flavour is exactly analogous, only the bottom quiver here has uneven end flavour (with one still 1) and so also comes with two versions, as per the discussion.}
\label{removalAA}
\end{figure}
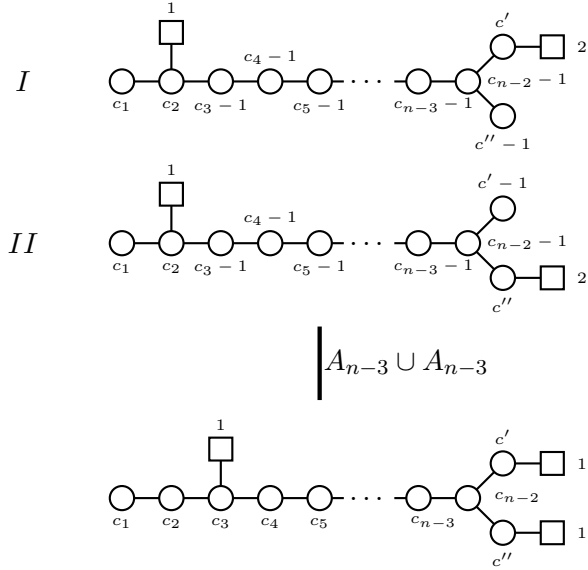

During quiver addition, the slice added to a given quiver must be such that its removal from the resulting quiver would give back the flavour arrangement of the quiver being added to. It is important to establish what happens to the flavour arrangement on the quivers when the slices are removed. Figures \ref{removalD} and \ref{removalAA} give two such examples.

The conclusion is that the $D_k$ and $A_{k}$ slices `push' a single flavour onto the first gauge node(s) attached to them which are not part of the slice. $A_{2k-1} \cup A_{2k-1}$ is a little more complicated, the transition still pushes flavour onto the first gauge nodes attached but not involved, however every time one of these slices is removed there is an option as to which leg of the Dynkin diagram is chosen. For example if a flavour node 1 is attached to both, the flavour on the one chosen gets pushed out of the diagram while the flavour on the one that wasn't chosen remains and is enhanced. 

\subsubsection{$D_4$ quivers and nilpotent varieties of $\kk{so}_8$}
$D_4$ Dynkin quivers will be used several times as examples throughout. Here the quivers which realise $\kk{so}_8$ nilpotent varieties are studied in detail.

The special slice Hasse diagram for $\kk{so}_8$ is:
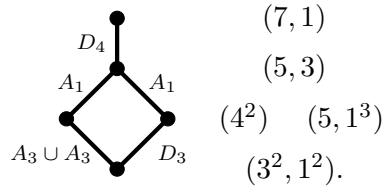
\begin{center}
	\begin{tikzpicture}[scale=2/3]
	
	\draw[\THICC] (0,7) -- (0,6) -- (1,5) -- (0,4) -- (-1,5) -- (0,6);
	
	\draw[fill=black] (0,4) circle (4pt);
	\draw[fill=black] (0,6) circle (4pt);
	\draw[fill=black] (0,7) circle (4pt);
	
	\draw[fill=black] (-1,5) circle (4pt);
	\draw[fill=black] (1,5) circle (4pt);
	
	\draw (-0.5,6.5) node {\sz$D_4$};
	
	\draw (-1.3,4.3) node {\sz$A_3 \cup A_3$};
	\draw (1.1,4.3) node {\sz$D_3$};
	
	\draw (-0.9,5.7) node {\sz$A_1$};
	\draw (0.9,5.7) node {\sz$A_1$};
	
	\begin{scope}[xshift = 3.5cm]
	\draw[fill=black] (0,4) node {$(3^2,1^2).$};
	\draw[fill=black] (0,6) node {$(5,3)$};
	\draw[fill=black] (0,7) node {$(7,1)$};
	
	\draw[fill=black] (-1,5) node {$(4^2)$};
	\draw[fill=black] (1,5) node {$(5,1^3)$};
	
	\end{scope}
	
	\end{tikzpicture}
\end{center}

\begin{figure}
	\begin{center}
		\begin{tikzpicture}[scale = 0.65]
		\def\q{7}
		\def\k{0.7071}
		\def\size{\tiny}
		
		\draw[thick] (3,0) -- (4,0) -- (4.7,0.7) (4,0) -- (4.7,-0.7) 
		
		;
		
		\draw[thick, fill=white] 
		(3,0) circle (\q pt)
		(4,0) circle (\q pt)
		(4+\k,\k) circle (\q pt)
		(4+\k,-\k) circle (\q pt)
		;
		
		\draw[thick, fill=white] 
		;

		\draw 
		

		(3,-0.5) node {\size $0$}
		(3.75,0.45) node {\size $0$}
		
		(4+\k,1.25) node {\size $0$}
		(4+\k,-1.25) node {\size $0$}
		;
		
		\draw[\THICC]
		(4,-1.5) -- (4,-2.5)
		(2.2,-4.3) -- (0.8,-5.2)
		(5.5,-5) -- (6.5,-6.5)
		(6.5,-9) -- (5.5,-10.5)
		(0.8,-10.7) -- (2.2,-11.8)
		
		;
		
		\draw
		(3.3,-2) node {$D_4$}
		(1.2,-4.3) node {$A_1$}
		(6.6,-5.5) node {$A_1$}
		(6.6,-10) node {$D_3$}
		
		(0.5,-11.6) node {$A_3 \cup A_3$}
		;

		\begin{scope}[yshift = -4cm]
		\draw[thick] (3,0) -- (4,0) -- (4.7,0.7) (4,0) -- (4.7,-0.7) 
		(4,0) -- (5,0)
		
		;
		
		\draw[thick, fill=white] 
		(3,0) circle (\q pt)
		(4,0) circle (\q pt)
		(4+\k,\k) circle (\q pt)
		(4+\k,-\k) circle (\q pt)
		;
		
		\draw[thick, fill=white] 
		(5 -\q/30,0 -\q/30) rectangle (5 +\q/30,0 +\q/30)
		;

		\draw 
		
		(5.6,0) node  {\size $1$}

		(3,-0.5) node {\size $1$}
		(3.75,0.45) node {\size $2$}
		
		(4+\k,1.25) node {\size $1$}
		(4+\k,-1.25) node {\size $1$}
		;
		\end{scope}

		\begin{scope}[yshift = -8.2cm, xshift = 4cm]
		\draw[thick] (3,0) -- (4,0) -- (4.7,0.7) (4,0) -- (4.7,-0.7) 
		(3,0) -- (3,1)
		
		;
		
		\draw[thick, fill=white] 
		(3,0) circle (\q pt)
		(4,0) circle (\q pt)
		(4+\k,\k) circle (\q pt)
		(4+\k,-\k) circle (\q pt)
		;
		
		\draw[thick, fill=white] 
		(3 -\q/30,1 -\q/30) rectangle (3 +\q/30,1 +\q/30)
		;

		\draw 
		(3,1.5) node {\size $2$}
		

		(3,-0.5) node {\size $2$}
		(3.75,0.45) node {\size $2$}
		
		(4+\k,1.25) node {\size $1$}
		(4+\k,-1.25) node {\size $1$}
		;
		\end{scope}

		\begin{scope}[yshift = -6.5cm,xshift = -5.5cm]
		\draw[thick] (3,0) -- (4,0) -- (4.7,0.7) (4,0) -- (4.7,-0.7) 
		
		(4+\k,\k) -- (5+\k,\k)
		;
		
		\draw[thick, fill=white] 
		(3,0) circle (\q pt)
		(4,0) circle (\q pt)
		(4+\k,\k) circle (\q pt)
		(4+\k,-\k) circle (\q pt)
		;
		
		\draw[thick, fill=white] 
		(5+\k -\q/30,\k -\q/30) rectangle (5+\k +\q/30,\k +\q/30)
		;

		\draw 
		
		(5.6 +\k,0+\k) node  {\size $2$}

		(3,-0.5) node {\size $1$}
		(3.75,0.45) node {\size $2$}
		
		(4+\k,1.25) node {\size $2$}
		(4+\k,-1.25) node {\size $1$}
		(1.5,0) node {$I$}
		;
		\end{scope}

		\begin{scope}[yshift = -9.5cm,xshift = -5.5cm]
		\draw[thick] (3,0) -- (4,0) -- (4.7,0.7) (4,0) -- (4.7,-0.7) 
		
		(4+\k,-\k) -- (5+\k,-\k)
		;
		
		\draw[thick, fill=white] 
		(3,0) circle (\q pt)
		(4,0) circle (\q pt)
		(4+\k,\k) circle (\q pt)
		(4+\k,-\k) circle (\q pt)
		;
		
		\draw[thick, fill=white] 
		(5+\k -\q/30,-\k -\q/30) rectangle (5+\k +\q/30,-\k +\q/30)
		;

		\draw 
		
		(5.6+\k,0-\k) node  {\size $2$}

		(3,-0.5) node {\size $1$}
		(3.75,0.45) node {\size $2$}
		
		(4+\k,1.25) node {\size $1$}
		(4+\k,-1.25) node {\size $2$}
		(1.5,0) node {$II$}
		;
		\end{scope}

		\begin{scope}[yshift = -12.2cm,xshift = 0.2cm]
		\draw[thick] (3,0) -- (4,0) -- (4.7,0.7) (4,0) -- (4.7,-0.7) 
		(3,0) -- (3,1)
		
		(4+\k,\k) -- (5+\k,\k)
		(4+\k,-\k) -- (5+\k,-\k)
		;
		
		\draw[thick, fill=white] 
		(3,0) circle (\q pt)
		(4,0) circle (\q pt)
		(4+\k,\k) circle (\q pt)
		(4+\k,-\k) circle (\q pt)
		;
		
		\draw[thick, fill=white] 
		(3 -\q/30,1 -\q/30) rectangle (3 +\q/30,1 +\q/30)
		(5+\k -\q/30,\k -\q/30) rectangle (5+\k +\q/30,\k +\q/30)
		(5+\k -\q/30,-\k -\q/30) rectangle (5+\k +\q/30,-\k +\q/30)
		;

		\draw 
		(3,1.5) node {\size $1$}
		
		(5.6 +\k,0+\k) node  {\size $1$}
		(5.6+\k,0-\k) node  {\size $1$}

		(3,-0.5) node {\size $2$}
		(3.75,0.45) node {\size $3$}
		
		(4+\k,1.25) node {\size $2$}
		(4+\k,-1.25) node {\size $2$}
		;
		\end{scope}

		\end{tikzpicture}
	\end{center}
	\caption{The Hasse diagram for the maximal special slice in $\kk{so}_8$ filled in with the $3d$ $\mathcal{N}=4$ quivers which realise the Slodowy slices corresponding to each node. The singularity structure of the Higgs branches of the quivers is therefore exactly all of the structure which dominates the node at which the quiver lives.}
	\label{D4quivers}
\end{figure}
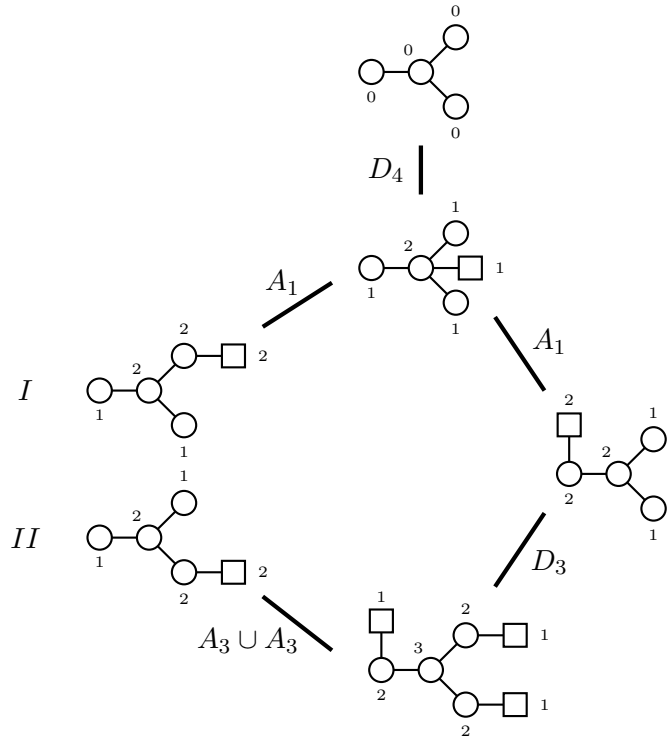

Beginning with a bare $D_4$ Dynkin quiver, the only slice which can push \textit{all} the flavour out of a $D_4$ quiver is a $D_4$ transition so this is the topmost singularity in figure \ref{D4quivers}. There are now three prior quivers which might have given this flavour arrangement by subtraction. At first they might all seem equivalent, however they can be distinguished by considering what may be added to them in subsequent steps. The full sequence is given in figure \ref{D4quivers}.
Quiver subtraction can now be used in order to determine the quivers for any subvariety of the maximal special slice in $\kk{so}_8$. An example of determining one such slice, namely the quiver with the Higgs branch $\mathcal{S}_{(3^2,1^2)} \cap \O_{(5,3)}$, is
\begin{center}
	\begin{tikzpicture}[scale= 0.65]
	\def\q{7}
	\def\k{0.7071}
	\def\size{\tiny}
	\draw (0.2,0) node {${\sf{Q}}_{\mathcal{H}}(\mathcal{S}_{(3^2,1^2)} \cap \O_{(5,3)}) = {\sf{Q}}_{\mathcal{H}}(\mathcal{S}_{(3^2,1^2)}) - {\sf{Q}}_{\mathcal{H}}(\mathcal{S}_{(5,3)})$};
	
	\draw (4.5,-2.5) node {$-$};
	\draw (-0.43,-2.5) node {$=$};
	\draw (-0.43,-6) node {$=$};
	
	\begin{scope}[yshift = -2.5cm, xshift = -2.5cm]
	\draw[thick] (3,0) -- (4,0) -- (4.7,0.7) (4,0) -- (4.7,-0.7) 
	(3,0) -- (3,1)
	
	(4+\k,\k) -- (5+\k,\k)
	(4+\k,-\k) -- (5+\k,-\k)
	;
	
	\draw[thick, fill=white] 
	(3,0) circle (\q pt)
	(4,0) circle (\q pt)
	(4+\k,\k) circle (\q pt)
	(4+\k,-\k) circle (\q pt)
	;
	
	\draw[thick, fill=white] 
	(3 -\q/30,1 -\q/30) rectangle (3 +\q/30,1 +\q/30)
	(5+\k -\q/30,\k -\q/30) rectangle (5+\k +\q/30,\k +\q/30)
	(5+\k -\q/30,-\k -\q/30) rectangle (5+\k +\q/30,-\k +\q/30)
	;

	\draw 
	(3,1.5) node {\size $1$}
	
	(5.6 +\k,0+\k) node  {\size $1$}
	(5.6+\k,0-\k) node  {\size $1$}

	(3,-0.5) node {\size $2$}
	(3.75,0.45) node {\size $3$}
	
	(4+\k,1.25) node {\size $2$}
	(4+\k,-1.25) node {\size $2$}
	;
	\end{scope}

	\begin{scope}[yshift = -2.5cm, xshift = 2.5cm]
	\draw[thick] (3,0) -- (4,0) -- (4.7,0.7) (4,0) -- (4.7,-0.7) 
	(4,0) -- (5,0)
	
	;
	
	\draw[thick, fill=white] 
	(3,0) circle (\q pt)
	(4,0) circle (\q pt)
	(4+\k,\k) circle (\q pt)
	(4+\k,-\k) circle (\q pt)
	;
	
	\draw[thick, fill=white] 
	(5 -\q/30,0 -\q/30) rectangle (5 +\q/30,0 +\q/30)
	;

	\draw 
	
	(5.6,0) node  {\size $1$}

	(3,-0.5) node {\size $1$}
	(3.75,0.45) node {\size $2$}
	
	(4+\k,1.25) node {\size $1$}
	(4+\k,-1.25) node {\size $1$}
	;
	\end{scope}

	\begin{scope}[yshift = -6cm, xshift = -2.5cm]
	\draw[thick] (3,0) -- (4,0) -- (4.7,0.7) (4,0) -- (4.7,-0.7) 
	(3,0) -- (3,1)
	
	(4+\k,\k) -- (5+\k,\k)
	(4+\k,-\k) -- (5+\k,-\k)
	;
	
	\draw[thick, fill=white] 
	(3,0) circle (\q pt)
	(4,0) circle (\q pt)
	(4+\k,\k) circle (\q pt)
	(4+\k,-\k) circle (\q pt)
	;
	
	\draw[thick, fill=white] 
	(3 -\q/30,1 -\q/30) rectangle (3 +\q/30,1 +\q/30)
	(5+\k -\q/30,\k -\q/30) rectangle (5+\k +\q/30,\k +\q/30)
	(5+\k -\q/30,-\k -\q/30) rectangle (5+\k +\q/30,-\k +\q/30)
	;

	\draw 
	(3,1.5) node {\size $1$}
	
	(5.6 +\k,0+\k) node  {\size $1$}
	(5.6+\k,0-\k) node  {\size $1$}

	(3,-0.5) node {\size $1$}
	(3.75,0.45) node {\size $1$}
	
	(4+\k,1.25) node {\size $1$}
	(4+\k,-1.25) node {\size $1$}
	;
	\end{scope}
	
	\end{tikzpicture}
\end{center}

The full table of such slices, analogous to the tables given in \cite{Rogers}, is given in figure \ref{D4table}. 
\begin{figure}
	\begin{center}
		\begin{tikzpicture}[scale = 0.46]
		\def\q{7}
		\def\k{0.7071}
		\def\size{\tiny}
		
		\begin{scope}[scale=1.4]
		\draw[thick] (1.5,3.1) -- (1.5,2.1) -- (0.5,2.1)  (1.5,3.1) -- (1.5,3.1) (1.5,2.1) -- (0.5,3.1) (8.5,-4) -- (12.1,-1.2) (12.1,-4) -- (8.5,-1.2) (8,-4) -- (14.5,-4) -- (14.5,3.1) (1.5,-8.6) -- (1.5,-12.2) (5.3,-8.6) -- (5.3,-12.2) -- (0.5,-12.2) -- (0.5,-8.6) -- (12.1,-8.6) -- (12.1,3.1) -- (12.1,-4) -- (0.5,-4) -- (5.3,-4) -- (5.3,-11.2) -- (0.5,-11.2) -- (0.5,3.1) -- (12,3.1) -- (5.3,3.1) -- (5.3,-8.6) -- (0.5,-8.6) -- (0.5,-1.2) -- (16,-1.2) -- (16,3.1) -- (8.5,3.1) -- (8.5,-11.2) -- (1.5,-11.2) -- (1.5,2.1) -- (16,2.1);
		
		\draw (1.2,2.8) node {$\mu$};
		\draw (0.8,2.4) node {$\nu$};
		
		\draw (1,0.45) node[rotate = 90] {\scriptsize$(3^2,1^2)$};
		\draw (1,-2.65) node[rotate = 90] {\scriptsize$(5,1^3)$};
		\draw (1,-6.3) node[rotate = 90] {\scriptsize$(4^2)$};
		\draw (1,-9.8) node[rotate = 90] {\scriptsize$(5,3)$};
		\draw (1,-11.7) node[rotate = 90] {\scriptsize$(7,1)$};
		
		
		\draw (3.4,2.6) node {\scriptsize$(7,1)$};
		\draw (6.9,2.6) node {\scriptsize$(5,3)$};
		\draw (10.3,2.6) node {\scriptsize$(4^2)$};
		\draw (13.3,2.6) node {\scriptsize$(5,1^3)$};
		\draw (15.25,2.6) node {\scriptsize$(3^2,1^2)$};
		\end{scope}

		\begin{scope}[yshift = 0.6cm, xshift = 0.3cm]
		\draw[thick] (3,0) -- (4,0) -- (4.7,0.7) (4,0) -- (4.7,-0.7) 
		(3,0) -- (3,1)
		
		(4+\k,\k) -- (5+\k,\k)
		(4+\k,-\k) -- (5+\k,-\k)
		;
		
		\draw[thick, fill=white] 
		(3,0) circle (\q pt)
		(4,0) circle (\q pt)
		(4+\k,\k) circle (\q pt)
		(4+\k,-\k) circle (\q pt)
		;
		
		\draw[thick, fill=white] 
		(3 -\q/30,1 -\q/30) rectangle (3 +\q/30,1 +\q/30)
		(5+\k -\q/30,\k -\q/30) rectangle (5+\k +\q/30,\k +\q/30)
		(5+\k -\q/30,-\k -\q/30) rectangle (5+\k +\q/30,-\k +\q/30)
		;

		\draw 
		(3,1.5) node {\size $1$}
		
		(5.6 +\k,0+\k) node  {\size $1$}
		(5.6+\k,0-\k) node  {\size $1$}

		(3,-0.5) node {\size $2$}
		(3.75,0.45) node {\size $3$}
		
		(4+\k,1.25) node {\size $2$}
		(4+\k,-1.25) node {\size $2$}
		;
		\end{scope}

		\begin{scope}[yshift = 0.6cm, xshift = 5.2cm]
		\draw[thick] (3,0) -- (4,0) -- (4.7,0.7) (4,0) -- (4.7,-0.7) 
		(3,0) -- (3,1)
		
		(4+\k,\k) -- (5+\k,\k)
		(4+\k,-\k) -- (5+\k,-\k)
		;
		
		\draw[thick, fill=white] 
		(3,0) circle (\q pt)
		(4,0) circle (\q pt)
		(4+\k,\k) circle (\q pt)
		(4+\k,-\k) circle (\q pt)
		;
		
		\draw[thick, fill=white] 
		(3 -\q/30,1 -\q/30) rectangle (3 +\q/30,1 +\q/30)
		(5+\k -\q/30,\k -\q/30) rectangle (5+\k +\q/30,\k +\q/30)
		(5+\k -\q/30,-\k -\q/30) rectangle (5+\k +\q/30,-\k +\q/30)
		;

		\draw 
		(3,1.5) node {\size $1$}
		
		(5.6 +\k,0+\k) node  {\size $1$}
		(5.6+\k,0-\k) node  {\size $1$}

		(3,-0.5) node {\size $1$}
		(3.75,0.45) node {\size $1$}
		
		(4+\k,1.25) node {\size $1$}
		(4+\k,-1.25) node {\size $1$}
		;
		\end{scope}

		\begin{scope}[yshift = -3.8cm, xshift = 0.3cm]
		\draw[thick] (3,0) -- (4,0) -- (4.7,0.7) (4,0) -- (4.7,-0.7) 
		(3,0) -- (3,1)
		
		;
		
		\draw[thick, fill=white] 
		(3,0) circle (\q pt)
		(4,0) circle (\q pt)
		(4+\k,\k) circle (\q pt)
		(4+\k,-\k) circle (\q pt)
		;
		
		\draw[thick, fill=white] 
		(3 -\q/30,1 -\q/30) rectangle (3 +\q/30,1 +\q/30)
		;

		\draw 
		(3,1.5) node {\size $2$}
		

		(3,-0.5) node {\size $2$}
		(3.75,0.45) node {\size $2$}
		
		(4+\k,1.25) node {\size $1$}
		(4+\k,-1.25) node {\size $1$}
		;
		\end{scope}

		\begin{scope}[yshift = -7.3cm,xshift = 0.3cm]
		\draw[thick] (3,0) -- (4,0) -- (4.7,0.7) (4,0) -- (4.7,-0.7) 
		
		(4+\k,\k) -- (5+\k,\k)
		;
		
		\draw[thick, fill=white] 
		(3,0) circle (\q pt)
		(4,0) circle (\q pt)
		(4+\k,\k) circle (\q pt)
		(4+\k,-\k) circle (\q pt)
		;
		
		\draw[thick, fill=white] 
		(5+\k -\q/30,\k -\q/30) rectangle (5+\k +\q/30,\k +\q/30)
		;

		\draw 
		
		(5.6 +\k,0+\k) node  {\size $2$}

		(3,-0.5) node {\size $1$}
		(3.75,0.45) node {\size $2$}
		
		(4+\k,1.25) node {\size $2$}
		(4+\k,-1.25) node {\size $1$}
		;
		\end{scope}

		\begin{scope}[yshift = -10.3cm,xshift = 0.3cm]
		\draw[thick] (3,0) -- (4,0) -- (4.7,0.7) (4,0) -- (4.7,-0.7) 
		
		(4+\k,-\k) -- (5+\k,-\k)
		;
		
		\draw[thick, fill=white] 
		(3,0) circle (\q pt)
		(4,0) circle (\q pt)
		(4+\k,\k) circle (\q pt)
		(4+\k,-\k) circle (\q pt)
		;
		
		\draw[thick, fill=white] 
		(5+\k -\q/30,-\k -\q/30) rectangle (5+\k +\q/30,-\k +\q/30)
		;

		\draw 
		
		(5.6+\k,0-\k) node  {\size $2$}

		(3,-0.5) node {\size $1$}
		(3.75,0.45) node {\size $2$}
		
		(4+\k,1.25) node {\size $1$}
		(4+\k,-1.25) node {\size $2$}
		;
		\end{scope}

		\begin{scope}[yshift = -13.8cm,xshift = 0.3cm]
		\draw[thick] (3,0) -- (4,0) -- (4.7,0.7) (4,0) -- (4.7,-0.7) 
		(4,0) -- (5,0)
		
		;
		
		\draw[thick, fill=white] 
		(3,0) circle (\q pt)
		(4,0) circle (\q pt)
		(4+\k,\k) circle (\q pt)
		(4+\k,-\k) circle (\q pt)
		;
		
		\draw[thick, fill=white] 
		(5 -\q/30,0 -\q/30) rectangle (5 +\q/30,0 +\q/30)
		;

		\draw 
		
		(5.6,0) node  {\size $1$}

		(3,-0.5) node {\size $1$}
		(3.75,0.45) node {\size $2$}
		
		(4+\k,1.25) node {\size $1$}
		(4+\k,-1.25) node {\size $1$}
		;
		\end{scope}

		\begin{scope}[yshift = 8.35cm, xshift = 9.9cm]

		\begin{scope}[yshift = -7.3cm,xshift = 0.3cm]
		\draw[thick] (3,0) -- (4,0) -- (4.7,0.7) 
		(3,0) -- (3,1)
		
		(4+\k,\k) -- (5+\k,\k)
		;
		
		\draw[thick, fill=white] 
		(3,0) circle (\q pt)
		(4,0) circle (\q pt)
		(4+\k,\k) circle (\q pt)
		;
		
		\draw[thick, fill=white] 
		(3 -\q/30,1 -\q/30) rectangle (3 +\q/30,1 +\q/30)
		(5+\k -\q/30,\k -\q/30) rectangle (5+\k +\q/30,\k +\q/30)
		;

		\draw 
		(3,1.5) node {\size $1$}
		
		(5.6 +\k,0+\k) node  {\size $1$}

		(2.5,0) node {\size $1$}
		(3.75,0.45) node {\size $1$}
		
		(4+\k,1.25) node {\size $1$}
		;
		\end{scope}

		\begin{scope}[yshift = -8.2cm,xshift = 0.3cm]
		\draw[thick] (3,0) -- (4,0)  (4,0) -- (4.7,-0.7) 
		(3,0) -- (3,-1)
		
		(4+\k,-\k) -- (5+\k,-\k)
		;
		
		\draw[thick, fill=white] 
		(3,0) circle (\q pt)
		(4,0) circle (\q pt)
		(4+\k,-\k) circle (\q pt)
		;
		
		\draw[thick, fill=white] 
		(3 -\q/30,-1 -\q/30) rectangle (3 +\q/30,-1 +\q/30)
		(5+\k -\q/30,-\k -\q/30) rectangle (5+\k +\q/30,-\k +\q/30)
		;

		\draw 
		(3,-1.5) node {\size $1$}
		
		(5.6+\k,0-\k) node  {\size $1$}

		(2.5,0) node {\size $1$}
		(3.75,-0.5) node {\size $1$}
		
		(4+\k,-1.25) node {\size $1$}
		;
		\end{scope}
		\end{scope}

		\begin{scope}[yshift = 0.6cm, xshift = 13.6cm]
		\draw[thick] (4,0) -- (4.7,0.7) (4,0) -- (4.7,-0.7) 
		
		(4+\k,\k) -- (5+\k,\k)
		(4+\k,-\k) -- (5+\k,-\k)
		;
		
		\draw[thick, fill=white] 
		(4,0) circle (\q pt)
		(4+\k,\k) circle (\q pt)
		(4+\k,-\k) circle (\q pt)
		;
		
		\draw[thick, fill=white] 
		(5+\k -\q/30,\k -\q/30) rectangle (5+\k +\q/30,\k +\q/30)
		(5+\k -\q/30,-\k -\q/30) rectangle (5+\k +\q/30,-\k +\q/30)
		;

		\draw 
		
		(5.6 +\k,0+\k) node  {\size $1$}
		(5.6+\k,0-\k) node  {\size $1$}

		(3.75,0.45) node {\size $1$}
		
		(4+\k,1.25) node {\size $1$}
		(4+\k,-1.25) node {\size $1$}
		;
		\end{scope}

		\begin{scope}[yshift = -4cm,xshift = 5.2cm]
		\draw[thick] 
		(3,0) -- (3,1)
		
		;
		
		\draw[thick, fill=white] 
		(3,0) circle (\q pt)
		;
		
		\draw[thick, fill=white] 
		(3 -\q/30,1 -\q/30) rectangle (3 +\q/30,1 +\q/30)
		;

		\draw 
		(3,1.5) node {\size $2$}
		

		(3,-0.5) node {\size $1$}
		
		;
		\end{scope}

		\begin{scope}[xshift = 4.7cm]
		\begin{scope}[yshift = -7.3cm,xshift = 0.3cm]
		\draw[thick] 
		
		(4+\k,-\k) -- (5+\k,-\k)
		;
		
		\draw[thick, fill=white] 
		(4+\k,-\k) circle (\q pt)
		;
		
		\draw[thick, fill=white] 
		(5+\k -\q/30,-\k -\q/30) rectangle (5+\k +\q/30,-\k +\q/30)
		;

		\draw 
		
		(5.6+\k,0-\k) node  {\size $2$}

		
		(4+\k,-1.25) node {\size $1$}
		;
		\end{scope}

		\begin{scope}[yshift = -10.3cm,xshift = 0.3cm]
		\draw[thick] 
		
		(4+\k,\k) -- (5+\k,\k)
		;
		
		\draw[thick, fill=white] 
		(4+\k,\k) circle (\q pt)
		;
		
		\draw[thick, fill=white] 
		(5+\k -\q/30,\k -\q/30) rectangle (5+\k +\q/30,\k +\q/30)
		;

		\draw 
		
		(5.6 +\k,0+\k) node  {\size $2$}

		
		(4+\k,1.25) node {\size $1$}
		;
		\end{scope}
		\end{scope}

		\end{tikzpicture}
	\end{center}
	\caption{A table in the style of \cite{Rogers} that gives the $D_4$  Dynkin quivers for all of the subvarieties of the maximal special slice of $\kk{so}_8$. The partitions which define the transverse slices $\O_\mu \cap \es_\nu$, that appear as the Higgs branches are given as column and row headings. The left hand column contains those theories which realise Slodowy slices and so appear in figure \ref{D4quivers}. Almost all of the remaining quivers are quivers for singularities, the only one that isn't is the quiver $\Q_{\Hg}(\es_{(3^2,1^2)} \cap \O_{(5,3)})$. Trivial theories have been left blank and boxes which don't correspond to a possible variety have been crossed. Unlike the linear case, the subvarieties of the maximal special slice of $\kk{so}_{2n}$ fall into a small number of different types and so writing down more general quivers for each of these types supersedes the need to enumerate quivers for small algebras in tables. } 
	\label{D4table}
\end{figure}

\noindent \textbf{An alternative derivation}

So far, all of the $D_4$ theories which realise $\kk{so}_8$ Slodowy slices as their Higgs branches have been balanced, and their Coulomb branch dimension (calculated easily as the sum of the gauge node ranks) has been the dimension of the nilpotent orbit in $\kk{so}_8$ for the partition which is the Lusztig-Spaltenstein map of the slice partition.

\begin{figure}
	\begin{center}
		\begin{tikzpicture}[scale = 0.85]
		\def\q{7}
		\def\k{0.7071}
		\def\size{\normalsize}
		
		\draw[thick] (3,0) -- (4,0) -- (4.7,0.7) (4,0) -- (4.7,-0.7)
		(3,0) -- (3,1)
		(4,0) -- (5,0)
		
		(4+\k,\k) -- (5+\k,\k)
		(4+\k,-\k) -- (5+\k,-\k)
		;
		
		\draw[thick, fill=white] 
		(3,0) circle (\q pt)
		(4,0) circle (\q pt)
		(4+\k,\k) circle (\q pt)
		(4+\k,-\k) circle (\q pt)
		;
		
		\draw[thick, fill=white] 
		(3 -\q/30,1 -\q/30) rectangle (3 +\q/30,1 +\q/30)
		(5 -\q/30,0 -\q/30) rectangle (5 +\q/30,0 +\q/30)
		(5+\k -\q/30,\k -\q/30) rectangle (5+\k +\q/30,\k +\q/30)
		(5+\k -\q/30,-\k -\q/30) rectangle (5+\k +\q/30,-\k +\q/30)
		;

		\draw 
		(3,1.5) node {\size $f_{1}$}
		(5.6,0) node  {\size $f_{2}$}
		(5.6 +\k,0+\k) node  {\size $f_3$}
		(5.6+\k,0-\k) node  {\size $f_4$}

		(3,-0.5) node {\size $g_{1}$}
		(3.75,0.45) node {\size $g_{2}$}
		
		(4+\k,1.25) node {\size $g_3$}
		(4+\k,-1.25) node {\size $g_4$}
		;

		\end{tikzpicture}
	\end{center}
	\caption{A completely general $D_4$ Dynkin quiver.}
	\label{quiver}
\end{figure}
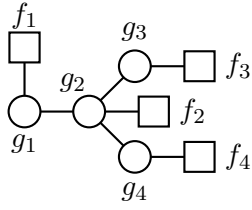

A completely general $D_4$ Dynkin quiver is given in figure \ref{quiver}. The gauge group is $U(g_1) \times U(g_2) \times U(g_3) \times U(g_4)$ with flavours $f_i$ for each gauge group. 
The quaternionic dimension of the Coulomb branch of this quiver is the sum of the values of the gauge nodes,
\begin{equation}\label{Cdim}
\bs{c} = \dim_{\mathbb{H}}(\mathcal{C}) = \sum_{i=1}^4 g_i.
\end{equation} 
When the quiver is balanced, the dimension of the Higgs branch $\mathcal{H}$ is
\begin{equation}\label{Hdim}
\bs{h} =  \dim_{\mathbb{H}}(\mathcal{H}) = \frac{1}{2}\sum_{i=1}^4 f_ig_i.
\end{equation}
Balance relations that $g_i$, $f_i$, $\bs{h}$ and $\bs{c}$ must satisfy, along with the known dimensions of Slodowy slices and nilpotent orbits, will be used to show that some moduli space dimensions are impossible using balanced quivers. The balance requirements for each one of the gauge nodes are
\begin{equation}\label{bal}
\begin{split}2g_i =& f_i + g_2 \qquad \textrm{for} \qquad i \in \{1,3,4\} \\
2g_2 =& f_2 + g_1+g_2+g_3.
\end{split}
\end{equation}
Rewriting the $g_2$ node balance relation using (\ref{Cdim}) gives $f_2 = 3g_2 - \bs{c}$. Expressing all $f$'s in terms of $g$'s in (\ref{Hdim}) gives $
2g_2^2 - \bs{c}g_2 + g_1^2 + g_3^2 + g_4^2 - \bs{h} = 0$. Solving for $g_2$ gives
\begin{equation}\label{method}
g_2 = \frac{1}{4}\bs{c} \pm \frac{1}{4} \sqrt{\bs{c}^2 + 8\bs{h} - 8(g_1^2+g_3^2+g_4^2)}.
\end{equation}
Given $\bs{c}$ and $\bs{h}$, which will be chosen to be the dimensions for a desired orbit and slice, $g_i$ must all be positive integers. If one of them equals zero the quiver is no longer Dynkin type $D_4$. Finally all of the $f_i$ must be non-negative integers. This will be the stumbling point to a few potential constructions. 

\noindent \textbf{Reproduction of known quivers}

These requirements can be used to reproduce the known appropriate $D_4$ Dynkin quivers. The first one will be the balanced quiver with $\mathcal{H} = \mathcal{S}_{(5,3)}$ and $\mathcal{C} = \O_{(2^2,1^4)}$. $\dim_{\mathbb{H}}(\mathcal{S}_{(5,3)}) = \bs{h} = 1$ and $\dim_{\mathbb{H}}(\O_{(2^2,1^4)}) = \bs{c} = 5$. Eq. (\ref{method}) becomes
\begin{equation}
g_2 = \frac{5}{4} \pm \frac{1}{4} \sqrt{33 - 8(g_1^2+g_3^2+g_4^2)}.
\end{equation}
for $g_i$ all positive integers. This requires that $0 < 33-8(g_1^2+g_3^2+g_4^2) = x^2$ for some $x = 3\mod 4$ (for the $+$) or  $x = 1\mod 4$ (for the $-$). For the $+$ scenario the only value of $x$ satisfying this is $3$. This restricts us to $g_1^2+g_3^2+g_4^2 = 3$ and hence $g_1=g_3=g_4 =1$. This gives $g_2 = 2$. Plugging these back into (\ref{bal}) gives $f_2 = 1$ and $f_1=f_3=f_4=0$. This is exactly the correct quiver.
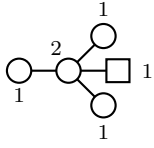
\begin{center}
	\begin{tikzpicture}[scale = 0.65]
	\def\q{7}
	\def\k{0.7071}
	\def\size{\sz}
	
	\draw[thick] (3,0) -- (4,0) -- (4.7,0.7) (4,0) -- (4.7,-0.7)
	(4,0) -- (5,0)
	
	;
	
	\draw[thick, fill=white] 
	(3,0) circle (\q pt)
	(4,0) circle (\q pt)
	(4+\k,\k) circle (\q pt)
	(4+\k,-\k) circle (\q pt)
	;
	
	\draw[thick, fill=white] 
	(5 -\q/30,0 -\q/30) rectangle (5 +\q/30,0 +\q/30)
	;

	\draw 
	(5.6,0) node  {\size $1$}

	(3,-0.5) node {\size $1$}
	(3.75,0.45) node {\size $2$}
	
	(4+\k,1.25) node {\size $1$}
	(4+\k,-1.25) node {\size $1$}
	;
	
	\end{tikzpicture}
\end{center}
The $-$ scenario is restricted to either $g_1^2+g_3^2+g_4^2 = 4$ or $g_1^2+g_3^2+g_4^2 = 1$ for $x=1$ and 5 respectively. Neither 1 or 4 can be written as the sum of three non-zero squares and so the procedure stops and no other solution yields viable quivers with appropriately dimensioned Higgs and Coulomb branches.

The next quivers need $\bs{h} = 2$ and $\bs{c} = 6$, there are multiple quivers which satisfy these dimensions, indeed there is a freedom in the solution which allows us to write all of them. These moduli space branch dimensions give
\begin{equation}
g_2 = \frac{3}{2} \pm \frac{1}{4} \sqrt{52 - 8(g_1^2+g_3^2+g_4^2)}.
\end{equation}
Working through the options gives values of $2$,$1$ and $1$ for $g_i$ for $i\in\{1,3,4\}$, the choice of which one takes the value 2 gives us the multiple viable quivers. 
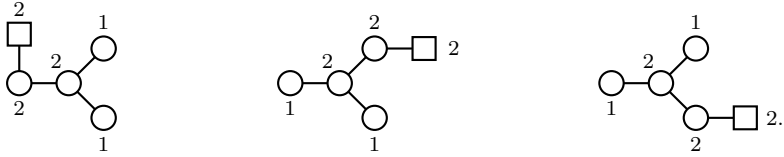
\begin{center}
	\begin{tikzpicture}[scale = 0.65]
	\def\q{7}
	\def\k{0.7071}
	\def\size{\sz}
	
	\draw[thick] (3,0) -- (4,0) -- (4.7,0.7) (4,0) -- (4.7,-0.7)
	(3,0) -- (3,1)
	
	;
	
	\draw[thick, fill=white] 
	(3,0) circle (\q pt)
	(4,0) circle (\q pt)
	(4+\k,\k) circle (\q pt)
	(4+\k,-\k) circle (\q pt)
	;
	
	\draw[thick, fill=white] 
	(3 -\q/30,1 -\q/30) rectangle (3 +\q/30,1 +\q/30)
	;

	\draw 
	(3,1.5) node {\size $2$}

	(3,-0.5) node {\size $2$}
	(3.75,0.45) node {\size $2$}
	
	(4+\k,1.25) node {\size $1$}
	(4+\k,-1.25) node {\size $1$}
	;
	
	\draw (1.8,0) node {$~$};
	
	\end{tikzpicture}$\qquad \qquad \quad$
	\begin{tikzpicture}[scale = 0.65]
	\def\q{7}
	\def\k{0.7071}
	\def\size{\sz}
	
	\draw[thick] (3,0) -- (4,0) -- (4.7,0.7) (4,0) -- (4.7,-0.7)
	
	(4+\k,\k) -- (5+\k,\k)
	;
	
	\draw[thick, fill=white] 
	(3,0) circle (\q pt)
	(4,0) circle (\q pt)
	(4+\k,\k) circle (\q pt)
	(4+\k,-\k) circle (\q pt)
	;
	
	\draw[thick, fill=white] 
	(5+\k -\q/30,\k -\q/30) rectangle (5+\k +\q/30,\k +\q/30)
	;

	\draw 
	(5.6 +\k,0+\k) node  {\size $2$}

	(3,-0.5) node {\size $1$}
	(3.75,0.45) node {\size $2$}
	
	(4+\k,1.25) node {\size $2$}
	(4+\k,-1.25) node {\size $1$}
	;
	
	\end{tikzpicture}$\qquad \qquad$
	\begin{tikzpicture}[scale = 0.65]
	\def\q{7}
	\def\k{0.7071}
	\def\size{\sz}
	
	\draw[thick] (3,0) -- (4,0) -- (4.7,0.7) (4,0) -- (4.7,-0.7)
	
	(4+\k,-\k) -- (5+\k,-\k)
	;
	
	\draw[thick, fill=white] 
	(3,0) circle (\q pt)
	(4,0) circle (\q pt)
	(4+\k,\k) circle (\q pt)
	(4+\k,-\k) circle (\q pt)
	;
	
	\draw[thick, fill=white] 
	(5+\k -\q/30,-\k -\q/30) rectangle (5+\k +\q/30,-\k +\q/30)
	;

	\draw 
	(5.6+\k,0-\k) node  {\size $2.$}

	(3,-0.5) node {\size $1$}
	(3.75,0.45) node {\size $2$}
	
	(4+\k,1.25) node {\size $1$}
	(4+\k,-1.25) node {\size $2$}
	;
	
	\end{tikzpicture}
\end{center}

The result for $\bs{h} = 3$, $\bs{c}=9$ gives
\begin{equation}
g_2 = \frac{9}{4} \pm \frac{1}{4} \sqrt{105 - 8(g_1^2+g_3^2+g_4^2)}.
\end{equation}
and so $105 - 8(g_1^2+g_3^2+g_4^2) \in \{1,9,25,49,81\}$. Setting this to 9 requires $g_1^2+g_3^2+g_4^2 = 12$ and so $g_1=g_3=g_4 = 2$ and the quiver is reproduced.
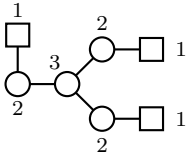
\begin{center}
	\begin{tikzpicture}[scale = 0.65]
	\def\q{7}
	\def\k{0.7071}
	\def\size{\sz}
	
	\draw[thick] (3,0) -- (4,0) -- (4.7,0.7) (4,0) -- (4.7,-0.7)
	(3,0) -- (3,1)
	
	(4+\k,\k) -- (5+\k,\k)
	(4+\k,-\k) -- (5+\k,-\k)
	;
	
	\draw[thick, fill=white] 
	(3,0) circle (\q pt)
	(4,0) circle (\q pt)
	(4+\k,\k) circle (\q pt)
	(4+\k,-\k) circle (\q pt)
	;
	
	\draw[thick, fill=white] 
	(3 -\q/30,1 -\q/30) rectangle (3 +\q/30,1 +\q/30)
	(5+\k -\q/30,\k -\q/30) rectangle (5+\k +\q/30,\k +\q/30)
	(5+\k -\q/30,-\k -\q/30) rectangle (5+\k +\q/30,-\k +\q/30)
	;

	\draw 
	(3,1.5) node {\size $1$}
	(5.6 +\k,0+\k) node  {\size $1$}
	(5.6+\k,0-\k) node  {\size $1$}

	(3,-0.5) node {\size $2$}
	(3.75,0.45) node {\size $3$}
	
	(4+\k,1.25) node {\size $2$}
	(4+\k,-1.25) node {\size $2$}
	;
	
	\end{tikzpicture}
\end{center}
The other possibilities aren't consistent.

\noindent \textbf{Impossibility of further quivers}

The remaining nilpotent varieties in $\kk{so}_8$ which might be the Higgs and Coulomb branches of a balanced $D_4$ Dynkin quiver have dimensions $(\bs{h}, \bs{c}) \in \{(6,10), (7,11), (12,12)\}$. The following calculations amount to an exhaustive checking of the viability of all possibilities in each case. Almost all fail because the necessary flavour nodes have a negative label and so the quiver isn't viable. Some fail earlier in the face of Legendre's three square theorem. That is, if an option for the dimensions of the moduli space branches requires $g_1^2+ g_3^2 + g_4^2$ to be of the form $4^a(8b+7)$ for $a$ and $b$ integers (in the set $7,15,23,28,\dots$) it necessarily cannot be written as the sum of three non-zero squares. 

$(\bs{h}, \bs{c}) = (6,10)$ requires that $148 - 8(g_1^2+g_3^2+g_4^2) = x^2$ for positive $x = 2\mod4$ which requires $(g_1,g_3,g_4)  \in \{(2,1,1),(1,2,3),(4,1,1), \textrm{permutations}\}$. The first (and perms.) yields $g_2 = 0$ but $g_2$ must be positive for the quiver to stay $D_4$, or $g_2 = 5$, which requires negative flavours. The second (and perms.) yields $g_2 = 1$ which requires negative $f_2$, or $g_2=4$ which requires negative flavour. The third (and perms.) yields $g_2 = 2$ which requires negative $f_2$, or $g_2=3$ which requires negative flavour.

$(\bs{h}, \bs{c}) = (7,11)$ requires that $177 - 8(g_1^2+g_3^2+g_4^2) = x^2$ for positive $x = 1\mod4$ or $x = 3\mod4$ and so $x^2 \in \{1,25,81,169\}$ or $x^2 \in \{9,49,121\}$. These seven options, in order, require $g_1^2 + g_3^2 + g_4^2 \in \{22,19,12,1,21,16,7\}$. For this to be 1 or 16 requires zero rank gauge nodes, 7 fails in the face of Legendre's three-square theorem, the rest of the options fail on the grounds of flavour in the same way as the previous example. 

$(\bs{h}, \bs{c}) = (12,12)$ requires that $240 - 8(g_1^2+g_3^2+g_4^2) = x^2$ for positive $x = 0\mod4$ and so $x^2 \in \{0,16,64,144\}$. These four options, in order, require $g_1^2 + g_3^2 + g_4^2 \in \{30,28,22,12\}$. Here 28 fails in the face of Legendre's three-square theorem, the rest of the options fail on the grounds of flavour in the same way as the previous examples.

\subsubsection{Quivers for maximal special slices}
\begin{figure}
	\begin{center}
		\begin{tikzpicture}[scale = 0.65]
		\def\q{7}
		\def\k{0.7071}
		\def\size{\sz}
		
		\draw[thick] (-1,0) -- (4,0) -- (4.7,0.7) (4,0) -- (4.7,-0.7) 
		(-1,0) -- (-1,1)
		
		(4+\k,\k) -- (5+\k,\k)
		(4+\k,-\k) -- (5+\k,-\k)
		;
		
		\draw[thick, fill=white] 
		(-1,0) circle (\q pt)
		(0,0) circle (\q pt)
		(1,0) circle (\q pt)
		(2,0) circle (\q pt)
		(3,0) circle (\q pt)
		(4,0) circle (\q pt)
		(4+\k,\k) circle (\q pt)
		(4+\k,-\k) circle (\q pt)
		;
		
		\draw[thick, fill=white] 
		(-1 -\q/30,1 -\q/30) rectangle (-1 +\q/30,1 +\q/30)
		(5+\k -\q/30,\k -\q/30) rectangle (5+\k +\q/30,\k +\q/30)
		(5+\k -\q/30,-\k -\q/30) rectangle (5+\k +\q/30,-\k +\q/30)
		;

		\draw 
		(-1,1.5) node {\size $1$}
		
		(5.6 +\k,0+\k) node  {\size $1$}
		(5.6+\k,0-\k) node  {\size $1$}
		
		(-1,-0.5) node {\size $2$}
		(0,-0.5) node {\size $3$}
		(1,-0.5) node {\size $4$}
		(2,-0.5) node {\size $g_3$}
		(3.3,-0.5) node {\size $2m-2$}
		(5,0) node {\size $2m-1$}
		
		(4+\k,1.25) node {\size $m$}
		(4+\k,-1.25) node {\size $m$}

		;
		
		\draw[white, fill=white] (1.5,-1) rectangle (2.5,2);
		
		\draw (2.05,0) node {$\dots$};
		
		\draw (-4.5,0) node {${\sf{Q}}_{\mathcal{H}}(\mathcal{S}_{(2m-1^2,1^2)}) ~~~= $};

		\begin{scope}[yshift=-4cm]
		
		\draw[thick] (-1,0) -- (4,0) -- (4.7,0.7) (4,0) -- (4.7,-0.7) 
		(-1,0) -- (-1,1)
		
		(4+\k,\k) -- (5+\k,\k)
		;
		
		\draw[thick, fill=white] 
		(-1,0) circle (\q pt)
		(0,0) circle (\q pt)
		(1,0) circle (\q pt)
		(2,0) circle (\q pt)
		(3,0) circle (\q pt)
		(4,0) circle (\q pt)
		(4+\k,\k) circle (\q pt)
		(4+\k,-\k) circle (\q pt)
		;
		
		\draw[thick, fill=white] 
		(-1 -\q/30,1 -\q/30) rectangle (-1 +\q/30,1 +\q/30)
		(5+\k -\q/30,\k -\q/30) rectangle (5+\k +\q/30,\k +\q/30)
		;

		\draw 
		(-1,1.5) node {\size $1$}
		
		(5.6 +\k,0+\k) node  {\size $2$}
		
		(-1,-0.5) node {\size $2$}
		(0,-0.5) node {\size $3$}
		(1,-0.5) node {\size $4$}
		(2,-0.5) node {\size $g_3$}
		(3.3,-0.5) node {\size $2m-1$}
		(3.75,0.45) node {\size $2m$}
		
		(4+\k,1.25) node {\size $m+1$}
		(4+\k,-1.25) node {\size $m$}
		
		(7.5,0) node {$I$}
		;
		
		\draw[white, fill=white] (1.5,-1) rectangle (2.5,2);
		
		\draw (2.05,0) node {$\dots$};
		\end{scope}
		
		\begin{scope}[yshift=-7cm]
		
		\draw[thick] (-1,0) -- (4,0) -- (4.7,0.7) (4,0) -- (4.7,-0.7) 
		(-1,0) -- (-1,1)
		
		(4+\k,-\k) -- (5+\k,-\k)
		;
		
		\draw[thick, fill=white] 
		(-1,0) circle (\q pt)
		(0,0) circle (\q pt)
		(1,0) circle (\q pt)
		(2,0) circle (\q pt)
		(3,0) circle (\q pt)
		(4,0) circle (\q pt)
		(4+\k,\k) circle (\q pt)
		(4+\k,-\k) circle (\q pt)
		;
		
		\draw[thick, fill=white] 
		(-1 -\q/30,1 -\q/30) rectangle (-1 +\q/30,1 +\q/30)
		(5+\k -\q/30,-\k -\q/30) rectangle (5+\k +\q/30,-\k +\q/30)
		;

		\draw 
		(-1,1.5) node {\size $1$}
		
		(5.6+\k,0-\k) node  {\size $2$}
		
		(-1,-0.5) node {\size $2$}
		(0,-0.5) node {\size $3$}
		(1,-0.5) node {\size $4$}
		(2,-0.5) node {\size $g_3$}
		(3.3,-0.5) node {\size $2m-1$}
		(3.75,0.45) node {\size $2m$}
		
		(4+\k,1.25) node {\size $m$}
		(4+\k,-1.25) node {\size $m+1$}

		(7.5,0) node { $II$}
		;
		
		\draw[white, fill=white] (1.5,-1) rectangle (2.5,2);
		
		\draw (2.05,0) node {$\dots$};

		\draw (-4,2) node {${\sf{Q}}_{\mathcal{H}}(\mathcal{S}_{(2m^2,1^2)}) ~= $};
		\end{scope}
		
		\end{tikzpicture}
	\end{center}
\caption{The $D_n$ Dynkin quivers which realise the maximal special slices of $\kk{so}_{4m}$ and $\kk{so}_{4m+2}$ respectively as their Higgs branches.}
\label{MaxD}
\end{figure}
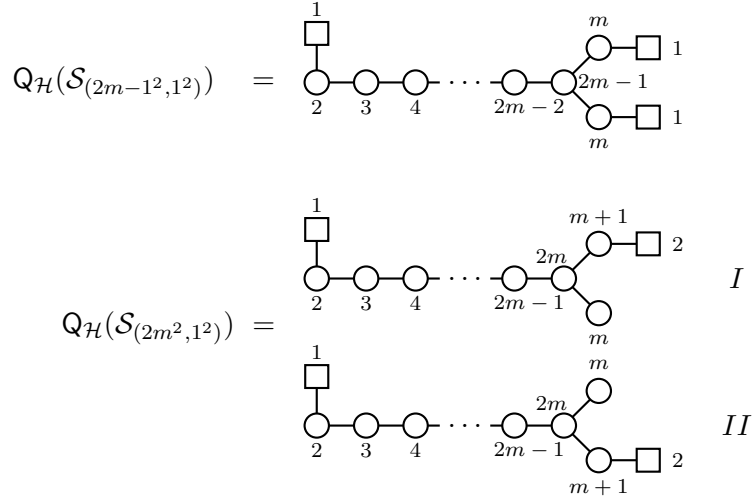

When writing down the general quiver for a maximal special slice there are two options arising from the two different Hasse diagrams given in figure \ref{MSSHasse}, $\mathcal{S}_{(2m-1^2,1^2)} \in \kk{so}_{4m}$ and $\mathcal{S}_{(2m^2,1^2)} \in \kk{so}_{4m+2}$. The quivers associated to each of these general slices are given in figure \ref{MaxD}. 
These can be established by adding singularity quivers down any given route in the Hasse diagrams in figure \ref{MSSHasse}, and can be checked in numerous ways. The first check is balance, for a gauge node labelled $c_i$ with flavour $f_i$ balance means
\begin{equation}
2c_i = f_i + \sum_{c_j \textrm{ linked to } c_i} c_j.
\end{equation}
It is easy to see that the quivers $\mathsf{Q}_\Hg(\mathcal{S}_{(2m-1^2,1^2)})$ and $\mathsf{Q}_\Hg(\mathcal{S}_{(2m^2,1^2)})$ in figure \ref{MaxD} fulfil this. Another test is Higgs branch dimension. For a general balanced unitary quiver, the Higgs branch dimension is given by \[\dim_{\mathbb{H}}(\mathcal{H}(\textsf{Q}_{\textrm{bal}})) = \frac {1}{2}\sum_i c_if_i.\] $\dim_{\mathbb{H}}(\mathcal{S}_{(2m^2,1^2)}) = m+2$ and $\dim_{\mathbb{H}}(\mathcal{S}_{(2m-1^2,1^2)}) = m+1$, which the quivers for the maximal special slices also satisfy.

\subsubsection{Quivers for special slice nilpotent varieties in $\kk{so}_{2n}$}
All the $D_n$ Dynkin quivers concerned here\footnote{$\Q(\mathcal{V})$ for $\mathcal{V}$ a subvariety of the maximal special slice $\mathcal{S}_{(2m-1^2,1^2)}$ (resp. $\mathcal{S}_{(2m^2,1^2)}$), enumerated in (\ref{so4vars}) and (\ref{so42vars}).} are descendants\footnote{$\Q_1$ \textit{descends} from another theory, $\Q_2$, if $\mathcal{H}(\Q_1) \subset \mathcal{H}(\Q_2)$.} of $\Q_{\mathcal{H}}(\mathcal{S}_{(2m-1^2,1^2)})$ (resp. $\Q_{\mathcal{H}}(\mathcal{S}_{(2m^2,1^2)})$), given in figure \ref{MaxD}. Instead of giving tables for small algebras as in \cite{Rogers}, it is possible to write the general form for any theory one would expect to find as a descendant of $\Q_{\mathcal{H}}(\mathcal{S}_{(2m-1^2,1^2)})$ or $\Q_{\mathcal{H}}(\mathcal{S}_{(2m^2,1^2)})$. 

\begin{figure}
	\begin{center}
		\begin{tikzpicture}[scale = 0.65]
		\def\q{7}
		\def\k{0.7071}
		\def\size{\tiny}
		
		\draw[thick] (-5,0) -- (-2.5,0) (-1.5,0) -- (1.5,0) (2.5,0) -- (4,0) -- (4.7,0.7) (4,0) -- (4.7,-0.7) 
		(0,0) -- (0,1)
		
		;
		
		\draw[thick, fill=white] 
		(-5,0) circle (\q pt)
		(-4,0) circle (\q pt)
		(-3,0) circle (\q pt)
		(-1,0) circle (\q pt)
		(0,0) circle (\q pt)
		(1,0) circle (\q pt)
		(3,0) circle (\q pt)
		(4,0) circle (\q pt)
		(4+\k,\k) circle (\q pt)
		(4+\k,-\k) circle (\q pt)
		;
		
		\draw[thick, fill=white] 
		(0 -\q/30,1 -\q/30) rectangle (0 +\q/30,1 +\q/30)
		;

		\draw 
		(0,1.5) node {\size $1$}
		
		
		(-7,0) node {$A(p,q) ~ = $}
		
		(-5,-0.5) node {\size $1$}
		(-4,-0.5) node {\size $2$}
		(-3,-0.5) node {\size $3$}
		(-1,0.5) node {\size $2p-1$}
		(0,-0.5) node {\size $2p$}
		(1,0.5) node {\size $2p$}
		(3,0.5) node {\size $2p$}
		(3.75,0.45) node {\size $2p$}
		
		(4+\k,1.25) node {\size $p$}
		(4+\k,-1.25) node {\size $p$}
		
		(2.5,-1) node {$\underbrace{~~~~~~~~~~~~~~~~~~}_{q}$}
		;
		
		\draw (2.05,0) node {$\dots$};
		\draw (-1.95,0) node {$\dots$};
		

		\begin{scope}[yshift = -4cm]
		
		\draw[thick] (-7,0) -- (-6.5,0) (-5.5,0) -- (-2.5,0) (-1.5,0) -- (1.5,0) (2.5,0) -- (4,0) -- (4.7,0.7) (4,0) -- (4.7,-0.7)
		(-7,0) -- (-7,1) 
		(0,0) -- (0,1)
		
		;
		
		\draw[thick, fill=white] 
		(-7,0) circle (\q pt)
		(-5,0) circle (\q pt)
		(-4,0) circle (\q pt)
		(-3,0) circle (\q pt)
		(-1,0) circle (\q pt)
		(0,0) circle (\q pt)
		(1,0) circle (\q pt)
		(3,0) circle (\q pt)
		(4,0) circle (\q pt)
		(4+\k,\k) circle (\q pt)
		(4+\k,-\k) circle (\q pt)
		;
		
		\draw[thick, fill=white] 
		(-7 -\q/30,1 -\q/30) rectangle (-7 +\q/30,1 +\q/30)
		(0 -\q/30,1 -\q/30) rectangle (0 +\q/30,1 +\q/30)
		;

		\draw 
		(-7,1.5) node {\size $1$}
		(0,1.5) node {\size $1$}
		
		
		(-9,0) node {$B(p,q) ~ = $}
		
		(-7,-0.5) node {\size $1$}
		(-5,-0.5) node {\size $1$}
		(-4,-0.5) node {\size $2$}
		(-3,-0.5) node {\size $3$}
		(-1,-0.5) node {\size $2p-1$}
		(0,-0.5) node {\size $2p$}
		(1,0.5) node {\size $2p$}
		(3,0.5) node {\size $2p$}
		(3.75,0.45) node {\size $2p$}
		
		(4+\k,1.25) node {\size $p$}
		(4+\k,-1.25) node {\size $p$}
		
		(2.5,-1) node {$\underbrace{~~~~~~~~~~~~~~~~~~}_{q}$}
		;
		
		\draw (2.05,0) node {$\dots$};
		\draw (-1.95,0) node {$\dots$};
		\draw (-5.95,0) node {$\dots$};
		
		\end{scope}
\end{tikzpicture}
\end{center}
\caption{The general forms for the $D_n$ Dynkin quivers which realise $\kk{so}_{2n}$ nilpotent varieties as their Higgs branches. The $B(p,q)$ quiver always has $2m$ (resp. $2m+1$) nodes for varieties in $\kk{so}_{4m}$ (resp. $\kk{so}_{4m+2}$). The conventions are that, in both quivers, when $q=-1$ this corresponds to having a flavour of 1 on both end nodes, and when $q=-2$ this corresponds to having a flavour 2 on one end node.}
\label{genquivs}
\end{figure}
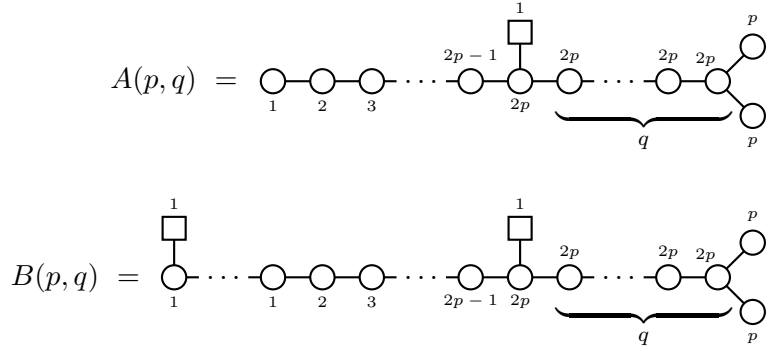

Because of (\ref{matchingvar}), one needs to establish quivers for only 7 of the varieties listed in (\ref{so4vars}) and (\ref{so42vars}). $\O_{(2m^2)} \cap \es_{(2m-1^2,1^2)} = A_{2m-1} \cup A_{2m-1}$ has been discussed so only 6 general quivers need to be found. All six can be written as one of two general forms given in figure \ref{genquivs}. There are some conventions for reading these quivers also given in figure \ref{genquivs}.
The quivers for the six general nilpotent varieties that are left to find take the forms
\begin{equation}\label{genforms}
\begin{split}
\Q_\Hg (\O_{\psi_j} \cap \es_{(2m^2)}) & = A(m-j,-2) \\
\Q_\Hg (\O_{\psi_j'} \cap \es_{(2m^2,1^2)}) & = B(m-j+1,2m+1) \\
\Q_\Hg (\O_{\psi_j} \cap \es_{\psi_k}) & = A(k-j,2m-2k-2) \\
\Q_\Hg (\O_{\varphi_j} \cap \es_{\varphi_k}) & = A(k-j,2m-2k-3) \\
\Q_\Hg (\O_{\psi_j} \cap \es_{\varphi_k}) & = B(k-j+1,2m-2k-3) \\
\Q_\Hg (\O_{\psi_j'} \cap \es_{\varphi_k'}) & = B(k-j,2m-2k-4).
\end{split}
\end{equation}
These have been established using quiver addition and using as a guide the Hasse subdiagrams of figure \ref{MSSHasse} which correspond to each subvariety. The claim that these are all the $\kk{so}_{2n}$ nilpotent varieties appearing as moduli space branches of $D_n$ Dynkin quivers is confirmed in the next section. There, the singularity structure of general good $D_n$ Dynkin quivers is found and a comparison with the known singularity structure of any further $\kk{so}_{2n}$ nilpotent varieties is made. This shows that none of the $D_n$ Dynkin quivers have further $\kk{so}_{2n}$ varieties as moduli space branches. 
 
\noindent$\bs{D_4}$ \textbf{example} $\quad$ Eq. (\ref{genforms}) can be used to reproduce the results for $D_4$ quivers relatively easily. There were only four $D_4$ Dynkin quivers which realised nilpotent varieties of $\kk{so}_{4m}$ (so $m=2$) which weren't also singularities. These were the quivers
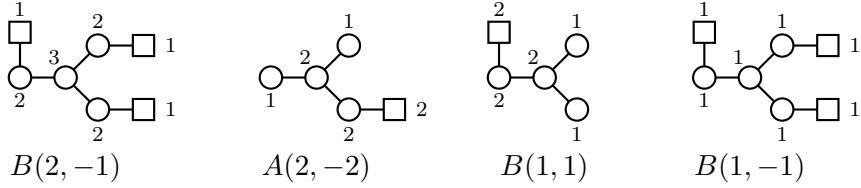
\begin{center}
	\begin{tikzpicture}[scale = 0.6]
	\def\q{7}
	\def\k{0.7071}
	\def\size{\sz}
	\begin{scope}[xshift = 15.5cm]
	\draw[thick] (3,0) -- (4,0) -- (4.7,0.7) (4,0) -- (4.7,-0.7)
	(3,0) -- (3,1)
	
	;
	
	\draw[thick, fill=white] 
	(3,0) circle (\q pt)
	(4,0) circle (\q pt)
	(4+\k,\k) circle (\q pt)
	(4+\k,-\k) circle (\q pt)
	;
	
	\draw[thick, fill=white] 
	(3 -\q/30,1 -\q/30) rectangle (3 +\q/30,1 +\q/30)
	;

	\draw 
	(3,1.5) node {\size $2$}

	(3,-0.5) node {\size $2$}
	(3.75,0.45) node {\size $2$}
	
	(4+\k,1.25) node {\size $1$}
	(4+\k,-1.25) node {\size $1$}
	
	(4,-2) node {$B(1,1)$}
	;
	
	\end{scope}
	\begin{scope}[xshift = 20cm]
	\draw[thick] (3,0) -- (4,0) -- (4.7,0.7) (4,0) -- (4.7,-0.7)
	(3,0) -- (3,1)
	
	(4+\k,\k) -- (5+\k,\k)
	(4+\k,-\k) -- (5+\k,-\k)
	;
	
	\draw[thick, fill=white] 
	(3,0) circle (\q pt)
	(4,0) circle (\q pt)
	(4+\k,\k) circle (\q pt)
	(4+\k,-\k) circle (\q pt)
	;
	
	\draw[thick, fill=white] 
	(3 -\q/30,1 -\q/30) rectangle (3 +\q/30,1 +\q/30)
	(5+\k -\q/30,\k -\q/30) rectangle (5+\k +\q/30,\k +\q/30)
	(5+\k -\q/30,-\k -\q/30) rectangle (5+\k +\q/30,-\k +\q/30)
	;

	\draw 
	(3,1.5) node {\size $1$}
	(5.6 +\k,0+\k) node  {\size $1$}
	(5.6+\k,0-\k) node  {\size $1$}

	(3,-0.5) node {\size $1$}
	(3.75,0.45) node {\size $1$}
	
	(4+\k,1.25) node {\size $1$}
	(4+\k,-1.25) node {\size $1$}
	
	(4,-2) node {$B(1,-1)$}
	;
	
	\end{scope}
	\begin{scope}[xshift= 10.5cm]
	\draw[thick] (3,0) -- (4,0) -- (4.7,0.7) (4,0) -- (4.7,-0.7)
	
	(4+\k,-\k) -- (5+\k,-\k)
	;
	
	\draw[thick, fill=white] 
	(3,0) circle (\q pt)
	(4,0) circle (\q pt)
	(4+\k,\k) circle (\q pt)
	(4+\k,-\k) circle (\q pt)
	;
	
	\draw[thick, fill=white] 
	(5+\k -\q/30,-\k -\q/30) rectangle (5+\k +\q/30,-\k +\q/30)
	;

	\draw 
	(5.6+\k,0-\k) node  {\size $2$}

	(3,-0.5) node {\size $1$}
	(3.75,0.45) node {\size $2$}
	
	(4+\k,1.25) node {\size $1$}
	(4+\k,-1.25) node {\size $2$}
	
	(4,-2) node {$A(2,-2)$}
	;

	\end{scope}

	\begin{scope}[xshift = 5cm]
	\draw[thick] (3,0) -- (4,0) -- (4.7,0.7) (4,0) -- (4.7,-0.7)
	(3,0) -- (3,1)
	
	(4+\k,\k) -- (5+\k,\k)
	(4+\k,-\k) -- (5+\k,-\k)
	;
	
	\draw[thick, fill=white] 
	(3,0) circle (\q pt)
	(4,0) circle (\q pt)
	(4+\k,\k) circle (\q pt)
	(4+\k,-\k) circle (\q pt)
	;
	
	\draw[thick, fill=white] 
	(3 -\q/30,1 -\q/30) rectangle (3 +\q/30,1 +\q/30)
	(5+\k -\q/30,\k -\q/30) rectangle (5+\k +\q/30,\k +\q/30)
	(5+\k -\q/30,-\k -\q/30) rectangle (5+\k +\q/30,-\k +\q/30)
	;

	\draw 
	(3,1.5) node {\size $1$}
	(5.6 +\k,0+\k) node  {\size $1$}
	(5.6+\k,0-\k) node  {\size $1$}

	(3,-0.5) node {\size $2$}
	(3.75,0.45) node {\size $3$}
	
	(4+\k,1.25) node {\size $2$}
	(4+\k,-1.25) node {\size $2$}
	
	(4,-2) node {$B(2,-1)$}
	;
	\end{scope}
	
	\end{tikzpicture}
\end{center}
whose appearance as the quivers which realised $\kk{so}_8$ nilpotent varieties as their Higgs branches can be checked:
\begin{equation}
\begin{split}
\Q_\Hg(\O_{(7,1)} \cap \es_{(3^2,1^2)}) & = \Q_\Hg(\O_{\psi_0}\cap\es_{\varphi_1}) = B(2,-1) \\
\Q_\Hg(\O_{(7,1)} \cap \es_{(4^2)}) & = \Q_\Hg(\O_{\psi_0}\cap\es_{(2m^2)}) = A(2,-2) \\
\Q_\Hg(\O_{(7,1)} \cap \es_{(5,1^3)}) & = \Q_\Hg(\O_{\psi_0}\cap\es_{\varphi_0}) = B(1,1) \\
\Q_\Hg(\O_{(5,3)} \cap \es_{(3^2,1^2)}) & = \Q_\Hg(\O_{\psi_1}\cap\es_{\varphi_1}) = B(1,-1).
\end{split}
\end{equation}
Here the variety is written in the form of (\ref{so4vars}), the type of quiver read from (\ref{genforms}) and figure \ref{genquivs} used to draw the quiver. This reproduces the familiar $D_4$ results.

\section{$\bs{D_n}$ Dynkin quivers}
Which varieties \textit{do} $D_n$ Dynkin quivers  realise with their moduli space branches? And is there a way of writing down a simple, complete description of every possible good $D_n$ Dynkin quiver which also provides a simple means by which to extract moduli space information\footnote{In analogy to the way the moduli space varieties of a linear ($A_n$ Dynkin) quiver and the name of that quiver are related, reviewed in (\cite{Rogers}(3.1),(3.2))}? This section tackles both problems simultaneously. The full singularity structure of the moduli space varieties of $D_n$ Dynkin quivers is provided and a classification based on that structure established. The analysis is performed without explicit reference to a brane construction in order to inform generalisations to quivers with no such description.

When discussing $D_n$ Dynkin quivers with nilpotent varieties of $\kk{so}_{2n}$ as their moduli space branches, essential building blocks were those quivers which correspond to the minimal degenerations, or singularities. The singularities are $A_k$, $a_k$, $A_{k} \cup A_{k}$ and $D_k$. These are the basic building blocks used to investigate the local structure of the moduli spaces for $D_n$ Dynkin quivers. In conjuncture with this structure, there is a natural characterisation of all good $D_n$ Dynkin quivers. The simplest subclass to tackle is that of \textit{balanced} $D_n$ Dynkin quivers, which are investigated first.

\subsection{Balanced $D_n$ Dynkin quivers}
At the level of the quiver, balance is the requirement that the sum of all the nodes connected to a given gauge node by an edge (flavours and other gauge nodes) is exactly double the rank of the given gauge node. This is referred to as the gauge node having zero \textit{excess}. Consider a general simply-laced quiver\footnote{This quiver needn't necessarily be in the shape of a Dynkin diagram.} with gauge nodes with positive rank $g^i$, where the upper index refers to some ordering of the gauge nodes. Attached to these gauge nodes are flavour nodes with non-negative label $f^i$. For each gauge node $g^i$, consider the set of gauge nodes which are connected to $g^i$ via an edge. This will be a collection of gauge nodes labelled with indices in the set $j_i$. The condition of balance is then
\begin{equation}\label{balanceeq}
e^i =  f^i - 2g^i +\sum_{k \in j_i} g^k  = 0,
\end{equation}
where $e^i$ is the excess for each node.

Balance imposes a general restriction on $D_n$ Dynkin quivers. The difference in the flavour attached to each of the end nodes must be even. For the quiver in figure \ref{Dynkinquiv} this means that $f'-f'' \equiv 0\mod2$. For concreteness take $f''\geq f'$. Consider the situation where the difference in flavour is odd, that is, $f'' = f'+2b+1$. Balancing the end nodes requires that $g' = \frac{1}{2}(f'+g_{n-2})$ and so $f'+g_{n-2}$ must be even, but also that $g'' = \frac{1}{2}(f'+2b+1+g_{n-2})$, so $f'+g_{n-2}$ must be odd. An odd difference in flavour is therefore unbalanceable. This will become important later for good $D_n$ Dynkin quivers, but for now all it means is that the balanced quivers must have $f''-f'$ even. Using nomenclature that will be introduced in more detail in the discussion of good quivers, this is the restriction that all balanced $D_n$ Dynkin quivers must be of \textit{even type}. 

The local moduli space analysis and classification of balanced $D_n$ Dynkin quivers arises from the giving of poset structure to the set of balanced quivers. In this analysis, this structure will be based on the inclusion relations of the Higgs branches of the theories. This structure will be illustrated using a Hasse diagram built up using quiver addition. 

The premise of quiver addition is that, for a given quiver, there exists at least one `larger' quiver from which the first quiver could have been found via the removal of a singularity from the larger quiver by quiver subtraction. Using the realisation of singularities as (sub)quivers, one can find all of the larger quivers for a given quiver by `adding back' the singularities. If one does this starting with some minimal quiver, and does so while insisting on maintaining balance at all times, one can recover \textit{any} balanced $D_n$ Dynkin quiver. This procedure also gives the set of balanced $D_n$ Dynkin quivers the desired poset structure, which is illustrated by the Hasse diagram one constructs by adding the slices back. Furthermore, since this procedure is based off of the transverse slice structure of the Higgs branches of the theories, one has automatically generated the Hasse diagram representing the singularity structure of the varieties that these quivers realise as moduli space branches. Once this poset structure is uncovered, the classification of balanced $D_n$ Dynkin quivers follows simply. As each node in the Hasse diagram represents a unique balanced $D_n$ Dynkin quiver, and every quiver is in the diagram, a unique label for every node gives a unique label for every quiver.

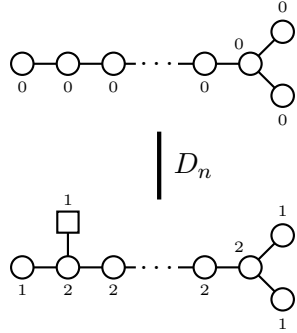
\begin{figure}
	\begin{center}
		\begin{tikzpicture}[scale = 0.6]
		\def\q{7}
		\def\k{0.7071}
		\def\size{\tiny}
		
		\draw[thick] (-1,0) -- (4,0) -- (4.7,0.7) (4,0) -- (4.7,-0.7) 
		
		;
		
		\draw[thick, fill=white] 
		(-1,0) circle (\q pt)
		(0,0) circle (\q pt)
		(1,0) circle (\q pt)
		(2,0) circle (\q pt)
		(3,0) circle (\q pt)
		(4,0) circle (\q pt)
		(4+\k,\k) circle (\q pt)
		(4+\k,-\k) circle (\q pt)
		;
		

		\draw 
		
		
		(-1,-0.5) node {\size $0$}
		(0,-0.5) node {\size $0$}
		(1,-0.5) node {\size $0$}
		(2,-0.5) node {\size $0$}
		(3,-0.5) node {\size $0$}
		(3.75,0.45) node {\size $0$}
		
		(4+\k,1.25) node {\size $0$}
		(4+\k,-1.25) node {\size $0$}
		;
		
		\draw[white, fill=white] (1.5,2) rectangle (2.4,-1);
		\draw (2,0) node {$\dots$};
		
		\draw[ultra thick] (2,-1.5) -- (2,-3);
		\draw (2.75,-2.25) node {$D_n$};
		
		\begin{scope}[yshift = -4.5cm]
		\draw[thick] (-1,0) -- (4,0) -- (4.7,0.7) (4,0) -- (4.7,-0.7) 
		(0,0) -- (0,1)
		
		;
		
		\draw[thick, fill=white] 
		(-1,0) circle (\q pt)
		(0,0) circle (\q pt)
		(1,0) circle (\q pt)
		(2,0) circle (\q pt)
		(3,0) circle (\q pt)
		(4,0) circle (\q pt)
		(4+\k,\k) circle (\q pt)
		(4+\k,-\k) circle (\q pt)
		;
		
		\draw[thick, fill=white] 
		(0 -\q/30,1 -\q/30) rectangle (0 +\q/30,1 +\q/30)
		;

		\draw 
		(0,1.5) node {\size $1$}
		
		
		(-1,-0.5) node {\size $1$}
		(0,-0.5) node {\size $2$}
		(1,-0.5) node {\size $2$}
		(2,-0.5) node {\size $2$}
		(3,-0.5) node {\size $2$}
		(3.75,0.45) node {\size $2$}
		
		(4+\k,1.25) node {\size $1$}
		(4+\k,-1.25) node {\size $1$}
		;
		
		\draw[white, fill=white] (1.5,0.7) rectangle (2.4,-1);
		\draw (2,0) node {$\dots$};
		
		\end{scope}
		\end{tikzpicture}
	\end{center}
	\caption{The only possible balanced singularity to consistently (maintaining balance) add to a bare $D_n$ Dynkin quiver is the $D_n$ singularity. This is drawn below the bare quiver when constructing the Hasse diagram because the Higgs branches of the quivers realise Slodowy slices which correspond to edges in the Hasse diagram \textit{dominating} the node at which the quiver sits, as per the discussion.}
	\label{FIGUREB}
\end{figure}

The smallest balanced $D_n$ Dynkin quiver is the flavourless trivial quiver at the top of figure \ref{FIGUREB}. All gauge nodes are trivially balanced. Note that a gauge node of rank zero cannot be balanced unless all gauge nodes are zero. Therefore the only transverse slice it is possible to add whilst maintaining balance is $D_n$, this addition is given in figure \ref{FIGUREB}. Recall that the manipulation of the quivers in performed in the Higgs branch geometry so the Hasse diagram is drawn descending from the trivial theory at the top.

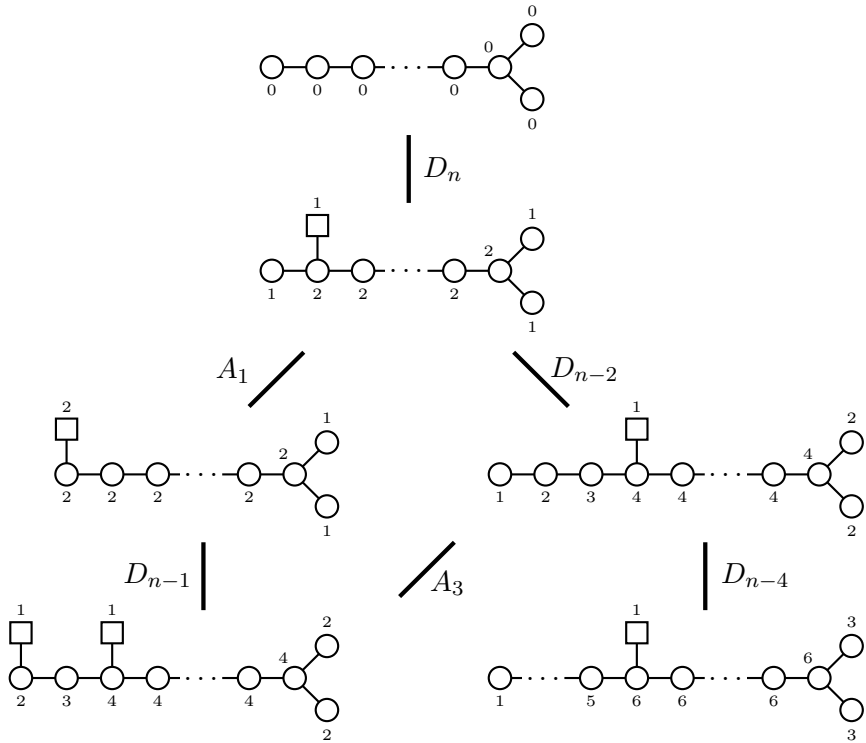
\begin{figure}
	\begin{center}
		\begin{tikzpicture}[scale = 0.6]
		\def\q{7}
		\def\k{0.7071}
		\def\size{\tiny}
		
		\draw[thick] (-1,0) -- (4,0) -- (4.7,0.7) (4,0) -- (4.7,-0.7) 
		
		;
		
		\draw[thick, fill=white] 
		(-1,0) circle (\q pt)
		(0,0) circle (\q pt)
		(1,0) circle (\q pt)
		(2,0) circle (\q pt)
		(3,0) circle (\q pt)
		(4,0) circle (\q pt)
		(4+\k,\k) circle (\q pt)
		(4+\k,-\k) circle (\q pt)
		;
		

		\draw 
		
		
		(-1,-0.5) node {\size $0$}
		(0,-0.5) node {\size $0$}
		(1,-0.5) node {\size $0$}
		(2,-0.5) node {\size $0$}
		(3,-0.5) node {\size $0$}
		(3.75,0.45) node {\size $0$}
		
		(4+\k,1.25) node {\size $0$}
		(4+\k,-1.25) node {\size $0$}
		;
		
		\draw[white, fill=white] (1.5,2) rectangle (2.4,-1);
		\draw (2,0) node {$\dots$};

		\begin{scope}[yshift = -4.5cm]
		\draw[thick] (-1,0) -- (4,0) -- (4.7,0.7) (4,0) -- (4.7,-0.7) 
		(0,0) -- (0,1)
		
		;
		
		\draw[thick, fill=white] 
		(-1,0) circle (\q pt)
		(0,0) circle (\q pt)
		(1,0) circle (\q pt)
		(2,0) circle (\q pt)
		(3,0) circle (\q pt)
		(4,0) circle (\q pt)
		(4+\k,\k) circle (\q pt)
		(4+\k,-\k) circle (\q pt)
		;
		
		\draw[thick, fill=white] 
		(0 -\q/30,1 -\q/30) rectangle (0 +\q/30,1 +\q/30)
		;

		\draw 
		(0,1.5) node {\size $1$}
		
		
		(-1,-0.5) node {\size $1$}
		(0,-0.5) node {\size $2$}
		(1,-0.5) node {\size $2$}
		(2,-0.5) node {\size $2$}
		(3,-0.5) node {\size $2$}
		(3.75,0.45) node {\size $2$}
		
		(4+\k,1.25) node {\size $1$}
		(4+\k,-1.25) node {\size $1$}
		;
		
		\draw[white, fill=white] (1.5,0.7) rectangle (2.4,-1);
		\draw (2,0) node {$\dots$};
		
		\end{scope}

		\begin{scope}[yshift = -9cm, xshift = 7cm]
		\draw[thick] (-3,0) -- (4,0) -- (4.7,0.7) (4,0) -- (4.7,-0.7) 
		(0,0) -- (0,1)
		
		;
		
		\draw[thick, fill=white] 
		(-3,0) circle (\q pt)
		(-2,0) circle (\q pt)
		(-1,0) circle (\q pt)
		(0,0) circle (\q pt)
		(1,0) circle (\q pt)
		(2,0) circle (\q pt)
		(3,0) circle (\q pt)
		(4,0) circle (\q pt)
		(4+\k,\k) circle (\q pt)
		(4+\k,-\k) circle (\q pt)
		;
		
		\draw[thick, fill=white] 
		(0 -\q/30,1 -\q/30) rectangle (0 +\q/30,1 +\q/30)
		;

		\draw 
		(0,1.5) node {\size $1$}
		
		(-3,-0.5) node {\size $1$}
		(-2,-0.5) node {\size $2$}
		(-1,-0.5) node {\size $3$}
		(0,-0.5) node {\size $4$}
		(1,-0.5) node {\size $4$}
		(2,-0.5) node {\size $4$}
		(3,-0.5) node {\size $4$}
		(3.75,0.45) node {\size $4$}
		
		(4+\k,1.25) node {\size $2$}
		(4+\k,-1.25) node {\size $2$}
		;
		
		\draw[white, fill=white] (1.5,0.7) rectangle (2.4,-1);
		\draw (2,0) node {$\dots$};
		
		\end{scope}

		\begin{scope}[yshift = -9cm, xshift = -4.5cm]
		\draw[thick] (-1,0) -- (4,0) -- (4.7,0.7) (4,0) -- (4.7,-0.7) 
		(-1,0) -- (-1,1)
		
		;
		
		\draw[thick, fill=white] 
		(-1,0) circle (\q pt)
		(0,0) circle (\q pt)
		(1,0) circle (\q pt)
		(2,0) circle (\q pt)
		(3,0) circle (\q pt)
		(4,0) circle (\q pt)
		(4+\k,\k) circle (\q pt)
		(4+\k,-\k) circle (\q pt)
		;
		
		\draw[thick, fill=white] 
		(-1 -\q/30,1 -\q/30) rectangle (-1 +\q/30,1 +\q/30)
		;

		\draw 
		(-1,1.5) node {\size $2$}
		
		
		(-1,-0.5) node {\size $2$}
		(0,-0.5) node {\size $2$}
		(1,-0.5) node {\size $2$}
		(2,-0.5) node {\size $2$}
		(3,-0.5) node {\size $2$}
		(3.75,0.45) node {\size $2$}
		
		(4+\k,1.25) node {\size $1$}
		(4+\k,-1.25) node {\size $1$}
		;
		
		\draw[white, fill=white] (1.5,0.7) rectangle (2.4,-1);
		\draw (2,0) node {$\dots$};
		
		\end{scope}

		\begin{scope}[yshift = -13.5cm, xshift = 7cm]
		\draw[thick] (-3,0) -- (4,0) -- (4.7,0.7) (4,0) -- (4.7,-0.7) 
		(0,0) -- (0,1)
		
		;
		
		\draw[thick, fill=white] 
		(-3,0) circle (\q pt)
		(-2,0) circle (\q pt)
		(-1,0) circle (\q pt)
		(0,0) circle (\q pt)
		(1,0) circle (\q pt)
		(2,0) circle (\q pt)
		(3,0) circle (\q pt)
		(4,0) circle (\q pt)
		(4+\k,\k) circle (\q pt)
		(4+\k,-\k) circle (\q pt)
		;
		
		\draw[thick, fill=white] 
		(0 -\q/30,1 -\q/30) rectangle (0 +\q/30,1 +\q/30)
		;

		\draw 
		(0,1.5) node {\size $1$}
		
		(-3,-0.5) node {\size $1$}
		(-2,-0.5) node {\size $2$}
		(-1,-0.5) node {\size $5$}
		(0,-0.5) node {\size $6$}
		(1,-0.5) node {\size $6$}
		(2,-0.5) node {\size $6$}
		(3,-0.5) node {\size $6$}
		(3.75,0.45) node {\size $6$}
		
		(4+\k,1.25) node {\size $3$}
		(4+\k,-1.25) node {\size $3$}
		;
		
		\draw[white, fill=white] (-2.5,0.7) rectangle (-1.6,-1);
		\draw (-2,0) node {$\dots$};
		
		\draw[white, fill=white] (1.5,0.7) rectangle (2.4,-1);
		\draw (2,0) node {$\dots$};
		
		\end{scope}

		\begin{scope}[yshift = -13.5cm, xshift = -4.5cm]
		\draw[thick] (-2,0) -- (4,0) -- (4.7,0.7) (4,0) -- (4.7,-0.7) 
		(-2,0) -- (-2,1)
		(0,0) -- (0,1)
		
		;
		
		\draw[thick, fill=white] 
		
		(-2,0) circle (\q pt)
		(-1,0) circle (\q pt)
		(0,0) circle (\q pt)
		(1,0) circle (\q pt)
		(2,0) circle (\q pt)
		(3,0) circle (\q pt)
		(4,0) circle (\q pt)
		(4+\k,\k) circle (\q pt)
		(4+\k,-\k) circle (\q pt)
		;
		
		\draw[thick, fill=white] 
		(-2 -\q/30,1 -\q/30) rectangle (-2 +\q/30,1 +\q/30)
		(0 -\q/30,1 -\q/30) rectangle (0 +\q/30,1 +\q/30)
		;

		\draw 
		(-2,1.5) node {\size $1$}
		(0,1.5) node {\size $1$}
		
		(-2,-0.5) node {\size $2$}
		(-1,-0.5) node {\size $3$}
		(0,-0.5) node {\size $4$}
		(1,-0.5) node {\size $4$}
		(2,-0.5) node {\size $4$}
		(3,-0.5) node {\size $4$}
		(3.75,0.45) node {\size $4$}
		
		(4+\k,1.25) node {\size $2$}
		(4+\k,-1.25) node {\size $2$}
		;
		
		\draw[white, fill=white] (1.5,0.7) rectangle (2.4,-1);
		\draw (2,0) node {$\dots$};
		
		\end{scope}

		\draw[ultra thick] (2,-1.5) -- (2,-3) (-0.3,-6.3) -- (-1.5,-7.5) (4.3,-6.3) -- (5.5,-7.5) (-2.5,-10.5) -- (-2.5,-12) (8.5,-10.5) -- (8.5,-12) (3,-10.5) -- (1.8,-11.7);
		\draw (2.75,-2.25) node {$D_n$};
		\draw (5.85,-6.7) node {$D_{n-2}$};
		\draw (-1.85,-6.7) node {$A_1$};
		
		\draw (9.6,-11.25) node {$D_{n-4}$};
		\draw (-3.5,-11.25) node {$D_{n-1}$};
		\draw (2.85,-11.4) node {$A_3$};
		\end{tikzpicture}
	\end{center}
	\caption{More of the Hasse diagram for completely balanced $D_n$ Dynkin quivers with the added singularities indicated by the labels of the edges. Again, because the quivers' Higgs branches are Slodowy slices, to be consistent with how singularity structure is read from Hasse diagrams, the larger quivers are placed lower. }
	\label{FIGURED}
\end{figure}

\begin{figure}
	\begin{center}
		\begin{tikzpicture}[xscale = 0.65, yscale = 0.65]
		\def\scriptsize{\tiny}
		
		
			\draw[ultra thick, red]
		(-0.5,2) -- (0.5,1)
		
		(0.5,1) -- (2,0)
		(0.5,0) -- (2,-1)
		
		(2,0) -- (4.5,-1)
		(2,-1) -- (4.5,-2)
		(2,-2) -- (4.5,-3)
		(2,-3) -- (3.5,-4)
		(2,-4) -- (3.5,-6)
		
		(4.5,-1) -- (8,-2)
		(4.5,-2) -- (8,-3)
		(4.5,-3) -- (8,-4)
		(5.5,-4) -- (9,-5)
		(3.5,-4) -- (7,-5)
		(4.5,-5) -- (7,-6)
		(5.5,-6) -- (9,-8)
		(3.5,-6) -- (7,-7.5)
		(4.5,-7) -- (7,-9)
		(4.5,-8) -- (7,-10.5)
		(4.5,-9) -- (7,-13)

		;
		
		\draw[ultra thick, red] (10,-9) -- (11,-9.4)
		(12,-10.35) -- (13,-10.7)
		(14,-11.6) -- (15,-11.9)
		;
		
		\begin{scope}[yshift = 1cm, xshift = -0.5cm]
		\filldraw[black] 
		(0,1) circle (4pt);
		\end{scope}
		
		\begin{scope}[yshift = 0cm, xshift = 0.5cm]
		\filldraw[black] (0,0) circle (4pt)
		(0,1) circle (4pt);
		
		\draw [ultra thick] (0,0) -- (0,1)
		
		(-0.4, 0.5) node {\scriptsize{$A_{1}$}};
		\end{scope}
		
		\draw (0.2,1.7) node[rotate = -45] {\scriptsize$D_n$}
		(1.4,0.75) node[rotate = -30] {\scriptsize$D_{n-2}$}
		(1.25,-0.2) node[rotate = -30] {\scriptsize$D_{n-1}$}
		
		(3.4,-0.3) node[rotate = -20] {\scriptsize$D_{n-4}$}
		(3.4,-1.3) node[rotate = -20] {\scriptsize$D_{n-3}$}
		(3.4,-2.3) node[rotate = -20] {\scriptsize$D_{n-2}$}
		(2.9,-3.3) node[rotate = -30] {\scriptsize$D_{n-2}$}
		(2.9,-4.7) node[rotate = -50] {\scriptsize$D_{n-1}$}
		
		(6.5,-1.3) node[rotate = -12] {\scriptsize$D_{n-6}$}
		(6.5,-2.3) node[rotate = -12] {\scriptsize$D_{n-5}$}
		(6.5,-3.3) node[rotate = -12] {\scriptsize$D_{n-4}$}
		(6.5,-4.05) node[rotate = -12] {\scriptsize$D_{n-3}$}
		(6.2,-4.55) node[rotate = -12] {\scriptsize$D_{n-4}$}
		(5.9,-5.3) node[rotate = -16] {\scriptsize$D_{n-3}$}
		(6.25,-6.15) node[rotate = -28] {\scriptsize$D_{n-2}$}
		(5.9,-6.75) node[rotate = -23] {\scriptsize$D_{n-3}$}
		(5.9,-7.8) node[rotate = -35] {\scriptsize$D_{n-2}$}
		(5.9,-9) node[rotate = -41] {\scriptsize$D_{n-2}$}
		(5.9,-10.75) node[rotate = -55] {\scriptsize$D_{n-1}$}
		;
		
		\begin{scope}[xshift=2cm, yshift = -4cm]
		\filldraw[black] (0,0) circle (4pt)
		(0,1) circle (4pt)
		(0,2) circle (4pt)
		(0,3) circle (4pt)
		(0,4) circle (4pt);
		
		\draw [ultra thick] (0,0) -- (0,4)
		(-0.3, 0.5) node {\scriptsize{$a_{3}$}}
		(-0.3, 1.5) node {\scriptsize{$a_{1}$}}
		(-0.3, 2.5) node {\scriptsize{$A_{1}$}}
		(0.35, 3.4) node {\scriptsize{$A_{3}$}}
		;
		\end{scope}
		
		\begin{scope}[xshift = 4.5cm, yshift = -9cm]
		\filldraw[black] (0,0) circle (4pt)
		(0,1) circle (4pt)
		(0,2) circle (4pt)
		(0,4) circle (4pt)
		(0,6) circle (4pt)
		(0,7) circle (4pt)
		(0,8) circle (4pt)
		(-1,3) circle (4pt)
		(1,3) circle (4pt)
		(-1,5) circle (4pt)
		(1,5) circle (4pt)
		;
		
		\draw [ultra thick] (0,0) -- (0,2) -- (-1,3) -- (0,4) -- (-1,5) -- (0,6) -- (0,8) -- (0,6) -- (1,5) -- (0,4) -- (1,3) -- (0,2)
		(-0.4, 0.5) node {\scriptsize{$a_{5}$}}
		(-0.4, 1.5) node {\scriptsize{$a_{3}$}}
		(0.33, 2.8) node {\scriptsize{$a_{1}$}}
		(-0.7, 2.3) node {\scriptsize{$a_{1}$}}
		(0.33, 3.33) node {\scriptsize{$a_{2}$}}
		(-0.7, 3.7) node {\scriptsize{$a_{2}$}}
		(0.6, 4.2) node {\scriptsize{$A_{2}$}}
		(-0.7, 4.3) node {\scriptsize{$A_{2}$}}
		(0.33, 5.33) node {\scriptsize{$A_{1}$}}
		(-0.7, 5.7) node {\scriptsize{$A_{1}$}}
		(0.4, 6.5) node {\scriptsize{$A_{3}$}}
		(0.4, 7.5) node {\scriptsize{$A_{5}$}}
		;
		\end{scope}
		
		\begin{scope}[xshift = 8cm, yshift = -16cm]

		\draw[ultra thick, red] (0,14) -- (1,13.85) (0,0) -- (0.5,-0.8);
		
		\filldraw[black] (0,0) circle (4pt)
		(0,1) circle (4pt)
		(0,2) circle (4pt)
		(0,12) circle (4pt)
		(0,13) circle (4pt)
		(0,14) circle (4pt)
		(1,3) circle (4pt)
		(1,4) circle (4pt)
		(1,5) circle (4pt)
		(1,6) circle (4pt)
		(1,7) circle (4pt)
		(1,8) circle (4pt)
		(1,9) circle (4pt)
		(1,10) circle (4pt)
		(1,11) circle (4pt)
		(-1,3) circle (4pt)
		(-1,4) circle (4pt)
		(-1,5.5) circle (4pt)
		(-1,7) circle (4pt)
		(-1,8.5) circle (4pt)
		(-1,10) circle (4pt)
		(-1,11) circle (4pt);

		\draw[ultra thick]  (0,0) -- (0,2) -- (-1,3) -- (-1,11) -- (0,12) -- (0,14) -- (0,12) -- (1,11) -- (1,3) -- (0,2) -- (1,3) -- (-1,4) -- (1,5) -- (1,6) -- (-1,7) -- (1,8) -- (1,9) -- (-1,10) -- (1,11) ;

		\begin{scope}[xscale = -1]
		\draw (0.4, 13.5) node {\scriptsize{$A_{7}$}}
		(0.4, 12.5) node {\scriptsize{$A_{5}$}}
		(0.4, 1.5) node {\scriptsize{$a_{5}$}}
		(0.4, 0.5) node {\scriptsize{$a_{7}$}}
		
		(0, 10.8) node {\scriptsize{$A_{2}$}}
		(0, 9.2) node {\scriptsize{$A_{2}$}}
		(0, 7.8) node {\scriptsize{$A_{1}$}}
		(0, 6.2) node {\scriptsize{$A_{1}$}}
		(0, 4.8) node {\scriptsize{$a_{2}$}}
		(0, 3.8) node {\scriptsize{$a_{2}$}}
		
		(-0.9, 2.3) node {\scriptsize{$a_{3}$}}
		(-1.3, 3.5) node {\scriptsize{$A_{1}$}}
		(-1.3, 4.5) node {\scriptsize{$a_{3}$}}
		(-1.3, 5.5) node {\scriptsize{$A_{1}$}}
		(-1.3, 6.5) node {\scriptsize{$A_{1}$}}
		(-1.3, 7.5) node {\scriptsize{$A_{1}$}}
		(-1.3, 8.5) node {\scriptsize{$A_{1}$}}
		(-1.3, 9.5) node {\scriptsize{$A_{3}$}}
		(-1.3, 10.5) node {\scriptsize{$A_{1}$}}
		(-0.9, 11.7) node {\scriptsize{$A_{3}$}}
		
		(0.8, 2.4) node {\scriptsize{$A_{1}$}}
		(0.7, 3.5) node {\scriptsize{$a_{4}$}}
		(1.3, 4.75) node {\scriptsize{$A_{2}$}}
		(1.3, 6.25) node {\scriptsize{$a_{3}$}}
		(1.3, 7.75) node {\scriptsize{$A_{3}$}}
		(0.7, 9.45) node {\scriptsize{$a_{2}$}}
		(0.65, 10.6) node {\scriptsize{$A_{4}$}}
		(0.6, 11.85) node {\scriptsize{$A_{1}$}}
		;
		\end{scope}
		\end{scope}

		\draw (11.5,-9.9) node {$\mathcal{P}(10)$}
		(13.5,-11.2) node {$\mathcal{P}(12)$}
		
		(15.7,-12) node[rotate = -13] {$\bs\dots$}
		;
		\end{tikzpicture}
	\end{center}
	\caption{The Hasse diagram resulting from balanced quiver addition for a generic $D_n$ Dynkin quiver. The diagram takes the form of linear subdiagrams of even magnitude partition Hasse diagrams (black) with $D_k$ singularity traversing structure (red). The Hasse diagram for the maximal special slice is readily identifiable as the top two lines of traversing structure. Clearly when a specific value of $n$ is chosen some of the traversing structure edges become undefined, this is dealt with momentarily with an editing prescription.}
	\label{FIGUREE}
\end{figure}
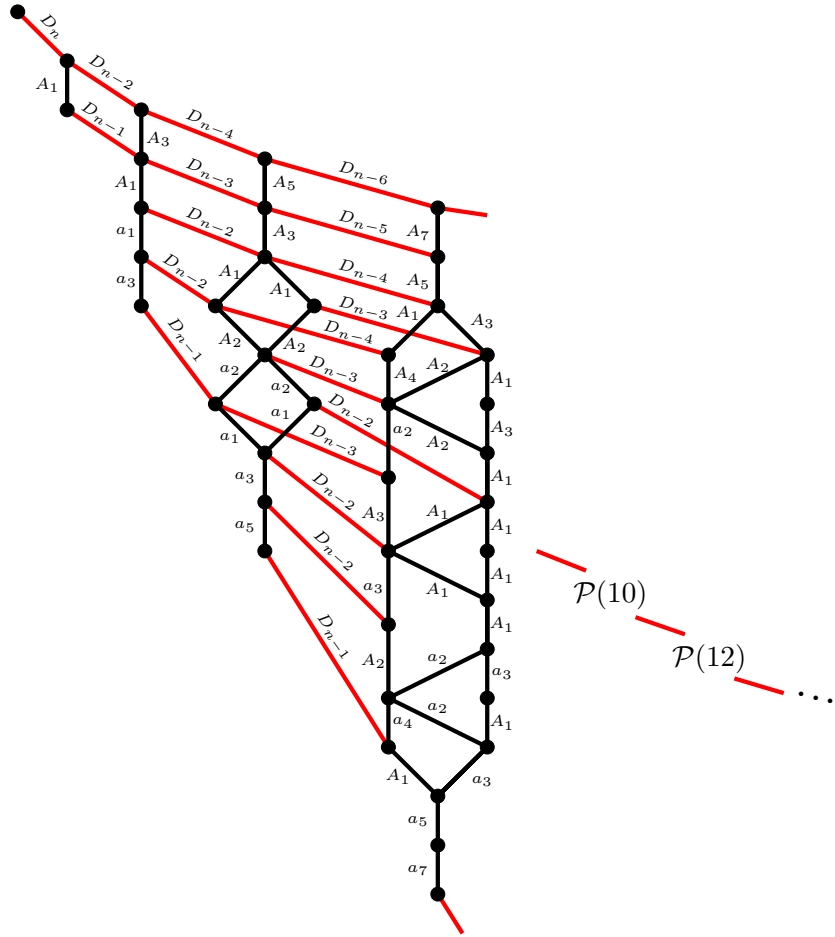

From here there are two options. The single flavour on the second node might have been the result of an $A_1$ transition in the end node of the tail, or a $D_{n-2}$ transition in the third through $n^\textrm{th}$ nodes. Both of these possibilities are added, figure \ref{FIGURED}. Now the only transverse slice that can be added to the right hand theory is $D_{n-1}$, whereas the left hand theory could have an $A_3$ or $D_{n-4}$ added to it. Note that adding $D_{n-1}$ to the right hand theory or $A_3$ to the left hand theory results in the same parent quiver. Each theory should be a single node in the Hasse diagram\footnote{There is a problem with over-naming for $D_4$ because it is more symmetrical, but we defer discussion of this to the future, for now we will treat the tail as different from the end nodes.} and so the structure drawn in figure \ref{FIGURED} reflects this. 
This process may be continued indefinitely but the structure is very regular. For general\footnote{What happens when a specific value of $n$ is chosen is explored momentarily.} $n$ the structure of the balanced $D_n$ Dynkin quiver Hasse diagram is given in figure \ref{FIGUREE}. 

The structure of the balanced Hasse diagram is that of Hasse diagrams for partitions of even integers with $D_k$ traversing structure. This provides the means to classify the balanced $D_n$ Dynkin diagrams very neatly. The problem of classification is now as follows: For a given $n$, how can every node of this balanced Hasse diagram be labelled uniquely? The interpretation of the Hasse diagram as even magnitude partition subdiagrams with traversing structure provides an easy answer. Assign to every node in figure \ref{FIGUREE} the partition, $\kappa$, which denotes the place of that node in its partition Hasse subdiagram\footnote{Importantly, these partitions are \textit{not} partitions for the nilpotent varieties of $\kk{so}_{2n}$. These even magnitude partitions are not otherwise restricted.}. In this way every balanced $D_n$ Dynkin quiver can be specified with the integer $n$ and the partition $\kappa$. The class of balanced $D_n$ Dynkin quivers is therefore denoted $D_\kappa(n)$.

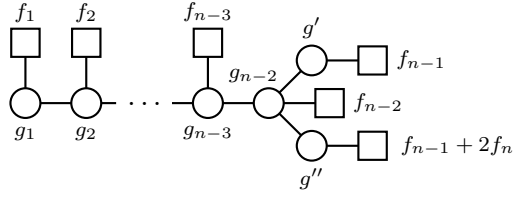
\begin{figure}
	\begin{center}
		\begin{tikzpicture}[scale = 0.8]
		\def\q{7}
		\def\k{0.7071}
		\def\size{\scriptsize}
		
		\draw[thick] (0,0) -- (4,0) -- (4.7,0.7) (4,0) -- (4.7,-0.7) (0,0) -- (0,1)
		(1,0) -- (1,1)
		(2,0) -- (2,1)
		(3,0) -- (3,1)
		(4,0) -- (5,0)
		
		(4+\k,\k) -- (5+\k,\k)
		(4+\k,-\k) -- (5+\k,-\k)
		;
		
		\draw[thick, fill=white] (0,0) circle (\q pt)
		(1,0) circle (\q pt)
		(2,0) circle (\q pt)
		(3,0) circle (\q pt)
		(4,0) circle (\q pt)
		(4+\k,\k) circle (\q pt)
		(4+\k,-\k) circle (\q pt)
		;
		
		\draw[thick, fill=white] 
		(0 -\q/30,1 -\q/30) rectangle (0 +\q/30,1 +\q/30)
		(1 -\q/30,1 -\q/30) rectangle (1 +\q/30,1 +\q/30)
		(2 -\q/30,1 -\q/30) rectangle (2 +\q/30,1 +\q/30)
		(3 -\q/30,1 -\q/30) rectangle (3 +\q/30,1 +\q/30)
		(5 -\q/30,0 -\q/30) rectangle (5 +\q/30,0 +\q/30)
		(5+\k -\q/30,\k -\q/30) rectangle (5+\k +\q/30,\k +\q/30)
		(5+\k -\q/30,-\k -\q/30) rectangle (5+\k +\q/30,-\k +\q/30)
		;

		\draw (0,1.5) node {\size $f_1$}
		(1,1.5) node {\size $f_2$}
		(2,1.5) node {\size $f_3$}
		(3,1.5) node {\size $f_{n-3}$}
		
		(5.8,0) node  {\size $f_{n-2}$}
		(5.8 +\k,0+\k) node  {\size $f_{n-1}$}
		(6.4+\k,0-\k) node  {\size $f_{n-1}+2f_n$}

		(0,-0.5) node {\size $g_1$}
		(1,-0.5) node {\size $g_2$}
		(2,-0.5) node {\size $g_3$}
		(3,-0.5) node {\size $g_{n-3}$}
		(3.75,0.45) node {\size $g_{n-2}$}
		
		(4+\k,1.25) node {\size $g'$}
		(4+\k,-1.25) node {\size $g''$}
		;
		
		\draw[white, fill=white] (1.5,2) rectangle (2.4,-1);
		\draw (2,0) node {$\dots$};

		\end{tikzpicture}
	\end{center}
	\caption{The necessary form of a balanced $D_n$ Dynkin quiver. This naming convention for the flavours is the one used for all the quivers of even type and allows for easy interpretation between a theory's quiver, name and moduli space singularity structure.  }
	\label{FIGUREG}
\end{figure}

Assigning these partitions to the theories has a natural interpretation in terms of the distribution of flavours across the quiver. By writing the flavours as in figure \ref{FIGUREG}, the partition to which the theory is assigned is
\begin{equation}\label{partitionbal}
\kappa  = (n^{f_n}, n-1^{f_{n-1}},\dots,2^{f_2},1^{f_1}).
\end{equation} 

Having labelled the nodes with these partitions the balanced Hasse diagram can be written more compactly as a sequence of partition Hasse diagrams along with an edge diagram capturing the traversing structure:
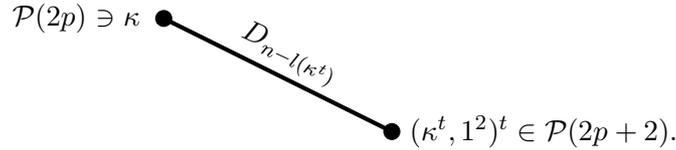
\begin{center}
	\begin{tikzpicture}
	\draw[fill=black] (0,0) circle (3pt)
	(3,-1.5) circle (3pt)
	;
	\draw[ultra thick] (0,0) -- (3,-1.5);
	
	\draw (-1.15,0) node {$\mathcal{P}(2p) \ni \kappa$}
	(5,-1.5) node {$(\kappa^t,1^2)^t \in \mathcal{P}(2p+2).$}
	(1.65,-0.475) node[rotate=-25] {$D_{n-l(\kappa^t)}$}
	;
	
	\end{tikzpicture}
\end{center}

It is necessary to consider when the construction given in figure \ref{FIGUREE} needs editing. Quiver addition can be performed forever. For a specific and finite $n$, there are obvious issues which arise. For a node $\kappa$ such that $n-l(\kappa^t) < 2$, the traversing edge which descends from the node represents a singularity which is not defined. In such circumstances one needs to edit the construction by removing or replacing some nodes and edges in a systematic way. This will be discussed in a moment.

We begin with a proof that the construction in figure \ref{FIGUREE} does indeed contain every balanced $D_n$ Dynkin quiver by proving that balancing the quiver in figure \ref{FIGUREG} requires the flavour to be distributed in the quiver such that the partition (\ref{partitionbal}) is of even magnitude. Since two different partitions necessarily give a different flavour distribution, such a proof demonstrates that every balanced $D_n$ quiver lives at a unique node in figure \ref{FIGUREE}, and so the classification for balanced quivers is complete.

\vspace{1mm}

\noindent \textbf{Proposition.} $\quad$ A balanced $D_n$ Dynkin quiver must take the form of figure \ref{FIGUREG} with an \textit{even} magnitude partition. Also $l(\kappa) = g_1$ and $|\kappa| = 2g''$.

\vspace{1mm}

\noindent \textbf{Corollary.} $\quad$ The class $D_\kappa(n)$ contains every balanced $D_n$ Dynkin quiver. 

\vspace{1mm}

\noindent \textbf{Proof.} $\quad$ Proving the proposition and corollary above requires demonstrating that balancing figure \ref{FIGUREG} requires $\kappa$ to have even magnitude, that is, $|\kappa| = \sum_{i=1}^nif_i=2x$ for some $x$. Also recall that $l(\kappa) = \sum_{i=1}^nf_i$.

The left hand gauge node of figure \ref{FIGUREG} is of rank $g_1$. Balancing this node requires $2g_1 = f_1+g_2$ and so $g_2 = 2g_1 - f_1$. Now consider balancing the second node, this requires that $2g_2 = g_1 + f_2 + g_3$ and so $g_3 = 2g_2 - g_1 - f_2 = 3g_1 - 2f_1 - f_2$. Balancing the nodes along the tail one at a time in this manner yields the balance criteria
\begin{equation}\label{gbalance}
g_k = kg_1 - \sum_{i=1}^{k-1}(k-i)f_i
\end{equation}
for $k\leq n-2$. Applying this to the $n-2^\textrm{th}$ node and rearranging terms yields
\begin{equation}\label{n2kappa}
\begin{split}
g_{n-2} & = (n-2)g_1 - \sum_{i=1}^{n-3}(n-2-i)f_i\\ 
& = (n-2)g_1 - (n-2)\sum_{i=1}^{n-3}f_i + |\kappa| - \sum_{i=n-2}^nif_i.
\end{split}
\end{equation}
This is one of two equations which give a balancing condition on $g_{n-2}$. The other comes from balancing $g_{n-2}$ directly such that $2g_{n-2} = g_{n-3} +f_{n-2} +g'+g''$. Using the balance requirements of $g'$ and $g''$, that is, $2g' = g_{n-2} + f_{n-1}$ and $2g'' = g_{n-2} + f_{n-1} + 2f_n$, gives
\begin{equation}\label{gn2balance}
g_{n-2} = g_{n-3} + \sum_{i=n-2}^nf_i
\end{equation}
Writing $g_{n-3}$ explicitly using (\ref{gbalance}), equating (\ref{n2kappa}) to (\ref{gn2balance}) and rearranging for $|\kappa|$ gives
\[
\begin{split}
|\kappa| & = -g_1 -(n-3)\sum_{i=1}^{n-4}f_i + \sum_{i=1}^{n-4}if_i + \sum_{i=n-2}^nf_i + (n-2)\sum_{i=1}^{n-3}f_i + \sum_{i=n-2}^nif_i \\
& = -g_1 + \sum_{i=1}^nif_i - (n-3)f_{n-3} -(n-3)\sum_{i=1}^{n-4}f_i + (n-3)\sum_{i=1}^{n-3}f_i + \sum_{i=1}^{n-3}f_i + \sum_{i=n-2}^nf_i \\
& = -g_1 + l(\kappa) + |\kappa|,
\end{split}
\]
so $g_1 = l(\kappa)$. When moving from lines one to two the third and sixth terms go to make the second and third terms. The first, second and fourth terms are pulled through and the fifth term is split into terms five and six. From lines two to three, term two is just $|\kappa|$, the last two terms come together to form $l(\kappa)$, the middle three terms cancel. Returning to (\ref{n2kappa}), rearranging for $|\kappa|$ and using $g_1 = l(\kappa)$ gives
\begin{equation}
\begin{split}\label{twogee}
|\kappa| & = g_{n-2} - (n-2)l(\kappa) - (n-2)\sum_{i=1}^{n-3}f_i + \sum_{i=n-2}^nif_i \\
& = g_{n-2} -(n-2)(f_{n-2}+f_{n-1}+f_n) + (n-2)f_{n-2} + (n-1)f_{n-1} + nf_n \\
& = 2f_n + f_{n-1} + g_{n-2} \\
& = 2g'',
\end{split}
\end{equation}
where the $g''$ balance requirement was used in the final line (the $g'$ balance requirement would have also worked). This completes the proof that balancing the quiver imposes that the magnitude of $\kappa$ is even, proving the proposition. Every node in the balanced Hasse diagram is a different quiver. Each quiver is determined completely by $n$ and $\kappa$. Therefore exactly one balanced quiver can be associated to each $(n,\kappa)$ pair, this is the quiver which appears in the balanced Hasse diagram, figure \ref{FIGUREE}. All balanced $D_n$ Dynkin quivers are present so the class $D_\kappa(n)$ contains all balanced $D_n$ Dynkin quivers exactly once. 

\vspace{2mm}

\begin{figure}
	\begin{center}
		\begin{tikzpicture}[scale = 0.6]
		\def\q{7}
		\def\k{0.7071}
		\def\size{\tiny}
		
		\draw[thick] (0,0) -- (4,0) -- (4.7,0.7) (4,0) -- (4.7,-0.7) (0,0) -- (0,1)
		(1,0) -- (1,1)
		(2,0) -- (2,1)
		(3,0) -- (3,1)
		(4,0) -- (5,0)
		
		(4+\k,\k) -- (5+\k,\k)
		(4+\k,-\k) -- (5+\k,-\k)
		;
		
		\draw[thick, fill=white] (0,0) circle (\q pt)
		(1,0) circle (\q pt)
		(2,0) circle (\q pt)
		(3,0) circle (\q pt)
		(4,0) circle (\q pt)
		(4+\k,\k) circle (\q pt)
		(4+\k,-\k) circle (\q pt)
		;
		
		\draw[thick, fill=white] 
		(0 -\q/30,1 -\q/30) rectangle (0 +\q/30,1 +\q/30)
		(1 -\q/30,1 -\q/30) rectangle (1 +\q/30,1 +\q/30)
		(2 -\q/30,1 -\q/30) rectangle (2 +\q/30,1 +\q/30)
		(3 -\q/30,1 -\q/30) rectangle (3 +\q/30,1 +\q/30)
		(5 -\q/30,0 -\q/30) rectangle (5 +\q/30,0 +\q/30)
		(5+\k -\q/30,\k -\q/30) rectangle (5+\k +\q/30,\k +\q/30)
		(5+\k -\q/30,-\k -\q/30) rectangle (5+\k +\q/30,-\k +\q/30)
		;

		\draw (0,1.5) node {\size $f_1$}
		(1,1.5) node {\size $f_2$}
		(2,1.5) node {\size $f_3$}
		(3,1.5) node {\size $f_{n-3}$}
		
		(5.8,0) node  {\size $f_{n-2}$}
		(5.8 +\k,0+\k) node  {\size $f_{n-1}$}
		(6.4+\k,0-\k) node  {\size $f_{n-1}+2f_n$}

		(0,-0.5) node {\size $g_1$}
		(1,-0.5) node {\size $g_2$}
		(2,-0.5) node {\size $g_3$}
		(3,-0.5) node {\size $g_{n-3}$}
		(3.75,0.45) node {\size $g_{n-2}$}
		
		(4+\k,1.25) node {\size $g'$}
		(4+\k,-1.25) node {\size $g''$}
		;
		
		\draw[white, fill=white] (-1,2) rectangle (2.4,-1);
		\draw (2,0) node {$\dots$};

		\draw (-6,0) node {$(n^{f_n}, n-1^{f_{n-1}}, n-2^{f_{n-2}}, \dots)$}
		(-6,-3.5) node {$(n^{f_n+1}, n-1^{f_{n-1}}, n-2^{f_{n-2}-1}, \dots )$}
		;

		\begin{scope}[yshift = -3.5cm]
		\draw[thick] (0,0) -- (4,0) -- (4.7,0.7) (4,0) -- (4.7,-0.7) (0,0) -- (0,1)
		(1,0) -- (1,1)
		(2,0) -- (2,1)
		(3,0) -- (3,1)
		(4,0) -- (5,0)
		
		(4+\k,\k) -- (5+\k,\k)
		(4+\k,-\k) -- (5+\k,-\k)
		;
		
		\draw[thick, fill=white] (0,0) circle (\q pt)
		(1,0) circle (\q pt)
		(2,0) circle (\q pt)
		(3,0) circle (\q pt)
		(4,0) circle (\q pt)
		(4+\k,\k) circle (\q pt)
		(4+\k,-\k) circle (\q pt)
		;
		
		\draw[thick, fill=white] 
		(0 -\q/30,1 -\q/30) rectangle (0 +\q/30,1 +\q/30)
		(1 -\q/30,1 -\q/30) rectangle (1 +\q/30,1 +\q/30)
		(2 -\q/30,1 -\q/30) rectangle (2 +\q/30,1 +\q/30)
		(3 -\q/30,1 -\q/30) rectangle (3 +\q/30,1 +\q/30)
		(5 -\q/30,0 -\q/30) rectangle (5 +\q/30,0 +\q/30)
		(5+\k -\q/30,\k -\q/30) rectangle (5+\k +\q/30,\k +\q/30)
		(5+\k -\q/30,-\k -\q/30) rectangle (5+\k +\q/30,-\k +\q/30)
		;

		\draw (0,1.5) node {\size $f_1$}
		(1,1.5) node {\size $f_2$}
		(2,1.5) node {\size $f_3$}
		(3,1.5) node {\size $f_{n-3}$}
		
		(6.2,0) node  {\size $f_{n-2}-1$}
		(5.8 +\k,0+\k) node  {\size $f_{n-1}$}
		(6.9+\k,0-\k) node  {\size $f_{n-1}+2f_n+2$}

		(0,-0.5) node {\size $g_1$}
		(1,-0.5) node {\size $g_2$}
		(2,-0.5) node {\size $g_3$}
		(3,-0.5) node {\size $g_{n-3}$}
		(3.75,0.45) node {\size $g_{n-2}$}
		
		(4+\k,1.25) node {\size $g'$}
		(4+\k,-1.25) node {\size $g''+1$}
		;
		
		\draw[white, fill=white] (-1,2) rectangle (2.4,-1);
		\draw (2,0) node {$\dots$};
		
		\end{scope}
		
		\draw (0,0) node {$\sim$};
		\draw (0.25,-3.5) node {$\sim$};
		
		\draw (-4,-1.75) node {$a_{f_{n-1}+2f_n+1}$};
		
		\draw[ultra thick] (-6,-0.75) -- (-6,-2.75);

		\begin{scope}[yshift = -11.5cm, xshift = 0cm]
		\draw[thick] (-2.5,0) -- (4,0) -- (4.7,0.7) (4,0) -- (4.7,-0.7) 
		(-1,0) -- (-1,1)
		(0,0) -- (0,1)
		
		(4+\k,\k) -- (5+\k,\k)
		(4+\k,-\k) -- (5+\k,-\k)
		;
		
		\draw[thick, fill=white] 
		
		(-2,0) circle (\q pt)
		(-1,0) circle (\q pt)
		(0,0) circle (\q pt)
		(1,0) circle (\q pt)
		(2,0) circle (\q pt)
		(3,0) circle (\q pt)
		(4,0) circle (\q pt)
		(4+\k,\k) circle (\q pt)
		(4+\k,-\k) circle (\q pt)
		;
		
		\draw[thick, fill=white] 
		(-1 -\q/30,1 -\q/30) rectangle (-1 +\q/30,1 +\q/30)
		(0 -\q/30,1 -\q/30) rectangle (0 +\q/30,1 +\q/30)
		(5+\k -\q/30,\k -\q/30) rectangle (5+\k +\q/30,\k +\q/30)
		(5+\k -\q/30,-\k -\q/30) rectangle (5+\k +\q/30,-\k +\q/30)
		;

		\draw 
		(-1.25,1.5) node {\size $f_{n-j-1}$}
		(0,1.5) node {\size $1$}
		
		(5.5 +\k,0+\k) node  {\size $1$}
		(6.15+\k,0-\k) node  {\size $2f_n+1$}
		(-2,0.5) node {\size $g_{n-j-2}$}
		(-1.85,-0.55) node[rotate = 10] {\size $g_{n-j-1}$}
		(0,-0.5) node[rotate = 0] {\size $g_{n-j}+1$}
		(2.2,-0.65) node[rotate = -10] {\size $g_{n-j+1}+1$}
		(3,0.5) node {\size $g_{n-3}+1$}
		(5.3,0) node[rotate = -0] {\size $g_{n-2}+1$}
		
		(4+\k,1.25) node {\size $g'+1$}
		(4+\k,-1.25) node {\size $g''$}
		;
		
		\draw[white, fill=white] (1.5,0.3) rectangle (2.4,-0.3);
		\draw (2,0) node {$\dots$};
		\draw (-3,0) node {$\dots$};
		
		\end{scope}
		
		\begin{scope}[yshift = -8cm, xshift = 0cm]
		\draw[thick] (-2.5,0) -- (4,0) -- (4.7,0.7) (4,0) -- (4.7,-0.7) 
		(-1,0) -- (-1,1)
		
		(4+\k,-\k) -- (5+\k,-\k)
		;
		
		\draw[thick, fill=white] 
		
		(-2,0) circle (\q pt)
		(-1,0) circle (\q pt)
		(0,0) circle (\q pt)
		(1,0) circle (\q pt)
		(2,0) circle (\q pt)
		(3,0) circle (\q pt)
		(4,0) circle (\q pt)
		(4+\k,\k) circle (\q pt)
		(4+\k,-\k) circle (\q pt)
		;
		
		\draw[thick, fill=white] 
		(-1 -\q/30,1 -\q/30) rectangle (-1 +\q/30,1 +\q/30)
		(5+\k -\q/30,-\k -\q/30) rectangle (5+\k +\q/30,-\k +\q/30)
		;

		\draw 
		(-1,1.5) node {\size $f_{n-j-1}+1$}
		
		(6.2+\k,0-\k) node  {\size $2f_n+2$}
		(-2,0.5) node {\size $g_{n-j-2}$}
		(-1,-0.5) node {\size $g_{n-j-1}$}
		(0,0.5) node {\size $g_{n-j}$}
		(1,-0.5) node {\size $g_{n-j+1}$}
		(3,-0.5) node {\size $g_{n-3}$}
		(3.75,0.45) node {\size $g_{n-2}$}
		
		(4+\k,1.25) node {\size $g'$}
		(4+\k,-1.25) node {\size $g''$}
		;
		
		\draw[white, fill=white] (1.5,0.7) rectangle (2.4,-0.3);
		\draw (2,0) node {$\dots$};
		\draw (-3,0) node {$\dots$};
		
		\end{scope}
		
		\begin{scope}[yshift = -8cm, xshift = -4.5cm]
		\draw (-6,0) node {$(n^{f_n+1}, n-j-1^{f_{n-j-1}+1}, \dots)$}
		(-6,-3.5) node {$(n^{f_n}, n-1,n-j, n-j-1^{f_{n-j-1}}, \dots )$}
		;
		\draw (-0.25,0) node {$\sim$};
		\draw (0.25,-3.5) node {$\sim$};
		
		\draw (-4.5,-1.75) node {$A_j \cup A_j$};
		
		\draw[ultra thick] (-6,-0.75) -- (-6,-2.75);
		
		\end{scope}
		
		\end{tikzpicture}
	\end{center}
	\caption{The editing prescription for the balanced $D_n$ Hasse diagram given in figure \ref{FIGUREE} when a specific value of $n$ is chosen and some edges become undefined. Once the offending edges and nodes have been removed, some edges must be added back into the Hasse diagram. This can be calculated from the point of view of the partitions assigned to the nodes between which the edges lie, or by appealing to the structure of the quiver in those circumstances, both are presented here.}
	\label{FIGUREH}
\end{figure}
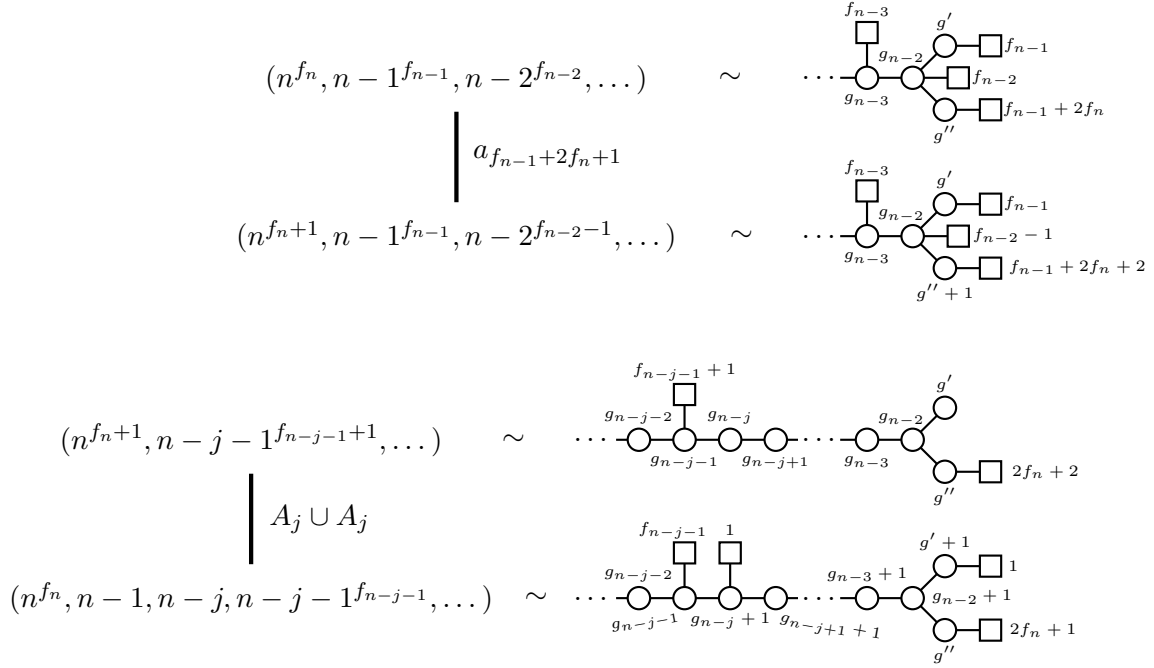

There remains the task of establishing a systematic editing of the figure \ref{FIGUREE} construction when a specific value of $n$ is chosen. This can be investigated in a number of corresponding ways. As mentioned already, for $\kappa$ with $n-l(\kappa^t) < 2$ some edge labels become undefined. It is also clear that there is no interpretation in the style of (\ref{partitionbal}) and figure \ref{FIGUREG} when the largest part of $\kappa$ is larger than $n$. For $n-l(\kappa^t) < 2$ to be true it must be the case that $\kappa$ contains parts larger than $n-2$. Therefore the traversing structure for the nodes labelled with $\kappa$ with no parts larger than $n-2$ is unaffected in the editing. There is also simply no theory corresponding to partitions with largest part larger than $n$. The editing prescription is then as follows:

\vspace{2mm}

\noindent \textbf{Editing prescription} $\quad$ To write down the balanced Hasse diagram for $D_n$ Dynkin quivers for some specific $n$, start with the general construction figure \ref{FIGUREE}. Identify in this construction all of the nodes with parts larger than $n$ and delete them and all the edges depending on them. Now identify the partitions with one or more parts equal to $n$ and/or $n-1$, change the edges coming from these nodes systematically using figure \ref{FIGUREH}. 

\vspace{2mm}

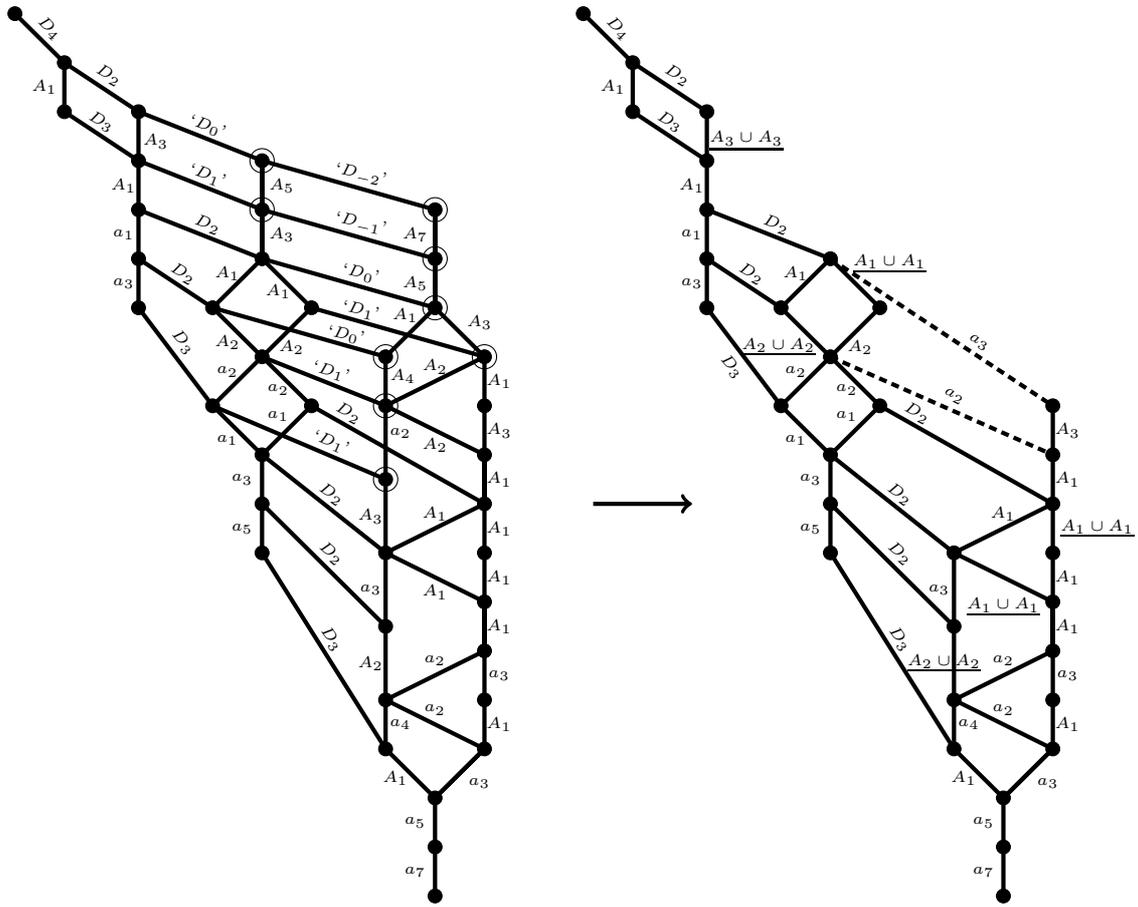
\begin{figure}
	\begin{center}
		\begin{tikzpicture}[xscale = 0.65, yscale = 0.65]
		\def\scriptsize{\tiny}
		
		
		\begin{scope}[yshift = 1cm, xshift = -0.5cm]
		\filldraw[black] 
		(0,1) circle (4pt);
		\end{scope}
		
		\begin{scope}[yshift = 0cm, xshift = 0.5cm]
		\filldraw[black] (0,0) circle (4pt)
		(0,1) circle (4pt);
		
		\draw [ultra thick] (0,0) -- (0,1)
		
		(-0.4, 0.5) node {\scriptsize{$A_{1}$}};
		\end{scope}
		
		\draw (0.2,1.7) node[rotate = -45] {\scriptsize$D_4$}
		(1.4,0.75) node[rotate = -30] {\scriptsize$D_{2}$}
		(1.25,-0.2) node[rotate = -30] {\scriptsize$D_{3}$}
		
		(3.4,-0.3) node[rotate = -20] {\scriptsize$`D_{0}\textrm{'}$}
		(3.4,-1.3) node[rotate = -20] {\scriptsize$`D_{1}\textrm{'}$}
		(3.4,-2.3) node[rotate = -20] {\scriptsize$D_{2}$}
		(2.9,-3.3) node[rotate = -30] {\scriptsize$D_{2}$}
		(2.9,-4.7) node[rotate = -50] {\scriptsize$D_{3}$}
		
		(6.5,-1.3) node[rotate = -12] {\scriptsize$`D_{-2}\textrm{'}$}
		(6.5,-2.3) node[rotate = -12] {\scriptsize$`D_{-1}\textrm{'}$}
		(6.5,-3.3) node[rotate = -12] {\scriptsize$`D_{0}\textrm{'}$}
		(6.5,-4.05) node[rotate = -12] {\scriptsize$`D_{1}\textrm{'}$}
		(6.2,-4.55) node[rotate = -12] {\scriptsize$`D_{0}\textrm{'}$}
		(5.9,-5.3) node[rotate = -16] {\scriptsize$`D_{1}\textrm{'}$}
		(6.25,-6.15) node[rotate = -28] {\scriptsize$D_{2}$}
		(5.9,-6.75) node[rotate = -23] {\scriptsize$`D_{1}\textrm{'}$}
		(5.9,-7.8) node[rotate = -35] {\scriptsize$D_{2}$}
		(5.9,-9) node[rotate = -41] {\scriptsize$D_{2}$}
		(5.9,-10.75) node[rotate = -55] {\scriptsize$D_{3}$}
		;
		
		\begin{scope}[xshift=2cm, yshift = -4cm]
		\filldraw[black] (0,0) circle (4pt)
		(0,1) circle (4pt)
		(0,2) circle (4pt)
		(0,3) circle (4pt)
		(0,4) circle (4pt);
		
		\draw [ultra thick] (0,0) -- (0,4)
		(-0.3, 0.5) node {\scriptsize{$a_{3}$}}
		(-0.3, 1.5) node {\scriptsize{$a_{1}$}}
		(-0.3, 2.5) node {\scriptsize{$A_{1}$}}
		(0.35, 3.4) node {\scriptsize{$A_{3}$}}
		;

		\end{scope}

		\begin{scope}[xshift = 4.5cm, yshift = -9cm]
		\filldraw[black] (0,0) circle (4pt)
		(0,1) circle (4pt)
		(0,2) circle (4pt)
		(0,4) circle (4pt)
		(0,6) circle (4pt)
		(0,7) circle (4pt) 
		(0,8) circle (4pt)
		(-1,3) circle (4pt)
		(1,3) circle (4pt)
		(-1,5) circle (4pt)
		(1,5) circle (4pt)
		;
		
		\draw[black] 
		(0,7) circle (7pt) 
		(0,8) circle (7pt)
		;
		
		\draw [ultra thick] (0,0) -- (0,2) -- (-1,3) -- (0,4) -- (-1,5) -- (0,6) -- (0,8) -- (0,6) -- (1,5) -- (0,4) -- (1,3) -- (0,2)
		(-0.4, 0.5) node {\scriptsize{$a_{5}$}}
		(-0.4, 1.5) node {\scriptsize{$a_{3}$}}
		(0.33, 2.8) node {\scriptsize{$a_{1}$}}
		(-0.7, 2.3) node {\scriptsize{$a_{1}$}}
		(0.33, 3.33) node {\scriptsize{$a_{2}$}}
		(-0.7, 3.7) node {\scriptsize{$a_{2}$}}
		(0.6, 4.2) node {\scriptsize{$A_{2}$}}
		(-0.7, 4.3) node {\scriptsize{$A_{2}$}}
		(0.33, 5.33) node {\scriptsize{$A_{1}$}}
		(-0.7, 5.7) node {\scriptsize{$A_{1}$}}
		(0.4, 6.5) node {\scriptsize{$A_{3}$}}
		(0.4, 7.5) node {\scriptsize{$A_{5}$}}
		;

		\end{scope}
		
		\begin{scope}[xshift = 8cm, yshift = -16cm]
		
		\filldraw[black] (0,0) circle (4pt)
		(0,1) circle (4pt)
		(0,2) circle (4pt)
		(0,12) circle (4pt)
		(0,13) circle (4pt)
		(0,14) circle (4pt)
		(1,3) circle (4pt)
		(1,4) circle (4pt)
		(1,5) circle (4pt)
		(1,6) circle (4pt)
		(1,7) circle (4pt)
		(1,8) circle (4pt)
		(1,9) circle (4pt)
		(1,10) circle (4pt)
		(1,11) circle (4pt)
		(-1,3) circle (4pt)
		(-1,4) circle (4pt)
		(-1,5.5) circle (4pt)
		(-1,7) circle (4pt)
		(-1,8.5) circle (4pt)
		(-1,10) circle (4pt)
		(-1,11) circle (4pt);
		
		\draw[black]
		(0,12) circle (7pt)
		(0,13) circle (7pt)
		(0,14) circle (7pt)
		(1,11) circle (7pt)
		(-1,8.5) circle (7pt)
		(-1,10) circle (7pt)
		(-1,11) circle (7pt);

		\draw[ultra thick] (0,0) -- (0,2) -- (-1,3) -- (-1,11) -- (0,12) -- (0,14) -- (0,12) -- (1,11) -- (1,3) -- (0,2) -- (1,3) -- (-1,4) -- (1,5) -- (1,6) -- (-1,7) -- (1,8) -- (1,9) -- (-1,10) -- (1,11);
		
		\begin{scope}[xscale = -1]
		\draw (0.4, 13.5) node {\scriptsize{$A_{7}$}}
		(0.4, 12.5) node {\scriptsize{$A_{5}$}}
		(0.4, 1.5) node {\scriptsize{$a_{5}$}}
		(0.4, 0.5) node {\scriptsize{$a_{7}$}}
		
		(0, 10.8) node {\scriptsize{$A_{2}$}}
		(0, 9.2) node {\scriptsize{$A_{2}$}}
		(0, 7.8) node {\scriptsize{$A_{1}$}}
		(0, 6.2) node {\scriptsize{$A_{1}$}}
		(0, 4.8) node {\scriptsize{$a_{2}$}}
		(0, 3.8) node {\scriptsize{$a_{2}$}}
		
		(-0.9, 2.3) node {\scriptsize{$a_{3}$}}
		(-1.3, 3.5) node {\scriptsize{$A_{1}$}}
		(-1.3, 4.5) node {\scriptsize{$a_{3}$}}
		(-1.3, 5.5) node {\scriptsize{$A_{1}$}}
		(-1.3, 6.5) node {\scriptsize{$A_{1}$}}
		(-1.3, 7.5) node {\scriptsize{$A_{1}$}}
		(-1.3, 8.5) node {\scriptsize{$A_{1}$}}
		(-1.3, 9.5) node {\scriptsize{$A_{3}$}}
		(-1.3, 10.5) node {\scriptsize{$A_{1}$}}
		(-0.9, 11.7) node {\scriptsize{$A_{3}$}}
		
		(0.8, 2.4) node {\scriptsize{$A_{1}$}}
		(0.7, 3.5) node {\scriptsize{$a_{4}$}}
		(1.3, 4.75) node {\scriptsize{$A_{2}$}}
		(1.3, 6.25) node {\scriptsize{$a_{3}$}}
		(1.3, 7.75) node {\scriptsize{$A_{3}$}}
		(0.7, 9.45) node {\scriptsize{$a_{2}$}}
		(0.65, 10.6) node {\scriptsize{$A_{4}$}}
		(0.6, 11.85) node {\scriptsize{$A_{1}$}}
		;
		\end{scope}
		\end{scope}
		
		\draw[ultra thick]
		(-0.5,2) -- (0.5,1)
		
		(0.5,1) -- (2,0)
		(0.5,0) -- (2,-1)
		
		(2,0) -- (4.5,-1)
		(2,-1) -- (4.5,-2)
		(2,-2) -- (4.5,-3)
		(2,-3) -- (3.5,-4)
		(2,-4) -- (3.5,-6)
		
		(4.5,-1) -- (8,-2)
		(4.5,-2) -- (8,-3)
		(4.5,-3) -- (8,-4)
		(5.5,-4) -- (9,-5)
		(3.5,-4) -- (7,-5)
		(4.5,-5) -- (7,-6)
		(5.5,-6) -- (9,-8)
		(3.5,-6) -- (7,-7.5)
		(4.5,-7) -- (7,-9)
		(4.5,-8) -- (7,-10.5)
		(4.5,-9) -- (7,-13)
		
		;
		
		\begin{scope}[xshift = 11.5cm]
		
		\begin{scope}[yshift = 1cm, xshift = -0.5cm]
		\filldraw[black] 
		(0,1) circle (4pt);
		\end{scope}
		
		\begin{scope}[yshift = 0cm, xshift = 0.5cm]
		\filldraw[black] (0,0) circle (4pt)
		(0,1) circle (4pt);
		
		\draw [ultra thick] (0,0) -- (0,1)
		
		(-0.4, 0.5) node {\scriptsize{$A_{1}$}};
		\end{scope}
		
		\draw (0.2,1.7) node[rotate = -45] {\scriptsize$D_4$}
		(1.4,0.75) node[rotate = -30] {\scriptsize$D_{2}$}
		(1.25,-0.2) node[rotate = -30] {\scriptsize$D_{3}$}
		
		(3.4,-2.3) node[rotate = -20] {\scriptsize$D_{2}$}
		(2.9,-3.3) node[rotate = -30] {\scriptsize$D_{2}$}
		(2.5,-5.2) node[rotate = -50] {\scriptsize$D_{3}$}
		
		(7.5,-4.7) node[rotate = -28] {\scriptsize$a_3$}
		(7,-5.8) node[rotate = -28] {\scriptsize$a_2$}
		(6.25,-6.15) node[rotate = -28] {\scriptsize$D_{2}$}
		(5.9,-7.8) node[rotate = -35] {\scriptsize$D_{2}$}
		(5.9,-9) node[rotate = -41] {\scriptsize$D_{2}$}
		(5.9,-10.75) node[rotate = -55] {\scriptsize$D_{3}$}
		;
		
		\begin{scope}[xshift=2cm, yshift = -4cm]
		\filldraw[black] (0,0) circle (4pt)
		(0,1) circle (4pt)
		(0,2) circle (4pt)
		(0,3) circle (4pt)
		(0,4) circle (4pt);
		
		\draw [ultra thick] (0,0) -- (0,4)
		(-0.3, 0.5) node {\scriptsize{$a_{3}$}}
		(-0.3, 1.5) node {\scriptsize{$a_{1}$}}
		(-0.3, 2.5) node {\scriptsize{$A_{1}$}}
		(0.8, 3.4) node {\underline{\scriptsize{$A_{3} \cup A_3$}}}
		;

		\end{scope}

		\begin{scope}[xshift = 4.5cm, yshift = -9cm]
		\filldraw[black] (0,0) circle (4pt)
		(0,1) circle (4pt)
		(0,2) circle (4pt)
		(0,4) circle (4pt)
		(0,6) circle (4pt)
		(-1,3) circle (4pt)
		(1,3) circle (4pt)
		(-1,5) circle (4pt)
		(1,5) circle (4pt)
		;
		
		\draw [ultra thick] (0,0) -- (0,2) -- (-1,3) -- (0,4) -- (-1,5) -- (0,6) -- (1,5) -- (0,4) -- (1,3) -- (0,2)
		(-0.4, 0.5) node {\scriptsize{$a_{5}$}}
		(-0.4, 1.5) node {\scriptsize{$a_{3}$}}
		(0.33, 2.8) node {\scriptsize{$a_{1}$}}
		(-0.7, 2.3) node {\scriptsize{$a_{1}$}}
		(0.33, 3.33) node {\scriptsize{$a_{2}$}}
		(-0.7, 3.7) node {\scriptsize{$a_{2}$}}
		(0.6, 4.2) node {\scriptsize{$A_{2}$}}
		(-1.05, 4.2) node {\scriptsize{\underline{$A_{2}\cup A_2$}}}
		(1.2, 5.9) node {\scriptsize{\underline{$A_{1}\cup A_1$}}}
		(-0.7, 5.7) node {\scriptsize{$A_{1}$}}
		;

		\end{scope}
		
		\begin{scope}[xshift = 8cm, yshift = -16cm]
		
		\filldraw[black] (0,0) circle (4pt)
		(0,1) circle (4pt)
		(0,2) circle (4pt)
		(1,3) circle (4pt)
		(1,4) circle (4pt)
		(1,5) circle (4pt)
		(1,6) circle (4pt)
		(1,7) circle (4pt)
		(1,8) circle (4pt)
		(1,9) circle (4pt)
		(1,10) circle (4pt)
		(-1,3) circle (4pt)
		(-1,4) circle (4pt)
		(-1,5.5) circle (4pt)
		(-1,7) circle (4pt);

		\draw[ultra thick] (0,0) -- (0,2) -- (-1,3) -- (-1,7)  (1,10) -- (1,3) -- (0,2) -- (1,3) -- (-1,4) -- (1,5) -- (1,6) -- (-1,7) -- (1,8) ;
		
		\begin{scope}[xscale = -1]
		\draw 
		(0.4, 1.5) node {\scriptsize{$a_{5}$}}
		(0.4, 0.5) node {\scriptsize{$a_{7}$}}
		
		(0, 7.8) node {\scriptsize{$A_{1}$}}
		(0, 5.9) node {\scriptsize{\underline{$A_{1}\cup A_1$}}}
		(0, 4.8) node {\scriptsize{$a_{2}$}}
		(0, 3.8) node {\scriptsize{$a_{2}$}}
		
		(-0.9, 2.3) node {\scriptsize{$a_{3}$}}
		(-1.3, 3.5) node {\scriptsize{$A_{1}$}}
		(-1.3, 4.5) node {\scriptsize{$a_{3}$}}
		(-1.3, 5.5) node {\scriptsize{$A_{1}$}}
		(-1.3, 6.5) node {\scriptsize{$A_{1}$}}
		(-1.9, 7.5) node {\scriptsize{\underline{$A_{1}\cup A_1$}}}
		(-1.3, 8.5) node {\scriptsize{$A_{1}$}}
		(-1.3, 9.5) node {\scriptsize{$A_{3}$}}
		
		(0.8, 2.4) node {\scriptsize{$A_{1}$}}
		(0.7, 3.5) node {\scriptsize{$a_{4}$}}
		(1.2, 4.75) node {\scriptsize{\underline{$A_{2}\cup A_2$}}}
		(1.3, 6.25) node {\scriptsize{$a_{3}$}}
		;
		\end{scope}
		\end{scope}
		
		\draw[ultra thick]
		(-0.5,2) -- (0.5,1)
		
		(0.5,1) -- (2,0)
		(0.5,0) -- (2,-1)
		
		(2,-2) -- (4.5,-3)
		(2,-3) -- (3.5,-4)
		(2,-4) -- (3.5,-6)
		
		(5.5,-6) -- (9,-8)
		(4.5,-7) -- (7,-9)
		(4.5,-8) -- (7,-10.5)
		(4.5,-9) -- (7,-13)
		
		;
		
		\draw[ultra thick, densely dashed]
		(4.5,-3) -- (9,-6)
		(4.5,-5) -- (9,-7)

		;
		
		
		\end{scope}
		
		\draw[\THICC, ->] (11.2,-8) -- (13.2,-8);

		\end{tikzpicture}
	\end{center}
	\caption{An example of the editing prescription for the balanced Hasse diagram to explicitly find the diagram for $D_\kappa(4)$ for $|\kappa|\leq8$. On the left we draw the general construction from figure \ref{FIGUREE} with $n=4$. The nodes corresponding to partitions with parts that are too big are circled and the labels of those edges which are no longer viable are in inverted commas. On the right the nodes with parts that are too large and edges with undefined labels have been removed as per the prescription and the new edges have been edited and added as per figure \ref{FIGUREH}. This structure can be explicitly verified using quiver addition on $D_4$ Dynkin quivers. At the very top of the structure, the top five nodes take on the form of the maximal special slice Hasse diagram for $D_4$. As observed in figure \ref{FIGUREE}, the top two lines of traversing structure of the general case gives the Hasse subdiagram for the maximal special slice of the corresponding $\kk{so}_{2n}$ algebra. When a specific value of $n$ is chosen, much of this structure is edited away. Since this subdiagram was the only part which appeared in both the Hasse diagram for nilpotent orbits of $\kk{so}_8$ and in the balanced Hasse diagram for $D_4$ Dynkin quivers, this once again shows that $D_4$ Dynkin quivers cannot realise nilpotent varieties of $\kk{so}_8$ outside the maximal special slice (The $A_1$ from partitions $(3,1)$ to $(2^2)$ also appears in $\kk{so}_8$, but is incidental here and is not a feature repeated for other algebras). }
	\label{FIGUREI}
\end{figure}

An illustrative example of the editing necessary for the Hasse diagram for $D_\kappa(4)$ for $|\kappa|\leq8$ is given in figure \ref{FIGUREI}. Notice how those $D_4$ Dynkin quivers which realise $\kk{so}_{8}$ nilpotent varieties are the tiny number living at the very top of the final $D_4$ balanced Hasse diagram. Figure \ref{FIGUREI} is limited to $|\kappa|\leq 8$ for practical purposes, the full $D_4$ balanced Hasse diagram can be expanded forever using quiver addition.

\subsubsection{Dimension matching for balanced theories}
Calculating the dimension of the moduli space branches of balanced $D_n$ Dynkin quivers gives a useful check of the construction and analysis in figure \ref{FIGUREE}. The simplest calculation is that of the dimension of the Higgs branch of $D_{\kappa}(n)$ using figure \ref{FIGUREE}. This requires picking a route from the node corresponding to the theory up to the top of figure \ref{FIGUREE}.

The simple route to choose is to go from the node labelled $\kappa$ up to the top of the linear subdiagram of nodes (the $\kk{sl}_n$ Slodowy slice to the $\kk{sl}_n$ nilpotent orbit $\O_{\kappa}$) then along the $D_k$ transitions at the top of the diagram. This gives
\begin{equation}
\dimh(\Hg(D_{\kappa}(n))) = 
\dimh(\es_{\kappa}) + \sum_{j=0}^{|\kappa|-2}\dimh(D_{n-j})
=  \frac{1}{2}\left( \sum_i(\kappa_i^t)^2 - |\kappa|\right) + \frac{1}{2}|\kappa| = \frac{1}{2} \sum_i(\kappa_i^t)^2,
\end{equation}
where the sum over $i$ is the sum over all the nonzero parts of the partition in each summation. The dimension of $\kk{sl}_n$ nilpotent varieties can be found in \cite{CollMc}.

A general calculation requires a general route up through figure \ref{FIGUREE} from the node $\kappa$ to the top. This route will go from $\kappa$ up to some node in $\mathcal{P}(|\kappa|)$ which dominates $\kappa$ and is of the form $(\eta^t, 1^2)^t$ (this is required by the traversing structure). There the route traverses up to the node $\eta \in \mathcal{P}(|\kappa|-2)$ and then up to some other node in $\mathcal{P}(|\kappa|-2)$ of appropriate form and across again, and so on up to the top of figure \ref{FIGUREE}. 

Denote the \textit{lowest} visited node in each linear subsystem $\mathcal{P}(d)$ as $\lambda_d$. The highest node in $\mathcal{P}(d)$ which the route passes through is then specified via the traversing structure by $\lambda_{d-2}$. Under this notation $\lambda_{|\kappa|} = \kappa$ and $\lambda_0 = (0)$. Using this notation the dimension of a general route up through figure \ref{FIGUREE} is
\begin{equation}\label{genhigcalc}
\begin{split}
\dimh(\Hg(D_\kappa(n))) & = \sum_{ j=0, ~ \mathrm{ even}}^{|\kappa|-2} \left[ \dimh\left( D_{n-l(\lambda_{|\kappa|-j-2}^t)}\right) + \dimh \left( \es_{\lambda_{|\kappa|-j}} \cap \O_{(\lambda_{|\kappa|-j-2}^t,1^2)^t}\right) \right] \\ & = \frac{1}{2}|\kappa| + \sum_{ j=0, ~ \mathrm{ even}}^{|\kappa|-2} \frac{1}{2}\left[ \sum_i ((\lambda_{|\kappa|-j|}^t)_i)^2 - \sum_i((\lambda_{|\kappa|-j-2}^t)_i)^2 - 2\right] \\ & = \frac{1}{2}\left( \sum_i((\lambda_{|\kappa|}^t)_i)^2 - \sum_i((\lambda_0^t)_i)^2\right) \\ & = \frac{1}{2}\sum_i(\kappa_i^t)^2.
\end{split}
\end{equation}
Once again, sums over $i$ mean sums over all the nonzero parts in the partition and the nilpotent varieties are those found in $\kk{sl}_n$ algebras. The general route agrees with the first, simpler, calculation, as expected since the dimensions ought to be route-independent.

The quaternionic dimension of the Coulomb branch for a unitary quiver gauge theory is  the sum of the ranks of the gauge nodes, therefore the dimension of the Coulomb branch for the $D_p$ singularity quiver is
\begin{equation}
\dimh(\mathcal{C}(\Q_{\Hg}(D_p))) = 2p-3  = \dimh(d_p),
\end{equation}
where $d_p$ is the mirror variety to $D_p$.

The construction in figure \ref{FIGUREE} can be checked by calculating the dimension of the Coulomb branch of a $D_n$ Dynkin quiver in two different ways. The first way is to sum over the ranks of all the nodes of a general balanced $D_n$ Dynkin quiver, figure \ref{FIGUREG}. The second way is to sum over the mirror varieties of a general route up through figure \ref{FIGUREE}. For $\kk{sl}_n$ nilpotent varieties the mirror of $\es_\rho \cap \O_\sigma$ is $\es_{\sigma^t} \cap \O_{\rho^t}$. 

Before proving the equality of the results of these methods it is worth recalling a small result which will form an essential step. Given a general partition $\eta = (z^{y_z},\dots,1^{y_1})$,
\begin{equation}\label{smallproof}
\sum_i(((\eta,1^a)^t)_i)^2  = \left( \sum_{m=1}^zy_m+a\right) ^2 + \sum_{q=2}^z\left( \sum_{m=q}^z y_m\right) ^2  = a^2 + 2l(\eta)a + \sum_i(\eta^t_i)^2
\end{equation}
generalizing the result used in \cite{Rogers}. 

\noindent \textbf{Proposition.} $\quad$ Figure \ref{FIGUREE} passes the Coulomb branch dimension check, that is,
\begin{equation}\label{dimprop}
\begin{split}
& \dimh(\mathcal{C}(D_\kappa(n)))   = \sum_{k=1}^{n-2}\left[ k g_1 - \sum_{i=1}^{k-1}(k-i)f_i \right] +g'+g'' \\ &  = \sum_{ j=0, ~ \mathrm{ even}}^{|\kappa|-2}\Bigg[ \dimh\left( \mathcal{C}\left( \Q_\Hg(D_{n-l(\lambda_{|\kappa|-j-2}^t)})\right) \right)  + \dimh\left( \textrm{Mirror}(\es_{\lambda_{|\kappa|-j}} \cap \O_{(\lambda_{|\kappa|-j-2}^t,1^2)^t})\right) \Bigg] .
\end{split}
\end{equation}
In fact
\begin{equation}\label{neweq}
\dimh(\mathcal{C}(D_\kappa(n))) = \frac{1}{2}|\kappa|(2n-1) - \frac{1}{2}\sum_i(\kappa_i)^2.
\end{equation}

\noindent \textbf{Proof.} $\quad$ Both lines on the right of (\ref{dimprop}) equal the right side of (\ref{neweq}). Firstly,
\begin{equation}\label{threetofour}
\begin{split}
& \sum_{ j=0, ~ \mathrm{ even}}^{|\kappa|-2}\left[ \dimh\left( \mathcal{C}\left( \Q_\Hg(D_{n-l(\lambda_{|\kappa|-j-2}^t)})\right) \right)  + \dimh\left( \textrm{Mirror}(\es_{\lambda_{|\kappa|-j}} \cap \O_{(\lambda_{|\kappa|-j-2}^t,1^2)^t})\right) \right] \\ & = 
\sum_{ j=0, ~ \mathrm{ even}}^{|\kappa|-2} \left[ 2(n-l(\lambda_{|\kappa|-j-2}^t)) - 3 + \dimh\left( \O_{\lambda_{|\kappa|-j}^t} \cap \es_{(\lambda^t_{|\kappa|-j-2},1^2)}\right) \right]
\\ & = \sum_{ j=0, ~ \mathrm{ even}}^{|\kappa|-2}\left[ 2n-3 - 2l(\lambda_{|\kappa|-j-2}^t) + \frac{1}{2}\left( \sum_i (((\lambda_{|\kappa|-j-2}^t,1^2)^t)_i)^2 - \sum_i((\lambda_{|\kappa|-j})_i)^2\right) \right]
\\ & = \sum_{ j=0, ~ \mathrm{ even}}^{|\kappa|-2}\Bigg{[} 2n-3 - 2l(\lambda_{|\kappa|-j-2}^t) - \frac{1}{2}\left( \sum_i ((\lambda_{|\kappa|-j})_i)^2\right)  \\ & \qquad \qquad ~~~~ + \frac{1}{2}\left( 4+4l(\lambda_{|\kappa|-j-2}^t) + \sum_i((\lambda_{|\kappa|-j-2})_i)^2\right)\Bigg{]} \\ & = \sum_{ j=0, ~ \mathrm{ even}}^{|\kappa|-2}\left[ 2n-1 + \frac{1}{2}\sum_i((\lambda_{|\kappa|-j-2})_i)^2 - \frac{1}{2}\sum_i((\lambda_{|\kappa|-j})_i)^2\right] \\ & = \frac{1}{2}|\kappa|(2n-1) + \frac{1}{2}\sum_i((\lambda_0)_i)^2 - \frac{1}{2}\sum_i((\lambda_{|\kappa|})_i)^2 \\ & = \frac{1}{2}|\kappa|(2n-1) - \frac{1}{2}\sum_i(\kappa_i)^2.
\end{split}
\end{equation}
The most important steps in (\ref{threetofour}) are from lines three to four where (\ref{smallproof}) was used and from lines five to six where the sum over $j$ is assessed, which results in massive cancellation between the $i$ sums. 

For the second part, begin with the realisation that
\begin{equation}\label{twotofourone}
\begin{split}
\sum_{k=1}^{n-2}\left[ k g_1 - \sum_{i=1}^{k-1}(k-i)f_i \right]  +g'+g''  & = g_1 \sum_{k=1}^{n-2}k - \sum_{k=1}^{n-2}\sum_{i=1}^k (k-i)f_i + g' +g'' \\ & = \frac{1}{2}g_1(n-2)(n-1)+g'+g'' - \sum_{l=1}^{n-3}\left[ f_l \sum_{j=1}^{n-2-l}j\right]
\end{split}
\end{equation}
which can be found by expanding the second term of the right of the first line directly, and rearranging. Assessing the $j$ sum and substituting the values of $g'$ and $g''$ in terms of $\kappa$ and $f_n$ known from balancing yields
\begin{equation}\label{twotofourtwo}
\begin{split}
&\frac{1}{2}g_1(n-2)(n-1) + |\kappa| - f_n - \sum_{l=1}^{n-3} \frac{1}{2}f_i(n-2-l)(n-1-l)
\\ & = \frac{1}{2}\left( \sum_{l=1}^nf_l\right) (n-2)(n-1) + |\kappa| - f_n \\ & ~~~ - \frac{1}{2}\left( \sum_{l=1}^{n-3}f_l\right) (n-2)(n-1) + \frac{1}{2}(2n-3)\sum_{l=1}^{n-3}lf_l - \frac{1}{2}\sum_{l=1}^{n-3}l^2f_l \\ & = 
\frac{1}{2}\left( \sum_{l=n-2}^nf_l\right) (n-2)(n-1) + |\kappa| - f_n -  \frac{1}{2}(2n-3)\sum_{l=1}^{n-3}lf_l - \frac{1}{2} \sum_{l=1}^{n-3}l^2f_l \\ & =
\frac{1}{2}|\kappa|(2n-1) + \frac{1}{2}\left[ (n^2-3n+2)\left( \sum_{l=n-2}^nf_l\right)  -(2n-3)\left( \sum_{l=n-2}^nlf_l\right)-2f_n - \sum_{l=1}^{n-3}l^2f_l\right] \\ & = \frac{1}{2}|\kappa|(2n-1) + \frac{1}{2}\left[- \sum_{l=1}^n l^2f_l\right]  \\ & = \frac{1}{2}|\kappa|(2n-1) - \frac{1}{2}\sum_i(\kappa_i)^2,
\end{split}
\end{equation}
as required, completing the proof. From lines one to two the $(n-2-l)(n-1-l)$ term was expanded and $g_1$, $g'$ and $g''$ written in terms of flavours and $\kappa$. From line two to three, terms one and four mostly cancel with one another. From lines three to four
\begin{equation}
\sum_{l=1}^{n-3}lf_l = \sum_{l=1}^n l f_l - \sum_{n-2}^n lf_l = |\kappa| - \sum_{l=n-2}^n lf_l.
\end{equation}
was used. From lines four to five the two sums over $\{n-2,n-1,n\}$ were assessed and the terms in the square brackets simplify considerably.

\subsection{From balanced to good quivers}
The discussion so far has concentrated on balanced $D_n$ Dynkin quivers and their moduli space singularity structure. The class $D_\kappa(n)$ of balanced quivers can be ordered into a Hasse diagram, figure \ref{FIGUREE}, by appealing to the moduli space inclusion relations. Classification of all \textit{good} $D_n$ Dynkin quivers can now also be performed. Whereas the gauge node excesses of balanced quivers must be zero, the excesses for good quivers need only be non-negative. Balanced quivers are therefore a subset of good quivers for a given gauge node topology. 

Given a complete set of balanced quivers for a class of quivers with a given gauge node topology, one can construct a set of good quivers fairly easily using the quiver subtraction introduced in \cite{QuivSub}. A quiver, $\Q_1$, can only be subtracted from another quiver, $\Q_2$, to give a third quiver, $\Q_3$, if $\Q_1$ and $\Q_2$ have the same gauge node topology and if the gauge nodes in $\Q_3$ have non-negative rank\footnote{Note that the rank of some gauge nodes in $\Q_3$ may be zero. This changes the gauge node topology since the rank zero nodes are effectively absent. This includes the possibility that $\Q_3$ is a disjoint quiver.}. 

Consider two balanced quivers $\Q^1_{\textrm{bal}}$ and $\Q^2_{\textrm{bal}}$ with gauge nodes with positive rank $g_i^1$ and $g_i^2$ respectively\footnote{Where the lower index refers to some ordering of the gauge nodes which is maintained across quivers.}. Attached to these gauge nodes are non-negative flavour nodes with label $f_i^1$ and $f_i^2$. Proving that $\Q^1_{\textrm{bal}} - \Q^2_{\textrm{bal}} = \Q^3$ is a good quiver is straightforward. Consider that quiver subtraction requires $g_i^1 \geq g_i^2$ for all $i$. For each gauge node $g_i$, consider the set of nodes which are connected to $g_i$ via an edge. This will be a collection of gauge nodes labelled with indices in the set $j_i$. The condition of balance then imposes that the excess $e$ of every gauge node is zero,
\begin{equation}\label{balanceeq}
e_i^a =  f_i^a - 2g_i^a +\sum_{k \in j_i} g_k^a  = 0,
\end{equation}
where $a \in \{1,2\}$. From quiver subtraction, $g_i^3 = g_i^1 - g_i^2 \geq 0$ and $f_i^3 = f_i^1$. The excess on the gauge nodes in $\Q_3$ is then,
\begin{equation}\label{excessflav}
\begin{split}
e_i^3 & =   f_i^3 - 2g_i^3 +\sum_{k \in j_i} g_k^3 \\ &
=   f_i^1 - 2(g_i^1 - g_i^2) +\sum_{k \in j_i} (g_k^1 - g_k^2)\\ & = 2g_i^2 - \sum_{k \in j_i} g_k^2 \\ & = f_i^2.
\end{split}
\end{equation}
Flavours are non-negative and so the excess is larger than zero, the result of quiver subtraction amongst balanced quivers is always a good quiver. As we will discuss in the next subsection, those good quivers that are constructable as a difference of balanced quivers are not necessarily all possible good quivers.

\subsection{Good $D_n$ Dynkin quivers}
This section lays out the characterisation and local moduli space analysis of good $D_n$ Dynkin quivers. Any two balanced quivers where one can be subtracted from the other yield a good quiver. Having already characterised all of the balanced $D_n$ Dynkin quivers, a large number of good quivers can be found by examining all possible subtractions. In the linear ($A_n$ Dynkin) quiver case this encapsulated all the possible good quivers, however for the $D_n$ case it does not. When looking at balanced quivers one quickly excluded any quivers with an odd difference in flavour on the end nodes as being unbalanceable. However loosening the restriction to the class of \textit{good} quivers means that they have to be included. The trouble is that the difference of two balanced quivers must necessarily have an even difference of end flavours, and good quivers with an odd difference in end flavours cannot possibly be found as balanced subtractions. The classification of good $D_n$ Dynkin quivers therefore necessarily divides into two parts: Those with an even difference in end flavours, \textit{even} type, and those with an odd difference in end flavours, \textit{odd} type.

It will be shown that all good $D_n$ Dynkin quivers of even type can be found as the subtraction of two balanced $D_n$ Dynkin quivers. The moduli space singularity structure can be expressed as a \textit{run} from one node down to another node on the balanced Hasse diagram constructed in figure \ref{FIGUREE}. This is shown by premising an arbitrary good quiver of even type and describing a general method by which the two relevant balanced quivers can be found. This will allow a classification of all good even $D_n$ Dynkin quivers using two even partitions (not necessarily of the same magnitude) and the integer $n$. Attention is then turned to good quivers of odd type. This classification requires very similar methods to even type. Poset structure is established for a class of quivers from which all good odd quivers can be found using quiver subtraction and the completeness of the classification verified using similar methods to the even case.

\subsubsection{$D_n$ quivers of even type}
\begin{figure}
	\begin{center}
		\begin{tikzpicture}[scale = 0.9]
		\def\q{7}
		\def\k{0.7071}
		\def\size{\tiny}
		
		\draw[thick] (0,0) -- (1.5,0) (2.5,0) -- (4,0) -- (4.7,0.7) (4,0) -- (4.7,-0.7) (0,0) -- (0,1)
		(1,0) -- (1,1)
		(3,0) -- (3,1)
		(4,0) -- (5,0)
		
		(4+\k,\k) -- (5+\k,\k)
		(4+\k,-\k) -- (5+\k,-\k)
		;
		
		\draw[thick, fill=white] (0,0) circle (\q pt)
		(1,0) circle (\q pt)
		(3,0) circle (\q pt)
		(4,0) circle (\q pt)
		(4+\k,\k) circle (\q pt)
		(4+\k,-\k) circle (\q pt)
		;
		
		\draw[thick, fill=white] 
		(0 -\q/30,1 -\q/30) rectangle (0 +\q/30,1 +\q/30)
		(1 -\q/30,1 -\q/30) rectangle (1 +\q/30,1 +\q/30)
		(3 -\q/30,1 -\q/30) rectangle (3 +\q/30,1 +\q/30)
		(5 -\q/30,0 -\q/30) rectangle (5 +\q/30,0 +\q/30)
		(5+\k -\q/30,\k -\q/30) rectangle (5+\k +\q/30,\k +\q/30)
		(5+\k -\q/30,-\k -\q/30) rectangle (5+\k +\q/30,-\k +\q/30)
		;

		\draw (0,1.5) node {\size $f_1$}
		(1,1.5) node {\size $f_2$}
		(3,1.5) node {\size $f_{n-3}$}
		
		(5.6,0) node  {\size $f_{n-2}$}
		(5.7 +\k,0+\k) node  {\size $f_{n-1}$}
		(6.1+\k,0-\k) node  {\size $f_{n-1}+2f_n$}

		(0,-0.45) node {\size $(g_1, e_1)$}
		(1,-0.45) node {\size $(g_2, e_2)$}
		(2,0) node {$\dots$}
		(3,-0.45) node {\size $(g_{n-3}, e_{n-3})$}
		(3.75,0.7) node {\size $(g_{n-2},$} (3.9,0.45) node {\size $e_{n-2})$}
		
		(4+\k,1.2) node {\size $(g', e_{n-1})$}
		(4+\k,-1.2) node {\size $(g'', e_{n-1}+2e_n)$}
		;
		
		\draw (-1.45,0) node {$~$};

		\end{tikzpicture}
	\end{center}
	\caption{The general form of a good $D_n$ Dynkin quiver of even type. Note that as well as the difference in flavours on the end nodes having to be even, the difference in excess of the end nodes also has to be even. }
	\label{FIGUREK}
\end{figure}
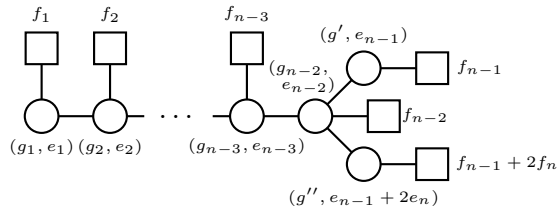

Each gauge node in a good quiver has associated to it a flavour, a rank and an excess. Even quivers have an even difference of flavours on the end nodes. It is simple to establish that the difference in the excess for the end nodes must also be even for an even quiver.
\begin{figure}
	\begin{center}
		\begin{tikzpicture}[scale = 0.9]
		\def\q{7}
		\def\k{0.7071}
		\def\size{\tiny}
		
		\draw[thick] (0,0) -- (1.5,0) (2.5,0) -- (4,0) -- (4.7,0.7) (4,0) -- (4.7,-0.7) (0,0) -- (0,1)
		(1,0) -- (1,1)
		(3,0) -- (3,1)
		(4,0) -- (5,0)
		
		(4+\k,\k) -- (5+\k,\k)
		(4+\k,-\k) -- (5+\k,-\k)
		;
		
		\draw[thick, fill=white] (0,0) circle (\q pt)
		(1,0) circle (\q pt)
		(3,0) circle (\q pt)
		(4,0) circle (\q pt)
		(4+\k,\k) circle (\q pt)
		(4+\k,-\k) circle (\q pt)
		;
		
		\draw[thick, fill=white] 
		(0 -\q/30,1 -\q/30) rectangle (0 +\q/30,1 +\q/30)
		(1 -\q/30,1 -\q/30) rectangle (1 +\q/30,1 +\q/30)
		(3 -\q/30,1 -\q/30) rectangle (3 +\q/30,1 +\q/30)
		(5 -\q/30,0 -\q/30) rectangle (5 +\q/30,0 +\q/30)
		(5+\k -\q/30,\k -\q/30) rectangle (5+\k +\q/30,\k +\q/30)
		(5+\k -\q/30,-\k -\q/30) rectangle (5+\k +\q/30,-\k +\q/30)
		;

		\draw (0,1.5) node {\size $f_1$}
		(1,1.5) node {\size $f_2$}
		(3,1.5) node {\size $f_{n-3}$}
		
		(5.6,0) node  {\size $f_{n-2}$}
		(5.5 +\k,0+\k) node  {\size $f'$}
		(5.5+\k,0-\k) node  {\size $f''$}

		(0,-0.45) node {\size $(g_1, e_1)$}
		(1,-0.45) node {\size $(g_2, e_2)$}
		(2,0) node {$\dots$}
		(3,-0.45) node {\size $(g_{n-3}, e_{n-3})$}
		(3.75,0.7) node {\size $(g_{n-2},$} (3.9,0.45) node {\size $e_{n-2})$}
		
		(4+\k,1.2) node {\size $(g', e')$}
		(4+\k,-1.2) node {\size $(g'', e'')$}
		;

		\end{tikzpicture}
	\end{center}
	\caption{The general form for a good $D_n$ Dynkin quiver. Each gauge node is now labelled with a rank and a non-negative \textit{excess}. Balanced quivers are the subset of good quivers where the excess on all the gauge nodes is zero. }
	\label{FIGUREJ}
\end{figure}
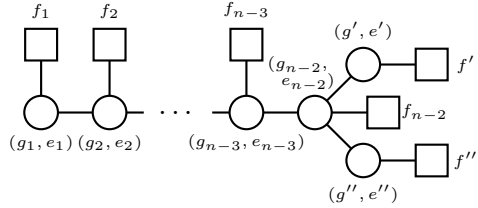
Consider figure \ref{FIGUREJ}, the `balance' (while regarding the excess) of the lower node requires $g_{n-2} + f_{n-1}+2f_n = 2g'' + e''$, and the balance of the upper node requires $g_{n-2}+f_{n-1}=2g'+e'$, putting these together requires $e''-e' = 2(g'+f_n-g'')$ so the difference is even. A general good $D_n$ Dynkin quiver of even type is given in figure \ref{FIGUREK}. 
Using figure \ref{FIGUREK}, define the following two partitions
\begin{equation}
\begin{split}\label{goodevenpart}
\kappa & = (n^{f_n}, n-1^{f_{n-1}},\dots,2^{f_2},1^{f_1}) \\
\lambda & = (n^{e_n}, n-1^{e_{n-1}},\dots,2^{e_2},1^{e_1}).
\end{split}
\end{equation}
In direct analogy with (\ref{gbalance}), it is simply established, by 'balancing' whilst taking into account the excess, that
\begin{equation}\label{newnodes}
g_k = kg_1 - \sum_{i=1}^{k-1}(k-i)(f_i-e_i).
\end{equation}
for $k\leq n-2$. Repeating the analysis of (\ref{gbalance}) - (\ref{twogee}) with this extra complication yields the analogous results
\begin{equation}\label{miracles}
\begin{split}
g_1 & = l(\kappa) - l(\lambda)\\
2g'' & = |\kappa|-|\lambda|.
\end{split}
\end{equation}
Note that all of this analysis reduces to the balanced case when we take $\lambda$ to be the zero partition which is equivalent to there being zero excess on every node. When it comes to examining the moduli space singularity structure this is the realisation that balanced quivers correspond to a run of edges and nodes in figure \ref{FIGUREE} from the very top to some node $\kappa$ whereas a good quiver corresponds to a run from a node $\lambda$ down to a node $\kappa$.

For a generic good, even $D_n$ Dynkin quiver one can find two balanced $D_n$ Dynkin quivers which give the good quiver under quiver subtraction. The larger quiver in quiver subtraction and the resulting quiver have the same flavour. The larger of the balanced quivers is therefore going to have flavour dictated by the partition $\kappa$. The clue for the smaller balanced quiver comes from (\ref{excessflav}). The flavour of a given node on the smaller quiver is exactly the excess of the good quiver under construction. The flavour of the smaller quiver is therefore dictated by the partition $\lambda$. 

A balanced quiver's flavour must be dictated by a partition of even magnitude, however there is nothing about good quivers which restricts (\ref{goodevenpart}) to be of even magnitude. This must be addressed. Note that the second relationship in (\ref{miracles}) tells us that either both $|\kappa|$ and $|\lambda|$ must be even, or both must be odd. 

The result is simple when both $\kappa$ and $\lambda$ are of even magnitude. In this case one \textit{can} construct balanced quivers using these partitions. The result when both are even is therefore
\begin{equation}\label{partitiondes}
\Q_{\textrm{good, even}}^{|\kappa|, |\lambda| \textrm{ even}} = \Q(D_{\kappa}(n)) - \Q(D_{\lambda}(n)).
\end{equation} 

When both $|\kappa|$ and $|\lambda|$ are odd there seems to be an impasse as the analogously defined 'balanced' quivers are unbalanceable. However the gauge node topology of the resulting quiver in quiver subtraction needn't be precisely that of the quivers involved in the subtraction. Nodes of rank zero might result from quiver subtraction which would change the quiver topology. The partitions which correspond to flavour arrangement are considered from the end of the tail of the quiver, premising another gauge node of rank $g_0=0$ on the good quiver changes the magnitude of the defining partitions considerably.

For $\kappa$ and $\lambda$ to be of odd magnitude, they must have an odd number of odd parts with odd multiplicity. Adding the zero rank gauge node to the good quiver has the effect of increasing all the parts by one, the odd number of odd parts with odd multiplicity is changed to an odd number of \textit{even} parts with odd multiplicity, which always gives an even number. However the previously even parts have now been shifted to being odd parts. If $\kappa$, for example, had an odd number of even parts with odd multiplicity, the new partition has an odd number of odd parts with odd multiplicity and so is odd. An odd magnitude partition with an odd number of even parts is necessarily of even length. The flavour on the new node can be arbitrary. When $\kappa$ is odd and $l(\kappa)$ is even choose the flavour, $f_0$, of the new zero node to be odd, and when $\kappa$ is odd and $l(\kappa)$ odd, choose $f_0$ even. This guarantees that the partition associated to the flavour for the new good quiver, $\kappa'$, is even. From here construct the quiver of excesses in the normal way. The excess of the new zero node is necessarily $g_1+f_0 = l(\kappa) - l(\lambda) +f_0$. The way $f_0$ was chosen now guarantees that the new partition on the quiver of excesses is also always even. To see this, note that when $l(\kappa)$ was odd $f_0$ was chosen even, therefore if $l(\lambda)$ was odd (and hence $\lambda$ had an even number of even parts with odd multiplicity) the flavour on the zeroth node in the excess quiver was even and the magnitude of $\lambda'$ is even. And if $l(\lambda)$ was even (and hence $\lambda$ had an odd number of even parts with odd multiplicity) the flavour on this node is odd and so the magnitude of $\lambda'$ is even again. In conclusion, when $\kappa$ and $\lambda$ as found from figure \ref{FIGUREK} are of odd magnitude, the good quiver is realised as
\begin{equation}
\Q_{\textrm{good, even}}^{|\kappa|, |\lambda| \textrm{ odd}} = \Q(D_{\kappa'}(n+1)) - \Q(D_{\lambda'}(n+1)).
\end{equation}
where
\begin{equation}
\begin{split}
\kappa' & = (n+1^{f_n}, n^{f_{n-1}},\dots,2^{f_1},1^{f_0}) \\
\lambda' & = (n+1^{e_n}, n^{e_{n-1}},\dots,2^{e_1},1^{l(\kappa)-l(\lambda)+f_0}).
\end{split}
\end{equation}
An important check to make on this construction is that the extra node added does indeed achieve a rank of zero after quiver subtraction. Using that the rank of the first node is the length of the partition for the balanced quivers constructed, the rank of the first node of their difference is
\begin{equation}\label{lengthdiff}
\begin{split}
g_1^{\kappa'}  - g_1^{\lambda'} & = l(\kappa') - l(\lambda') \\ & = \sum_{i=0}^nf_i - \sum_{i=1}^ne_i - l(\kappa) + l(\lambda) - f_0 \\ & = l(\kappa) + f_0 - l(\lambda) - l(\kappa) + l(\lambda) - f_0 \\ & = 0
\end{split}
\end{equation}
as required. 

\subsubsection{An alternative for even theories with odd partitions}
There is an alternative construction which allows an easier reading of the moduli space singularity structure when $|\kappa|$ and $|\lambda|$ are odd as compared to when the quiver is described as the difference of balanced quivers.

\begin{figure}
	\begin{center}
		\begin{tikzpicture}[scale=0.65]
		
		\def\scriptsize{\tiny}
		
		\filldraw[black] (0,18) circle (4pt)
		
		;
		
		\begin{scope}[yshift = 15cm, xshift = 1cm]
		\filldraw[black] (0,0) circle (4pt)
		(0,1) circle (4pt)
		(0,2) circle (4pt);
		
		\draw[\THICC] (0,0) -- (0,2)
		(-0.4, 0.5) node {\scriptsize{$a_{2}$}}
		(-0.4, 1.5) node {\scriptsize{$A_{2}$}}
		;
		\end{scope}
		
		\begin{scope}[yshift = 10cm, xshift = 3cm]
		\filldraw[black] (0,0) circle (4pt)
		(0,1) circle (4pt)
		(0,2) circle (4pt)
		(0,3) circle (4pt)
		(0,4) circle (4pt)
		(0,5) circle (4pt)
		(0,6) circle (4pt);
		
		\draw[\THICC] (0,0) -- (0,6)
		(-0.4, 0.5) node {\scriptsize{$a_{4}$}}
		(-0.4, 1.5) node {\scriptsize{$a_{2}$}}
		(-0.4, 2.5) node {\scriptsize{$a_{1}$}}
		(0.4, 3.5) node {\scriptsize{$A_{1}$}}
		(0.4, 4.5) node {\scriptsize{$A_{2}$}}
		(0.4, 5.5) node {\scriptsize{$A_{4}$}}
		;
		
		\end{scope}
		
		\begin{scope}[yshift = 4cm, xshift = 6cm]
		
		\filldraw[black] (0,0) circle (4pt)
		(0,1) circle (4pt)
		(0,2) circle (4pt)
		(0,4) circle (4pt)
		(0,7) circle (4pt)
		(0,9) circle (4pt)
		(0,10) circle (4pt)
		(0,11) circle (4pt)
		(1,3) circle (4pt)
		(-1,3) circle (4pt)
		(-1,5.5) circle (4pt)
		(1,5) circle (4pt)
		(1,6) circle (4pt)
		(-1,8) circle (4pt)
		(1,8) circle (4pt);
		
		\draw[\THICC] (0,0) -- (0,2) -- (-1,3) -- (0,4) -- (-1,5.5) -- (0,7) -- (-1,8) -- (0,9) -- (0,11) -- (0,9) -- (1,8) -- (0,7) -- (1,6) -- (1,5) -- (0,4) -- (1,3) -- (0,2);
		\begin{scope}[xscale=-1]
		\draw
		(0.4, 0.5) node {\scriptsize{$a_{6}$}}
		(0.4, 1.5) node {\scriptsize{$a_{4}$}}
		(0.7, 2.3) node {\scriptsize{$A_{1}$}}
		(-0.7, 2.3) node {\scriptsize{$a_{2}$}}
		(0.7, 3.7) node {\scriptsize{$a_{3}$}}
		(-0.7, 3.7) node {\scriptsize{$a_{2}$}}
		(0.8, 4.7) node {\scriptsize{$A_{2}$}}
		(-0.7, 4.3) node {\scriptsize{$A_{1}$}}
		(0.9, 6.3) node {\scriptsize{$a_{2}$}}
		(-1.3, 5.5) node {\scriptsize{$A_{1}$}}
		(-0.76, 6.7) node {\scriptsize{$A_{1}$}}
		(0.4, 10.5) node {\scriptsize{$A_{6}$}}
		(0.4, 9.5) node {\scriptsize{$A_{4}$}}
		(-0.8, 7.4) node {\scriptsize{$A_{2}$}}
		(0.3, 7.8) node {\scriptsize{$A_{3}$}}
		(-0.8, 8.7) node {\scriptsize{$A_{2}$}}
		(0.8, 8.7) node {\scriptsize{$A_{1}$}}
		;
		\end{scope}
		\end{scope}
		
		\draw[\THICC] (0,18) -- (1,17) -- (3,16) -- (6,15) -- (8,14.7)
		(1,16) -- (3,15)
		(1,15) -- (3,13)
		(3,15) -- (6,14)
		(3,14) -- (6,13)
		(3,13) -- (5,12)
		(3,12) -- (6,11)
		(3,11) -- (5,9.5)
		(3,10) -- (5,7)
		(6,4) -- (7,2)
		
		(7.9,9.4) -- (8.9,9)
		(10.6,8.5) -- (12,8.1)
		(14,7.75) -- (15,7.6)
		;
		
		\draw (1,17.7) node {\tiny$D_{n-1}$}
		(2.2,16.7) node[rotate = -23] {\tiny$D_{n-3}$}
		(2.05,15.75) node[rotate = -23] {\tiny$D_{n-2}$}
		(1.2,14) node[rotate = 0] {\tiny$D_{n-1}$}
		
		(4.5,15.8) node[rotate = -17] {\tiny$D_{n-5}$}
		(4.5,14.8) node[rotate = -17] {\tiny$D_{n-4}$}
		(4.5,13.8) node[rotate = -17] {\tiny$D_{n-3}$}
		(4.2,12.75) node[rotate = -23] {\tiny$D_{n-3}$}
		(4.4,11.8) node[rotate = -17] {\tiny$D_{n-2}$}
		(4.05,10.55) node[rotate = -31] {\tiny$D_{n-2}$}
		(3.6,8) node[rotate = 0] {\tiny$D_{n-1}$}
		
		(9.75,8.75) node {$\mathcal{P}(9)$}
		(13,8) node {$\mathcal{P}(11)$}
		
		(16,7.5) node[rotate = -7] {$\dots$}
		;
		
		\end{tikzpicture}
	\end{center}
	\caption{The general Hasse diagram for even theories with odd magnitude partitions. This Hasse diagram is similar to figure \ref{FIGUREE}. Cutting either this or figure \ref{FIGUREE} off at an arbitrary point yields a finite Hasse diagram. Finite Hasse subdiagrams of arbitrary size for one can be found \textit{somewhere} in the other. Finding arbitrarily large subdiagrams inside figure \ref{FIGUREE} is the same as the statement that even type theories with odd partitions can be found as the difference of two balanced theories (whose partitions must necessarily be even), which we  have already seen. }
	\label{FIGUREMUE}
\end{figure}
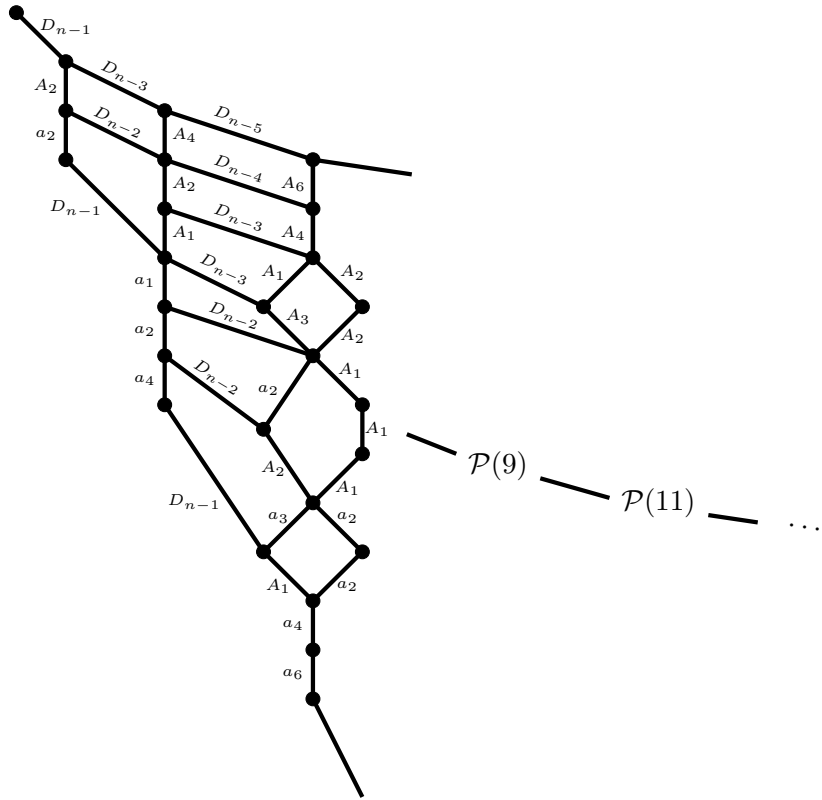

In figure \ref{FIGUREE}, the nodes were all labelled with even magnitude partitions. This can be viewed as arising because the diagram started with a trivial, flavourless quiver with partition $(0)$, and the traversing structure could only shift partition magnitude by an even amount. Starting with the trivial theory with partition (1) (at the top of figure \ref{ADDING}) and employing the same style of traversing structure yields a diagram analogous to figure \ref{FIGUREE} but built entirely of good, even theories with \textit{odd} partitions dictating flavour and whose only non-negative excess appears on the end node of the tail, figure \ref{FIGUREMUE}. This exactly gives the even type theories with odd partitions and hence allows a short-cut to their moduli space singularity analysis. As good and even theories with even partition magnitudes can be found as runs in figure \ref{FIGUREE}, good and even theories with odd partitions can be found as runs in figure \ref{FIGUREMUE}. Along with the previous discussion regarding identifying odd partitioned theories within figure \ref{FIGUREE} this shows that an arbitrarily sized section of figure \ref{FIGUREMUE} can always be found sufficiently far into figure \ref{FIGUREE}. Reversing the previous discussion also implies the opposite way round. 
\begin{figure}
	\begin{center}
		\begin{tikzpicture}[scale = 0.5]
		\def\q{7}
		\def\k{0.7071}
		\def\size{\tiny}
		
		\begin{scope}[xshift = 0.5cm]
		\draw[thick] (-2,0) -- (4,0) -- (4.7,0.7) (4,0) -- (4.7,-0.7) 
		(-2,0) -- (-2,1)
		
		;
		
		\draw[thick, fill=white] 
		(-2,0) circle (\q pt)
		(-1,0) circle (\q pt)
		(0,0) circle (\q pt)
		(1,0) circle (\q pt)
		(2,0) circle (\q pt)
		(3,0) circle (\q pt)
		(4,0) circle (\q pt)
		(4+\k,\k) circle (\q pt)
		(4+\k,-\k) circle (\q pt)
		;
		
		\draw[thick, fill=white] 
		(-2 -\q/30,1 -\q/30) rectangle (-2 +\q/30,1 +\q/30)
		;

		\draw 
		(-2,1.5) node {\size $1$}
		
		
		(-2,-0.5) node {\size $0$}
		(-1,-0.5) node {\size $0$}
		(0,-0.5) node {\size $0$}
		(1,-0.5) node {\size $0$}
		(2,-0.5) node {\size $0$}
		(3,-0.5) node {\size $0$}
		(3.75,0.45) node {\size $0$}
		
		(4+\k,1.25) node {\size $0$}
		(4+\k,-1.25) node {\size $0$}
		;
		
		\draw[white, fill=white] (1.5,2) rectangle (2.4,-1);
		\draw (2,0) node {$\dots$};
		\end{scope}

		\begin{scope}[yshift = -4.5cm, xshift = 0.5cm]
		\draw[thick] (-2,0) -- (4,0) -- (4.7,0.7) (4,0) -- (4.7,-0.7) 
		(0,0) -- (0,1)
		
		;
		
		\draw[thick, fill=white] 
		(-2,0) circle (\q pt)
		(-1,0) circle (\q pt)
		(0,0) circle (\q pt)
		(1,0) circle (\q pt)
		(2,0) circle (\q pt)
		(3,0) circle (\q pt)
		(4,0) circle (\q pt)
		(4+\k,\k) circle (\q pt)
		(4+\k,-\k) circle (\q pt)
		;
		
		\draw[thick, fill=white] 
		(0 -\q/30,1 -\q/30) rectangle (0 +\q/30,1 +\q/30)
		;

		\draw 
		(0,1.5) node {\size $1$}
		
		
		(-2,-0.5) node {\size $0$}
		(-1,-0.5) node {\size $1$}
		(0,-0.5) node {\size $2$}
		(1,-0.5) node {\size $2$}
		(2,-0.5) node {\size $2$}
		(3,-0.5) node {\size $2$}
		(3.75,0.45) node {\size $2$}
		
		(4+\k,1.25) node {\size $1$}
		(4+\k,-1.25) node {\size $1$}
		;
		
		\draw[white, fill=white] (1.5,0.7) rectangle (2.4,-1);
		\draw (2,0) node {$\dots$};
		
		\end{scope}

		\begin{scope}[yshift = -9cm, xshift = 7.5cm]
		\draw[thick] (-4,0) -- (4,0) -- (4.7,0.7) (4,0) -- (4.7,-0.7) 
		(0,0) -- (0,1)
		
		;
		
		\draw[thick, fill=white] 
		(-4,0) circle (\q pt)
		(-3,0) circle (\q pt)
		(-2,0) circle (\q pt)
		(-1,0) circle (\q pt)
		(0,0) circle (\q pt)
		(1,0) circle (\q pt)
		(2,0) circle (\q pt)
		(3,0) circle (\q pt)
		(4,0) circle (\q pt)
		(4+\k,\k) circle (\q pt)
		(4+\k,-\k) circle (\q pt)
		;
		
		\draw[thick, fill=white] 
		(0 -\q/30,1 -\q/30) rectangle (0 +\q/30,1 +\q/30)
		;

		\draw 
		(0,1.5) node {\size $1$}
		
		(-4,-0.5) node {\size $0$}
		(-3,-0.5) node {\size $1$}
		(-2,-0.5) node {\size $2$}
		(-1,-0.5) node {\size $3$}
		(0,-0.5) node {\size $4$}
		(1,-0.5) node {\size $4$}
		(2,-0.5) node {\size $4$}
		(3,-0.5) node {\size $4$}
		(3.75,0.45) node {\size $4$}
		
		(4+\k,1.25) node {\size $2$}
		(4+\k,-1.25) node {\size $2$}
		;
		
		\draw[white, fill=white] (1.5,0.7) rectangle (2.4,-1);
		\draw (2,0) node {$\dots$};
		
		\end{scope}

		\begin{scope}[yshift = -9cm, xshift = -4.5cm]
		\draw[thick] (-2,0) -- (4,0) -- (4.7,0.7) (4,0) -- (4.7,-0.7) 
		(-2,0) -- (-2,1)
		(-1,0) -- (-1,1)
		
		;
		
		\draw[thick, fill=white] 
		(-2,0) circle (\q pt)
		(-1,0) circle (\q pt)
		(0,0) circle (\q pt)
		(1,0) circle (\q pt)
		(2,0) circle (\q pt)
		(3,0) circle (\q pt)
		(4,0) circle (\q pt)
		(4+\k,\k) circle (\q pt)
		(4+\k,-\k) circle (\q pt)
		;
		
		\draw[thick, fill=white] 
		(-2 -\q/30,1 -\q/30) rectangle (-2 +\q/30,1 +\q/30)
		(-1 -\q/30,1 -\q/30) rectangle (-1 +\q/30,1 +\q/30)
		;

		\draw 
		(-2,1.5) node {\size $1$}
		(-1,1.5) node {\size $1$}
		
		(-2,-0.5) node {\size $1$}
		(-1,-0.5) node {\size $2$}
		(0,-0.5) node {\size $2$}
		(1,-0.5) node {\size $2$}
		(2,-0.5) node {\size $2$}
		(3,-0.5) node {\size $2$}
		(3.75,0.45) node {\size $2$}
		
		(4+\k,1.25) node {\size $1$}
		(4+\k,-1.25) node {\size $1$}
		;
		
		\draw[white, fill=white] (1.5,0.7) rectangle (2.4,-1);
		\draw (2,0) node {$\dots$};
		
		\end{scope}

		\begin{scope}[yshift = -13.5cm, xshift = 14cm]
		\draw[thick] (-4,0) -- (4,0) -- (4.7,0.7) (4,0) -- (4.7,-0.7) 
		(0,0) -- (0,1)
		
		;
		
		\draw[thick, fill=white] 
		(-4,0) circle (\q pt)
		(-3,0) circle (\q pt)
		(-2,0) circle (\q pt)
		(-1,0) circle (\q pt)
		(0,0) circle (\q pt)
		(1,0) circle (\q pt)
		(2,0) circle (\q pt)
		(3,0) circle (\q pt)
		(4,0) circle (\q pt)
		(4+\k,\k) circle (\q pt)
		(4+\k,-\k) circle (\q pt)
		;
		
		\draw[thick, fill=white] 
		(0 -\q/30,1 -\q/30) rectangle (0 +\q/30,1 +\q/30)
		;

		\draw 
		(0,1.5) node {\size $1$}
		
		(-4,-0.5) node {\size $0$}
		(-3,-0.5) node {\size $1$}
		(-2,-0.5) node {\size $2$}
		(-1,-0.5) node {\size $5$}
		(0,-0.5) node {\size $6$}
		(1,-0.5) node {\size $6$}
		(2,-0.5) node {\size $6$}
		(3,-0.5) node {\size $6$}
		(3.75,0.45) node {\size $6$}
		
		(4+\k,1.25) node {\size $3$}
		(4+\k,-1.25) node {\size $3$}
		;
		
		\draw[white, fill=white] (-2.5,0.7) rectangle (-1.6,-1);
		\draw (-2,0) node {$\dots$};
		
		\draw[white, fill=white] (1.5,0.7) rectangle (2.4,-1);
		\draw (2,0) node {$\dots$};
		
		\end{scope}

		\begin{scope}[yshift = -13.5cm, xshift = -7cm]
		\draw[thick] (-3,0) -- (4,0) -- (4.7,0.7) (4,0) -- (4.7,-0.7) 
		(-3,0) -- (-3,1)
		
		;
		
		\draw[thick, fill=white] 
		(-3,0) circle (\q pt)
		(-2,0) circle (\q pt)
		(-1,0) circle (\q pt)
		(0,0) circle (\q pt)
		(1,0) circle (\q pt)
		(2,0) circle (\q pt)
		(3,0) circle (\q pt)
		(4,0) circle (\q pt)
		(4+\k,\k) circle (\q pt)
		(4+\k,-\k) circle (\q pt)
		;
		
		\draw[thick, fill=white] 
		(-3 -\q/30,1 -\q/30) rectangle (-3 +\q/30,1 +\q/30)
		;

		\draw 
		(-3,1.5) node {\size $3$}
		
		(-3,-0.5) node {\size $2$}
		(-2,-0.5) node {\size $2$}
		(-1,-0.5) node {\size $2$}
		(0,-0.5) node {\size $2$}
		(1,-0.5) node {\size $2$}
		(2,-0.5) node {\size $2$}
		(3,-0.5) node {\size $2$}
		(3.75,0.45) node {\size $2$}
		
		(4+\k,1.25) node {\size $1$}
		(4+\k,-1.25) node {\size $1$}
		;
		
		\draw[white, fill=white] (1.5,0.7) rectangle (2.4,-1);
		\draw (2,0) node {$\dots$};
		
		\end{scope}

		\begin{scope}[yshift = -13.5cm, xshift = 3cm]
		\draw[thick] (-3,0) -- (4,0) -- (4.7,0.7) (4,0) -- (4.7,-0.7) 
		(-3,0) -- (-3,1)
		(0,0) -- (0,1)
		
		;
		
		\draw[thick, fill=white] 
		(-3,0) circle (\q pt)
		(-2,0) circle (\q pt)
		(-1,0) circle (\q pt)
		(0,0) circle (\q pt)
		(1,0) circle (\q pt)
		(2,0) circle (\q pt)
		(3,0) circle (\q pt)
		(4,0) circle (\q pt)
		(4+\k,\k) circle (\q pt)
		(4+\k,-\k) circle (\q pt)
		;
		
		\draw[thick, fill=white] 
		(-3 -\q/30,1 -\q/30) rectangle (-3 +\q/30,1 +\q/30)
		(0 -\q/30,1 -\q/30) rectangle (0 +\q/30,1 +\q/30)
		;

		\draw 
		(-3,1.5) node {\size $1$}
		(0,1.5) node {\size $1$}
		
		(-3,-0.5) node {\size $1$}
		(-2,-0.5) node {\size $2$}
		(-1,-0.5) node {\size $3$}
		(0,-0.5) node {\size $4$}
		(1,-0.5) node {\size $4$}
		(2,-0.5) node {\size $4$}
		(3,-0.5) node {\size $4$}
		(3.75,0.45) node {\size $4$}
		
		(4+\k,1.25) node {\size $2$}
		(4+\k,-1.25) node {\size $2$}
		;
		
		\draw[white, fill=white] (1.5,0.7) rectangle (2.4,-1);
		\draw (2,0) node {$\dots$};
		
		\end{scope}

		\draw[ultra thick] 
		(2,-1.5) -- (2,-3) 
		(-0.3,-6.3) -- (-1.5,-7.5) 
		(4.3,-6.3) -- (5.5,-7.5) 
		(-3.5,-10.5) -- (-5,-12)
		(9.5,-10.5) -- (11,-12) 
		(3,-11.5) -- (1.5,-10.5)
		(4,-11.5) -- (5.5,-10.5)
		;
		\draw (2.9,-2.25) node {\sz$D_{n-1}$};
		\draw (5.85,-6.7) node {\sz$D_{n-3}$};
		\draw (11.35,-11.25) node {\sz$D_{n-5}$};
		\draw (-1.85,-6.7) node {\sz$A_2$};
		
		\draw (1.5,-11.25) node {\sz$D_{n-2} $};
		\draw (-3.5,-11.25) node {\sz$a_2$};
		\draw (5.5,-11.3) node {\sz$ A_4$};
		\end{tikzpicture}
	\end{center}
	\caption{The beginning of the Hasse diagram for quiver addition of even theories with an odd magnitude partition. These can be realised as the difference of two balanced quivers with even magnitude partitions, but an easier way to read off the moduli space singularity structure is to perform quiver addition having premised that the node at the end of the tail has an excess of one. This arises because the traversing structure for even theories necessarily changes the magnitude of the partitions assigned to the theory by an even amount. To cover all possible assignations therefore, it is proper to construct two Hasse diagrams, one starting at the partition (0) and the other at the partition (1). This observation will be important for the $\wt{D}_n$ Dynkin quivers, but we defer discussion to future work. }
	\label{ADDING}
\end{figure}
The position of a theory in a partition subdiagram of figure \ref{FIGUREMUE} corresponds to the partitions $\kappa$ and $\lambda$ that can be extracted from the quiver in the usual manner via (\ref{goodevenpart}). The editing prescription for figure \ref{FIGUREMUE} is the same as for figure \ref{FIGUREE}.

\subsubsection{Good, even classification}
All good, even $D_n$ Dynkin quivers can be considered as the difference between two balanced quivers. While it is possible to find all good, even quivers when restricted to even magnitude partitions only and using figure \ref{FIGUREE}, the moduli space analysis is more direct when allowing odd magnitude partitions as well using figure \ref{FIGUREMUE}. All good even type $D_n$ Dynkin quivers are classified using two partitions (not necessarily of equal magnitude), say $\kappa$ and $\lambda$, and a value $n$. The class is denoted $D_{\kappa}^{\lambda}(n)_{e}$, where $e$ denotes even type, $\kappa$ and $\lambda$ are restricted in a number of ways such that they are compatible with the value of $n$ (no part is larger than $n$) and with the need for quiver subtraction to not produce gauge nodes with negative rank. 

The moduli space singularity structure for good quivers can then be read off of figure \ref{FIGUREE} almost immediately by considering \textit{runs} of nodes and edges. A run on the Hasse diagram is simply a pair of nodes between which there is a Hasse subdiagram. Take a very simple example, the pair of partitions $(6,1^2),(3^2,1^2) \in \mathcal{P}(8)$ form a run as there is a Hasse subdiagram suspended between them, whereas $(5,1^3)$ and $(4^2)$ do not form a run. Runs corresponding to partition Hasse subdiagrams exactly correspond to pairs of partitions where one dominates the other in the usual partition dominance ordering sense. When the Hasse diagram is more complicated the notion of dominance is maintained. Dominance in figures \ref{FIGUREE} and \ref{FIGUREMUE} also needs to act \textit{between} partition Hasse subdiagrams. For $\lambda \in \mathcal{P}(2p)$ and $\kappa \in \mathcal{P}(2p+2j)$ to define a run corresponding to a variety that is realised as a $D_n$ Dynkin quiver Higgs branch, it is required that $ (\lambda^t,1^{2j})^t > \kappa$. This extension of dominance to figures \ref{FIGUREE} and \ref{FIGUREMUE} depends on the traversing structure. This is just the usual partition dominance ordering when $j=0$. The relationship for $\lambda$ and $\kappa$ is exactly the condition that quiver subtraction needs in order to be well defined (none of the gauge nodes become negative). The edits it is necessary to perform on figure \ref{FIGUREE} correspond to the restrictions on the partitions from the value of $n$. 

\subsubsection{Dimension matching for good, even theories}
A number of aspects of the analysis can be checked by performing further moduli space dimension calculations. We start with the calculation of the Higgs branch dimensions for a good, even theory. These theories are constructed as differences of balanced theories, so it is expected that
\begin{equation}\label{higgsdiff}
\begin{split}
\dimh(\Hg(D_{\nu}^{\mu}(n)_e)) & = \dimh(\Hg(D_\nu(n)_e - D_\mu(n)_e)) \\ & = \dimh(\Hg(D_\nu(n)_e)) - \dimh(\Hg(D_\mu(n)_e)) \\ & = \frac{1}{2}\left( \sum_i(\nu_i^t)^2 - \sum_i(\mu_i^t)^2\right) 
\end{split}
\end{equation}
which is indeed the case. It may be confirmed by summing over $j$ up to $|\nu|-|\mu|-2$, instead of $|\nu|-2$ as in (\ref{genhigcalc}),
\begin{equation}
\begin{split}
\dimh(\Hg(D_{\nu}^{\mu}(n)_e)) & = \sum_{ j=0, ~ \mathrm{ even}}^{|\nu|-|\mu|-2} \left[ \dimh\left( D_{n-l(\lambda_{|\kappa|-j-2}^t)}\right) + \dimh \left( \es_{\lambda_{|\kappa|-j}} \cap \O_{(\lambda_{|\kappa|-j-2}^t,1^2)^t}\right) \right] \\ & = \frac{1}{2}(|\nu|-|\mu|) + \frac{1}{2}\left( \sum_i((\lambda_{|\nu|}^t)_i)^2 - \sum_i((\lambda_{|\mu|}^t)_i)^2 \right) - \frac{1}{2}(|\nu|-|\mu|) \\ & = \frac{1}{2}\left( \sum_i(\nu_i^t)^2 - \sum_i(\mu_i^t)^2\right).
\end{split}
\end{equation}

The Coulomb branch dimension check is relatively simple. On the one side the dimension is calculated by a sum of the dimensions of individual edges of figure \ref{FIGUREE}, this was done for balanced quivers in (\ref{threetofour}). For good theories, replace the sum up to $|\nu|-2$ with a sum to $|\nu|-|\mu|-2$, and skip immediately to line 5 of (\ref{threetofour}) replacing the sum's limits appropriately, 
\begin{equation}\label{coulombgen}
\begin{split}
\dimh(\mathcal{C}(D_\nu^\mu(n)_e)) & = \sum_{ j=0, ~ \mathrm{ even}}^{|\nu|-|\mu|-2}\left[ 2n-1 + \frac{1}{2}\sum_i((\lambda_{|\kappa|-j-2})_i)^2 - \frac{1}{2}\sum_i((\lambda_{|\kappa|-j})_i)^2\right] \\ & = \frac{1}{2}(|\nu|-|\mu|)(2n-1) - \frac{1}{2}\left( \sum_i(\nu_i)^2 - \sum_i(\mu_i)^2\right).
\end{split}
\end{equation}
When $\mu=(0)$, this simplifies to the balanced case exactly as expected.

The other computation of the Coulomb branch dimension check for balanced quivers was (\ref{twotofourone}) and (\ref{twotofourtwo}). For good quivers, consider the extra factor of excess as in (\ref{newnodes}), and the analysis is essentially the same. Begin with the generalization of (\ref{twotofourone}),
\begin{equation}
\sum_{k=1}^{n-2}\left[ k g_1 - \sum_{i=1}^{k-1}(k-i)(f_i-e_i) \right]  +g'+g''   = \frac{1}{2}g_1(n-2)(n-1)+g'+g'' - \sum_{l=1}^{n-3}\left[ (f_l-e_l) \sum_{j=1}^{n-2-l}j\right],
\end{equation}
and follow the same analysis as (\ref{twotofourtwo})
\begin{equation}
\begin{split}
& \frac{1}{2}g_1(n-2)(n-1) + |\kappa| - (f_n-e_n) - \sum_{l=1}^{n-3} \frac{1}{2}(f_i-e_i)(n-2-l)(n-1-l)
\\ & = \frac{1}{2}(|\nu|-|\mu|)(2n-1) + \frac{1}{2}\left[- \sum_{l=1}^n l^2(f_l-e_l)\right] \\ & = \frac{1}{2}(|\nu|-|\mu|)(2n-1) - \frac{1}{2}\left( \sum_i(\nu_i)^2 - \sum_i(\mu_i)^2\right).
\end{split}
\end{equation}
This matches the result in (\ref{coulombgen}) and so the Coulomb branch dimension check is passed for good quivers.

Explicit calculation of moduli space dimension also allows a consistency check when identifying the theories with an odd partition magnitude as a difference of balanced (and hence even partitioned) theories. The claim is that for $|\nu|$ and $|\mu|$ odd,
\begin{equation}\label{odddiff}
D_\nu^\mu(n)_e = D_{\nu'}(n+1) - D_{\mu'}(n+1).
\end{equation}
with
\begin{equation}\label{newpartitions}
\begin{split}
\nu' & = (n+1^{f_n}, n^{f_{n-1}},\dots,2^{f_1},1^{f_0}) \\
\mu' & = (n+1^{e_n}, n^{e_{n-1}},\dots,2^{e_1},1^{l(\nu)-l(\mu)+f_0}).
\end{split}
\end{equation}
Alternatively these theories could be read from figure \ref{FIGUREMUE}. This alternative construction allows a similar analysis as the one performed for figure \ref{FIGUREE}. (Figure \ref{FIGUREMUE} is of the same form, using different partitions. The only change to make is that the $j$ sum is taken over odd values instead of even values). Therefore, from the alternative construction in figure \ref{FIGUREMUE} on one hand, and from (\ref{higgsdiff}) and (\ref{odddiff}) on the other, consistency requires
\begin{equation}
\dimh(\Hg(D_{\nu}^{\mu}(n)_e))  = \frac{1}{2}\left( \sum_i(\nu_i^t)^2 - \sum_i(\mu_i^t)^2\right) = \frac{1}{2}\left( \sum_i((\nu')_i^t)^2 - \sum_i((\mu')_i^t)^2\right).
\end{equation}
for $\nu$ and $\mu$ given in the standard way and $\nu'$ and $\mu'$ given by (\ref{newpartitions}). Note that for $\nu'$ we have the relation
\begin{equation}
\begin{split}
\sum_i((\nu')_i^t)^2 & = \sum_{q=1}^n\left( \sum_{j=q}^n f_j\right)^2 + \left( \sum_{j=0}^n f_j\right)^2 \\ & = \sum_i(\nu_i^t)^2 + (l(\nu)+f_0)^2.
\end{split}
\end{equation}
Thus 
\begin{equation}
\begin{split}
\frac{1}{2}\left( \sum_i((\nu')_i^t)^2 - \sum_i((\mu')_i^t)^2\right) & = \frac{1}{2}\left( \sum_i(\nu_i^t)^2 - \sum_i(\mu_i^t)^2\right) \\ & ~~~~+ (l(\nu)+f_0)^2 - (l(\mu)+l(\nu)-l(\mu) + f_0)^2 \\ & =  \frac{1}{2}\left( \sum_i(\nu_i^t)^2 - \sum_i(\mu_i^t)^2\right)
\end{split}
\end{equation}
as required. A second check of the odd partition even theories is a Coulomb branch dimension consideration which requires
\begin{equation}\label{coulombodd}
\begin{split}
\frac{1}{2}(|\nu|-|\mu|)&(2n-1)  - \frac{1}{2}\left( \sum_i(\nu_i)^2 - \sum_i(\mu_i)^2\right) \\ & = \frac{1}{2}(|\nu'|-|\mu'|)(2(n+1)-1) - \frac{1}{2}\left( \sum_i((\nu')_i)^2 - \sum_i((\mu')_i)^2\right).
\end{split}
\end{equation}
Note that $|\nu'| = \sum_{j=0}^n (j+1)f_j = |\nu|+l(\nu) +f_0$ and $|\mu'| = \sum_{j=0}^n (j+1)e_j = |\mu|+l(\mu) +(l(\nu) - l(\mu) +f_0)$ and so it is plain that $|\nu'|-|\mu'| = |\nu|-|\mu|$. Also (writing $e_0 = l(\nu)-l(\mu)+f_0$), 
\begin{equation}
\begin{split}
-\frac{1}{2}\left( \sum_i((\nu')_i)^2 - \sum_i((\mu')_i)^2\right) & = - \frac{1}{2}\left( \sum_{j=0}^n(j+1)^2(f_j-e_j)\right)  \\& = -\frac{1}{2}\left( \sum_{j=0}^nj^2(f_j-e_j) + 2\sum_{j=0}^nj(f_j-e_j) +l(\nu') - l(\mu')\right) \\ & = -(|\nu|-|\mu|) - \frac{1}{2}\left( \sum_i(\nu_i)^2 - \sum_i(\mu_i)^2\right),
\end{split}
\end{equation}
which completes the equality in (\ref{coulombodd}). (\ref{lengthdiff}) was used in moving from line two to three.

\subsubsection{Recovering $\kk{so}_{2n}$ nilpotent varieties}
The two general quivers given in figure \ref{genquivs} are those quivers realising nilpotent varieties of $\kk{so}_{2n}$. Recasting these quivers under the classification via moduli space given here is straight-forward. Realising $\kk{so}_{2n}$ nilpotent varieties as $D_n$ quivers gives
\begin{equation}\label{recovery}
\begin{split}
A(p,q) & = D_{(n-q-2)}^{(n-q-2p-2)}(n)_e = D_{(2p)}(2p+q+2)_e \\
B(p,q) & = D_{(n-q-2,1)}^{(n-q-2p-1)}(n)_e
\end{split}
\end{equation}
where, in the context of the previous discussion, $n=m$ or $m+1$.

\subsubsection{Odd $D_n$ quiver Hasse diagram}
\begin{figure}
	\begin{center}
		\begin{tikzpicture}[scale = 1]
		\def\q{7}
		\def\k{0.7071}
		\def\size{\sz}
		
		\draw[thick] (0,0) -- (1.5,0) (2.5,0) --  (4,0) -- (4.7,0.7) (4,0) -- (4.7,-0.7) (0,0) -- (0,1)
		(1,0) -- (1,1)
		(3,0) -- (3,1)
		(4,0) -- (5,0)
		
		(4+\k,\k) -- (5+\k,\k)
		(4+\k,-\k) -- (5+\k,-\k)
		;
		
		\draw[thick, fill=white] (0,0) circle (\q pt)
		(1,0) circle (\q pt)
		(3,0) circle (\q pt)
		(4,0) circle (\q pt)
		(4+\k,\k) circle (\q pt)
		(4+\k,-\k) circle (\q pt)
		;
		
		\draw[thick, fill=white] 
		(0 -\q/30,1 -\q/30) rectangle (0 +\q/30,1 +\q/30)
		(1 -\q/30,1 -\q/30) rectangle (1 +\q/30,1 +\q/30)
		(3 -\q/30,1 -\q/30) rectangle (3 +\q/30,1 +\q/30)
		(5 -\q/30,0 -\q/30) rectangle (5 +\q/30,0 +\q/30)
		(5+\k -\q/30,\k -\q/30) rectangle (5+\k +\q/30,\k +\q/30)
		(5+\k -\q/30,-\k -\q/30) rectangle (5+\k +\q/30,-\k +\q/30)
		;

		\draw (0,1.5) node {\size $f_1$}
		(1,1.5) node {\size $f_2$}
		(3,1.5) node {\size $f_{n-3}$}
		
		(5.8,0) node  {\size $f_{n-2}$}
		(5.8 +\k,0+\k) node  {\size $f_{n-1}$}
		(6.4+\k,0-\k) node  {\size $f_{n-1}+2f_n+1$}

		(0,-0.5) node {\size $g_1$}
		(1,-0.5) node {\size $g_2$}
		(3,-0.5) node {\size $g_{n-3}$}
		(3.75,0.45) node {\size $g_{n-2}$}
		
		(4+\k,1.25) node {\size $g'$}
		(4+\k,-1.25) node {\size $g''$}
		;
		
		\draw (2,0) node {$\dots$};

		\end{tikzpicture}
	\end{center}
	\caption{The general structure of a $D_n$ Dynkin quiver of \textit{odd} type. Note that this quiver is necessarily unbalanceable like in the alternative construction for good, even theories of odd partition magnitude. Because the difference between the end node flavours is odd, it is implicit that for every quiver of odd type there are the options $I$ and $II$ discussed earlier. For simplicity these won't both be written from here on, however one should always recall that there are two quivers with the same field content at every point.}
	\label{FIGUREM}
\end{figure}
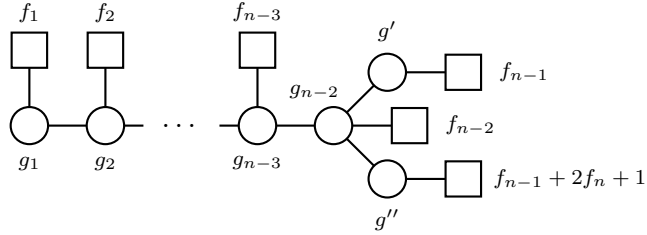

It is time to consider all of the good $D_n$ quivers which are not captured by $D_\nu^\mu(n)_e$. These are of odd type. Odd $D_n$ Dynkin quivers have an odd difference in the flavour on the end nodes, and are therefore always unbalanceable. Without loss of generality one can assume the flavours of the end nodes are $f_{n-1}$ and $f_{n-1}+2f_n+1$ respectively, figure \ref{FIGUREM}.
It can be shown that odd flavour difference requires an odd excess difference for the end nodes.

In order to give the class of good, odd theories an appropriate poset structure, and so build a Hasse diagram like figure \ref{FIGUREE}, quiver addition is employed once again. This will yield all possible good $D_n$ Dynkin quivers of odd type under quiver subtraction in the same manner as even theories were determined.

The positions of flavours on $D_n$ Dynkin quivers can be associated to partitions as (\ref{goodevenpart}) as in figure \ref{FIGUREM}. Note that this time there is an extra flavour on one end node. No singularity changes the fact that the difference of flavour is odd. Therefore the Hasse diagrams from odd theories never connect with the Hasse diagram for even theories. The lowest rank quiver, at the top of the Hasse diagram, is the quiver corresponding to the zero partition. Implicit for odd quivers is the option to swap the flavours and ranks of the end nodes, which formally gives two options, $I$ and $II$, for every node in the Hasse diagram. This extra notation is dropped from here for simplicity.

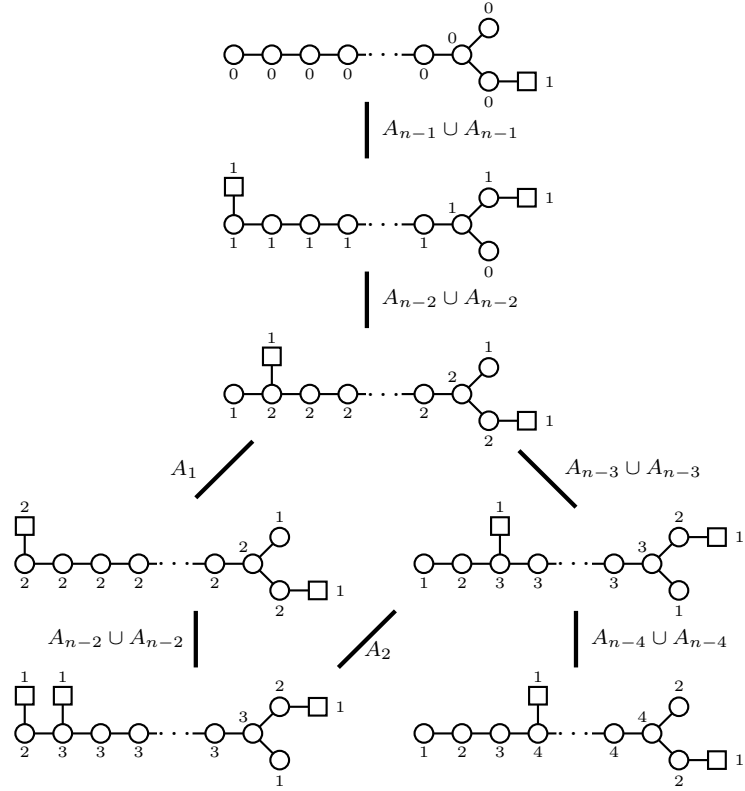
\begin{figure}
	\begin{center}
		\begin{tikzpicture}[scale = 0.5]
		\def\q{7}
		\def\k{0.7071}
		\def\size{\tiny}
		
		\begin{scope}[xshift = 0.5cm]
		\draw[thick] (-2,0) -- (4,0) -- (4.7,0.7) (4,0) -- (4.7,-0.7) 
		
		(4+\k,-\k) -- (5+\k,-\k)
		;
		
		\draw[thick, fill=white] 
		(-2,0) circle (\q pt)
		(-1,0) circle (\q pt)
		(0,0) circle (\q pt)
		(1,0) circle (\q pt)
		(2,0) circle (\q pt)
		(3,0) circle (\q pt)
		(4,0) circle (\q pt)
		(4+\k,\k) circle (\q pt)
		(4+\k,-\k) circle (\q pt)
		;
		
		\draw[thick, fill=white] 
		(5+\k -\q/30,-\k -\q/30) rectangle (5+\k +\q/30,-\k +\q/30)
		;

		\draw 
		
		(5.6+\k,0-\k) node  {\size $1$}
		
		(-2,-0.5) node {\size $0$}
		(-1,-0.5) node {\size $0$}
		(0,-0.5) node {\size $0$}
		(1,-0.5) node {\size $0$}
		(2,-0.5) node {\size $0$}
		(3,-0.5) node {\size $0$}
		(3.75,0.45) node {\size $0$}
		
		(4+\k,1.25) node {\size $0$}
		(4+\k,-1.25) node {\size $0$}
		;
		
		\draw[white, fill=white] (1.5,2) rectangle (2.4,-1);
		\draw (2,0) node {$\dots$};
		\end{scope}

		\begin{scope}[yshift = -4.5cm, xshift = 0.5cm]
		\draw[thick] (-2,0) -- (4,0) -- (4.7,0.7) (4,0) -- (4.7,-0.7) 
		(-2,0) -- (-2,1)
		
		(4+\k,\k) -- (5+\k,\k)
		;
		
		\draw[thick, fill=white] 
		(-2,0) circle (\q pt)
		(-1,0) circle (\q pt)
		(0,0) circle (\q pt)
		(1,0) circle (\q pt)
		(2,0) circle (\q pt)
		(3,0) circle (\q pt)
		(4,0) circle (\q pt)
		(4+\k,\k) circle (\q pt)
		(4+\k,-\k) circle (\q pt)
		;
		
		\draw[thick, fill=white] 
		(-2 -\q/30,1 -\q/30) rectangle (-2 +\q/30,1 +\q/30)
		(5+\k -\q/30,\k -\q/30) rectangle (5+\k +\q/30,\k +\q/30)
		;

		\draw 
		(-2,1.5) node {\size $1$}
		
		(5.6 +\k,0+\k) node  {\size $1$}
		
		(-2,-0.5) node {\size $1$}
		(-1,-0.5) node {\size $1$}
		(0,-0.5) node {\size $1$}
		(1,-0.5) node {\size $1$}
		(2,-0.5) node {\size $1$}
		(3,-0.5) node {\size $1$}
		(3.75,0.45) node {\size $1$}
		
		(4+\k,1.25) node {\size $1$}
		(4+\k,-1.25) node {\size $0$}
		;
		
		\draw[white, fill=white] (1.5,0.7) rectangle (2.4,-1);
		\draw (2,0) node {$\dots$};
		
		\end{scope}

		\begin{scope}[yshift = -9cm, xshift = 0.5cm]
		\draw[thick] (-2,0) -- (4,0) -- (4.7,0.7) (4,0) -- (4.7,-0.7) 
		(-1,0) -- (-1,1)
		
		(4+\k,-\k) -- (5+\k,-\k)
		;
		
		\draw[thick, fill=white] 
		(-2,0) circle (\q pt)
		(-1,0) circle (\q pt)
		(0,0) circle (\q pt)
		(1,0) circle (\q pt)
		(2,0) circle (\q pt)
		(3,0) circle (\q pt)
		(4,0) circle (\q pt)
		(4+\k,\k) circle (\q pt)
		(4+\k,-\k) circle (\q pt)
		;
		
		\draw[thick, fill=white] 
		(-1 -\q/30,1 -\q/30) rectangle (-1 +\q/30,1 +\q/30)
		(5+\k -\q/30,-\k -\q/30) rectangle (5+\k +\q/30,-\k +\q/30)
		;

		\draw 
		(-1,1.5) node {\size $1$}
		
		(5.6+\k,0-\k) node  {\size $1$}
		
		(-2,-0.5) node {\size $1$}
		(-1,-0.5) node {\size $2$}
		(0,-0.5) node {\size $2$}
		(1,-0.5) node {\size $2$}
		(2,-0.5) node {\size $2$}
		(3,-0.5) node {\size $2$}
		(3.75,0.45) node {\size $2$}
		
		(4+\k,1.25) node {\size $1$}
		(4+\k,-1.25) node {\size $2$}
		;
		
		\draw[white, fill=white] (1.5,0.7) rectangle (2.4,-1);
		\draw (2,0) node {$\dots$};
		
		\end{scope}

		\begin{scope}[yshift = -13.5cm, xshift = -5cm]
		\draw[thick] (-2,0) -- (4,0) -- (4.7,0.7) (4,0) -- (4.7,-0.7) 
		(-2,0) -- (-2,1)
		
		(4+\k,-\k) -- (5+\k,-\k)
		;
		
		\draw[thick, fill=white] 
		(-2,0) circle (\q pt)
		(-1,0) circle (\q pt)
		(0,0) circle (\q pt)
		(1,0) circle (\q pt)
		(2,0) circle (\q pt)
		(3,0) circle (\q pt)
		(4,0) circle (\q pt)
		(4+\k,\k) circle (\q pt)
		(4+\k,-\k) circle (\q pt)
		;
		
		\draw[thick, fill=white] 
		(-2 -\q/30,1 -\q/30) rectangle (-2 +\q/30,1 +\q/30)
		(5+\k -\q/30,-\k -\q/30) rectangle (5+\k +\q/30,-\k +\q/30)
		;

		\draw 
		(-2,1.5) node {\size $2$}
		
		(5.6+\k,0-\k) node  {\size $1$}
		
		(-2,-0.5) node {\size $2$}
		(-1,-0.5) node {\size $2$}
		(0,-0.5) node {\size $2$}
		(1,-0.5) node {\size $2$}
		(2,-0.5) node {\size $2$}
		(3,-0.5) node {\size $2$}
		(3.75,0.45) node {\size $2$}
		
		(4+\k,1.25) node {\size $1$}
		(4+\k,-1.25) node {\size $2$}
		;
		
		\draw[white, fill=white] (1.5,0.7) rectangle (2.4,-1);
		\draw (2,0) node {$\dots$};
		
		\end{scope}

		\begin{scope}[yshift = -13.5cm, xshift = 5.5cm]
		\draw[thick] (-2,0) -- (4,0) -- (4.7,0.7) (4,0) -- (4.7,-0.7) 
		(0,0) -- (0,1)
		
		(4+\k,\k) -- (5+\k,\k)
		;
		
		\draw[thick, fill=white] 
		(-2,0) circle (\q pt)
		(-1,0) circle (\q pt)
		(0,0) circle (\q pt)
		(1,0) circle (\q pt)
		(2,0) circle (\q pt)
		(3,0) circle (\q pt)
		(4,0) circle (\q pt)
		(4+\k,\k) circle (\q pt)
		(4+\k,-\k) circle (\q pt)
		;
		
		\draw[thick, fill=white] 
		(0 -\q/30,1 -\q/30) rectangle (0 +\q/30,1 +\q/30)
		(5+\k -\q/30,\k -\q/30) rectangle (5+\k +\q/30,\k +\q/30)
		;

		\draw 
		(0,1.5) node {\size $1$}
		
		(5.6 +\k,0+\k) node  {\size $1$}
		
		(-2,-0.5) node {\size $1$}
		(-1,-0.5) node {\size $2$}
		(0,-0.5) node {\size $3$}
		(1,-0.5) node {\size $3$}
		(2,-0.5) node {\size $3$}
		(3,-0.5) node {\size $3$}
		(3.75,0.45) node {\size $3$}
		
		(4+\k,1.25) node {\size $2$}
		(4+\k,-1.25) node {\size $1$}
		;
		
		\draw[white, fill=white] (1.5,0.7) rectangle (2.4,-1);
		\draw (2,0) node {$\dots$};
		
		\end{scope}

		\begin{scope}[yshift = -18cm, xshift = -5cm]
		\draw[thick] (-2,0) -- (4,0) -- (4.7,0.7) (4,0) -- (4.7,-0.7) 
		(-2,0) -- (-2,1)
		(-1,0) -- (-1,1)
		
		(4+\k,\k) -- (5+\k,\k)
		;
		
		\draw[thick, fill=white] 
		(-2,0) circle (\q pt)
		(-1,0) circle (\q pt)
		(0,0) circle (\q pt)
		(1,0) circle (\q pt)
		(2,0) circle (\q pt)
		(3,0) circle (\q pt)
		(4,0) circle (\q pt)
		(4+\k,\k) circle (\q pt)
		(4+\k,-\k) circle (\q pt)
		;
		
		\draw[thick, fill=white] 
		(-2 -\q/30,1 -\q/30) rectangle (-2 +\q/30,1 +\q/30)
		(-1 -\q/30,1 -\q/30) rectangle (-1 +\q/30,1 +\q/30)
		(5+\k -\q/30,\k -\q/30) rectangle (5+\k +\q/30,\k +\q/30)
		;

		\draw 
		(-2,1.5) node {\size $1$}
		(-1,1.5) node {\size $1$}
		
		(5.6 +\k,0+\k) node  {\size $1$}
		
		(-2,-0.5) node {\size $2$}
		(-1,-0.5) node {\size $3$}
		(0,-0.5) node {\size $3$}
		(1,-0.5) node {\size $3$}
		(2,-0.5) node {\size $3$}
		(3,-0.5) node {\size $3$}
		(3.75,0.45) node {\size $3$}
		
		(4+\k,1.25) node {\size $2$}
		(4+\k,-1.25) node {\size $1$}
		;
		
		\draw[white, fill=white] (1.5,0.7) rectangle (2.4,-1);
		\draw (2,0) node {$\dots$};
		
		\end{scope}

		\begin{scope}[yshift = -18cm, xshift = 5.5cm]
		\draw[thick] (-2,0) -- (4,0) -- (4.7,0.7) (4,0) -- (4.7,-0.7) 
		(1,0) -- (1,1)
		
		(4+\k,-\k) -- (5+\k,-\k)
		;
		
		\draw[thick, fill=white] 
		(-2,0) circle (\q pt)
		(-1,0) circle (\q pt)
		(0,0) circle (\q pt)
		(1,0) circle (\q pt)
		(2,0) circle (\q pt)
		(3,0) circle (\q pt)
		(4,0) circle (\q pt)
		(4+\k,\k) circle (\q pt)
		(4+\k,-\k) circle (\q pt)
		;
		
		\draw[thick, fill=white] 
		(1 -\q/30,1 -\q/30) rectangle (1 +\q/30,1 +\q/30)
		(5+\k -\q/30,-\k -\q/30) rectangle (5+\k +\q/30,-\k +\q/30)
		;

		\draw 
		(1,1.5) node {\size $1$}
		
		(5.6+\k,0-\k) node  {\size $1$}
		
		(-2,-0.5) node {\size $1$}
		(-1,-0.5) node {\size $2$}
		(0,-0.5) node {\size $3$}
		(1,-0.5) node {\size $4$}
		(2,-0.5) node {\size $4$}
		(3,-0.5) node {\size $4$}
		(3.75,0.45) node {\size $4$}
		
		(4+\k,1.25) node {\size $2$}
		(4+\k,-1.25) node {\size $2$}
		;
		
		\draw[white, fill=white] (1.5,0.7) rectangle (2.4,-1);
		\draw (2,0) node {$\dots$};
		
		\end{scope}
		
		\draw[\THICC]
		(2,-1.25) -- (2,-2.75)
		(2,-5.75) -- (2,-7.25)
		(-1,-10.25) -- (-2.5,-11.75)
		(6,-10.5) -- (7.5,-12)
		(-2.5,-14.75) -- (-2.5,-16.25)
		(7.5,-14.75) -- (7.5,-16.25)
		(2.75,-14.75) -- (1.25,-16.25)
		;
		
		\draw (4.2,-2) node {\sz$A_{n-1}\cup A_{n-1}$}
		(4.2,-6.5) node {\sz$A_{n-2}\cup A_{n-2}$}
		(-4.6,-15.5) node {\sz$A_{n-2}\cup A_{n-2}$}
		(9.7,-15.5) node {\sz$A_{n-4}\cup A_{n-4}$}
		(9,-11) node {\sz$A_{n-3}\cup A_{n-3}$}
		(-2.8,-11) node {\sz$A_1$}
		(2.3,-15.8) node {\sz$A_2$}
		;
		
		\end{tikzpicture}
	\end{center}
	\caption{The beginning of the quiver addition for odd theories. Recall that at all stages there are options $I$ and $II$ as discussed previously. Note that because the difference of flavour on the end nodes is odd, the $D_k$ traversing structure is never possible. There is $A_k \cup A_k$ traversing structure only. However these transitions change the magnitude of the assigned partition by one each time, this means that all partitions are included in this Hasse diagram and there is no need to use two different starting theories to easily find all of the possible theories. In this sense the Hasse diagram structure for odd theories is simpler than for even theories. Note that whilst the end node with the excess of one is always as assigned at the top of the diagram, the `extra flavour' flips back and forth when only one of the two options is written.}
	\label{FIGUREO}
\end{figure}

The only first option under quiver addition is to add a single $A_{n-1}\cup A_{n-1}$ singularity as in figure \ref{FIGUREO}. Note that whilst the `extra' flavour has in a sense swapped nodes, the `extra' excess remains on the lower node. 

\begin{figure}
	\begin{center}
		\begin{tikzpicture}[scale=0.65]
		
		\def\scriptsize{\tiny}
		
		\begin{scope}[yshift = -8cm]
		
		\filldraw[black] (0,22) circle (4pt)
		(1,21) circle (4pt)
		;
		
		\begin{scope}[xshift = 2cm, yshift = 19cm]
		\filldraw[black] (0,0) circle (4pt)
		(0,1) circle (4pt);
		
		\draw[\THICC] (0,0) -- (0,1)
		(-0.4, 0.5) node {\scriptsize{$A_{1}$}};
		
		\end{scope}

		\begin{scope}[xshift = 4cm, yshift = 17cm]
		\filldraw[black] (0,0) circle (4pt)
		(0,1) circle (4pt)
		(0,2) circle (4pt);
		
		\draw[\THICC] (0,0) -- (0,2)
		(-0.4, 0.5) node {\scriptsize{$a_{2}$}}
		(-0.4, 1.5) node {\scriptsize{$A_{2}$}}
		;
		\end{scope}

		\begin{scope}[xshift = 6cm, yshift = 14cm]
		\filldraw[black] (0,0) circle (4pt)
		(0,1) circle (4pt)
		(0,2) circle (4pt)
		(0,3) circle (4pt)
		(0,4) circle (4pt);
		
		\draw[\THICC] (0,0) -- (0,4)
		(-0.4, 0.5) node {\scriptsize{$a_{3}$}}
		(0.4, 1.5) node {\scriptsize{$A_{1}$}}
		(-0.4, 2.5) node {\scriptsize{$A_{1}$}}
		(0.4, 3.5) node {\scriptsize{$A_{3}$}}
		;
		\end{scope}

		\begin{scope}[xshift = 9cm, yshift = 11cm]
		\filldraw[black] (0,0) circle (4pt)
		(0,1) circle (4pt)
		(0,2) circle (4pt)
		(0,3) circle (4pt)
		(0,4) circle (4pt)
		(0,5) circle (4pt)
		(0,6) circle (4pt);
		
		\draw[\THICC] (0,0) -- (0,6)
		(-0.4, 0.5) node {\scriptsize{$a_{4}$}}
		(-0.4, 1.6) node {\scriptsize{$a_{2}$}}
		(-0.4, 2.35) node {\scriptsize{$a_{1}$}}
		(-0.4, 3.5) node {\scriptsize{$A_{1}$}}
		(0.35, 4.55) node {\scriptsize{$A_{2}$}}
		(-0.4, 5.5) node {\scriptsize{$A_{4}$}}
		;
		\end{scope}

		\begin{scope}[xshift = 12cm, yshift = 8cm]
		\filldraw[black] (0,0) circle (4pt)
		(0,1) circle (4pt)
		(0,2) circle (4pt)
		(0,4) circle (4pt)
		(0,6) circle (4pt)
		(0,7) circle (4pt)
		(0,8) circle (4pt)
		(-1,3) circle (4pt)
		(1,3) circle (4pt)
		(-1,5) circle (4pt)
		(1,5) circle (4pt)
		;
		
		\draw[\THICC] (0,0) -- (0,2) -- (-1,3) -- (0,4) -- (-1,5) -- (0,6) -- (0,8) -- (0,6) -- (1,5) -- (0,4) -- (1,3) -- (0,2)
		
		(0.4, 0.5) node {\scriptsize{$a_{5}$}}
		(0.4, 1.5) node {\scriptsize{$a_{3}$}}
		(0.7, 2.3) node {\scriptsize{$a_{1}$}}
		(-0.7, 2.3) node {\scriptsize{$a_{1}$}}
		(0.7, 3.7) node {\scriptsize{$a_{2}$}}
		(-0.7, 3.7) node {\scriptsize{$a_{2}$}}
		(0.7, 4.3) node {\scriptsize{$A_{2}$}}
		(-0.2, 4.7) node {\scriptsize{$A_{2}$}}
		(0.7, 5.7) node {\scriptsize{$A_{1}$}}
		(-0.7, 5.7) node {\scriptsize{$A_{1}$}}
		(0.4, 6.5) node {\scriptsize{$A_{3}$}}
		(0.4, 7.5) node {\scriptsize{$A_{5}$}}
		;
		\end{scope}
		\end{scope}
		
		\draw[\THICC]
		(0,14) -- (2,12)
		(2,12) -- (4,11) -- (6,10) -- (9,9) -- (12,8) -- (14,7.5)
		(2,11) -- (4,10) -- (6,9) -- (9,8) -- (12,7)
		(4,9) -- (6,7)
		(6,8) -- (9,7) -- (12,6)
		(6,7) -- (9,6) -- (11,5)
		(6,6) -- (9,4) -- (11,3)
		(9,5) -- (12,4)
		(9,3) -- (12,1)
		(12,0) -- (13,-1)
		
		(13.2,3.6) -- (14.1,3.3)
		(15.8,2.95) -- (16.8,2.7)
		(18.2,2.45) -- (19.2,2.3)
		;
		
		\draw (15,3) node {$\mathcal{P}(7)$}
		(17.5,2.5) node {$\mathcal{P}(8)$}
		(20,2.2) node[rotate = -5] {$\dots$}
		;
		
		\draw (1.8,13.7) node {\tiny$A_{n-1}\cup A_{n-1}$}
		(2.8,12.7) node {\tiny$A_{n-2}\cup A_{n-2}$}
		(4.3,11.7) node {\tiny$A_{n-3}\cup A_{n-3}$}
		(1.8,10.3) node {\tiny$A_{n-2}\cup A_{n-2}$}
		
		(6.3,10.7) node {\tiny$A_{n-4}\cup A_{n-4}$}
		(7.6,9.75) node[rotate = -17] {\tiny$A_{n-5}\cup A_{n-5}$}
		(10.6,8.75) node[rotate = -17] {\tiny$A_{n-6}\cup A_{n-6}$}
		(10.6,7.75) node[rotate = -17] {\tiny$A_{n-5}\cup A_{n-5}$}
		(10.6,6.75) node[rotate = -17] {\tiny$A_{n-4}\cup A_{n-4}$}
		(10,6.2) node[rotate = -22] {\tiny$A_{n-4}$}
		(10.2,5.7) node[rotate = -22] {\tiny$\cup A_{n-4}$}
		(10.3,4.8) node[rotate = -19] {\tiny$A_{n-3}\cup A_{n-3}$}
		(10,4.2) node[rotate = -22] {\tiny$A_{n-3}$}
		(10.2,3.7) node[rotate = -22] {\tiny$\cup A_{n-3}$}
		(10.4,2.4) node[rotate = -33] {\tiny$A_{n-2}\cup A_{n-2}$}
		
		(4.8,10.25) node[rotate = -22] {\tiny$A_{n-3}$}
		(5.2,9.65) node[rotate = -22] {\tiny$\cup A_{n-3}$}
		
		(3.5,7.9) node {\tiny$A_{n-2}\cup A_{n-2}$}
		
		(7.45,8.2) node[rotate = -17] {\tiny$A_{n-4}\cup A_{n-4}$}
		(7.45,7.8) node[rotate = -17] {\tiny$A_{n-3}\cup A_{n-3}$}
		(7.45,6.2) node[rotate = -17] {\tiny$A_{n-3}\cup A_{n-3}$}
		(7.4,4.7) node[rotate = -34] {\tiny$A_{n-2}\cup A_{n-2}$}
		
		;

		\end{tikzpicture}
	\end{center}
	\caption{The general structure of the quiver addition Hasse diagram for theories of odd type. This is used in the same way as figures \ref{FIGUREE} and \ref{FIGUREMUE} were used for even quivers to deduce the moduli space singularity structure of any \textit{good} $D_n$ theory of odd type. }
	\label{FIGUREP}
\end{figure}
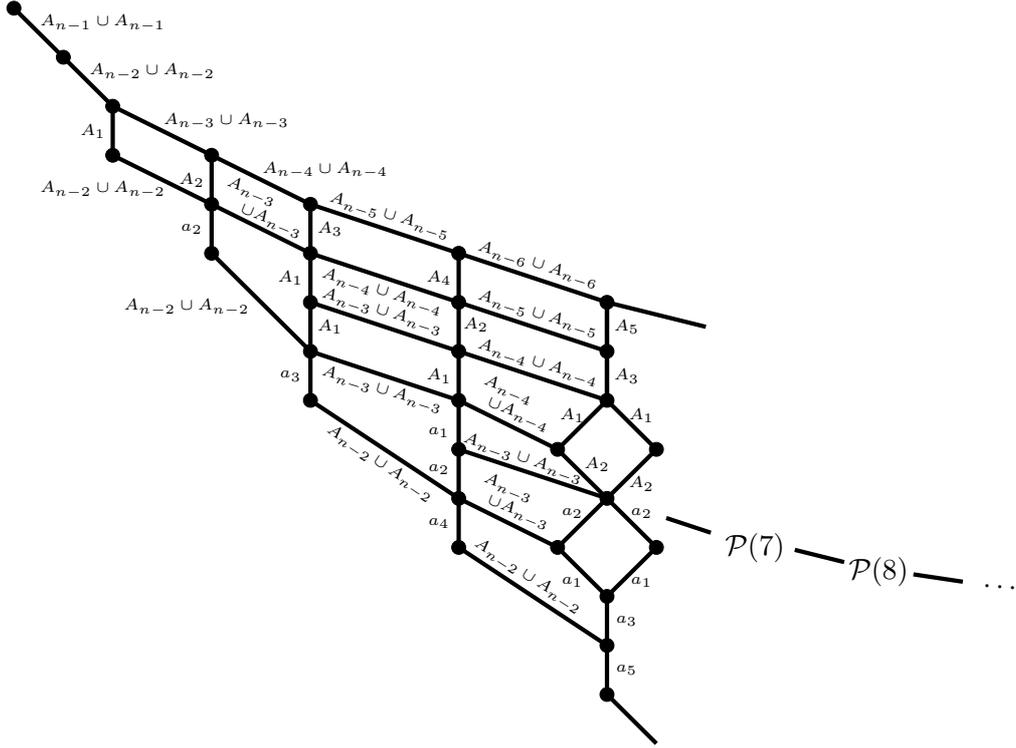

The full picture is given in figure \ref{FIGUREP}.
Once again a pattern of partition Hasse subdiagram emerges whereby the partition associated to the node in the subdiagram is the same as the partition associated to the flavours in the quiver. For balanced cases the structure traversing from a partition subdiagram to another consisted of $D_k$ transverse slices which changed the magnitude of the partitions by two. For figure \ref{FIGUREP} the traversing structure consists of $A_k\cup A_k$ singularities which change the partition magnitude by one at a time. This can once again be encapsulated as an edge diagram:
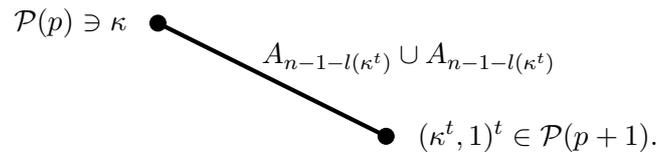
\begin{center}
	\begin{tikzpicture}
	\draw[fill=black] (0,0) circle (3pt)
	(3,-1.5) circle (3pt)
	;
	\draw[ultra thick] (0,0) -- (3,-1.5);
	
	\draw (-1.15,0) node {$\mathcal{P}(p) \ni \kappa$}
	(5,-1.5) node {$(\kappa^t,1)^t \in \mathcal{P}(p+1).$}
	(3.3,-0.475) node[rotate=0] {$A_{n-1 - l(\kappa^t)}\cup A_{n-1 - l(\kappa^t)}$}
	;
	
	\end{tikzpicture}
\end{center}

Observe that when the partition is of odd magnitude, the node with extra flavour and the node with excess are opposite, whereas when the partition is of even magnitude, they are on the same node. Since quiver addition doesn't change the excess of the nodes one can also observe that the end node with excess remains the only node with excess. By repeating the analysis (\ref{gbalance}) - (\ref{twogee}) it can be shown that indeed when the excess and flavour are on opposite nodes, $|\kappa| = 2g''-1$. When they are on the same node $|\kappa| = 2g''$. 

Like in the balanced case, the choice of a concrete $n$ will inevitably necessitate editing of the general structure presented in figure \ref{FIGUREP}. Once again this can be determined in a systematic way by observing which quivers and transitions are defined in the Hasse diagram or are possible at the level of the quivers and exploring what happens in the fringe cases.
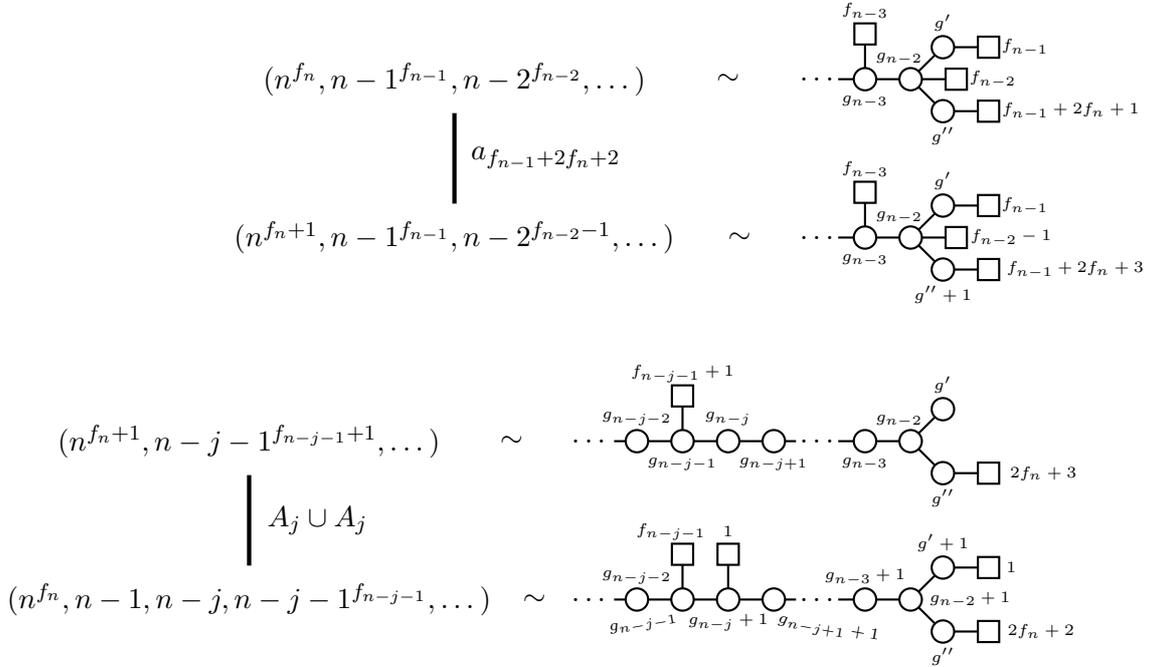
\begin{figure}
	\begin{center}
		\begin{tikzpicture}[scale = 0.6]
		\def\q{7}
		\def\k{0.7071}
		\def\size{\tiny}
		
		\draw[thick] (0,0) -- (4,0) -- (4.7,0.7) (4,0) -- (4.7,-0.7) (0,0) -- (0,1)
		(1,0) -- (1,1)
		(2,0) -- (2,1)
		(3,0) -- (3,1)
		(4,0) -- (5,0)
		
		(4+\k,\k) -- (5+\k,\k)
		(4+\k,-\k) -- (5+\k,-\k)
		;
		
		\draw[thick, fill=white] (0,0) circle (\q pt)
		(1,0) circle (\q pt)
		(2,0) circle (\q pt)
		(3,0) circle (\q pt)
		(4,0) circle (\q pt)
		(4+\k,\k) circle (\q pt)
		(4+\k,-\k) circle (\q pt)
		;
		
		\draw[thick, fill=white] 
		(0 -\q/30,1 -\q/30) rectangle (0 +\q/30,1 +\q/30)
		(1 -\q/30,1 -\q/30) rectangle (1 +\q/30,1 +\q/30)
		(2 -\q/30,1 -\q/30) rectangle (2 +\q/30,1 +\q/30)
		(3 -\q/30,1 -\q/30) rectangle (3 +\q/30,1 +\q/30)
		(5 -\q/30,0 -\q/30) rectangle (5 +\q/30,0 +\q/30)
		(5+\k -\q/30,\k -\q/30) rectangle (5+\k +\q/30,\k +\q/30)
		(5+\k -\q/30,-\k -\q/30) rectangle (5+\k +\q/30,-\k +\q/30)
		;

		\draw (0,1.5) node {\size $f_1$}
		(1,1.5) node {\size $f_2$}
		(2,1.5) node {\size $f_3$}
		(3,1.5) node {\size $f_{n-3}$}
		
		(5.8,0) node  {\size $f_{n-2}$}
		(5.8 +\k,0+\k) node  {\size $f_{n-1}$}
		(6.8+\k,0-\k) node  {\size $f_{n-1}+2f_n+1$}

		(0,-0.5) node {\size $g_1$}
		(1,-0.5) node {\size $g_2$}
		(2,-0.5) node {\size $g_3$}
		(3,-0.5) node {\size $g_{n-3}$}
		(3.75,0.45) node {\size $g_{n-2}$}
		
		(4+\k,1.25) node {\size $g'$}
		(4+\k,-1.25) node {\size $g''$}
		;
		
		\draw[white, fill=white] (-1,2) rectangle (2.4,-1);
		\draw (2,0) node {$\dots$};

		\draw (-6,0) node {$(n^{f_n}, n-1^{f_{n-1}}, n-2^{f_{n-2}}, \dots)$}
		(-6,-3.5) node {$(n^{f_n+1}, n-1^{f_{n-1}}, n-2^{f_{n-2}-1}, \dots )$}
		;

		\begin{scope}[yshift = -3.5cm]
		\draw[thick] (0,0) -- (4,0) -- (4.7,0.7) (4,0) -- (4.7,-0.7) (0,0) -- (0,1)
		(1,0) -- (1,1)
		(2,0) -- (2,1)
		(3,0) -- (3,1)
		(4,0) -- (5,0)
		
		(4+\k,\k) -- (5+\k,\k)
		(4+\k,-\k) -- (5+\k,-\k)
		;
		
		\draw[thick, fill=white] (0,0) circle (\q pt)
		(1,0) circle (\q pt)
		(2,0) circle (\q pt)
		(3,0) circle (\q pt)
		(4,0) circle (\q pt)
		(4+\k,\k) circle (\q pt)
		(4+\k,-\k) circle (\q pt)
		;
		
		\draw[thick, fill=white] 
		(0 -\q/30,1 -\q/30) rectangle (0 +\q/30,1 +\q/30)
		(1 -\q/30,1 -\q/30) rectangle (1 +\q/30,1 +\q/30)
		(2 -\q/30,1 -\q/30) rectangle (2 +\q/30,1 +\q/30)
		(3 -\q/30,1 -\q/30) rectangle (3 +\q/30,1 +\q/30)
		(5 -\q/30,0 -\q/30) rectangle (5 +\q/30,0 +\q/30)
		(5+\k -\q/30,\k -\q/30) rectangle (5+\k +\q/30,\k +\q/30)
		(5+\k -\q/30,-\k -\q/30) rectangle (5+\k +\q/30,-\k +\q/30)
		;

		\draw (0,1.5) node {\size $f_1$}
		(1,1.5) node {\size $f_2$}
		(2,1.5) node {\size $f_3$}
		(3,1.5) node {\size $f_{n-3}$}
		
		(6.2,0) node  {\size $f_{n-2}-1$}
		(5.8 +\k,0+\k) node  {\size $f_{n-1}$}
		(6.9+\k,0-\k) node  {\size $f_{n-1}+2f_n+3$}

		(0,-0.5) node {\size $g_1$}
		(1,-0.5) node {\size $g_2$}
		(2,-0.5) node {\size $g_3$}
		(3,-0.5) node {\size $g_{n-3}$}
		(3.75,0.45) node {\size $g_{n-2}$}
		
		(4+\k,1.25) node {\size $g'$}
		(4+\k,-1.25) node {\size $g''+1$}
		;
		
		\draw[white, fill=white] (-1,2) rectangle (2.4,-1);
		\draw (2,0) node {$\dots$};
		
		\end{scope}
		
		\draw (0,0) node {$\sim$};
		\draw (0.25,-3.5) node {$\sim$};
		
		\draw (-4,-1.75) node {$a_{f_{n-1}+2f_n+2}$};
		
		\draw[ultra thick] (-6,-0.75) -- (-6,-2.75);

		\begin{scope}[yshift = -11.5cm, xshift = 0cm]
		\draw[thick] (-2.5,0) -- (4,0) -- (4.7,0.7) (4,0) -- (4.7,-0.7) 
		(-1,0) -- (-1,1)
		(0,0) -- (0,1)
		
		(4+\k,\k) -- (5+\k,\k)
		(4+\k,-\k) -- (5+\k,-\k)
		;
		
		\draw[thick, fill=white] 
		
		(-2,0) circle (\q pt)
		(-1,0) circle (\q pt)
		(0,0) circle (\q pt)
		(1,0) circle (\q pt)
		(2,0) circle (\q pt)
		(3,0) circle (\q pt)
		(4,0) circle (\q pt)
		(4+\k,\k) circle (\q pt)
		(4+\k,-\k) circle (\q pt)
		;
		
		\draw[thick, fill=white] 
		(-1 -\q/30,1 -\q/30) rectangle (-1 +\q/30,1 +\q/30)
		(0 -\q/30,1 -\q/30) rectangle (0 +\q/30,1 +\q/30)
		(5+\k -\q/30,\k -\q/30) rectangle (5+\k +\q/30,\k +\q/30)
		(5+\k -\q/30,-\k -\q/30) rectangle (5+\k +\q/30,-\k +\q/30)
		;

		\draw 
		(-1.25,1.5) node {\size $f_{n-j-1}$}
		(0,1.5) node {\size $1$}
		
		(5.5 +\k,0+\k) node  {\size $1$}
		(6.15+\k,0-\k) node  {\size $2f_n+2$}
		(-2,0.5) node {\size $g_{n-j-2}$}
		(-1.85,-0.55) node[rotate = 10] {\size $g_{n-j-1}$}
		(0,-0.5) node[rotate = 0] {\size $g_{n-j}+1$}
		(2.2,-0.65) node[rotate = -10] {\size $g_{n-j+1}+1$}
		(3,0.5) node {\size $g_{n-3}+1$}
		(5.3,0) node[rotate = -0] {\size $g_{n-2}+1$}
		
		(4+\k,1.25) node {\size $g'+1$}
		(4+\k,-1.25) node {\size $g''$}
		;
		
		\draw[white, fill=white] (1.5,0.3) rectangle (2.4,-0.3);
		\draw (2,0) node {$\dots$};
		\draw (-3,0) node {$\dots$};
		
		\end{scope}
		
		\begin{scope}[yshift = -8cm, xshift = 0cm]
		\draw[thick] (-2.5,0) -- (4,0) -- (4.7,0.7) (4,0) -- (4.7,-0.7) 
		(-1,0) -- (-1,1)
		
		(4+\k,-\k) -- (5+\k,-\k)
		;
		
		\draw[thick, fill=white] 
		
		(-2,0) circle (\q pt)
		(-1,0) circle (\q pt)
		(0,0) circle (\q pt)
		(1,0) circle (\q pt)
		(2,0) circle (\q pt)
		(3,0) circle (\q pt)
		(4,0) circle (\q pt)
		(4+\k,\k) circle (\q pt)
		(4+\k,-\k) circle (\q pt)
		;
		
		\draw[thick, fill=white] 
		(-1 -\q/30,1 -\q/30) rectangle (-1 +\q/30,1 +\q/30)
		(5+\k -\q/30,-\k -\q/30) rectangle (5+\k +\q/30,-\k +\q/30)
		;

		\draw 
		(-1,1.5) node {\size $f_{n-j-1}+1$}
		
		(6.2+\k,0-\k) node  {\size $2f_n+3$}
		(-2,0.5) node {\size $g_{n-j-2}$}
		(-1,-0.5) node {\size $g_{n-j-1}$}
		(0,0.5) node {\size $g_{n-j}$}
		(1,-0.5) node {\size $g_{n-j+1}$}
		(3,-0.5) node {\size $g_{n-3}$}
		(3.75,0.45) node {\size $g_{n-2}$}
		
		(4+\k,1.25) node {\size $g'$}
		(4+\k,-1.25) node {\size $g''$}
		;
		
		\draw[white, fill=white] (1.5,0.7) rectangle (2.4,-0.3);
		\draw (2,0) node {$\dots$};
		\draw (-3,0) node {$\dots$};
		
		\end{scope}
		
		\begin{scope}[yshift = -8cm, xshift = -4.5cm]
		\draw (-6,0) node {$(n^{f_n+1}, n-j-1^{f_{n-j-1}+1}, \dots)$}
		(-6,-3.5) node {$(n^{f_n}, n-1,n-j, n-j-1^{f_{n-j-1}}, \dots )$}
		;
		\draw (-0.25,0) node {$\sim$};
		\draw (0.25,-3.5) node {$\sim$};
		
		\draw (-4.5,-1.75) node {$A_j \cup A_j$};
		
		\draw[ultra thick] (-6,-0.75) -- (-6,-2.75);
		
		\end{scope}
		
		\end{tikzpicture}
	\end{center}
	\caption{The editing prescription for the quiver addition Hasse diagram for theories of odd type presented in the same manner as figure \ref{FIGUREH}.}
	\label{FIGURER}
\end{figure}

\noindent \textbf{Editing prescription} $\quad$ To write down the Hasse diagram for good, odd $D_n$ Dynkin quivers for some specific $n$ one starts with the general construction in figure \ref{FIGUREP}, identifies in this construction all of the nodes with parts larger than $n$ and deletes them. Also delete any badly defined traversing edges. The final step is to put in edges following figure \ref{FIGURER}.

\begin{figure}
	\begin{center}
		\begin{tikzpicture}[scale=0.65]
		
		\def\scriptsize{\tiny}
		
		\begin{scope}[yshift = -8cm]
		
		\filldraw[black] (0,22) circle (4pt)
		(1,21) circle (4pt)
		;
		
		\begin{scope}[xshift = 2cm, yshift = 19cm]
		\filldraw[black] (0,0) circle (4pt)
		(0,1) circle (4pt);
		
		\draw[\THICC] (0,0) -- (0,1)
		(-0.4, 0.5) node {\scriptsize{$A_{1}$}};
		
		\end{scope}

		\begin{scope}[xshift = 4cm, yshift = 17cm]
		\filldraw[black] (0,0) circle (4pt)
		(0,1) circle (4pt)
		(0,2) circle (4pt);
		
		\draw[\THICC] (0,0) -- (0,2)
		(-0.4, 0.5) node {\scriptsize{$a_{2}$}}
		(-0.4, 1.5) node {\scriptsize{$A_{2}$}}
		;
		\end{scope}

		\begin{scope}[xshift = 6cm, yshift = 14cm]
		\filldraw[black] (0,0) circle (4pt)
		(0,1) circle (4pt)
		(0,2) circle (4pt)
		(0,3) circle (4pt)
		(0,4) circle (4pt);
		
		\draw[\THICC] (0,0) -- (0,4)
		(-0.4, 0.5) node {\scriptsize{$a_{3}$}}
		(0.4, 1.5) node {\scriptsize{$A_{1}$}}
		(-0.4, 2.5) node {\scriptsize{$A_{1}$}}
		(0.4, 3.5) node {\scriptsize{$A_{3}$}}
		;
		\end{scope}

		\begin{scope}[xshift = 9cm, yshift = 11cm]
		\filldraw[black] (0,0) circle (4pt)
		(0,1) circle (4pt)
		(0,2) circle (4pt)
		(0,3) circle (4pt)
		(0,4) circle (4pt)
		(0,5) circle (4pt)
		(0,6) circle (4pt);
		
		\draw (0,6) circle (7pt);
		
		\draw[\THICC] (0,0) -- (0,6)
		(-0.4, 0.5) node {\scriptsize{$a_{4}$}}
		(-0.4, 1.6) node {\scriptsize{$a_{2}$}}
		(-0.4, 2.35) node {\scriptsize{$a_{1}$}}
		(-0.4, 3.5) node {\scriptsize{$A_{1}$}}
		(0.35, 4.55) node {\scriptsize{$A_{2}$}}
		(-0.4, 5.5) node {\scriptsize{$A_{4}$}}
		;
		\end{scope}
		\end{scope}
		
		\draw[\THICC]
		(0,14) -- (2,12)
		(2,12) -- (4,11) -- (6,10) -- (9,9)
		(2,11) -- (4,10) -- (6,9) -- (9,8) 
		(4,9) -- (6,7)
		(6,8) -- (9,7) 
		(6,7) -- (9,6) 
		(6,6) -- (9,4)

		;

		\draw (1.6,13.7) node {\tiny$A_{3}\cup A_{3}$}
		(2.6,12.7) node {\tiny$A_{2}\cup A_{2}$}
		(4.1,11.7) node {\tiny$A_{1}\cup A_{1}$}
		(1.8,10.3) node {\tiny$A_{2}\cup A_{2}$}
		
		(5.9,10.7) node {\tiny$`A_{0}\cup A_{0}\textrm{'}$}
		(7.6,9.75) node[rotate = -17] {\tiny$`A_{-1}\cup A_{-1}\textrm{'}$}

		(4.9,9.95) node[rotate = -22] {\tiny$A_{1}\cup A_{1}$}
		
		(4,7.9) node {\tiny$A_{2}\cup A_{2}$}
		
		(7.45,8.2) node[rotate = -17] {\tiny$`A_{0}\cup A_{0}\textrm{'}$}
		(7.45,7.8) node[rotate = -17] {\tiny$A_{1}\cup A_{1}$}
		(7.45,6.2) node[rotate = -17] {\tiny$A_{1}\cup A_{1}$}
		(7.4,4.7) node[rotate = -34] {\tiny$A_{2}\cup A_{2}$}
		
		;

		\begin{scope}[xshift = 11cm]
		\begin{scope}[yshift = -8cm]
		
		\filldraw[black] (0,22) circle (4pt)
		(1,21) circle (4pt)
		;
		
		\begin{scope}[xshift = 2cm, yshift = 19cm]
		\filldraw[black] (0,0) circle (4pt)
		(0,1) circle (4pt);
		
		\draw[\THICC] (0,0) -- (0,1)
		(-0.4, 0.5) node {\scriptsize{$A_{1}$}};
		
		\end{scope}

		\begin{scope}[xshift = 4cm, yshift = 17cm]
		\filldraw[black] (0,0) circle (4pt)
		(0,1) circle (4pt)
		(0,2) circle (4pt);
		
		\draw[\THICC] (0,0) -- (0,2)
		(-0.4, 0.5) node {\scriptsize{$a_{2}$}}
		(-0.4, 1.5) node {\scriptsize{$A_{2}$}}
		;
		\end{scope}

		\begin{scope}[xshift = 6cm, yshift = 14cm]
		\filldraw[black] (0,0) circle (4pt)
		(0,1) circle (4pt)
		(0,2) circle (4pt)
		(0,3) circle (4pt)
		(0,4) circle (4pt);
		
		\draw[\THICC] (0,0) -- (0,4)
		(-0.4, 0.5) node {\scriptsize{$a_{3}$}}
		(0.4, 1.5) node {\scriptsize{$A_{1}$}}
		(-0.4, 2.5) node {\scriptsize{$A_{1}$}}
		(1, 3.5) node {\underline{\scriptsize{$A_{3}\cup A_3$}}}
		;
		\end{scope}

		\begin{scope}[xshift = 9cm, yshift = 11cm]
		\filldraw[black] (0,0) circle (4pt)
		(0,1) circle (4pt)
		(0,2) circle (4pt)
		(0,3) circle (4pt)
		(0,4) circle (4pt)
		(0,5) circle (4pt);
		
		\draw[\THICC] (0,0) -- (0,5)
		(-0.4, 0.5) node {\scriptsize{$a_{4}$}}
		(-0.4, 1.6) node {\scriptsize{$a_{2}$}}
		(-0.4, 2.35) node {\scriptsize{$a_{1}$}}
		(-0.4, 3.5) node {\scriptsize{$A_{1}$}}
		(1, 4.5) node {\underline{\scriptsize{$A_{2}\cup A_2$}}}
		;
		\end{scope}
		\end{scope}
		
		\draw[\THICC]
		(0,14) -- (2,12)
		(2,12) -- (4,11) 
		(2,11) -- (4,10) -- (6,9)
		(4,9) -- (6,7)
		(6,8) -- (9,7) 
		(6,7) -- (9,6) 
		(6,6) -- (9,4) 
		;
		
		\draw (1.6,13.7) node {\tiny$A_{3}\cup A_{3}$}
		(2.6,12.7) node {\tiny$A_{2}\cup A_{2}$}
		(3,11.2) node[rotate = -22] {\tiny$A_{1}\cup A_{1}$}
		(1.8,10.3) node {\tiny$A_{2}\cup A_{2}$}
		
		(4.9,9.2) node[rotate = -22] {\tiny$A_{1}\cup A_{1}$}
		
		(4,7.9) node {\tiny$A_{2}\cup A_{2}$}
		
		(7.45,7.2) node[rotate = -17] {\tiny$A_{1}\cup A_{1}$}
		(7.45,6.2) node[rotate = -17] {\tiny$A_{1}\cup A_{1}$}
		(7.4,4.7) node[rotate = -34] {\tiny$A_{2}\cup A_{2}$}
		
		(4,12) node {\tiny$a_2$}
		(8,8.6) node {\tiny$a_2$}
		
		;
		
		\draw[\THICC, densely dashed] (2,12) edge[bend left] (6,10);
		\draw[\THICC, densely dashed] (4,10) -- (9,8);
		\end{scope}
		
		\draw[\THICC, ->] (11,7) -- (14,7);
		
		\end{tikzpicture}
	\end{center}
	\caption{An example of the application of the odd type editing prescription to explicitly find the Hasse diagram for odd $D_4$ Dynkin quivers with $|\kappa| \leq 5$. This can be checked explicitly using quiver arithmetic. }
	\label{FIGUREXX}
\end{figure}
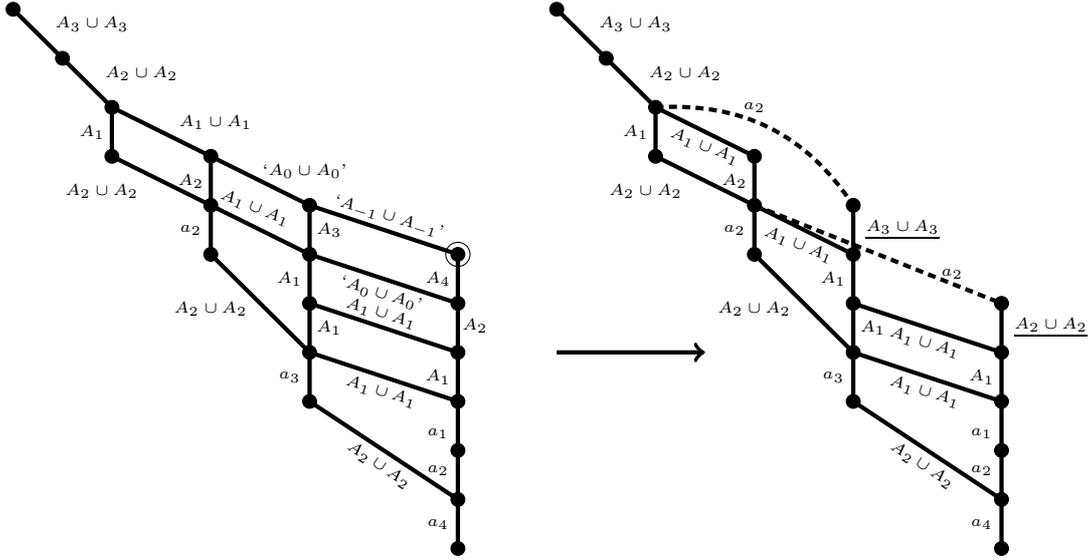

An example of performing this editing for $n=4$ theories and partitions $\kappa$ with $|\kappa| \leq 5$ in given in figure \ref{FIGUREXX}.

Theories living at the nodes of figure \ref{FIGUREP}, while not balanced, play the same role as balanced quivers in the even case and may be classified using just one partition and the number $n$, $D_\kappa(n)_o$. The moduli space singularity structure for the theory $D_\kappa(n)_o$ is given by the run on the general construction, after editing, from the very top to the node labelled with the partition $\kappa$. The difference between two quivers in figure \ref{FIGUREP} is a \textit{good}, odd $D_n$ Dynkin quiver. Taking differences of quivers in figure \ref{FIGUREP} will encompass all good, odd $D_n$ Dynkin quivers, as we will discuss now. 

\subsubsection{Good $D_n$ quivers of odd type}
\textit{Any} good $D_n$ Dynkin quiver of odd type can be realised as the difference between two quivers living at nodes in figure \ref{FIGUREP} and hence we need to know the two partitions $\mu$ and $\nu$ (not necessarily of equal magnitude) and the integer $n$. 

\begin{figure}
	\begin{center}
		\begin{tikzpicture}[scale = 0.9]
		\def\q{7}
		\def\k{0.7071}
		\def\size{\tiny}
		
		\draw[thick] (0,0) -- (1.5,0) (2.5,0) -- (4,0) -- (4.7,0.7) (4,0) -- (4.7,-0.7) (0,0) -- (0,1)
		(1,0) -- (1,1)
		(3,0) -- (3,1)
		(4,0) -- (5,0)
		
		(4+\k,\k) -- (5+\k,\k)
		(4+\k,-\k) -- (5+\k,-\k)
		;
		
		\draw[thick, fill=white] (0,0) circle (\q pt)
		(1,0) circle (\q pt)
		(3,0) circle (\q pt)
		(4,0) circle (\q pt)
		(4+\k,\k) circle (\q pt)
		(4+\k,-\k) circle (\q pt)
		;
		
		\draw[thick, fill=white] 
		(0 -\q/30,1 -\q/30) rectangle (0 +\q/30,1 +\q/30)
		(1 -\q/30,1 -\q/30) rectangle (1 +\q/30,1 +\q/30)
		(3 -\q/30,1 -\q/30) rectangle (3 +\q/30,1 +\q/30)
		(5 -\q/30,0 -\q/30) rectangle (5 +\q/30,0 +\q/30)
		(5+\k -\q/30,\k -\q/30) rectangle (5+\k +\q/30,\k +\q/30)
		(5+\k -\q/30,-\k -\q/30) rectangle (5+\k +\q/30,-\k +\q/30)
		;

		\draw (0,1.5) node {\size $f_1$}
		(1,1.5) node {\size $f_2$}
		(3,1.5) node {\size $f_{n-3}$}
		
		(5.6,0) node  {\size $f_{n-2}$}
		(5.7 +\k,0+\k) node  {\size $f_{n-1}$}
		(6.4+\k,0-\k) node  {\size $f_{n-1}+2f_n+1$}

		(0,-0.45) node {\size $(g_1, e_1)$}
		(1,-0.45) node {\size $(g_2, e_2)$}
		(2,0) node {$\dots$}
		(3,-0.45) node {\size $(g_{n-3}, e_{n-3})$}
		(3.75,0.7) node {\size $(g_{n-2},$} (3.9,0.45) node {\size $e_{n-2})$}
		
		(4+\k,1.2) node {\size $(g', e_{n-1})$}
		(4+\k,-1.2) node {\size $(g'', e_{n-1}+2e_n+1)$}
		;
		
		\draw (-1.45,0) node {$~$};

		\end{tikzpicture}
	\end{center}
	\caption{The general form of a good, odd $D_n$ Dynkin quiver. These can all be found as the difference of two odd $D_n$ Dynkin quivers from figure \ref{FIGUREP} and their moduli space singularity structure is given by the appropriate run on figure \ref{FIGUREP}.  }
	\label{FIGURES}
\end{figure}
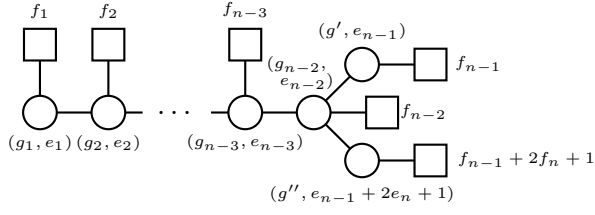

The general good $D_n$ Dynkin quiver of odd type is given in figure \ref{FIGURES}, however there is a subtlety that must be addressed. Since there exist two equivalent quivers at every node, for quiver subtraction to work in the desired way, the extra single end flavour must be on the same node in the subtraction. This is always possible since there are always two options, $I$ and $II$, for odd quivers. 

It was previously recognised that the `extra' flavour and the extra excess needn't be on the same node. But in figure \ref{FIGURES} it is drawn such that they are both associated to the bottom node. This is allowed because of the extra freedom in the general good case. A quiver like figure \ref{FIGURES} but with the `extra' end flavour on the upper node would in fact be of the form of figure \ref{FIGURES} with $f_{n-1} \rightarrow f_{n-1}+1$ and $f_n \rightarrow f_n -1$. When $f_n=0$ this transform isn't possible which sets the `extra' flavour and excess as having to be on the same node, so all cases are covered. 

The partitions associated to the general good, odd $D_n$ Dynkin quiver are
\begin{equation}
\begin{split}\label{goododdpart}
\kappa & = (n^{f_n}, n-1^{f_{n-1}},\dots,2^{f_2},1^{f_1}) \\
\lambda & = (n^{e_n}, n-1^{e_{n-1}},\dots,2^{e_2},1^{e_1}),
\end{split}
\end{equation}
in the usual manner. From here it follows that we have
\begin{equation}
\Q_{\textrm{good, odd}} = \Q(D_\kappa(n)_o) - \Q(D_\lambda(n)_o).
\end{equation}

A general good $D_n$ Dynkin quiver of odd type can therefore be encapsulated by two partitions, say, $\mu$ and $\nu$ with $(\lambda^t, 1^j)^t > \kappa$, and an integer $n$. The class can therefore be written $D_\nu^\mu(n)_o$. 

The singularity structure of the Higgs branch of these theories, $\Hg(D_\nu^\mu(n)_o)$, is given by the run in figure \ref{FIGUREP}, after editing, from a node $\mu$ down to a node $\nu$.

\subsubsection{Dimension matching for good, odd theories}
The calculations for odd theories are similar to those for even theories, only the routes have to be defined on figure \ref{FIGUREP}. This is a matter of replacing the sum over even values of $j$ to a sum over all values and replacing the manner in which partitions for nodes in different partition subdiagrams are determined by one another in order to be commensurate with the $A_k \cup A_k$ traversing structure. Otherwise the construction is the same. The Higgs branch calculation for good, odd theories is
\begin{equation}
\begin{split}
&\dimh(\Hg(D_\nu^\mu(n)_o)) \\ & = \sum_{j=0}^{|\nu|-|\mu|-1} \left( \dimh (A_{n-1-l(\lambda^t_{|\nu|-j-1})} \cup A_{n-1-l(\lambda^t_{|\nu|-j-1})}) + \dimh(\es_{\lambda_{|\nu|-j}} \cap \O_{(\lambda^t_{|\nu|-j-1},1)^t})\right) \\ & = \sum_{j=0}^{|\nu|-|\mu|-1} \left( \frac{1}{2} + \frac{1}{2}\left[ \sum_i ((\lambda_{|\nu|-j}^t)_i)^2 -  \sum_i ((\lambda_{|\nu|-j-1}^t)_i)^2 -1\right] \right) \\ & = \frac{1}{2}\left( \sum_i(\nu_i^t)^2 - \sum_i(\mu_i^t)^2\right).
\end{split}
\end{equation}

For the Coulomb branch calculation one can observe that an odd and even theory with the same partition data have the same ranks on the gauge nodes and so should have the same Coulomb branch dimension. For the even case the partitions either had to be both even or both odd, however for odd theories this needn't be the case. The calculation is
\begin{equation}\label{thisone}
\begin{split}
&\dimh(\mathcal{C}(\Q_\Hg(D_\nu^\mu(n)_o))) \\ & = \sum_{j=0}^{|\nu|-|\mu|-1} \left( \dimh(\mathcal{C}(\Q_\Hg (A_{n-1-l(\lambda^t_{|\nu|-j-1})} \cup A_{n-1-l(\lambda^t_{|\nu|-j-1})}) + \dimh(\es_{(\lambda^t_{|\nu|-j-1},1)} \cap \O_{\lambda_{|\nu|-j}^t})\right) \\ & = \sum_{j=0}^{|\nu|-|\mu|-1}\Bigg( n-1-l(\lambda^t_{|\nu|-j-1}) - \frac{1}{2}\sum_i((\lambda_{|\nu|-j})_i)^2 \\ & ~~~~~~~~~~~~~~~~+ \frac{1}{2}\left[ 1+2l(\lambda^t_{|\nu|-j-1}) + \sum_i((\lambda_{|\nu|-j-1})_i)^2\right]  \Bigg) \\ & = \sum_{j=0}^{|\nu|-|\mu|-1}\left( n-\frac{1}{2} + \frac{1}{2}\left[ \sum_i((\lambda_{|\nu|-j-1})_i)^2 \sum_i((\lambda_{|\nu|-j})_i)^2\right] \right) \\ & = \frac{1}{2}(2n-1)(|\nu|-|\mu|) - \frac{1}{2}\left( \sum_i(\nu_i)^2 - \sum_i(\mu_i)^2\right).
\end{split}
\end{equation}
In the case of even theories, since both of the partitions had to be odd, or both even, this result was guaranteed to be an integer, since the differences were always even. For odd theories, however, there can be one odd and one even partition. In this case, the first term is clearly not an integer. However an odd magnitude partition must contain an odd number of odd parts with odd multiplicity, since odd numbers square to odd numbers. The sum of the squares of the parts has an odd number of odd numbers in it and so is odd. If one term in (\ref{thisone}) is a half integer, the other must be a half integer and so the total is an integer.

\section{Conclusions and outlook}
This work has fully characterised the singularity structure for the moduli space branches of good $D_n$ Dynkin quivers. Along with this local analysis came a natural classification of $D_n$ Dynkin quivers in terms of four pieces of data; an integer $n \geq 2$, the letter $p \in \{e,o\}$ distinguishing even and odd cases, whose end flavours differ by an even or odd amount; and partitions $\mu$ and $\nu$ obeying certain relations dependant on $p$. The theories are therefore denoted $D_\nu^\mu (n)_p$. Balanced theories are exactly the subclass $D_\nu(n)_e \subset D_\nu^\mu (n)_p$, that is, even theories with $\mu=(0)$, the zero partition. The $D_n$ Dynkin quivers which realise $\kk{so}_{2n}$ special Slodowy slices as their Higgs branches are the subclass of balanced quivers where $\nu \in \mathcal{P}(p)$ takes the form\footnote{Recall these partitions come from the set of all partitions, not the restricted set associated to the classification of nilpotent varieties in $\kk{so}_{2n}$.} $(p)$ or $(p-1,1)$ and which are more fully realised within this classification by (\ref{recovery}). 

The primary method used to classify the $D_n$ Dynkin quivers and construct the local analysis of their moduli space branches was \textit{quiver arithmetic}. Quiver subtraction, first discussed formally in \cite{QuivSub}, has been complemented here by the corresponding quiver \textit{addition}. Whereas in \cite{QuivSub} the subtraction of two quivers was used to identify the transverse slice between their moduli space varieties, the addition of quivers requires the identification of those slices which it is possible to add to a given quiver in a consistent manner. For a given quiver the slices could be realised as the difference between that quiver and some larger quiver of the same class, say balanced $D_n$ Dynkin quivers. One starts with the smallest (lowest rank) quiver of the chosen class, for example the balanced zero quiver where all gauge nodes are flavourless and have rank zero. Quivers corresponding to singularities are then added, building the structure of the moduli space from the ground up while also imbuing the class concerned with poset structure in a natural manner. The quiver addition, and hence the poset structure, can be illustrated using a Hasse diagram, as in figures \ref{FIGUREE}, \ref{FIGUREMUE} and \ref{FIGUREP}. Uniquely labelling every node in one of the resulting Hasse diagrams then gives a manner to name every theory constructed. The most useful way of doing this is the identification of Hasse subdiagrams corresponding to sets of integer partitions. At the level of the quiver, these subdiagrams are associated to linear subquivers following the relationship between the moduli space branches of linear ($A_n$ Dynkin) quiver gauge theories and nilpotent orbits of the $\kk{sl}_n$. The exact manner in which these partition subdiagrams are incorporated into the overall Hasse diagram depends on the linear subquiver's relationship with the non-linear aspects of the overall quiver, hence the marked difference between the poset structure for even and odd theories.

\vspace{1mm}

There are three immediately apparent avenues for generalization. In \cite{Rogers}, a generalization from linear to circular quivers was performed, the singularity structure of circular quiver moduli spaces analysed and circular quiver gauge theories classified. In the language of Dynkin quivers, the class of linear quiver gauge theories was \textit{affinized} by the inclusion of further gauge groups and fields such that the quiver went from an $A_n$ Dynkin quiver to an $\wt{A}_n$ Dynkin quiver. A precisely analogous generalization is possible here. The Dynkin diagram for $\wt{D}_n$ is given in figure \ref{DynkinwtD}.
\begin{figure}
	\begin{center}
				\begin{tikzpicture}
				\draw[thick] (0.3,-0.7) -- (1,0) -- (2.5,0) (0.3,0.7) -- (1,0) -- (2.5,0)  (3.5,0) -- (5,0) -- (5.7,0.7) (5,0) -- (5.7,-0.7);
		
				\draw[ thick, fill=white] (1,0) circle (6pt);
				\draw[ thick, fill=white] (2,0) circle (6pt);
				\draw[thick, fill=white] (4,0) circle (6pt);
			\draw[ thick, fill=white] (5,0) circle (6pt);
				\begin{scope}[xshift=5cm]
				\draw[ thick, fill=white] (45:1) circle (6pt);
				\draw[ thick, fill=white] (-45:1) circle (6pt);
				\end{scope}
				\begin{scope}[xshift=1cm]
				\draw[ thick, fill=white] (135:1) circle (6pt);
				\draw[ thick, fill=white] (-135:1) circle (6pt);
				\end{scope}
				\draw (3,0.6) node {$\overbrace{~~\qquad\qquad\qquad\qquad\qquad}^{n-3}$}
		
				(3,0) node {$\dots$};
				\end{tikzpicture}
	\end{center}
	\caption{The Dynkin diagram for $\wt{D}_n$.}
	\label{DynkinwtD}
\end{figure}
Progress on the classification and moduli space analysis of $\wt{D}_n$ Dynkin quivers utilising the techniques used here, and a full discussion, will be published elsewhere. 

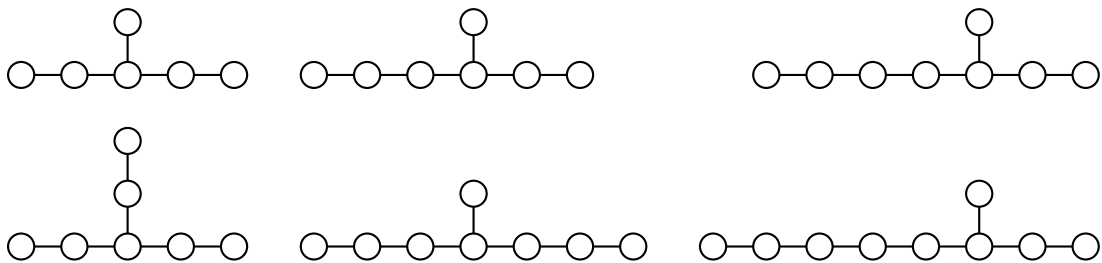
\begin{figure}
	\begin{center}
		\begin{tikzpicture}[xscale = -1, scale=0.7]
		
		\begin{scope}[xshift=-1cm]
		\draw[thick] (1,0) -- (7,0) (3,0) -- (3,1);
		
		\draw[thick, fill=white] 
		(1,0) circle (7pt)
		(2,0) circle (7pt)
		(3,0) circle (7pt)
		(4,0) circle (7pt)
		(5,0) circle (7pt)
		(6,0) circle (7pt)
		(7,0) circle (7pt)
		
		(3,1) circle (7pt);
		
		\end{scope}
		
		\begin{scope}[xshift=8.5cm]
		\draw[thick] (1,0) -- (6,0) (3,0) -- (3,1);
		
		\draw[thick, fill=white] 
		(1,0) circle (7pt)
		(2,0) circle (7pt)
		(3,0) circle (7pt)
		(4,0) circle (7pt)
		(5,0) circle (7pt)
		(6,0) circle (7pt)
		
		(3,1) circle (7pt);
		
		\end{scope}
		
		\begin{scope}[xshift=15cm]
		\draw[thick] (1,0) -- (5,0) (3,0) -- (3,1);
		
		\draw[thick, fill=white] 
		(1,0) circle (7pt)
		(2,0) circle (7pt)
		(3,0) circle (7pt)
		(4,0) circle (7pt)
		(5,0) circle (7pt)
		
		(3,1) circle (7pt);
		
		\end{scope}
		
		\begin{scope}[yshift = -3.25cm]
		
		\begin{scope}[xshift=-1cm]
		\draw[thick] (1,0) -- (8,0) (3,0) -- (3,1);
		
		\draw[thick, fill=white] 
		(1,0) circle (7pt)
		(2,0) circle (7pt)
		(3,0) circle (7pt)
		(4,0) circle (7pt)
		(5,0) circle (7pt)
		(6,0) circle (7pt)
		(7,0) circle (7pt)
		(8,0) circle (7pt)
		
		(3,1) circle (7pt);
		
		\end{scope}
		
		\begin{scope}[xshift=8.5cm]
		\draw[thick] (0,0) -- (6,0) (3,0) -- (3,1);
		
		\draw[thick, fill=white] 
		(0,0) circle (7pt)
		(1,0) circle (7pt)
		(2,0) circle (7pt)
		(3,0) circle (7pt)
		(4,0) circle (7pt)
		(5,0) circle (7pt)
		(6,0) circle (7pt)
		
		(3,1) circle (7pt);
		
		\end{scope}
		
		\begin{scope}[xshift=15cm]
		\draw[thick] (1,0) -- (5,0) (3,0) -- (3,2);
		
		\draw[thick, fill=white] 
		(1,0) circle (7pt)
		(2,0) circle (7pt)
		(3,0) circle (7pt)
		(4,0) circle (7pt)
		(5,0) circle (7pt)
		
		(3,1) circle (7pt)
		(3,2) circle (7pt);
		
		\end{scope}
		
		\end{scope}

		\end{tikzpicture}
	\end{center}
	\caption{The Dynkin diagrams for $E_6$, $E_7$ and $E_8$ (top row) and $\wt{E}_6$, $\wt{E}_7$ and $\wt{E}_8$ (bottom row). }
	\label{ealgs}
\end{figure}

An alternative generalization is to the remaining simply-laced non-affine Dynkin diagrams, namely $E_6$, $E_7$ and $E_8$. As $D_n$ brought with it the $D_k$ and $A_k \cup A_k$ singularities, the $E_n$ family come with their own, singularities. The quivers associated to these new singularities have the $E_n$ Du Val singularities as Higgs branches and the closures of the minimal nilpotent orbits in the corresponding $\kk{e}_n$ algebras as Coulomb branches. They are also amenable to the quiver addition techniques used here. Furthermore, each has an affinization all of which are \textit{also} simply-laced, figure \ref{ealgs}.

The third obvious generalisation is to turn to the other classical and exceptional, non-affine, Dynkin diagrams, $B_n$, $C_n$, $F_4$ and $G_2$ and their affine partners. These are set apart from $ADE$ diagrams by the presence of non-simply-laced edges and so carry complications in their interpretation field theoretically. Further investigations might then elucidate the connections between Dynkin quivers of different shapes, especially via the phenomenon of \textit{folding}, by which (typically simply-laced) diagrams are folded to yield non-simply-laced diagrams. Initial constructions of Hasse diagrams for BCFG quivers via quiver addition give numerous suggestions of an intimate link between families of Dynkin quivers related diagrammatically by folding. 
Folding of affine Lie algebras yields an even wider class of diagrams associated to \textit{twisted} affine Lie algebras.

The present work did not depend on or require an interpretation of the quivers as describing the low energy dynamics of a brane construction in string theory. Many constructions are known which have such a description \cite{Kasputin}. The interpretation of the present discussion into an explicitly string-theoretic context in this manner would provide a further interesting line of investigation. However many quiver gauge theories do not currently have an interpretation as the low energy dynamics of a brane configuration. Since the present work does not explicitly depend on manipulation at the level of a brane configuration,  the generalization to quivers of different shapes is more readily available. 

\vspace{1mm}

More speculative future directions are also available. The present work and the work in \cite{Rogers} share a crucial property: the analysis of the moduli space of vacua is \textit{local}, it does not depend on, nor does it easily provide, a global description of the moduli space of the theories which are analysed. The description of a variety as a Hasse diagram of its transverse singularities is a detailed and constraining one and the promoting of a full local description of the singularity structure of a given variety\footnote{The moduli space of vacua of a Dynkin quiver gauge theory in this context, although not necessarily.} into a complete global description is an interesting mathematical challenge.

A fundamental limit to the methods here is the identification of singularity quivers to add. The only gauge node topologies that are balanceable with no flavour are the affine Dynkin diagrams. The singularities added during quiver addition are balanced and look like lowest rank balanced affine Dynkin diagrams with the `extra' node acting as flavour. It seems that the singularities arising in nilpotent varieties which are realisable as moduli space branches (and close relations such as considering the $A_l \cup A_l$ singularity to include even $l$\footnote{Only singularities with odd $l$ appear naturally in nilpotent varieties of Lie algebras.}) are all the singularities one has to play with. The full singularity structure of the nilpotent varieties of the exceptional Lie algebras was only studied relatively recently \cite{fu}. There the authors identify numerous singularities which have no known quiver interpretation, nor are even particularly well understood geometrically. 

It is impossible to construct balanced Hasse diagrams of non-Dynkin quiver gauge theories analogous to figures \ref{FIGUREE}, \ref{FIGUREMUE} and \ref{FIGUREP}. However, good, non-Dynkin quiver gauge theories which yield to deconstruction one singularity quiver at a time assuredly exist and their moduli spaces of vacua are locally analysable in a sense. The techniques developed here have the potential to provide crucial insight into understanding the moduli spaces of simply laced non-Dynkin, or even generic, $3d$ $\mathcal{N}=4$ quiver gauge theories.

\appendix

\section{$\bs{A_n}$ and $\bs{D_n}$ nilpotent orbit Hasse diagrams} \label{appendix}
\subsection{$\kk{sl}_{n}$ for $n=2, \dots, 9$}
\begin{center}
	\begin{tikzpicture}[xscale = 0.65, yscale = 0.65]
	\filldraw[black] (0,0) circle (4pt)
	(0,1) circle (4pt);
	
	\draw (0,-2.5) node {};
	
	\draw[ultra thick] (0,0) -- (0,1);
	
	\draw[ultra thick] (-1, 1.5) -- (3.5,1.5);
	
	\draw (0,2.25) node {\scriptsize{Hasse}}
	(0,1.85) node {\scriptsize{Diagram}}
	(2.5,2.05) node {\scriptsize{Partition}}
	(1.25,3.3) node {\Large{$\kk{sl}_2$}}
	(2.5,1) node {$(2)$}
	(2.5,0) node {$(1^2)$}
	(0.4, 0.5) node {\scriptsize{$A_{1}$}}
	;
	\end{tikzpicture}
	$ \qquad $
	\begin{tikzpicture}[xscale = 0.65, yscale = 0.65]
	\filldraw[black] (0,0) circle (4pt)
	(0,1) circle (4pt)
	(0,2) circle (4pt);
	
	\draw[ultra thick] (0,0) -- (0,2);
	
	\draw (0,-2) node {};
	
	\draw[ultra thick] (-1, 2.5) -- (3.5,2.5);
	
	\draw (0,3.25) node {\scriptsize{Hasse}}
	(0,2.85) node {\scriptsize{Diagram}}
	(2.5,3.05) node {\scriptsize{Partition}}
	(1.25,4.3) node {\Large{$\kk{sl}_3$}}
	(2.5,2) node {$(3)$}
	(2.5,1) node {$(2,1)$}
	(2.5,0) node {$(1^3)$}
	(0.4, 0.5) node {\scriptsize{$a_{2}$}}
	(0.4, 1.5) node {\scriptsize{$A_{2}$}}
	;
	\end{tikzpicture}
	$ \qquad $
	\begin{tikzpicture}[xscale = 0.65, yscale = 0.65]
	\filldraw[black] (0,0) circle (4pt)
	(0,1) circle (4pt)
	(0,2) circle (4pt)
	(0,3) circle (4pt)
	(0,4) circle (4pt);
	
	\draw (0,-1.5) node {};
	
	\draw[ultra thick] (0,0) -- (0,4);
	
	\draw[ultra thick] (-1, 4.5) -- (3.5,4.5);
	
	\draw (0,5.25) node {\scriptsize{Hasse}}
	(0,4.85) node {\scriptsize{Diagram}}
	(2.5,5.05) node {\scriptsize{Partition}}
	(1.25,6.3) node {\Large{$\kk{sl}_4$}}
	(2.5,4) node {$(4)$}
	(2.5,3) node {$(3,1)$}
	(2.5,2) node {$(2^2)$}
	(2.5,1) node {$(2,1^2)$}
	(2.5,0) node {$(1^4)$}
	(0.4, 0.5) node {\scriptsize{$a_{3}$}}
	(0.4, 1.5) node {\scriptsize{$A_{1}$}}
	(0.4, 2.5) node {\scriptsize{$A_{1}$}}
	(0.4, 3.5) node {\scriptsize{$A_{3}$}}
	;
	\end{tikzpicture}
	$\qquad$
	\begin{tikzpicture}[xscale = 0.65, yscale = 0.65]
	\filldraw[black] (0,0) circle (4pt)
	(0,1) circle (4pt)
	(0,2) circle (4pt)
	(0,3) circle (4pt)
	(0,4) circle (4pt)
	(0,5) circle (4pt)
	(0,6) circle (4pt);
	
	\draw[ultra thick] (0,0) -- (0,6);
	
	\draw[ultra thick] (-1, 6.5) -- (3.5,6.5);
	
	\draw (0,7.25) node {\scriptsize{Hasse}}
	(0,6.85) node {\scriptsize{Diagram}}
	(2.5,7.05) node {\scriptsize{Partition}}
	(1.25,8.3) node {\Large{$\kk{sl}_5$}}
	(2.5,6) node {$(5)$}
	(2.5,5) node {$(4,1)$}
	(2.5,4) node {$(3,2)$}
	(2.5,3) node {$(3,1^2)$}
	(2.5,2) node {$(2^2,1)$}
	(2.5,1) node {$(2,1^3)$}
	(2.5,0) node {$(1^5)$}
	(0.4, 0.5) node {\scriptsize{$a_{4}$}}
	(0.4, 1.5) node {\scriptsize{$a_{2}$}}
	(0.4, 2.5) node {\scriptsize{$A_{1}$}}
	(0.4, 3.5) node {\scriptsize{$A_{1}$}}
	(0.4, 4.5) node {\scriptsize{$A_{2}$}}
	(0.4, 5.5) node {\scriptsize{$A_{4}$}}
	;
	
	\end{tikzpicture}
	$\qquad \qquad $
	\begin{tikzpicture}[xscale = 0.65, yscale = 0.65]
	\filldraw[black] (0,0) circle (4pt)
	(0,1) circle (4pt)
	(0,2) circle (4pt)
	(0,4) circle (4pt)
	(0,6) circle (4pt)
	(0,7) circle (4pt)
	(0,8) circle (4pt)
	(-1,3) circle (4pt)
	(1,3) circle (4pt)
	(-1,5) circle (4pt)
	(1,5) circle (4pt)
	;
	
	\draw (0,-1.4) node {};
	
	\draw[ultra thick] (0,0) -- (0,2) -- (-1,3) -- (0,4) -- (-1,5) -- (0,6) -- (0,8) -- (0,6) -- (1,5) -- (0,4) -- (1,3) -- (0,2);
	
	\draw[ultra thick] (-1, 8.5) -- (5.5,8.5);
	
	\draw (0,9.25) node {\scriptsize{Hasse}}
	(0,8.85) node {\scriptsize{Diagram}}
	(4.5,9.05) node {\scriptsize{Partition}}
	(2.25,9.3) node {\Large{$\kk{sl}_6$}}
	(4.5,8) node {$(6)$}
	(4.5,7) node {$(5,1)$}
	(4.5,6) node {$(4,2)$}
	(3.5,5) node {$(4,1^2)$}
	(5.5,5) node {$(3^2)$}
	(4.5,4) node {$(3,2,1)$}
	(3.5,3) node {$(3,1^3)$} 
	(5.5,3) node {$(2^3)$}
	(4.5,2) node {$(2^2,1^2)$}
	(4.5,1) node {$(2,1^4)$}
	(4.5,0) node {$(1^6)$}
	
	(0.4, 0.5) node {\scriptsize{$a_{5}$}}
	(0.4, 1.5) node {\scriptsize{$a_{3}$}}
	(0.7, 2.3) node {\scriptsize{$A_{1}$}}
	(-0.7, 2.3) node {\scriptsize{$A_{1}$}}
	(0.7, 3.7) node {\scriptsize{$a_{2}$}}
	(-0.7, 3.7) node {\scriptsize{$a_{2}$}}
	(0.7, 4.3) node {\scriptsize{$A_{2}$}}
	(-0.7, 4.3) node {\scriptsize{$A_{2}$}}
	(0.7, 5.7) node {\scriptsize{$A_{1}$}}
	(-0.7, 5.7) node {\scriptsize{$A_{1}$}}
	(0.4, 6.5) node {\scriptsize{$A_{3}$}}
	(0.4, 7.5) node {\scriptsize{$A_{5}$}}
	;
	\end{tikzpicture}
	$\qquad \qquad $
	\begin{tikzpicture}[xscale = 0.65, yscale = 0.65]
	\begin{scope}[xscale=-1]
	\filldraw[black] (0,0) circle (4pt)
	(0,1) circle (4pt)
	(0,2) circle (4pt)
	(0,4) circle (4pt)
	(0,7) circle (4pt)
	(0,9) circle (4pt)
	(0,10) circle (4pt)
	(0,11) circle (4pt)
	(1,3) circle (4pt)
	(-1,3) circle (4pt)
	(-1,5.5) circle (4pt)
	(1,5) circle (4pt)
	(1,6) circle (4pt)
	(-1,8) circle (4pt)
	(1,8) circle (4pt);
	
	\draw[ultra thick] (0,0) -- (0,2) -- (-1,3) -- (0,4) -- (-1,5.5) -- (0,7) -- (-1,8) -- (0,9) -- (0,11) -- (0,9) -- (1,8) -- (0,7) -- (1,6) -- (1,5) -- (0,4) -- (1,3) -- (0,2);
	\end{scope}

	\draw[ultra thick] (-1, 11.5) -- (5.5,11.5);
	
	\draw (0,12.25) node {\scriptsize{Hasse}}
	(0,11.85) node {\scriptsize{Diagram}}
	(4.5,12.05) node {\scriptsize{Partition}}
	(2.25,12.3) node {\Large{$\kk{sl}_7$}}
	(4.5,11) node {$(7)$}
	(4.5,10) node {$(6,1)$}
	(4.5,9) node {$(5,2)$}
	(3.5,8) node {$(4,3)$}
	(5.5,8) node {$(5,1^2)$}
	(4.5,7) node {$(4,2,1)$}
	(3.5,6) node {$(3^2,1)$}
	(3.5,5) node {$(3,2^2)$}
	(5.5,5.5) node {$(4,1^3)$}
	(4.5,4) node {$(3,2,1^2)$}
	(3.5,3) node {$(2^3,1)$}
	(5.5,3) node {$(3,1^4)$}
	(4.5,2) node {$(2^2,1^3)$}
	(4.5,1) node {$(2,1^5)$}
	(4.5,0) node {$(1^7)$}
	
	(0.4, 0.5) node {\scriptsize{$a_{6}$}}
	(0.4, 1.5) node {\scriptsize{$a_{4}$}}
	(0.7, 2.3) node {\scriptsize{$A_{1}$}}
	(-0.7, 2.3) node {\scriptsize{$a_{2}$}}
	(0.7, 3.7) node {\scriptsize{$a_{3}$}}
	(-0.7, 3.7) node {\scriptsize{$a_{2}$}}
	(0.8, 4.7) node {\scriptsize{$A_{2}$}}
	(-0.7, 4.3) node {\scriptsize{$A_{1}$}}
	(0.9, 6.3) node {\scriptsize{$a_{2}$}}
	(-1.3, 5.5) node {\scriptsize{$A_{1}$}}
	(-0.76, 6.7) node {\scriptsize{$A_{1}$}}
	(0.4, 10.5) node {\scriptsize{$A_{6}$}}
	(0.4, 9.5) node {\scriptsize{$A_{4}$}}
	(-0.8, 7.4) node {\scriptsize{$A_{2}$}}
	(0.8, 7.4) node {\scriptsize{$A_{3}$}}
	(-0.8, 8.7) node {\scriptsize{$A_{2}$}}
	(0.8, 8.7) node {\scriptsize{$A_{1}$}}
	;
	
	\end{tikzpicture}
	
	\vspace{5mm}
	
	\begin{tikzpicture}[xscale = 0.65, yscale = 0.65]
	\begin{scope}[xscale=-1]
	\filldraw[black] (0,0) circle (4pt)
	(0,1) circle (4pt)
	(0,2) circle (4pt)
	(0,12) circle (4pt)
	(0,13) circle (4pt)
	(0,14) circle (4pt)
	(1,3) circle (4pt)
	(1,4) circle (4pt)
	(1,5) circle (4pt)
	(1,6) circle (4pt)
	(1,7) circle (4pt)
	(1,8) circle (4pt)
	(1,9) circle (4pt)
	(1,10) circle (4pt)
	(1,11) circle (4pt)
	(-1,3) circle (4pt)
	(-1,4) circle (4pt)
	(-1,5.5) circle (4pt)
	(-1,7) circle (4pt)
	(-1,8.5) circle (4pt)
	(-1,10) circle (4pt)
	(-1,11) circle (4pt);
	
	\draw[ultra thick] (0,0) -- (0,2) -- (-1,3) -- (-1,11) -- (0,12) -- (0,14) -- (0,12) -- (1,11) -- (1,3) -- (0,2) -- (1,3) -- (-1,4) -- (1,5) -- (1,6) -- (-1,7) -- (1,8) -- (1,9) -- (-1,10) -- (1,11);
	\end{scope}
	
	\draw (0,-2.5) node {};
	
	\draw[ultra thick] (-1, 14.5) -- (5.5,14.5);
	
	\draw (0,15.25) node {\scriptsize{Hasse}}
	(0,14.85) node {\scriptsize{Diagram}}
	(4.5,15.05) node {\scriptsize{Partition}}
	(2.25,15.3) node {\Large{$\kk{sl}_8$}}
	(4.5,14) node {$(8)$}
	(4.5,13) node {$(7,1)$}
	(4.5,12) node {$(6,2)$}
	(3.5,11) node {$(5,3)$}
	(3.5,10) node {$(4^2)$}
	(3.5,9) node {$(4,3,1)$}
	(3.5,8) node {$(4,2^2)$}
	(3.5,7) node {$(3^2,2)$}
	(3.5,6) node {$(3^2,1^2)$}
	(3.5,5) node {$(3,2^2,1)$}
	(3.5,4) node {$(2^4)$}
	(3.5,3) node {$(2^3,1^2)$}
	(5.5,11) node {$(6,1^2)$}
	(5.5,10) node {$(5,2,1)$}
	(5.5,8.5) node {$(5,1^3)$}
	(5.5,7) node {$(4,2,1^2)$}
	(5.5,5.5) node {$(4,1^4)$}
	(5.5,4) node {$(3,2,1^3)$}
	(5.5,3) node {$(3,1^5)$}
	(4.5,2) node {$(2^2,1^4)$}
	(4.5,1) node {$(2,1^6)$}
	(4.5,0) node {$(1^8)$}
	
	(0.4, 13.5) node {\scriptsize{$A_{7}$}}
	(0.4, 12.5) node {\scriptsize{$A_{5}$}}
	(0.4, 1.5) node {\scriptsize{$a_{5}$}}
	(0.4, 0.5) node {\scriptsize{$a_{7}$}}
	
	(0, 10.8) node {\scriptsize{$A_{2}$}}
	(0, 9.2) node {\scriptsize{$A_{2}$}}
	(0, 7.8) node {\scriptsize{$A_{1}$}}
	(0, 6.2) node {\scriptsize{$A_{1}$}}
	(0, 4.8) node {\scriptsize{$a_{2}$}}
	(0, 3.2) node {\scriptsize{$a_{2}$}}
	
	(-0.9, 2.3) node {\scriptsize{$a_{3}$}}
	(-1.3, 3.5) node {\scriptsize{$A_{1}$}}
	(-1.3, 4.5) node {\scriptsize{$a_{3}$}}
	(-1.3, 5.5) node {\scriptsize{$A_{1}$}}
	(-1.3, 6.5) node {\scriptsize{$A_{1}$}}
	(-1.3, 7.5) node {\scriptsize{$A_{1}$}}
	(-1.3, 8.5) node {\scriptsize{$A_{1}$}}
	(-1.3, 9.5) node {\scriptsize{$A_{3}$}}
	(-1.3, 10.5) node {\scriptsize{$A_{1}$}}
	(-0.9, 11.7) node {\scriptsize{$A_{3}$}}
	
	(0.9, 2.3) node {\scriptsize{$A_{1}$}}
	(1.3, 3.5) node {\scriptsize{$a_{4}$}}
	(1.3, 4.75) node {\scriptsize{$A_{2}$}}
	(1.3, 6.25) node {\scriptsize{$a_{3}$}}
	(1.3, 7.75) node {\scriptsize{$A_{3}$}}
	(1.3, 9.25) node {\scriptsize{$a_{2}$}}
	(1.3, 10.5) node {\scriptsize{$A_{4}$}}
	(0.9, 11.7) node {\scriptsize{$A_{1}$}}
	;
	\end{tikzpicture}
	$\qquad \qquad $
	\begin{tikzpicture}[xscale=0.65, yscale=0.65]
	\begin{scope}[xscale=-1]
	\filldraw[black] (0,0) circle (4pt)
	(0,1) circle (4pt)
	(0,2) circle (4pt)
	(0,16) circle (4pt)
	(0,17) circle (4pt)
	(0,18) circle (4pt)
	(0,6) circle (4pt)
	(0,8) circle (4pt)
	(0,10) circle (4pt)
	(0,12) circle (4pt)
	(-1,3) circle (4pt)
	(-1,5) circle (4pt)
	(-1,7) circle (4pt)
	(-1,9) circle (4pt)
	(-1,11) circle (4pt)
	(-1,13) circle (4pt)
	(-1,15) circle (4pt)
	(-1,4) circle (4pt)
	(-1,14) circle (4pt)
	(1,3) circle (4pt)
	(1,4) circle (4pt)
	(1,5) circle (4pt)
	(1,6) circle (4pt)
	(1,7) circle (4pt)
	(1,9) circle (4pt)
	(1,11) circle (4pt)
	(1,12) circle (4pt)
	(1,13) circle (4pt)
	(1,14) circle (4pt)
	(1,15) circle (4pt);
	
	\draw[ultra thick] (0,0) -- (0,2) -- (1,3) -- (1,15) -- (0,16) -- (0,18) -- (0,16) -- (-1,15) -- (-1,3) -- (0,2)
	(1,5) -- (-1,7)
	(0,6) -- (1,7) -- (0,8) -- (-1,7) -- (0,8) --(0,10) -- (-1,11) -- (0,10) -- (1,11) -- (0,12)
	(0,12) -- (1,13) -- (-1,11)
	(1,3) -- (-1,4) -- (1,5)
	(1,15) -- (-1,14) -- (1,13)
	;
	
	\end{scope}
	
	\draw[ultra thick] (-1, 18.5) -- (6.5,18.5);
	
	\draw (0,19.25) node {\scriptsize{Hasse}}
	(0,18.85) node {\scriptsize{Diagram}}
	(5.1,19.05) node {\scriptsize{Partition}}
	(2.5,19.3) node {\Large{$\kk{sl}_9$}};
	
	\begin{scope}[xscale = 0.9, xshift = 0.2cm]
	
	\draw
	(5.5,18) node {$(9)$}
	(5.5,17) node {$(8,1)$}
	(5.5,16) node {$(7,2)$}
	
	(3.5,15) node {$(6,3)$}
	(3.5,14) node {$(5,4)$}
	(3.5,13) node {$(5,3,1)$}
	(3.5,12) node {$(4^2,1)$}
	(3.5,11) node {$(4,3,2)$}
	(3.5,9) node {$(3^3)$}
	(3.5,7) node {$(3^2,2,1)$}
	(3.5,6) node {$(3,2^3)$}
	(3.5,5) node {$(3,2^2,1^2)$}
	(3.5,4) node {$(2^4,1)$}
	(3.5,3) node {$(2^3,1^3)$}
	
	(5.5,12) node {$(5,2^2)$}
	(5.5,10) node {$(4,3,1^2)$}
	(5.5,8) node {$(4,2^2,1)$}
	(5.5,6) node {$(3^2,1^3)$}
	
	(7.5,15) node {$(7,1^2)$}
	(7.5,14) node {$(6,2,1)$}
	(7.5,13) node {$(6,1^3)$}
	(7.5,11) node {$(5,2,1^2)$}
	(7.5,9) node {$(5,1^4)$}
	(7.5,7) node {$(4,2,1^3)$}
	(7.5,5) node {$(4,1^5)$}
	(7.5,4) node {$(3,2,1^4)$}
	(7.5,3) node {$(3,1^6)$}
	
	(5.5,2) node {$(2^2,1^5)$}
	(5.5,1) node {$(2,1^7)$}
	(5.5,0) node {$(1^9)$}
	;
	
	\end{scope}
	
	\draw
	(0.4, 0.5) node {\scriptsize{$a_{8}$}}
	
	(0.4, 1.5) node {\scriptsize{$a_{6}$}}
	
	(0.8, 2.3) node {\scriptsize{$A_{1}$}}
	(-0.8, 2.3) node {\scriptsize{$a_{4}$}}
	
	(-1.4, 3.5) node {\scriptsize{$a_{2}$}}
	(0, 3.8) node {\scriptsize{$a_{2}$}}
	(1.4, 3.5) node {\scriptsize{$a_{5}$}}
	
	(-1.4, 4.5) node {\scriptsize{$a_{3}$}}
	(0, 4.8) node {\scriptsize{$a_{3}$}}
	(1.4, 4.5) node {\scriptsize{$A_{2}$}}
	
	(-1.4, 5.5) node {\scriptsize{$A_{1}$}}
	(-0.25, 5.35) node {\scriptsize{$A_{1}$}}
	(1.4, 6) node {\scriptsize{$a_{4}$}}
	
	(-1.4, 6.5) node {\scriptsize{$a_{2}$}}
	(-0.35, 6.8) node {\scriptsize{$a_{2}$}}
	(0.35, 6.8) node {\scriptsize{$A_{1}$}}
	
	(-1.4, 8) node {\scriptsize{$A_{2}$}}
	(-0.6, 7.8) node {\scriptsize{$A_{1}$}}
	(0.6, 7.8) node {\scriptsize{$a_{2}$}}
	(1.4, 8) node {\scriptsize{$A_{3}$}}

	(0.4, 9) node {\scriptsize{$A_{1}$}}
	
	;
	
	\begin{scope}
	\draw[yshift = 18cm, yscale=-1]
	(0.4, 0.5) node {\scriptsize{$A_{8}$}}
	
	(0.4, 1.5) node {\scriptsize{$A_{6}$}}
	
	(0.8, 2.3) node {\scriptsize{$A_{1}$}}
	(-0.8, 2.3) node {\scriptsize{$A_{4}$}}
	
	(-1.4, 3.5) node {\scriptsize{$A_{2}$}}
	(0, 3.8) node {\scriptsize{$A_{2}$}}
	(1.4, 3.5) node {\scriptsize{$A_{5}$}}
	
	(-1.4, 4.5) node {\scriptsize{$A_{3}$}}
	(0, 4.8) node {\scriptsize{$A_{3}$}}
	(1.4, 4.5) node {\scriptsize{$a_{2}$}}
	
	(-1.4, 5.5) node {\scriptsize{$A_{1}$}}
	(-0.25, 5.35) node {\scriptsize{$A_{1}$}}
	(1.4, 6) node {\scriptsize{$A_{4}$}}
	
	(-1.4, 6.5) node {\scriptsize{$A_{2}$}}
	(-0.35, 6.8) node {\scriptsize{$A_{2}$}}
	(0.35, 6.8) node {\scriptsize{$A_{1}$}}
	
	(-1.4, 8) node {\scriptsize{$a_{2}$}}
	(-0.6, 7.8) node {\scriptsize{$A_{1}$}}
	(0.6, 7.8) node {\scriptsize{$A_{2}$}}
	(1.4, 8) node {\scriptsize{$a_{3}$}}
	;
	\end{scope}

	\end{tikzpicture}
	
\end{center}
	
\subsection{$\kk{so}_{2n}$ for $n=2,\dots,6$}

\begin{center}
\begin{tikzpicture}[scale = 0.65]
\begin{scope}[yshift=1cm]
\draw[ultra thick] (-1, 1.5) -- (3.5,1.5);

\draw (0,2.25) node {\scriptsize{Hasse}}
(0,1.85) node {\scriptsize{Diagram}}
(2.5,2.05) node {\scriptsize{Partition}}
(1.25,3.3) node {\Large{$\kk{so}_4$}};
\end{scope}

\draw[fill=black] (0,0) circle (4pt);
\draw[fill=black] (0,1) circle (4pt);
\draw[fill=black] (0,2) circle (4pt);

\draw[\THICC] (0,0) -- (0,2);

\draw (2.5,0) node {$(1^4)$};
\draw (2.5,1) node {$(2^2)$};
\draw (2.5,2) node {$(3,1)$};

\draw (-0.5,1.5) node {\scriptsize$A_1$};
\draw (-1,0.5) node {\scriptsize$A_1 \cup A_1$};

\draw (0,-5.27) node {};
\end{tikzpicture}$\qquad$
\begin{tikzpicture}[scale = 0.65]
		\begin{scope}[yshift=3cm]
		\draw[ultra thick] (-1, 1.5) -- (3.5,1.5);
		
		\draw (0,2.25) node {\scriptsize{Hasse}}
		(0,1.85) node {\scriptsize{Diagram}}
		(2.5,2.05) node {\scriptsize{Partition}}
		(1.25,3.3) node {\Large{$\kk{so}_6$}};
		\end{scope}
		
		\draw[fill=black] (0,0) circle (4pt);
		\draw[fill=black] (0,1) circle (4pt);
		\draw[fill=black] (0,2) circle (4pt);
		\draw[fill=black] (0,3) circle (4pt);
		\draw[fill=black] (0,4) circle (4pt);
		
		\draw[\THICC] (0,0) -- (0,4);
		
		\draw (2.5,0) node {$(1^6)$};
		\draw (2.5,1) node {$(2^2,1^2)$};
		\draw (2.5,2) node {$(3,1^3)$};
		\draw (2.5,3) node {$(3^2)$};
		\draw (2.5,4) node {$(5,1)$};
		
		\draw (-0.5,3.5) node {\scriptsize$D_3$};
		\draw (-0.5,2.5) node {\scriptsize$A_1$};
		\draw (-0.5,1.5) node {\scriptsize$A_1$};
		\draw (-0.5,0.5) node {\scriptsize$d_3$};
		
		\draw (0,-3.27) node {};
		\end{tikzpicture}$\qquad$
		\begin{tikzpicture}[scale=0.65]
		\begin{scope}[yshift=6cm]
		\draw[ultra thick] (-1, 1.5) -- (4.5,1.5);
		
		\draw (0,2.25) node {\scriptsize{Hasse}}
		(0,1.85) node {\scriptsize{Diagram}}
		(3.5,2.05) node {\scriptsize{Partition}}
		(1.75,3.3) node {\Large{$\kk{so}_8$}};
		\end{scope}
		
		\draw[\THICC] (0,0) -- (0,1) -- (1,2) -- (0,3) -- (0,4) -- (1,5) -- (0,6) -- (0,7) (0,6) -- (-1,5) -- (0,4) (0,3) -- (-1,2) -- (0,1);
		
		\draw[fill=black] (0,0) circle (4pt);
		\draw[fill=black] (0,1) circle (4pt);
		\draw[fill=black] (0,3) circle (4pt);
		\draw[white, fill=white] (0,3) circle (1.8pt);
		\draw[fill=black] (0,4) circle (4pt);
		\draw[fill=black] (0,6) circle (4pt);
		\draw[fill=black] (0,7) circle (4pt);
		
		\draw[fill=black] (-1,2) circle (4pt);
		\draw[fill=black] (1,2) circle (4pt);
		\draw[fill=black] (-1,5) circle (4pt);
		\draw[fill=black] (1,5) circle (4pt);
		
		\begin{scope}[xshift = 3.5cm]
		\draw[fill=black] (0,0) node {$(1^8)$};
		\draw[fill=black] (0,1) node {$(2^2,1^4)$};
		\draw[fill=black] (0,3) node {$(3,2^2,1)$};
		\draw[fill=black] (0,4) node {$(3^2,1^2)$};
		\draw[fill=black] (0,6) node {$(5,3)$};
		\draw[fill=black] (0,7) node {$(7,1)$};
		
		\draw[fill=black] (-1,2) node {$(2^4)$};
		\draw[fill=black] (1,2) node {$(3,1^5)$};
		\draw[fill=black] (-1,5) node {$(4^2)$};
		\draw[fill=black] (1,5) node {$(5,1^3)$};
		
		\end{scope}
		
		\draw (-0.5,0.5) node {\scriptsize$d_4$};
		\draw (-0.5,3.5) node {\scriptsize$A_1$};
		\draw (-0.5,6.5) node {\scriptsize$D_4$};
		
		\draw (-1.3,1.3) node {\scriptsize$A_1 \cup A_1$};
		\draw (0.9,1.3) node {\scriptsize$A_1$};
		
		\draw (-0.9,2.7) node {\scriptsize$c_2$};
		\draw (0.9,2.7) node {\scriptsize$b_2$};
		
		\draw (-1.3,4.3) node {\scriptsize$A_3 \cup A_3$};
		\draw (0.9,4.3) node {\scriptsize$D_3$};
		
		\draw (-0.9,5.7) node {\scriptsize$A_1$};
		\draw (0.9,5.7) node {\scriptsize$A_1$};
		
		\end{tikzpicture}
		
	\end{center}
	\begin{center}
		\begin{tikzpicture}[scale=0.65]
		\begin{scope}[yshift=10cm]
		\draw[ultra thick] (-1, 1.5) -- (5.5,1.5);
		
		\draw (0,2.25) node {\scriptsize{Hasse}}
		(0,1.85) node {\scriptsize{Diagram}}
		(4.5,2.05) node {\scriptsize{Partition}}
		(2.25,2.3) node {\Large{$\kk{so}_{10}$}};
		\end{scope}
		
		\draw[\THICC] (0,0) -- (0,1) -- (1,2) -- (0,3) -- (0,4) -- (1,5) -- (1,7) -- (0,8) -- (1,9) -- (0,10) -- (0,11) (0,1) -- (-1,2) -- (0,3) (0,10) -- (-1,9) -- (0,8) -- (-1,7) -- (-1,5) -- (0,4) (-1,7) -- (1,6);
		
		\draw[fill=black] (0,0) circle (4pt);
		\draw[fill=black] (0,1) circle (4pt);
		\draw[fill=black] (0,3) circle (4pt);
		\draw[white, fill=white] (0,3) circle (1.8pt);
		\draw[fill=black] (0,4) circle (4pt);
		\draw[fill=black] (0,8) circle (4pt);
		\draw[fill=black] (0,10) circle (4pt);
		\draw[fill=black] (0,11) circle (4pt);
		
		\draw[fill=black] (-1,2) circle (4pt);
		\draw[fill=black] (-1,5) circle (4pt);
		\draw[fill=black] (-1,7) circle (4pt);
		\draw[white, fill=white] (-1,7) circle (1.8pt);
		\draw[fill=black] (-1,9) circle (4pt);
		
		\draw[fill=black] (1,2) circle (4pt);
		\draw[fill=black] (1,5) circle (4pt);
		\draw[fill=black] (1,6) circle (4pt);
		\draw[fill=black] (1,7) circle (4pt);
		\draw[fill=black] (1,9) circle (4pt);
		
		\begin{scope}[xshift = 4.5cm]
		\draw[fill=black] (0,0) node{$(1^{10})$};
		\draw[fill=black] (0,1) node{$(2^2,1^6)$};
		\draw[fill=black] (0,3) node{$(3,2^2,1^3)$};
		\draw[fill=black] (0,4) node{$(3^2,1^4)$};
		\draw[fill=black] (0,8) node{$(5,3,1^2)$};
		\draw[fill=black] (0,10) node{$(7,3)$};
		\draw[fill=black] (0,11) node{$(9,1)$};
		
		\draw[fill=black] (-1,2) node{$(2^4,1^2)$};
		\draw[fill=black] (-1,5) node{$(5,1^5)$};
		\draw[fill=black] (-1,7) node{$(5,2^2,1)$};
		\draw[fill=black] (-1,9) node{$(5^2)$};
		
		\draw[fill=black] (1,2) node{$(3,1^7)$};
		\draw[fill=black] (1,5) node{$(3^2,2^2)$};
		\draw[fill=black] (1,6) node{$(3^3,1)$};
		\draw[fill=black] (1,7) node{$(4^2,1^2)$};
		\draw[fill=black] (1,9) node{$(7,1^3)$};
		\end{scope}
		
		\draw (-0.9,1.3) node {\scriptsize$d_3$};
		\draw (0.9,1.3) node {\scriptsize$A_1$};
		
		\draw (-0.9,2.7) node {\scriptsize$c_2$};
		\draw (0.9,2.7) node {\scriptsize$b_3$};
		
		\draw (-0.9,4.3) node {\scriptsize$D_3$};
		\draw (1.3,4.3) node {\scriptsize$A_1 \cup A_1$};
		
		\draw (-0.9,7.7) node {\scriptsize$A_1$};
		\draw (0.9,7.7) node {\scriptsize$A_1$};
		
		\draw (-0.9,8.3) node {\scriptsize$A_3$};
		\draw (0.9,8.3) node {\scriptsize$D_4$};
		
		\draw (-0.9,9.7) node {\scriptsize$D_3$};
		\draw (0.9,9.7) node {\scriptsize$A_1$};
		
		\draw (-0.5,0.5) node {\scriptsize$d_5$};
		\draw (-0.5,3.5) node {\scriptsize$A_1$};
		\draw (-0.5,10.5) node {\scriptsize$D_5$};
		\draw (0.1,6.8) node[rotate=-20] {\scriptsize$A_1$};
		
		\draw (-1.5,6) node {\scriptsize$b_2$};
		
		\draw (1.5,5.5) node {\scriptsize$A_1$};
		\draw (1.5,6.5) node {\scriptsize$A_1$};
		
		\draw (0,-4.27) node {};
		\end{tikzpicture}$\qquad$
		\begin{tikzpicture}[scale=0.65]
		\begin{scope}[yshift=14cm]
		\draw[ultra thick] (-1, 1.5) -- (6.5,1.5);
		
		\draw (0,2.25) node {\scriptsize{Hasse}}
		(0,1.85) node {\scriptsize{Diagram}}
		(5.5,2.05) node {\scriptsize{Partition}}
		(2.75,2.3) node {\Large{$\kk{so}_{12}$}};
		\end{scope}
		
		\draw[\THICC] (0,0) -- (0,1) -- (1,2) -- (1,13) -- (0,14) -- (0,15) (0,14) -- (-1,13) -- (-1,7) -- (0,6) -- (0,5) -- (-1,4) -- (-1,2) -- (0,1) (-1,2) -- (1,3) -- (-1,4) (0,5) -- (1,4) (1,7) -- (0,6) -- (0,7) -- (1,8) -- (-1,10) -- (1,11) (-1,11) -- (1,12) -- (-1,13) (0,9) -- (-1,8) -- (0,7)
		
		;
		
		\foreach \y in {0,1,5,6,7,9,14,15}{\draw[fill=black] (0,\y) circle (4pt);}
		\foreach \y in {2,3,4,7,8,9,10,11,12,13}{\draw[fill=black] (-1,\y) circle (4pt);}
		\foreach \y in {2,3,4,5.5,7,8,9.5,11,12,13}{\draw[fill=black] (1,\y) circle (4pt);}
		
		\draw[white, fill=white] (1,11) circle (1.8pt)
		(1,7) circle (1.8pt)
		(-1,4) circle (1.8pt)
		(1,3) circle (1.8pt)
		;
		
		\begin{scope}[xshift = 5.5cm]
		\draw (0,0) node {$(1^{12})$}
		(0,1) node {$(2^2,1^8)$}
		(0,5) node {$(3^2,2^2,1^2)$}
		(0,6) node {$(3^3,1^3)$}
		(0,7) node {$(4^2,1^4)$}
		(0,9) node {$(5,3,2^2)$}
		(0,14) node {$(9,3)$}
		(0,15) node {$(11,1)$}
		
		(-1.5,2) node {$(2^4,1^4)$}
		(-1.5,3) node {$(2^6)$}
		(-1.5,4) node {$(3,2^4,1)$}
		(-1.5,7) node {$(3^4)$}
		(-1.5,8) node {$(4^2,2^2)$}
		(-2.15,9) node {$(4^2,3,1)$}
		(-1.5,10) node {$(5,3^2,1)$}
		(-1.5,11) node {$(5^2,1^2)$}
		(-1.5,12) node {$(6^2)$}
		(-1.5,13) node {$(7,5)$}
		
		(1.5,2) node {$(3,1^9)$}
		(1.5,3) node {$(3,2^2,1^5)$}
		(1.5,4) node {$(3^2,1^6)$}
		(2,5.5) node {$(5,1^7)$}
		(2.15,7) node {$(5,2^2,1^3)$}
		(1.5,8) node {$(5,3,1^4)$}
		(2,9.5) node {$(7,1^5)$}
		(1.5,11) node {$(7,2^2,1)$}
		(1.5,12) node {$(7,3,1^2)$}
		(1.5,13) node {$(9,1^3)$}
		
		;
		\end{scope}
		
		\draw (-0.5,0.5) node {\scriptsize{$d_6$}}
		(-0.5,5.5) node {\scriptsize{$A_1$}}
		(-0.3,6.8) node {\scriptsize{$A_1$}}
		(-0.5,14.5) node {\scriptsize{$D_6$}}
		
		(-1.9,2.5) node {\scriptsize{$A_1 \cup A_1$}}
		(-1.5,3.5) node {\scriptsize{$c_3$}}
		
		(-1.9,7.5) node {\scriptsize{$A_1 \cup A_1$}}
		(-1.5,8.5) node {\scriptsize{$A_1$}}
		(-1.5,9.5) node {\scriptsize{$A_1$}}
		(-1.5,10.5) node {\scriptsize{$D_3$}}
		(-1.9,11.5) node {\scriptsize{$A_5 \cup A_5$}}
		(-1.5,12.5) node {\scriptsize{$A_1$}}

		(1.5,2.5) node {\scriptsize{$b_4$}}
		(1.5,3.5) node {\scriptsize{$A_1$}}
		(1.5,4.75) node {\scriptsize{$D_3$}}
		(1.5,6.25) node {\scriptsize{$b_3$}}
		(1.5,7.5) node {\scriptsize{$A_1$}}
		(1.5,8.75) node {\scriptsize{$D_4$}}
		(1.5,10.25) node {\scriptsize{$b_2$}}
		(1.5,11.5) node {\scriptsize{$A_1$}}
		(1.5,12.5) node {\scriptsize{$D_5$}}

		(-0.1,2.8) node[rotate = 30] {\scriptsize{$c_2$}}
		(-0.1,11.8) node[rotate = 30] {\scriptsize{$D_3$}}
		(-0.1,10.8) node[rotate = 30] {\scriptsize{$A_3$}}
		
		(0.1,3.8) node[rotate = -20] {\scriptsize{$b_2$}}
		(0.1,12.8) node[rotate = -20] {\scriptsize{$A_3$}}
		
		(-0.8,1.3) node {\scriptsize{$d_4$}}
		(0.8,1.3) node {\scriptsize{$A_1$}}
		(-0.7,4.8) node {\scriptsize{$c_2$}}
		(0.55,4.9) node {\scriptsize{$d_3$}}
		
		(-0.7,6.2) node {\scriptsize{$A_1$}}
		(0.6,6.2) node {\scriptsize{$A_1$}}
		(-0.25,7.6) node[rotate = 40] {\scriptsize{$A_1 \cup A_1$}}
		
		(0.6,7.25) node {\scriptsize{$A_1$}}
		(0.62,8.8) node {\scriptsize{$A_1$}}
		(-0.62,8.8) node {\scriptsize{$A_1$}}
		(-0.2,9.8) node {\scriptsize{$A_1$}}
		(-0.8,13.7) node {\scriptsize{$D_4$}}
		(0.8,13.7) node {\scriptsize{$A_1$}}
		
		;

		\end{tikzpicture}
	\end{center}

\begingroup

\endgroup

\end{document}